\documentclass[a4paper,11pt]{article}
\usepackage{jcappub}
\usepackage[caption = false]{subfig}
\usepackage{graphicx}
\usepackage{dcolumn}
\usepackage{bm}
\usepackage{hyperref}
\usepackage{color}
\usepackage{amsmath}

\def\prl{Phys. Rev. Lett.}

\def\jcap{J. Cosmol.  Astropart. Phys.}
\def\aap{A\&A}
\def\apj{ApJ}

\def\apjl{ApJ}
\def\mnras{MNRAS}
\def\araa{ARA\&A}
\def\aj{AJ}

\def\nat{Nature}

\newcommand{\rom}[1]{%
  \textup{\uppercase\expandafter{\romannumeral#1}}%
}

\newcommand{\msun}{{\,\rm M_\odot}}

\newcommand{\kms}{\,{\rm km}\,{\rm s}^{-1}}

\newcommand{\erg}{\,{\rm erg}}
\newcommand{\Gyr}{\,{\rm Gyr}}

\newcommand{\Mpc}{\,{\rm Mpc}}

\newcommand{\cpm}{\,{\rm cm}^2\,{\rm g}^{-1}}

\definecolor{cerulean}{rgb}{0., 0.52,0.65}

\begin{document}

\newpage

\title{SMBH Seeds from Dissipative Dark Matter}

\author{Huangyu Xiao$^1$}
\author{Xuejian Shen$^2$}
\author{Philip F. Hopkins$^2$}
\author{Kathryn M. Zurek$^3$}
        
\affiliation{$^1$ Department of Physics, University of Washington,  Seattle, WA 98195, USA}
\affiliation{$^2$ TAPIR, California Institute of Technology, Pasadena, CA 91125, USA}
\affiliation{$^3$ Walter Burke Institute for Theoretical Physics, California Institute of Technology, Pasadena, CA 91125, USA}
\date{\today}

\abstract{
The existence of supermassive black holes (SMBHs) with masses greater than $\sim 10^{9}M_{\odot}$ at high redshift ($z\gtrsim 7$) is difficult to accommodate in standard astrophysical scenarios. We study the possibility that (nearly) totally dissipative self-interacting dark matter (tdSIDM)--in rare, high density dark matter fluctuations in the early Universe--produces SMBH seeds through catastrophic collapse. We use a semi-analytic model, tested and calibrated by a series of N-body simulations of isolated dark matter halos, to compute the collapse criteria and timescale of tdSIDM halos, where dark matter loses nearly all of its kinetic energy in a single collision in the center-of-momentum frame. Applying this model to halo merger trees, we empirically assign SMBH seeds to halos and trace the formation and evolution history of SMBHs. We make predictions for the quasar luminosity function, the $M_{\rm BH}$-$\sigma_{\rm v}^{\ast}$ relation, and cosmic SMBH mass density at high redshift and compare them to observations. We find that a dissipative dark matter interaction cross-section of $\sigma/m \sim 0.05~\rm cm^2/g$ is sufficient to produce the SMBHs observed in the early Universe while remaining consistent with ordinary SMBHs in the late Universe.   
}
\maketitle

\section{Introduction}
Observations of quasars at $z\gtrsim 6$ indicate that supermassive black holes (SMBHs) with masses greater than $\sim 10^9 M_{\odot}$ formed in the early Universe ({\em e.g.} \cite{Matsuoka_2016,Ba_ados_2017,Wang_2018,Matsuoka_2019,Yang_2020}). The discovery of such SMBHs is puzzling in the current understanding of SMBHs, \textit{i.e., how did the first SMBHs grow so large so quickly? } 
One possible scenario is that the SMBHs were seeded by the remnants of the Population $\rom{3}$ (Pop $\rom{3}$) stars, which are expected to form in $\sim 10^{5-6}M_{\odot}$ dark matter minihalos through primordial gas undergoing molecular hydrogen cooling. Since the primordial gas is significantly warmer than the usual star-forming molecular clouds at low redshift, the cooling is less efficient, leading to inefficient fragmentation \cite{Madau2001,Abel2002,Bromm2002,OShea2007,Turk2009,Tanaka2009,Greif2012,Valiante2016}. Therefore, Pop $\rom{3}$ stars are expected to be more massive than stars in the Local Universe, and simulations have suggested a mass range of $10\lesssim M_{\star}/M_{\odot}\lesssim 10^3$ \cite{Hirano_2014}. If SMBH growth is dominated by Eddington-limited accretion, SMBH seeds will grow exponentially within an $e$-folding time $t_{\rm edd}\approx 50$ Myr, assuming a radiative efficiency $\epsilon_{\rm r} \approx 10\%$. In the Eddington-limit, a $100\,M_{\odot}$ Pop $\rom{3}$ seed will need $\approx 0.8\,\rm Gyr$ to reach a billion solar mass, a time greater than the age of the universe at $z=7$ even assuming a duty-cycle $D\approx 1$ over eight orders of magnitude growth in mass, making it impossible to explain the mass growth of SMBHs with masses $10^{9} M_{\odot}$  at $z=7$. A high duty-cycle ($D\approx 1$) is also disfavored by the feedback effects from accretion onto the SMBH, as well as displacement of the gas reservoir by UV radiation and supernovae explosions of the Pop $\rom{3}$ stars in the shallow gravitational potential of minihalos \cite{10.1111/j.1365-2966.2006.11275.x,Whalen_2008,Milosavljevi__2009,Alvarez_2009}. 

Several different scenarios have been proposed to ease the timescale constraints (see \cite{Inayoshi_2020} for a review of the assembly of SMBHs at high redshift). Generally, one can increase either the SMBH seed mass or the growth rate. One possibility is that a small fraction of SMBH seeds in rare massive halos may be able to sustain Eddington accretion over most of the history of the Universe or even grow at a super-Eddington rate \cite{brightman2019breaking}. Super-Eddington accretion at a few times the Eddington-limited rate could be maintained with duty-cycles $\sim$ $20-30\%$ in some accretion disk models (e.g. \cite{Madau_2014}), which could explain the existence of billion solar mass SMBHs at $z \gtrsim 7$. Another popular scenario relies on the formation of massive SMBH seeds with mass $\approx 10^{4-6} M_{\odot}$ formed through collapse of chemically pristine primordial gas in so-called ``atomic cooling halos" with virial temperature $T_{\rm vir}\sim 10^4\,{\rm K}$ at $z \simeq 15\operatorname{-}20$ \cite{Bromm2003,Koushiappas2004,Begelman2006,Lodato:2006,Ferrara2014,Pacucci2015,Valiante2016}. 
However, even in these models, an Eddington-limit accretion has to be sustained for most of the lifetime of the seeds, which implies a very high duty-cycle of SMBHs in the early Universe. 
Thus, such a scenario is hard to reconcile with some of the massive quasars at $z \gtrsim 6$ with low measured Eddington ratios \cite{Matsuoka2019,Onoue2019} as well as the short quasar lifetimes ($\sim 10^{4-5}\,{\rm yr}$) found in observations of quasar proximity zones at $z\sim 6$ \cite{Eilers:2020htq,2017ApJ...840...24E,Eilers_2018,Davies_2019,Andika_2020}. 

Self-interacting dark matter (SIDM) halos have the potential to seed massive SMBHs in a much more accelerated way through the ``gravothermal catastrophe'' \cite{Balberg:2002ue,Pollack:2014rja,Hu_2006,padilla2020corehalo,Koda2011}. 
Finite self-gravitating systems ({\em e.g.} dark matter halos, globular clusters) have a negative heat capacity and the heat conduction will eventually lead to the ``gravothermal catastrophe'' of the system ({\em e.g.} \cite{Bell1968,Bell1980}). In a halo with elastic dark matter self-interactions, effective heat conduction is realized by collisions between dark matter particles and the SIDM halo cores could ultimately experience run-away collapse into compact objects ({\em e.g.} \cite{Burkert2000,Kochanek2000,Balberg:2002ue,Colin2002,Koda2011,Vogelsberger2012}). However, such elastic self-interacting dark matter (eSIDM) requires a cross-section $\sigma/m=5~\rm cm^2/g$ to seed SMBHs with masses $10^6 M_{\odot}$ at $z\sim 10$ \cite{Balberg:2002ue}, which is now ruled out by observations of galaxy cluster collisions \cite{Randall_2008}. Those constraints are derived at relative velocities 1000-2000 km/s, while the dark matter halos we are interested in have virial velocities 200-2000 km/s. If the cross-section is velocity dependent, those constraints might be avoided and a large cross-section that can seed SMBHs efficiently is allowed, which we have not studied quantitatively. To accelerate the ``gravothermal catastrophe'', hybrid dark matter models were proposed where the bulk of dark matter does not have any self-interaction, but a small fraction is SIDM with a large cross-section \cite{Pollack:2014rja,Choquette:2018lvq}. Alternatively, the presence of baryons in protogalaxies has also been shown to accelerate the gravothemal collapse of eSIDM halos \cite{feng2020seeding} with a smaller cross-section. 

If the self-interaction is totally inelastic (hit-and-stick), the collapse timescale can be two orders of magnitude shorter than the prediction in elastic SIDM \cite{Huo:2019yhk,Choquette:2018lvq,Essig2019}. Therefore, totally dissipative self-interacting dark matter (tdSIDM) can greatly accelerate the catastrophic collapse of halos, which leads to the formation of SMBHs in the early universe. Our study is motivated by the analysis of dark nuggets in Refs.~\cite{Gresham:2018anj,Gresham:2017cvl,Gresham:2017zqi}, based on the model of Refs.~\cite{Wise:2014jva,Wise:2014ola} featuring hit-and-stick interactions that are crucial for accelerating the catastrophic collapse of SIDM halos. Other dissipative dark matter models, such as atomic dark matter, exciting dark matter, and composite strongly interacting dark matter \cite{Fan:2013tia,Fan:2013yva,Foot_2015,Boddy_2014,Kaplan_2010,CyrRacine:2012fz,Cline:2013pca,Finkbeiner:2014sja,Boddy:2016bbu,Schutz:2014nka,Das:2017fyl,jo2020exploring}, feature a constant kinetic energy loss in the center-of-momentum frame, which needs to be tuned to accelerate the catastrophic collapse efficiently.
The proposal of Ref.~\cite{Gresham:2018anj} was to consider {\em rare, high density} fluctuations of dissipative dark matter which features hit-and-stick interactions as the seeds of SMBHs at high redshift. The goal of this paper is to test this hypothesis in detail, and determine whether a hit-and-stick dissipative dark matter model that produces SMBHs through this mechanism can explain the SMBH abundance in the early Universe while remaining consistent with observations of dark matter halos (and their SMBHs) in the late Universe. 

Though the timescale of seeding SMBHs in an isolated tdSIDM halo was well-studied numerically using N-body simulations \cite{Choquette:2018lvq,Huo:2019yhk}, the cosmological abundance of SMBHs in the early Universe has never been calculated. For the first time, we compute the cosmological abundance of SMBHs seeded from tdSIDM halos and show this seeding mechanism can be consistent with both high redshift and low redshift observations of SMBHs. We study the formation of initial SMBH seeds in isolated dark matter halos using a series of N-body simulations and calibrate our semi-analytic model based on the simulation results. Then, we run Monte Carlo simulations to generate the merger trees of halos that can trace the growth of those SMBH seeds. Our model is also testable in the future when the abundance of SMBHs is better measured at different redshifts.

This paper is organized as follows. In Sec.~\ref{sec:simulation}, we discuss our semi-analytical model of the dissipation timescale and calibrate it with a series of N-body simulations in isolated NFW halos. In Sec.~\ref{sec:cosmological_evolution}, we generate the merger trees of halos and track the cosmological evolution of SMBH seeds in those halo progenitors, allowing us to compute the mass function of SMBHs and compare it with observables. In Sec.~\ref{sec:conssistency_check}, we show our tdSIDM model will not cause the formation of overly massive SMBHs at low redshift, remaining consistent with low redshift observations.

\section{Simulating black hole formation in isolated halos}\label{sec:simulation}
We performed N-body simulations of dark matter halos with the NFW density profile as the initial condition, using the code {\sc Gizmo} \cite{Hopkins_2015}. The initial condition is generated using the code {\sc pyICs} which was first used in \cite{2017MNRAS.470.4941H}. 15 N-body simulations are performed for 15 different NFW halos with a mass range $\sim10^{9}-10^{13}M_{\odot}$ which are completely isolated in each simulation box. There are $6\times 10^6$ particles in each simulation box and the gravitational softening length is taken to be $2d_0$ where $d_0$ is the particle mean separation within $0.07 r_{\rm s}$ at the beginning of the simulation. $r_s$ is the scale radius of different NFW halos and $0.07 r_{\rm s}$, as we will show later, is the universal collapsed radius.
{\sc Gizmo} is a multi-method gravity plus hydrodynamics code and is capable of simulating both gas and dissipative dark matter. Baryonic simulations are much more computationally
expensive, however, and the formation of SMBH seeds in our model is mainly driven by the dissipation in the dark sector. Therefore, we run dark-matter-only (DMO) simulations to study the formation of SMBH seeds from the catastrophic collapse of halos. The gravity of dark matter is solved with an improved version of the Tree-PM solver from GADGET-3 \cite{Springel2005}. Dark matter self-interactions are simulated in a Monte-Carlo fashion with the implementation in \cite{Rocha2013}. In the tdSIDM model, when two dark matter particles collide with each other, they lose a fraction $f$ of their kinetic energy in the center-of-momentum frame.  We focus on the case that nearly all the kinetic energy is dissipated in the interaction, $f \approx 1$.  This is a particular feature of the nugget fusion model presented in Refs.~\cite{Wise:2014jva,Wise:2014ola,Gresham:2017cvl,Gresham:2017zqi,Gresham:2018anj}, not shared in general by other dissipative dark matter models.  

As explained in the introduction, we are interested in SMBH seed formation in massive, rare halos in the mass range $10^{9}-10^{13}M_{\odot}$, which can produce SMBH seeds in the mass range $10^{6}-10^{10}M_{\odot}$ (if the SMBH-to-halo mass ratio is about $10^{-3}$ as we will show later in the simulation results). In order to sample such rare structures in cosmological simulations, a comoving boxsize of order ${\rm Gpc}^{3}$ is required. Meanwhile, the physical size of the collapsed region is about $0.07 r_{\rm s}$, as we will show later in the simulation results, where $r_{\rm s}$ is the scale radius. This poses a challenge to cosmological simulations due to limitations on mass and spatial resolution. For example, if we are interested in SMBH formation in a rare dark matter halo at high redshift with mass $10^{12}M_{\odot}$, the particle number in the central region within $0.07 r_{\rm s}$ has to be larger than $\sim 200$ \cite{Power2003} to resolve the dense core. Assuming the virial radius is $4r_{\rm s}$, the particle number in this halo is $\sim 2\times 10^4$, which requires a mass resolution of $5\times 10^7M_{\odot}$ in the simulation box. However, we need a boxsize of $\gtrsim {\rm Gpc}^{3}$ to simulate the structure formation from extremely rare fluctuations. Therefore, the particle number in the simulation box has to be $\sim 10^{12}$, which is at least $100$ times larger than the particle number $\lesssim 10^{10}$ in state-of-the-art cosmological simulations ({\em e.g.} \cite{MBK2009,Klypin2011,Hopkins2018,Pillepich2018}). Therefore, it is very challenging to simulate the formation of SMBH seeds with cosmological simulations, even employing a zoom-in technique. An alternative strategy is to simulate individual isolated halos with various halo parameters and test the formation of SMBH seeds separately. Large scale structure with moderate dark matter self-interactions will not differ significantly from the CDM case \cite[{\em e.g.},][]{Vogelsberger2012,Rocha2013,Vogelsberger2016,Lovell2018}. The calibration from the isolated halo simulations can then be used to study the cosmological population of SMBHs with semi-analytic approaches.

\subsection{Semi-Analytic Model}
Before introducing simulations, it is useful to develop an analytic model that can predict the collapse timescale of the dissipative dark matter model. The analytic predictions can then be compared to and calibrated by the simulation results. In this section, we will discuss the analytic model that predicts the collapse timescale and show that it agrees well with our simulation results after calibrating the result by a universal ${\cal O}(1)$ prefactor in the formula. 
We focus on SMBH seed formation in rare, massive halos at high redshift with high central dark matter density and thus high dissipative dark matter self-interaction rates, following Ref.~\cite{Gresham:2018anj}. Dark matter halos formed from rare fluctuations are the ideal environments for seeding SMBHs, as such halos form at higher redshift relative to typical halos, where the background density of the universe is larger, implying a higher halo central density. 
The collapse rate of tdSIDM halos is characterized by $\rho v \sigma/m$, where $\rho$ is the average density, $v$ is the velocity dispersion and $\sigma$ is the cross-section of dark matter self-interaction.  Thus higher densities shorten the dissipation timescale, as we will discuss in detail in Eq.~(\ref{eq:collapse_time}). We will take the average density and velocity dispersion to be those in the collapsing central region of the halo, characterized by a collapse radius $r_0$ determined by our N-body simulations. 

Because of our reliance on high density dark matter fluctuations to seed SMBHs, we must quantify the rareness of halos, which can be explicitly define through the variance of density fluctuation:
\begin{equation}
    \sigma_{\rm h}(M,z)= \nu_{\rm h}\sigma_0(M,z),
\end{equation}
where $\sigma_0(M,z)$ is the variance in the density fluctuation field smoothed over a top-hat filter of scale $R_{\rm s}=(3M/4\pi\bar{\rho})^{1/3}$, $\bar{\rho}$ is the average comoving background density and $\sigma_{\rm h}(M,z)$ is the variance of a local density fluctuation that can differ from the average fluctuation. Clearly $\nu_{\rm h}$ defines the rareness of the fluctuation and the halo that just formed from this fluctuation. One can also define the peak height $\nu$:
\begin{equation} \label{eq:nu_sigma}
    \nu(M,z)\equiv \frac{\delta_c}{\sigma_0(M,z)}=\frac{\delta_c}{\sigma_0(M,z=0)\, D(z)},
\end{equation}
where $\delta_{\rm c}=1.686$ is the critical overdensity for collapsed halos derived from the spherical top-hat model and $D(z)$ is the growth factor normalized to unity at $z=0$. In the spherical collapse model, a halo forms when the variance in the density fluctuation field satisfies $\sigma = \delta_c$, corresponding to when typical halos with $\nu = 1$ collapse. However, rare, high-variance halos may collapse earlier than typical halos, when $\nu=\nu_{\rm h}$.
In what follows, $5-\sigma$ ($3-\sigma$) halos are defined by $\nu(M,z)=5$ ($\nu(M,z)=3$) at $z = z_f$, where $z_f$ is determined by when a density perturbation reaches $\sigma(M,z_f) = \delta_c$. 
Note that different halos with different peak heights $\nu$ may collapse at the same redshift, though rare fluctuations correspond to more massive halos. 

We use the model of \cite{Macci__2008}, which is a modification of the Bullock model \cite{Bullock_2001}, to define the halo parameters such as characteristic density $\rho_0$, halo concentration $c_{\rm vir}$ and halo mass $M_{\rm vir}$ and their evolution with redshift.
The density profile of virialized dark matter halos are well described by the Navarro–Frenk–White profile \cite{Navarro_1996}
\begin{equation}
\rho(r)=\frac{\rho_0}{\frac{r}{r_{\rm s}}\left(1+\frac{r}{r_{\rm s}}\right)^2}.    \end{equation}
where $\rho_{0}$ is the scale density of the halo and $r_{\rm s}$ is the scale radius. $\rho_0$ characterizes the central density of a dark matter halo.
As is typical in a halo formation model, we can link the central density of a halo formed at redshift $z_{\rm f}$ to that of the critical density of the universe at $z_{\rm f}$. Therefore we define the mass of the halo at the formation redshift $z_{\rm f}$ via 
\begin{equation}\label{eq:mass_delta}
    M_{\rm f}=\frac{4}{3}\pi c_0^3 r_{\rm s}^3  \Delta(z_{\rm f}) \rho_{\rm crit}(z_{\rm f}),
\end{equation}
where $c_0$ is the halo concentration at the formation time, $\rho_{\rm crit}(z)$ is the critical density of the universe at redshift $z$, and $\Delta(z)$ is the overdensity of the halo with respect to the critical density. One common choice is to set $\Delta(z_f) =200$, which is motivated by the spherical collapse model. The only dependence on cosmology in this mass definition comes from $\rho_{\rm crit}(z)$. It has been universally found that the initial halo concentration at the moment of the first collapse satisfies $c_0\approx 4$ \cite{Zhao_2003}. Therefore the halo central density is $\rho_0\approx 200c_0^3 \rho_{\rm crit}(z_{\rm f})$, suggesting that a halo formed at high $z_{\rm f}$ has a large central density. The scale radius $r_{\rm s}$ is determined at the time of formation and does not evolve with time:
\begin{equation}\label{eq:determine_rs}
    r_{\rm s}= \left(\frac{3M_{\rm f}}{4\pi c_0^3\Delta(z_{\rm f})\rho_{\rm crit}(z_{\rm f})}\right)^{1/3}.
\end{equation}

The NFW profile is truncated at a virial radius that depends on redshift, which is defined as $R_{\rm vir}(z)\equiv c_{\rm vir}(z) r_{\rm s}$, where $c_{\rm vir}(z)$ is a redshift dependent concentration number. As the universe expands, the background density drops but the halo central density $\rho_{\rm s}$ should remain the same, leading to a larger concentration number and a larger virial radius.
Therefore the mass within the virial radius for an NFW profile should grow (logarithmically) as the universe evolves, which can be represented as
\begin{equation}\label{eq:mass_conc}
\begin{split}
        M_{\rm vir}(z)&= 4\pi \rho_0 r_{\rm s}^3f(c_{\rm vir}(z))=M_{\rm f}\frac{f(c_{\rm vir}(z))}{f(c_0)},
\end{split}
\end{equation}
where $f(c)={\rm ln}(1+c)-c/(c+1)$. This equation is obtained by integrating the NFW profile truncated at the virial radius.  
 The redshift dependence of the halo concentration is thus defined by
\begin{equation}\label{eq:concentration_z}
    \frac{c_{\rm vir}(z)^3}{f(c_{\rm vir}(z))} = \frac{c_{0}^3}{f(c_{0})}\frac{\Delta(z_{\rm f})\rho_{\rm col}(z_{\rm f})}{\Delta(z)\rho_{\rm crit}(z)},
\end{equation}
such that we see that $c_{\rm vir}(z) \propto 1+z$ in the limit of large concentration parameters and $\Delta(z) = \Delta(z_f)$.  Equivalently, we have a generalized form of Eq.~(\ref{eq:mass_delta}):
\begin{equation}\label{eq:mass_deltaz}
    M_{\rm vir}(z)=\frac{4}{3}\pi c_{\rm vir}(z)^3 r_{\rm s}^3  \Delta(z) \rho_{\rm crit}(z),
\end{equation}
where we assume $\Delta(z) = \Delta(z_f)$.
From these relations, one finds the characteristic density of dark matter halos
\begin{equation}\label{eq:determine_rho0}
    \rho_0=\frac{M_{\rm vir}(z)}{4\pi r_{\rm s}^3(z) f(c_{\rm vir}(z))}=\frac{c_{0}^3}{3f(c_{0})}\Delta(z_{\rm f})\rho_{\rm crit}(z_{\rm f}).
\end{equation}
This expression clearly states that the halo central density is directly determined by the background density of the universe at the redshift of formation.
The invariance of $\rho_0$ and $r_{\rm s}$ indicates that the inner profiles of dark matter halos do not change over time. The boundary of a halo, described by $R_{\rm vir}$, must move outwards owing to the decreasing background density $\rho_{\rm crit}(z)$, which is known as the ``pseudo-growth" of dark matter halos \cite{Diemer_2013}.

Now that we know how to determine the halo parameters from the halo mass and concentration number at the observation redshift, we can further study the behavior of halos that are made of dissipative self-interacting dark matter at high redshift. 
Dark matter particles dissipate their kinetic energy through self-interactions (referred to as ``collisions''). The average timescale for a particle to encounter one such collision in radius $r$ can be estimated as
\begin{equation}\label{eq:relaxation_time}
    t_{\rm relax}(r) = \dfrac{1}{\alpha \rho(r) \sigma_{\rm v}(r)}\frac{1}{\sigma/m},
\end{equation}
where $\rho(r)$ and $\sigma_{\rm v}(r)$ are the dark matter mass density and one-dimensional velocity dispersion at radius $r$, $\alpha=\sqrt{16/\pi}$ is a constant factor assuming hard-sphere-like scattering and a Maxwell-Boltzmann velocity distribution and $\sigma/m$ is the dissipative interaction cross-section per particle mass. If the velocity field of dark matter is isotropic (as found in \cite{Lemze2012,Sparre2012,Wojtak2013}), $\sigma_{\rm v}$(r) can be obtained by solving the spherical Jeans equation \cite{Lokas2001}
\begin{align}
    & \sigma_{\rm v}(r) = \sqrt{4\pi G \rho_0 r_{\rm s}^2 F(r/r_{\rm s})} ,\nonumber \\
    & F(x) = \dfrac{1}{2} x (1+x)^{2} \Big[ \pi^2 - \ln{(x)} - \dfrac{1}{x} \nonumber \\
    & - \dfrac{1}{(1+x)^2} - \dfrac{6}{1+x} + \Big( 1 + \dfrac{1}{ x^2} - \dfrac{4}{x} - \dfrac{2}{1+x}\Big) \nonumber \\ 
    & \times \ln{(1+x)} + 3 \ln^{2}{(1+x)} - 2{\rm Li}_{2}(-x) \Big], 
    \label{eq:velocity-dispersion}
\end{align}
where ${\rm Li}_{2}(x)$ is the dilogarithm. The timescale for local kinetic energy to dissipate through such collisions is
\begin{align}
    t_{\rm diss}(r) & = \dfrac{ 3 \rho(r) \sigma_{\rm v}^2 /2 }{ C f \rho^{2}(r) \sigma_{\rm v}^{3}(r) }\frac{1}{\sigma/m} \nonumber \\
    & = \dfrac{  1 }{ \beta f \rho(r) \sigma_{\rm v}(r)}\frac{1}{\sigma/m } ,
    \label{eq:dissipation_timescale}
\end{align}
where $f$ is the fraction of kinetic energy loss in the center-of-mass frame per collision, $C$ is a constant factor and $\beta=4\alpha/3$~\cite{Shen2021}, assuming a Maxwell-Boltzmann velocity distribution, a velocity independent cross-section and all the kinetic energy in the center-of-momentum frame is dissipated during a collision.

\begin{figure}
\centering
\includegraphics[width=10cm]{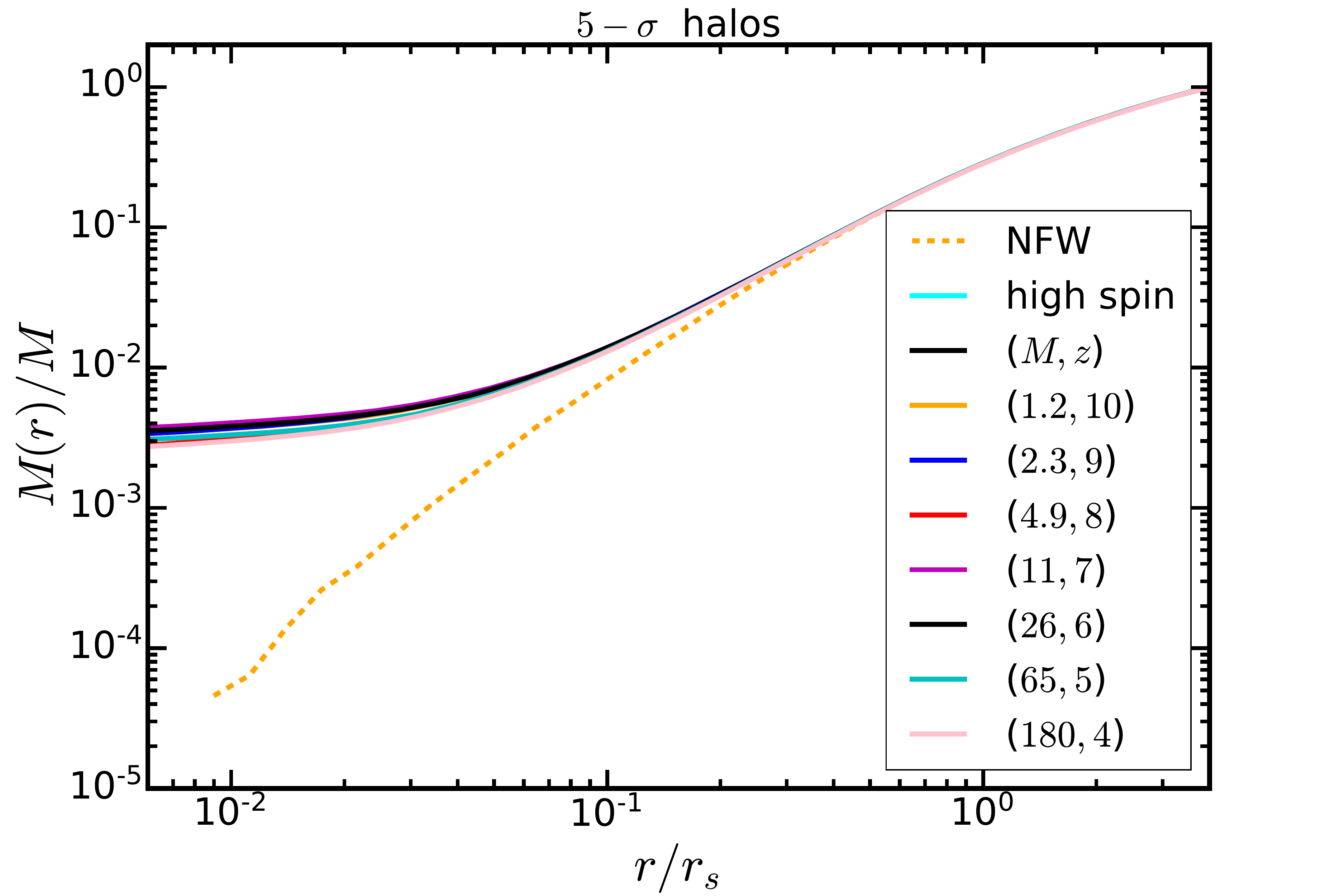}
\includegraphics[width=10cm]{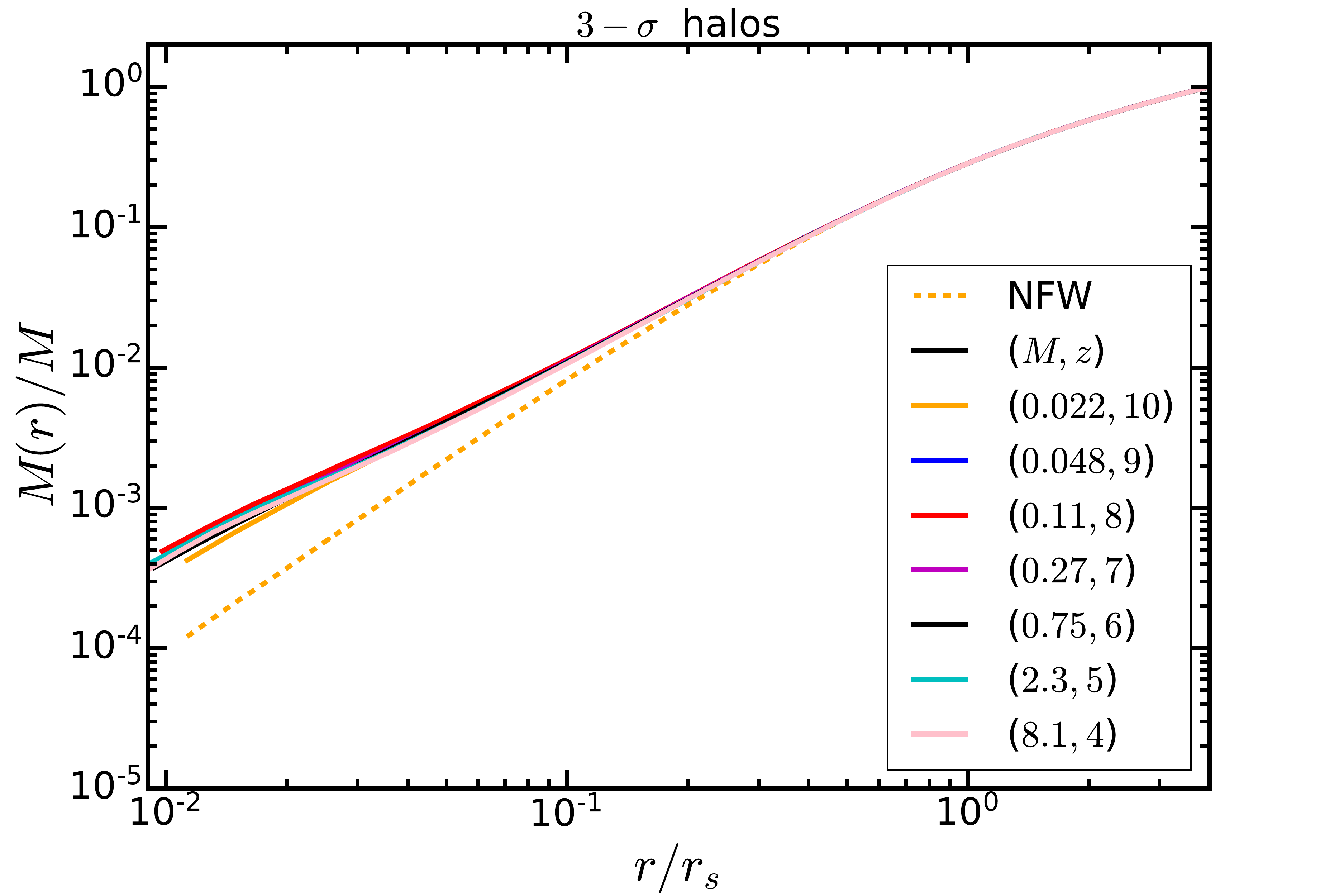}
\caption{Enclosed mass fraction as a function of radius (normalized to the scale radius $r_s$), for 8 and 7 different 5- and 3-$\sigma$ halos (upper and lower panels, respectively) including dark matter dissipation with cross-section $1\rm cm^2/g$. Different halos are labeled with the mass (in units of $10^{11}M_{\odot}$) and redshift. The high spin curve corresponds to $\lambda=0.1$ at $z=10$, while other halos have spin parameter $\lambda=0.03$.  In these figures, the more dense $5-\sigma$ halos show core collapse, indicated by the region of constant density at small radius, while $3-\sigma$ halos have not been destabilized, consistent with Fig.~\ref{fig:massz}.  In the collapsed halos, the collapse fraction is found universally to be $\sim 3\times 10^{-3}$.
}
\label{fig:emass}
\end{figure}

Rapid kinetic (thermal) energy dissipation will inevitably result in gravitational collapse of the central halo. The collapse timescale should be on the same order as the dissipation time, $t_{\rm col}= A t_{\rm diss}$, where the order one factor $A$ is determined by our simulations of isolated halos.
Collapse is expected to happen at radii where $t_{\rm col}(r) \ll t_{\rm life}$, where $t_{\rm life}$ is the lifetime of the system. 
It is hard to determine the collapse radius analytically, but our N-body simulations can give us the desired information. The details of our simulation results shall be discussed in the next subsection but we can briefly describe the findings. We run a series of dark-matter-only simulations for isolated NFW halos to study the evolution of their density profiles. We show the final stage of the cumulative mass profile $M(r)/M$ in Fig.~\ref{fig:emass}, where $M(r)/M$ is roughly a constant in the central region, indicating the formation of SMBHs. We studied the collapse of dark matter halos with different masses, all of which formed an SMBH with mass $\sim 3\times 10^{-3}M$, where $M$ is the halo mass. 
If an NFW halo within some radius $r_0$ collapses to an SMBH seed, the fraction of the initial mass in the seed is
\begin{equation}
    f_{\rm col}= \frac{{\rm ln}(1+r_0/r_{\rm s})-r_0/(r_{\rm s}+r_0)}{{\rm ln}(1+c)-c/(c+1)}.
\end{equation}
Furthermore, if we take a collapse radius $r_0=0.07 r_{\rm s}$ and a concentration $c=c_{0} = 4$ at formation time, this equation gives a collapse fraction $f_{\rm col }\approx 3\times10^{-3}$. 
Note that $c_0 = 4$ is a universal prediction for halos at formation \cite{Diemer_2019}, independent of their rarity $\nu$. A higher $\nu$ halo, of a given mass, will simply form at a higher redshift $z_{\rm f}$ relative to typical halos, and hence will have a higher concentration at lower redshift $z$, according to Eq.~(\ref{eq:concentration_z}). 
A large central density, as shown in Eq.~(\ref{eq:dissipation_timescale}), corresponds to a smaller dissipation time. 

As we will discuss in the next subsection, the simulation results, as shown in Fig.~\ref{fig:emass}, indicate that the collapse fraction of tdSIDM halos is universally $3\times 10^{-3}$ independent of halo mass and redshift. Therefore, we conclude that the collapse radius (the radius where dark matter particles will fall into the halo center and collapse) is about $ 0.07 r_{\rm s}$ independent of halo mass and redshift, corresponding to a collapse fraction of $3\times10^{-3}$. In Appendix \ref{append:col_frac}, we further confirm that the collapse fraction is independent of the cross-section and provide a theoretical explanation for the universality of the collapse fraction. 
We thus evaluate the collapse time within a collapse radius $0.07 r_{\rm s}$, which gives the timescale of SMBH formation at the halo center. The corresponding collapse timescale $t_{\rm col}(0.07r_{\rm s})$ is
\begin{equation}\label{eq:collapse_time}
\begin{split}
 t_{\rm col}& =  \frac{A}{f}\,\,  1.29\times 10^{11}{\rm yr}\left(c_{\rm vir}(z)^3\frac{\Delta(z)}{200}\frac{\rho_{\rm crit}(z)}{\Omega_{\rm m}\rho_{\rm crit}(0)}\right)^{-7/6} \\
 &\times\left[f(c_{\rm vir}(z)\right]^{3/2}\left(\frac{\rm 1cm^2/g}{\sigma/m}\right)\left(\frac{10^{15}M_{\odot}}{M_{\rm vir}}\right)^{1/3},
\end{split}
\end{equation}
where $\Omega_m$ is the matter density today.

\subsection{Simulations of halo collapse and black hole formation}
We ran series of DMO simulations with different initial conditions to calibrate the collapse timescale and determine the SMBH-to-halo mass ratio. The initial conditions are characterized by the NFW profile parameters $\rho_0$ and $r_{\rm s}$, which can be determined by the concentration number $c_{\rm vir}$, halo mass $M_{\rm vir}$ and the observation redshift $z$ by using Eq.~(\ref{eq:determine_rho0}) and Eq.~(\ref{eq:determine_rs}). We simulate the evolution of $5-\sigma$ and $3-\sigma$ rare halos whose mass can be determined from $\nu=$ 5,~3. The concentration of those halos can be determined from the models that give the relation between halo mass $M$ and peak height $\nu$ \cite{Diemer_2019,Diemer_2015,ishiyama2020uchuu,2001MNRAS.321..559B}, though for large $\nu$, the halo concentration is roughly 4, with the exact value weakly depending on redshift. Therefore we assume those halos have a concentration of 4 in our simulations. Selecting an observation redshift $z$ for $5-\sigma$ or $3-\sigma$ halos, we obtain the halo mass and concentration, from which we determine $\rho_0$, $r_{\rm s}$ needed to create initial conditions for our N-body simulations.

We expect the most massive halos at high redshift, corresponding to rare fluctuations, will have higher central dark matter density and thus smaller collapse timescales. The dark matter self-interaction cross-sections in our simulation is taken to be $\sigma_0/m=1~\rm cm^2/g$. The analytic formula in Eq.~(\ref{eq:collapse_time}) suggests that the collapse timescale is inversely proportional to the cross-section. Therefore we can easily apply the simulation results to other cross-sections. The gravitational softening length is chosen to be $2 d_0$ where $d_0$ is the mean separation for particles within radius $0.07~r_{\rm s}$. The particle number in the whole simulation box is $6\times10^6$. 

As the tdSIDM halo evolves, dissipation will drive radial contraction of the halo as well as a ``dark cooling flow'' found in recent cosmological simulations of tdSIDM \cite{Shen2021}. The contraction at a certain stage could be halted by centrifugal forces. However, if dark matter substructure torque, created by global gravitational instability or dark matter viscosity, efficiently transports angular momentum, the run-away collapse of the halo into an SMBH may occur. During this process, the central dark matter density is expected to very rapidly increase, causing the integration time-step to approach zero. {\sc Gizmo} uses adaptive time-stepping, which allows us to study the halo profiles at the moment of collapse. In the extreme case, we expect dark matter particles to lose all of their kinetic energy in the center-of-mass frame, typical in the dark nugget model \cite{Gresham:2018anj}. In the simulation, we choose $f=0.8$ to avoid numerical difficulties ({\em e.g.} particles cluster in the same position in phase space and blow out the integration time) but the results are nearly identical for $f\approx 1$ (if we correct for the dependence of $t_{\rm diss}$ on $f$). After the catastrophic collapse, the enclosed mass $M(r)$ is expected to be flat at the halo center, which agrees well with what we found in simulations of isolated 5-$\sigma$ halos, as shown in Fig.~\ref{fig:emass}. The NFW parameters of 5-$\sigma$ halos are $r_{\rm s}=$ 39.1, 23.2, 14.6, 9.6, 6.5, 4.6, 3.3 kpc, and $\rho_0=0.030,\,0.051,\,0.081,\,0.121,\,0.173,\,0.237,\,0.316\,M_{\odot}\rm pc^{-3}$ for redshift $z=4-10$. The NFW parameters $\rho_0$ and $r_{\rm s}$ can be obtained from Eqs.~(\ref{eq:determine_rho0}) and (\ref{eq:determine_rs}) once we fix the halo mass $M_{\rm vir}$ and the concentration number $c_{\rm vir}$ at a given observation redshift.
The halo spin parameter is taken to be $\lambda = 0.03$. For comparison, we also run a simulation with high spin parameter $\lambda = 0.1$. The halo collapses within the same timescale, suggesting that the centrifugal barrier discussed in Appendix \ref{append:barrier_fragmentation} is not important, such that we expect SMBHs to form if the halo mass is above the threshold.

After running simulations for isolated NFW halos, we calibrated our semi-analytic model and found the timescale for SMBH formation is
\begin{equation}\label{eq:collapse_time_simulated}
\begin{split}
 t_{\rm col} = 1.06 t_{\rm diss}.
\end{split}
\end{equation}
Therefore the collapse timescale can be determined from our analytic prediction of the dissipation timescale, after adding a calibration factor of 1.06. The collapse radius is universally found to be $\sim0.07 ~r_{\rm s}$ for 5-$\sigma$ halos at different redshift, corresponding to a collapse fraction $\sim 3\times 10^{-3}$, independent of halo mass. Different mass halos have slightly different calibration factors, which are found to be 1.02, 1.05, 1.05, 1.15, 1.12, 1.05, 1.01 for 5-$\sigma$ halos at redshift $z=4-10$. Even though there are some uncertainties, our semi-analytic formula in Eq.~(\ref{eq:collapse_time}) agrees well with the simulation results after adding a calibration factor. To confirm that less rare halos will not collapse, we also run simulations with $3-\sigma$ halos and stop the evolution at time $\sigma/\sigma_0 \epsilon t_{H}$, where $\sigma$ is the cross-section that will be appropriate for seeding SMBHs, $\sigma_0=\rm 1cm^2/g$ is the cross-section we are using in our simulation and $\epsilon$ is the parameter of seeding criterion discussed in Eq. (\ref{eq:collapse_condition_prac}). We will show later in Sec.~\ref{sec:cosmological_evolution} that $\sigma \sim 0.1 \rm cm^2/g$ is appropriate for seeding the high mass SMBHs at high redshift while not causing inconsistencies at low redshift.


\begin{figure*}
\begin{minipage}[b]{0.495\textwidth}
\includegraphics[width=1.21\textwidth]{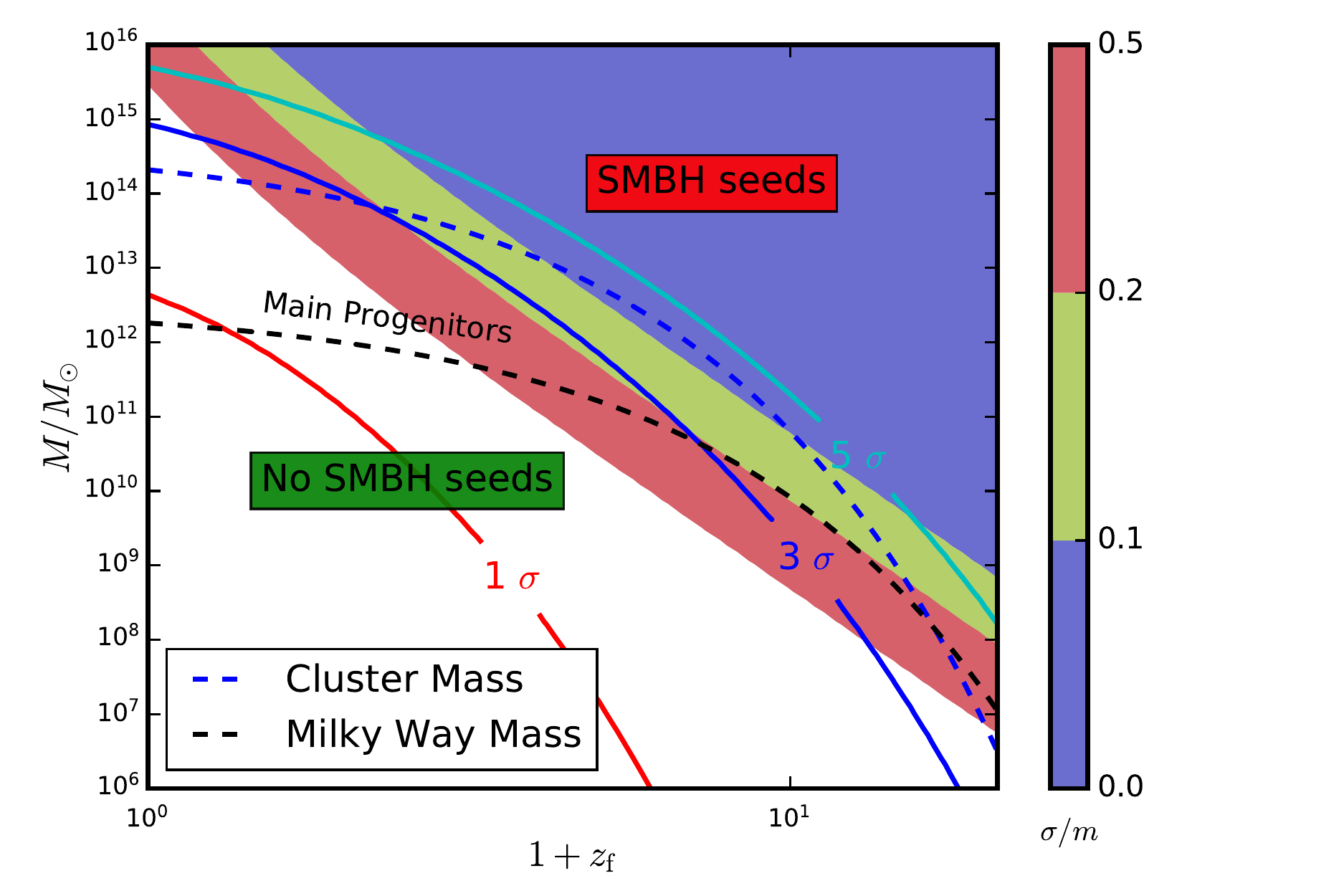}
\end{minipage}
\hfill
\begin{minipage}[b]{0.49\textwidth}
\includegraphics[width=1.21\textwidth]{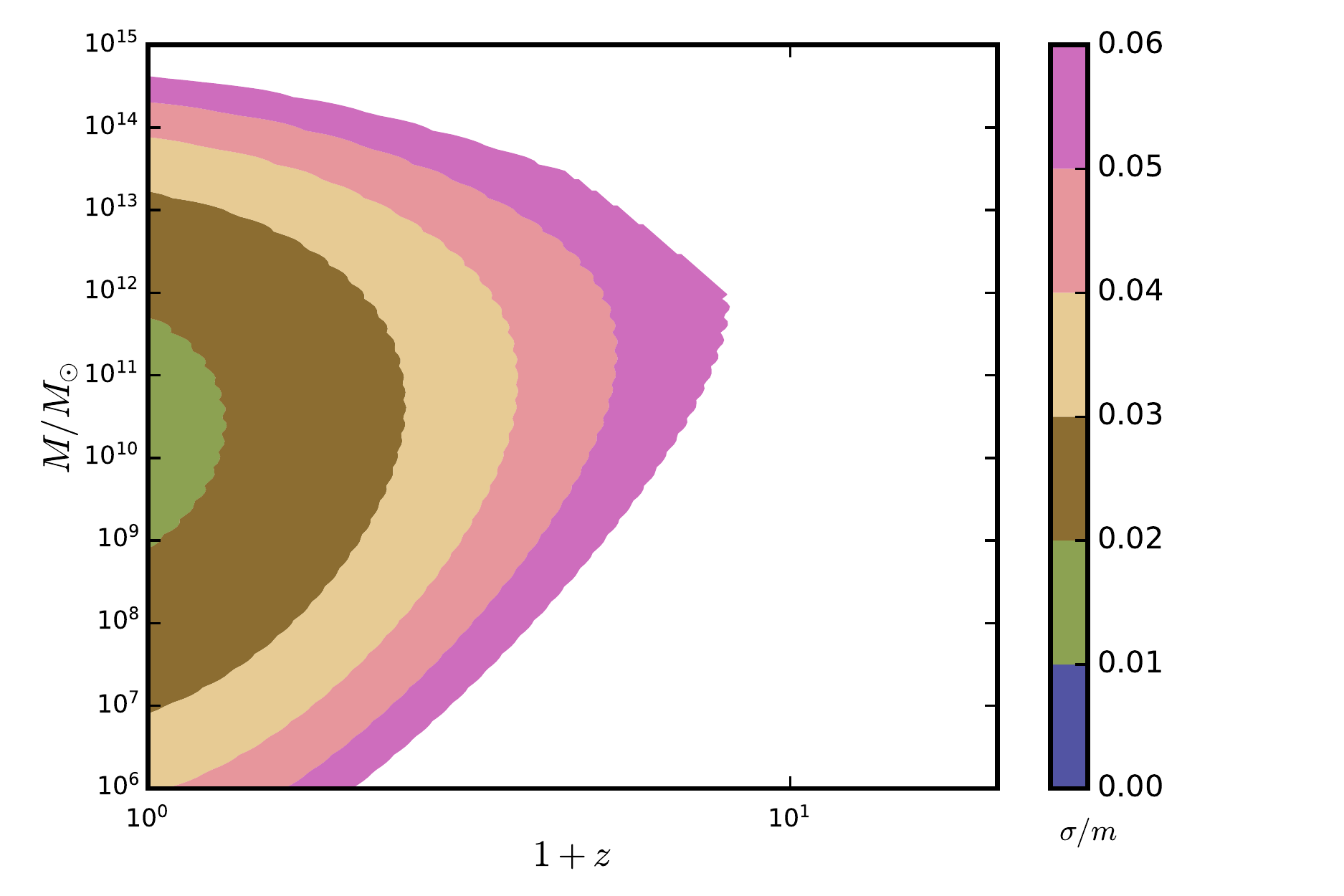}
\end{minipage}
\caption{
$\textit{left panel}$: Shown as shaded contours, minimum halo mass $M$ to seed an SMBH (labeled ``SMBH seeds'') immediately at redshift of halo formation $z_f$, for a fixed tdSIDM cross-section ($\sigma/m$ in units of $\mbox{cm}^2/\mbox{g}$).  The solid curves show the mass of a $\nu = 1,~3,~5$ halo in a spherical collapse model formed at redshift $z_f$.  A halo is available in the cosmological history to seed an SMBH (for a given cross-section) if the shaded region corresponding to that cross-section is above a solid curve.  For comparison, we show as dashed curves the cosmological history of the Main Progenitors of a Milky Way Mass and Cluster Mass galaxy, as given in Eq.~(\ref{eq:main_progenitor}).  Interestingly, the main progenitor of a $1-\sigma$ halo at $z=0$, can be a $3-\sigma$ halo at $z\sim 10$, which is more likely to form SMBHs. 
$\textit{right panel}$: 
Mass $M$ of a $\nu = 5$ halo, again for fixed cross-section corresponding to colored regions, that may seed an SMBH at a lower redshift $z\leq z_{\rm f}$.  We can see that rare halos that do not seed an SMBH immediately may do so later in the history of the Universe. 
During the evolution of these halos, we assume the central density and the halo mass are fixed; we will track the assembly history of halos more completely in  Sec.~\ref{sec:cosmological_evolution} utilizing merger trees.
}
\label{fig:massz}
\end{figure*}

\subsection{Mass Threshold of Black Hole Seeding}
In this subsection, we discuss the collapse criterion of dark matter halos analytically based on the collapse timescale calibrated by our simulations. Other criteria related to the halo spin parameter and halo dynamical timescale are discussed in Appendix \ref{append:barrier_fragmentation}, where we will show they are not relevant for the problem at hand. We also study the halo masses that lead to SMBH formation at different redshifts, assuming a median mass-concentration relation discussed in \cite{Diemer_2019}. There will, however, always be a scatter in the halo concentration, which is related to the halo assembly history. The complication is that halos may form early, but not merge until late. Such halos will have a very large central density, corresponding to scatter above the median mass-concentration relation. Another complication is that even though a certain halo is not massive or concentrated enough to seed an SMBH, one of its halo progenitors may have seeded an SMBH which subsequently falls to the halo center. We will fully address those questions in Sec.~\ref{sec:cosmological_evolution} with a merger tree. In this section, we only discuss SMBH formation based on median mass-concentration relations for a given halo mass $M_{\rm vir}$ and redshift $z$. This works well for rare halos that have large masses at high redshift because we do not expect much scatter in their concentration. Therefore the production of high mass SMBHs is well predicted in this section with purely analytic formulae.

For a halo with mass $M_{\rm vir}(z)$, the criterion that an SMBH seed form in the halo at redshift $z$ is approximately given by
\begin{equation}\label{eq:collapse_condition}
 t_{\rm col}(M_{\rm vir}(z), \sigma/m, z) \ll t_{\rm H}(z),
\end{equation}
where $t_{\rm H}(z)$ is the Hubble time at $z$. This indicates that seeding happens when the collapse time is significantly shorter than the lifetime of the system. Practically, we determine that an SMBH seed would form when
\begin{equation}\label{eq:collapse_condition_prac}
      t_{\rm col}(M_{\rm vir}(z), \sigma/m, z) = \epsilon t_{\rm H}(z),
\end{equation}
where $\epsilon$ is set to be $0.1$; this parameter is somewhat arbitrary, but also degenerate with a rescaling of the cross-section, such that $\epsilon$ can be viewed as the uncertainty on the cross-section. Therefore the only parameter that will determine the seeding of SMBHs is $\epsilon\sigma/m$. The choice of $\epsilon=0.1$ is reasonable because the time threshold of collapse and the Hubble time are expected to be roughly within the same order of magnitude. In our seeding model, the collapse timescale is greater than the halo dynamical timescale as discussed in Appendix~\ref{append:barrier_fragmentation} to avoid local fragmentation. An extremely small collapse timescale is disfavored if an SMBH is seeded in a halo instead of forming many local dark stars. 
The fraction of a dark matter halo that eventually collapses into a black hole is crucial for determining the mass function of the SMBH seeds. From the simulation, we know this collapsing fraction is about $3\times 10^{-3}$. Eq.~(\ref{eq:collapse_condition}) gives the lower bound of the halo mass that would lead to a collapsing halo. 
For high redshift where the cosmological constant is not important and the universe is dominated by matter, the critical density scales like $\rho_{\rm crit}\propto (1+z)^{3/2}$. We can further assume $\Delta(z) = 200$ regardless of redshift and the mass threshold for collapsed halos can be determined by combining Eq. (\ref{eq:collapse_time}) and Eq. (\ref{eq:collapse_condition_prac}): 
\begin{equation}\label{eq:minimum_mass}
\begin{split}
M > & M_0(c_{\rm vir}(z),z)\\
 &= 1.63\times 10^{17}
    M_{\odot} \left[{\rm ln}(1+c_{\rm vir}(z))-\frac{c_{\rm vir}(z)}{c_{\rm vir}(z)+1}\right]^{9/2}\\& \times\frac{1}{(c_{\rm vir}(z)/4)^{21/2}}\left(\frac{\Omega_m\rho_{\rm crit}(0)}{\rho_{\rm crit}(z)}\right)^2 \left(\frac{\rm 0.01cm^2/g}{\epsilon \sigma/m}\right)^3.
\end{split}
\end{equation}
For a given dark matter halo with mass $M$ at $z$, it has an SMBH seed with mass $f M$ in the halo center if $M>M_0(c,z)$, where $f \sim  3\times 10^{-3}$ is the collapse fraction of the dissipative dark matter halo.

Eq.~(\ref{eq:minimum_mass}) suggests that the minimum halo mass to seed an SMBH is much smaller at higher redshift, and furthermore, high concentration (rare) halos are more likely to form an SMBH at higher redshift.  This is shown in Fig.~\ref{fig:massz}, similar to the proposal and discussion in Ref.~\cite{Gresham:2018anj}. In the left panel of Fig.~\ref{fig:massz}, 
the minimum mass halo to form an SMBH is shown in shaded regions for different cross-sections.  To determine whether a halo is available in the cosmological history that meets this minimum mass requirement, these colored regions are compared against the solid lines corresponding to a $\nu=1,~3,~5$ halo of mass $M$ formed at redshift $z_f$, using Eq. (\ref{eq:minimum_mass}).  When a colored region is above a solid line of fixed $\nu$, halos of a given cosmological rarity $\nu$ are available to make SMBH seeds via dissipation at redshift $z$. We can thus see that rare halos can seed SMBHs at high redshift. 

In the right panel of Fig.~\ref{fig:massz}, we track the $5-\sigma$ halos to lower redshift $z$ to determine if these rare halos of mass $M$ can seed a black hole later in the history of the Universe.  The shaded contours, with the colors corresponding again to different cross-sections, indicate where SMBH seeds can form at lower redshifts.  Note that we assume the halo mass and central density remains fixed, while the concentration is given by Eq.~(\ref{eq:concentration_z}); this assumption is idealized since halos will accrete and merge with other halos.  Nevertheless, fixing the tdSIDM cross-section, this demonstrates how rare halos that cannot form an SMBH seed immediately may form one at lower redshift.  


Furthermore, if the halo is not massive enough to seed an SMBH at high redshift, it may still have lighter SMBH seeds because its progenitors may have formed black holes at higher redshift.  
We can thus see that the assembly history has to be determined to fully study SMBH formation at low redshift.  This will be discussed in detail in Sec.~\ref{sec:cosmological_evolution} using Monte Carlo simulations to generate the merger tree. However, there is still an analytic shortcut to describe the evolution of the most massive progenitors, known as main progenitors, during the assembly history. Empirically, the mass accretion histories for main progenitors, as observed at $z=0$, can be characterized by a simple function \cite{Wechsler_2002}
\begin{equation}\label{eq:main_progenitor}
    M(z)=M_{0}e^{-\alpha z},
\end{equation}
where $M_0$ is the halo mass at $z=0$ and $M(z)$ is the most massive progenitor in the merger tree. Although the mass accretion history of individual halos may deviate from this form, it provides a good characterization of the range of halo mass accretion trajectories, as we will show later in Sec.~\ref{sec:cosmological_evolution}. $\alpha$ is related to the halo mass at the observed time. The average $\alpha$ is $\approx 0.6$ for a typical halo with mass $M=10^{12}M_{\odot}$ at $z=0$, and $\approx 0.9$ for a rarer halo with mass $M=10^{14}M_{\odot}$ at $z=0$. We show two halo trajectories in Fig.~\ref{fig:massz} for masses $M=10^{12}$ and $M=10^{14}M_{\odot}$. We can see from Fig.~\ref{fig:massz} that the most massive progenitor of a typical ($1-\sigma$) halo at low redshift may instead be a rare $3-\sigma$ halo at high redshift. The rare progenitors, which formed relatively early, have a large central density even at low redshift before merging, and they can potentially seed an SMBH at the halo center. Thus, while the dissipative nature of dark matter helps us explain the most massive SMBHs at high redshift, one must further examine the merger history of halos to check consistency with observations of SMBHs at low-$z$ in Milky Way-like galaxies.
We will show in Sec.~\ref{sec:cosmological_evolution} and Sec.~\ref{sec:conssistency_check} that this suggests a range of cross-sections where high redshift SMBH formation could occur, while simultaneously remaining consistent with low-$z$ observations. 

\section{Cosmological evolution and abundance of SMBHs}\label{sec:cosmological_evolution}

In this section, we aim to make predictions for the cosmological abundance of SMBHs formed via direct collapse of tdSIDM halos and the observed luminosity functions of quasars formed via this mechanism. In contrast to the canonical seeding mechanisms for smaller SMBH seeds ({\em e.g.}, remnants of Pop \rom{3} stars with typical mass of $\sim 10\operatorname{-}10^{3}\,\msun$ \cite{Madau2001,Abel2002,Bromm2002,OShea2007,Turk2009,Tanaka2009,Greif2012,Valiante2016} or directly collapsed pristine gas clouds of mass $\sim 10^{4}\operatorname{-}10^{6}\,\msun$ \cite{Bromm2003,Koushiappas2004,Begelman2006,Lodato:2006,Ferrara2014,Pacucci2015,Valiante2016}), the mechanism in this paper could naturally explain the existence of massive quasars ($M_{\rm BH}\gtrsim 10^{9}\msun$) at $z\gtrsim 6$ discovered in recent years \cite{Mortlock2011,Venemans2013,Wu2015,Mazzucchelli2017,Banados2018,Wang2018,Matsuoka2019,Onoue2019,Yang2020}. According to the mass criterion for seeding in Eq. (\ref{eq:minimum_mass}) and the seed-to-host mass ratio $f_{\rm col}$ discussed in Sec.~\ref{sec:simulation}, $M_{\rm BH}\gtrsim 10^{9}\msun$ SMBHs at $z\sim 7$ will form in $M \gtrsim 10^{12}\msun$ halos with normal concentrations. However, it is still an open question whether this model can produce the correct cosmic abundance of the SMBHs and observed quasars at high redshift.

To investigate this aspect, the cosmological evolution of SMBH seeds and their host halos need to be tracked. In this model, halos with different masses and concentrations could be coupled to the seeding mechanism at very different cosmic times. The seeding should be considered as a continuous process rather than happening only in a short period of time. In addition, the decoupled seeds could further increase their masses through the accretion of baryonic matter, and the amount of such accretion depends on the evolutionary history of halos ({\em e.g.} a major galaxy merger could trigger such accretion). Furthermore, the seeding criterion has a strong dependence on the concentration of the halo, which in turn depends on the assembly history of the halo \cite[{\em e.g.},][]{Zhao_2009}, and is subject to various biases ({\em e.g.} environment of formation). A simple median mass-concentration relation may not be accurate enough to describe the seeding process of the entire cosmological population of dark matter halos.

Given the physical processes and uncertainties involved in the evolution of SMBH seeds, we employ halo merger trees to trace the merger history of halos and SMBH seeds and to evolve SMBHs using empirical prescriptions. The halo merger trees are generated using the {\sc SatGen}~\footnote{https://github.com/shergreen/SatGen} code~\cite{Jiang2020}, which is based on the Extended Press-Schechter (EPS) formalism~\cite{Lacey1993} and the algorithm introduced in Refs. \cite{Parkinson2008,Benson2017}. The virial mass and radius of halos in the merger trees are defined with the redshift-dependent $\Delta_{\rm vir}$ in Ref. \cite{Bryan1998}. When creating the merger trees, we uniformly sample $10$ halos per dex of halo mass ranging from $10^{8}$ to $10^{16.4} \msun$ at $z=4$ and trace their progenitors up to $z\simeq 20$, with a progenitor mass resolution $5$ ($6$ for trees more massive than $10^{15} \msun$) orders of magnitude lower than the final halo mass at $z=4$. The merger tree traces the mass and concentration of each halo from the time when it enters the tree (becomes more massive than the mass resolution of the tree) to the time when it merges into a more massive halo. The halo concentration is obtained from an empirical relation calibrated via simulations~\cite{Zhao_2009}, which relates the main branch (the branch that tracks the most massive progenitor) merging history to the concentration parameter by
\begin{equation}
    c_{\rm vir}(M_{\rm vir}, z) = \big[4^8 + \big(t(z)/t_{\rm 0.04}(M_{\rm vir}, z)\big)^{8.4} \big]^{1/8},
    \label{eq:concentration_Zhao09}
\end{equation}
where $t(z)$ is the cosmic time at redshift $z$ and $t_{\rm 0.04}$ is the cosmic time when the host halo has assembled $4\%$ of its instantaneous mass, $M_{\rm vir}(z)$. In principle, the gravitational impact of baryonic matter ({\em e.g.} adiabatic contraction of dark matter \cite{Blumenthal1986,Ryden1987}), star formation and subsequent feedback processes could potentially affect the structure of high redshift halos. However, self-consistently modelling the baryonic content of high redshift galaxies is beyond the scope of this paper, and we defer a detailed analysis of this aspect to follow-up work.

\begin{figure}[h!]
\centering
\includegraphics[width=0.7\textwidth]{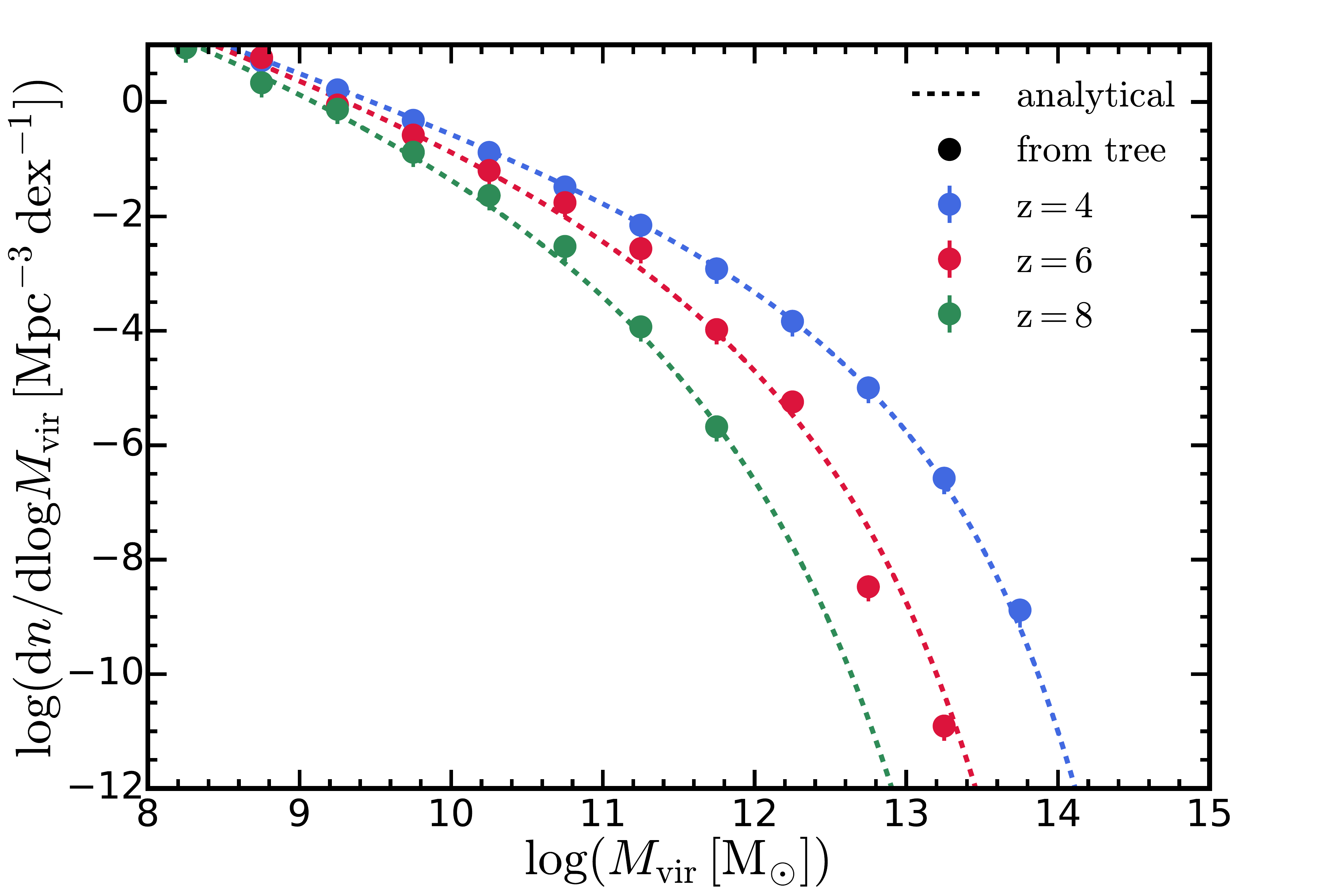}
\caption{\textbf{Halo mass function.} The reconstructed halo mass functions at $z=4,6,8$ based on the weighted abundance of halos in the merger trees (shown by circles of different colors). They are compared to the halo mass functions determined analytically using the {\sc hmf} code (shown by dashed lines), which itself is calibrated based on numerical cosmological simulations~\cite{Tinker2008}. The halo mass functions determined by the merger trees agree reasonably well with the analytic ones up to $10^{\rm 12}\msun$ ($10^{\rm 13.5}\msun$) at $z=8$ ($z=6$), which covers the mass range of quasar host halos of interest.}
\label{fig:hmf}
\end{figure}

All the progenitors of one merger tree are weighted by the number density of the final halo sampled at $z=4$, determined analytically by the halo mass function from the {\sc hmf} code~\cite{Murray2013}, which itself is calibrated based on numerical cosmological simulations~\cite{Tinker2008}. In Fig.~\ref{fig:hmf}, we show the halo mass functions at $z=4,6,8$ reproduced with the weighted abundance of halos in the merger trees. They are in agreement with the halo mass functions determined analytically up to $10^{12}\msun$ ($10^{13.5}\msun$) at $z=8$ ($z=6$), which covers the mass range of quasar host halos of interest. In the subsequent analysis, we will use the weighted results for any predictions in the cosmological context. 

\subsection{SMBH seeding and growth}

Based on the halo merger trees, we initialize and evolve the SMBH seeds with the following empirical prescriptions. An SMBH seed is initialized when the halo meets the seeding criterion introduced in Eq.  (\ref{eq:minimum_mass}). The initial mass of the seed is set as a constant fraction, $f_{\rm col}=3\times 10^{-3}$, of the instantaneous mass of the host halo, motivated by the simulation results in Sec. \ref{sec:simulation}. Subsequently, as long as the host halo still meets the seeding criterion, we maintain the seed-to-host mass ratio as $f_{\rm col}$ (referred to as the $reseeding$ $mechanism$). The treatment relies on the assumption that, after the initial collapse of the dark matter halo, the accretion of dark matter onto the central SMBH seed will continue until a dynamical equilibrium between the SMBH seed and the host halo is reached. This dynamical equilibrium results in the roughly constant seeding fraction found in the simulations, and should hold as long as dark matter can still be efficiently fed to the halo center via dissipative self-interactions. For halos that are coupled to the seeding mechanism, the $reseeding$ would effectively erase the unique growth history of the SMBHs and set a tight correlation between host halo mass and SMBH mass. However, for halos that no longer meet the seeding criterion, the subsequent growth of SMBHs they host will no longer be affected by dark matter physics but by hierarchical mergers of SMBHs during halo mergers and accretion of baryonic matter. During the merger of host halos, the dynamical friction against the dark matter background could drag the satellite SMBH towards the primary SMBH and a bound SMBH binary will form. We assume that this happens when the mass ratio of the two SMBH-plus-halo systems is larger than $0.3$, as suggested in Ref. \cite{Volonteri:2002vz}. For simplicity, we do not model the subsequent evolution of the binary and treat the bound binary as a single SMBH right after the merger. The typical timescale for a billion solar mass SMBH binary to go through the hardening stage to the final coalesce is of $\sim 1\Gyr$ (e.g. \cite{Volonteri:2002vz}). Therefore, in the early Universe, it is likely that binary SMBHs seeded through this mechanism are common. These binaries could have different accretion (quasar) activities from low redshift AGNs. In addition, SMBH triplets will likely form through hierarchical merger as well. The intruding SMBH can facilitate the coalesce of the binary through close three-body interactions and Kozai-Lidov oscillations (e.g. \cite{Iwasawa2006,Hoffman2007}). The lightest SMBHs are expected to be ejected from the galaxy center in about $40\%$ of the cases (e.g. \cite{Iwasawa2006,Hoffman2007}). Moreover, the recoil due to the gravitational wave emission after binary merger (e.g. \cite{Fitchett1983}) could also lead to the ejection of the remnant SMBH. These processes could introduce order unity correction factors to the SMBH occupation fractions and SMBH masses. Self-consistently modelling these processes is beyond the scope of this paper, thus our results should be treated as upper limits.

For the accretion of baryonic matter, we model the ``merger driven'' accretion of SMBHs, which has been adopted in previous studies of the cosmic evolution of SMBHs~\cite{Volonteri:2002vz,Volonteri2006,Volonteri2008,2014GReGr..46.1702N}. The efficient gas inflow triggered by galaxy mergers feeds both the accretion of SMBHs and the star-formation in galaxy bulges. We assume this feeding happens when the mass ratio between the two progenitor halos is larger than $0.1$ (defined as ``major merger''). The stellar/supernovae feedback from rapid star-formation and potential active galactic nucleus (AGN) feedback will eventually quench the gas inflow as well as further growth of the SMBH. The total amount of mass accreted during each major merger event is related to the complicated gas dynamics and feedback processes in the galaxy bulge. Hypothetically, it manifests as the observed statistical correlation between the SMBH mass and bulge velocity dispersion of its host galaxy (the $M_{\rm BH}-\sigma^{\ast}_{\rm v}$ relation \cite{Ferrarese:2002ct,Kormendy2013})
\begin{equation}
    M_{\rm BH} = (4.4 \pm 0.9) \times 10^7 \msun \, (\sigma^{\ast}_{\rm v}/150\kms)^{4.58\pm 0.52}.
\end{equation}
This motivates us to set the mass gain of an SMBH through accretion of baryonic matter during each merger event as
\begin{equation} 
    \Delta M_{\rm BH} = \Delta M_{\rm 0} (1-\epsilon_{\rm r}) \, (\sigma^{\ast}_{\rm v}/150\kms)^{4.58},
    \label{eq:dmbh-sigma}
\end{equation}
where $\sigma^{\ast}_{\rm v}$ is the bulge velocity dispersion of the merged galaxy, $\epsilon_{\rm r}$ is the radiative efficiency (assumed to be the canonical value $0.1$) and $\Delta M_{\rm 0}$ is a free normalization parameter, which has been set to $\sim 10^{4} \operatorname{-} 10^{7}\msun$ in previous studies of low-mass seeds \cite{Volonteri:2002vz,Volonteri2006,2014GReGr..46.1702N}. In observations, the bulge velocity dispersion is found to correlate with the asymptotic value of the halo circular velocity as \cite{Ferrarese:2002ct}
\begin{equation}
    \log{V_{\rm c}} = (0.892 \pm 0.041) \log{\sigma^{\ast}_{\rm v}} + (0.44\pm 0.09).
    \label{eq:Vc-sigma}
\end{equation}
And for the NFW profile, the maximum circular velocity of the host halo is related to the halo mass as~\cite{Klypin2001,Klypin2011}
\begin{align}
    & V_{\rm c} = \Big[G \dfrac{f(x_{\rm max})}{f(c_{\rm vir})}\dfrac{c_{\rm vir}}{x_{\rm max}} \Big(\dfrac{4\pi}{3} \Delta(z) \rho_{\rm crit}(z) \Big)^{1/3}  \Big]^{1/2} M_{\rm vir}^{1/3}, \nonumber \\
    & f(x) = \ln{(1+x)} - \dfrac{x}{1+x}, \,\, x_{\rm max}=2.15.
    \label{eq:Vc-Mhalo}
\end{align}

Combining Eqs. (\ref{eq:dmbh-sigma}), (\ref{eq:Vc-sigma}) and (\ref{eq:Vc-Mhalo}) above results in a link between $\Delta M_{\rm BH}$ and host halo parameters ($M_{\rm vir}$, $c_{\rm vir}$) at a given redshift. This forms an empirical prescription to model the mass growth of SMBHs during galaxy mergers tracked by halo merger trees, with the assumption that the statistical correlations between SMBHs and their host galaxies (halos) are maintained throughout cosmic time. Overall, the two free parameters of the SMBH catalog are the self-interaction cross-section per unit mass, $\sigma/m$, and the baryonic mass accretion constant, $\Delta M_{\rm 0}$. Similar to the host halos, the SMBHs are assigned with statistical weights corresponding to the number density of the final halo of the merger tree at its sampling redshift. For our fiducial model, we set $\Delta M_{\rm 0}=0$ to study the pure impact of dark matter physics and hierarchical mergers of SMBH seeds. In addition, we will try varying $\Delta M_{\rm 0}$ to $10^{7}\msun$ to study the ``maximum'' effect (since $10^{7}\msun$ is already close to the normalization of the local $M_{\rm BH}-\sigma^{\ast}_{\rm v}$ relation) that baryonic accretion can have on this population of SMBHs at high redshift.

\subsection{Predictions for high redshift quasars}

\begin{figure}[h!]
\centering
\includegraphics[width=0.6\textwidth]{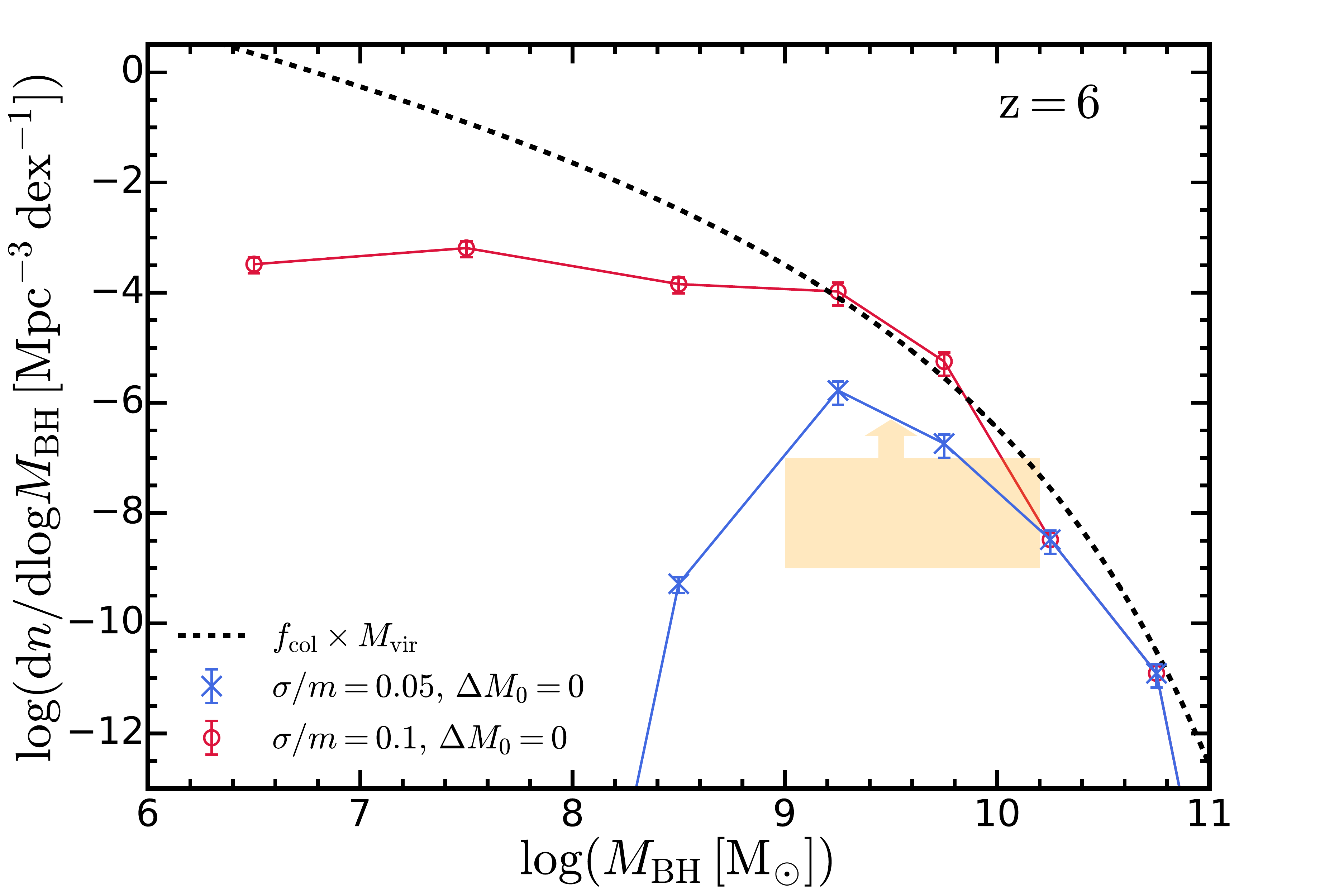}
\includegraphics[width=0.6\textwidth]{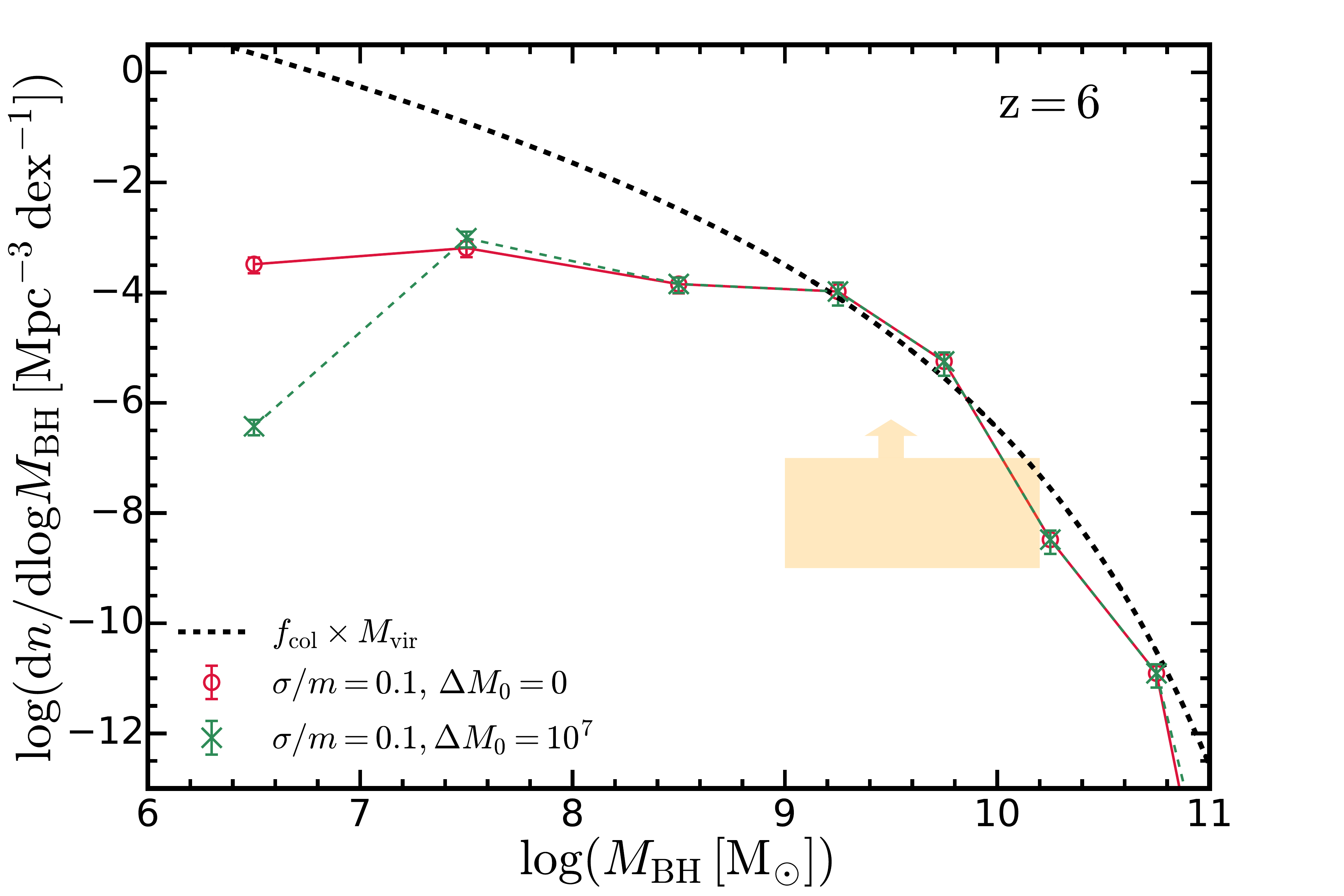}
\caption{\textbf{SMBH mass function.} {\it Top:} Number density of SMBHs as a function SMBH mass at $z=6$, calculated from the weighted abundance of SMBHs in the merger trees. The prediction assuming $\sigma/m = 0.1\,(0.05)\cpm$ and $\Delta M_{\rm 0}=0$ is shown and compared to the halo mass function multiplied by the collapse fraction $f_{\rm col}$. The massive end of the BHMF is coupled with the seeding mechanism, and the shape of the SMBH mass function resembles the exponential cut-off in the halo mass function. Low mass SMBHs has decoupled from the seeding mechanism and the low-mass end of the mass function deviates from the halo mass function. The choice of self-interaction cross-section does not affect the massive end but changes the characteristic mass where the SMBH mass function deviates from the halo mass function. The shaded region indicates the abundance of observed massive quasars ($M_{\rm BH} \gtrsim 10^{9}\msun$) at high redshift and the abundance of underlying SMBH population should at least be larger. {\it Bottom:} We show the SMBH mass functions in the model with $\sigma/m = 0.1\cpm$ and $\Delta M_{\rm 0}=0\,(10^{7})\msun$. The baryonic accretion arguably only has an impact at the low-mass end (shifting the lowest mass of the seeds produced by the mechanism), hardly changing the abundance of the most massive SMBHs. }
\label{fig:BHMF}
\end{figure}

In this section, we aim to make predictions for the abundance of luminous quasars at high redshift and explain the unexpectedly large masses of these quasars with the seeding model discussed in this paper. We will first derive predictions for the mass function of SMBHs seeded by tdSIDM, and then link it to the luminosity function of quasars. Binned estimations of SMBH mass functions are derived based on the weighted abundance of SMBHs in the merger trees and the results are shown in Fig.~\ref{fig:BHMF}. At the massive end, the shape of the SMBH mass function resembles the halo mass function with an exponential decrease, since the massive SMBHs are still coupled to the seeding mechanism with mass proportional to the host halo mass. At the low-mass end, SMBHs start to decouple from the seeding mechanism, so the SMBH mass function turns over and starts to decrease with lower $M_{\rm BH}$, as opposed to the behaviour of the halo mass function. Varying the self-interaction cross-section has almost no effects at the massive end while changing the characteristic lower mass when the SMBH mass function turns over. The model with $\sigma/m=0.1\cpm$ predicts a more extended tail of SMBHs at the low-mass end, compared to the model with $\sigma/m=0.05\cpm$, with no apparent mass cut-off. This is because the redshift range of seeding in the model with $\sigma/m=0.1\cpm$ is broader than the model with $\sigma/m=0.05\cpm$, as illustrated in the left panel of Fig.\ref{fig:massz}. SMBHs seeded and decoupled at higher redshift can populate the low-mass end of the SMBH mass function. Quasar surveys and theoretical modelling indicate that the number density of luminous high redshift quasars with $M_{\rm BH} \gtrsim 10^{9}\msun$ and $L_{\rm bol}\gtrsim 10^{46} \erg/{\rm s}$ is $10^{-9} \lesssim \Phi \lesssim 10^{-7} [\Mpc^{-3}\,{\rm dex}^{-1}]$ \cite{Inayoshi_2020,Trakhtenbrot2020,Shen:2020obl}, which sets a lower limit of the abundance of underlying SMBH population \footnote{The estimation is done using the bright UV-selected quasars at $z\gtrsim 6$. If actively accreting at (sub-)Eddington rate, a billion solar mass SMBH roughly gives bolometric radiation output $L_{\rm bol}\sim 10^{46-47} \erg/{\rm s}$, which corresponds to $M_{\rm UV} \sim -24$ after applying the bolometric corrections (e.g. \cite{Shen:2020obl}). The number density of bright UV-selected quasar with such luminosities is roughly the range quoted in the main text (e.g. \cite{Kulkarni2019,Wang2019b,Shen:2020obl}). Similar number density estimations were given in \cite{Trakhtenbrot2020,feng2020seeding}}. The predictions here are consistent with this limit. In the lower panel of Fig.~\ref{fig:BHMF}, we compare the model predictions with $\Delta M_{\rm 0}=0$ and $\Delta M_{\rm 0}=10^{7}$. Baryonic accretion during major mergers only has a weak impact at the low-mass end (shifting the lowest mass of the seeds produced by the mechanism) and can hardly affect the mass of the most massive SMBHs seeded through this mechanism. 

In order to relate the SMBH mass function to the quasar luminosity function, the fraction of SMBHs that are active (the duty-cycle $D$) and the luminosity of active quasars are required. The bolometric luminosity (luminosity integrated over the entire quasar spectrum and free from extinction) of a quasar, $L_{\rm bol}$, is often described as its ratio to the Eddington luminosity
\begin{align}
  L_{\rm bol} & = \lambda_{\rm edd} L_{\rm edd}=\lambda_{\rm edd}\dfrac{4\pi G m_{\rm p}c}{\sigma_{\rm T}}M_{\rm BH} \nonumber \\
  & = 1.26 \times 10^{47} {\rm erg}/{\rm s} \,\, \Big(\dfrac{\lambda_{\rm edd}}{1}\Big) \Big(\dfrac{M_{\rm BH}}{10^{9}\msun}\Big),
    \label{eq:edd}
\end{align}
where $m_{\rm p}$ is the proton mass and $\sigma_{\rm T}$ is the Thomson scattering cross-section for the electron. The ratio $\lambda_{\rm edd}$ is referred to as the Eddington ratio. For simplicity, we first adopt a log-normal Eddington ratio distribution function (ERDF)
\begin{equation}
    P_{\rm 1}(\log{\lambda_{\rm edd}}) = \dfrac{1}{\sqrt{2\pi} \sigma_{\rm edd}} e^{-(\log{\lambda_{\rm edd}}-\log{\lambda_{\rm c}})^{2}/2\sigma^{2}_{\rm edd}},
\end{equation}
with $\lambda_{\rm c}=0.6$ and $\sigma_{\rm edd}=0.3$, motivated by observational constraints of $z\sim 6$ quasars \cite{Willott2010}, as well as the extrapolation of models constrained at lower redshift \cite[e.g.,][]{Kelly2013,Tucci:2016tyc}. Such a log-normal ERDF implies that active SMBHs accrete at close to the Eddington limit. However, it is still possible that a substantial fraction of active SMBHs accrete at much lower rates and the observed massive quasars are only tip-of-the-iceberg of the SMBH population. Therefore, in addition to the log-normal ERDF, we also try using a cut-off power-law ERDF that extends to $\lambda_{\rm edd}=10^{-4}$
\begin{equation}
    P_{\rm 2}(\log{\lambda_{\rm edd}}) = N \Big( \dfrac{\lambda_{\rm edd}}{\lambda_{\rm c}} \Big)^{\alpha} e^{-\lambda_{\rm edd}/\lambda_{\rm c}},
    \label{eq:erdf-power-law}
\end{equation}
where $N$ is a normalization factor to keep the integrated probability at unity, $\lambda_{\rm c}=1.5$ \cite{Tucci:2016tyc} sets a cut-off in the super Eddington regime and $\alpha$ is the faint-end slope, which is free and can be tuned to match the prediction with the observed bolometric quasar luminosity function. The SMBH mass function can then be mapped to the bolometric quasar luminosity function through the convolution
\begin{align}
    & \phi_{\rm L}(\log{L_{\rm bol}}) = D \int_{-4}^{\infty } \phi_{\rm M}(\log{L_{\rm bol}}-\log{\lambda_{\rm edd}}-\log{C}) \nonumber \\ 
    & \hspace{4cm} P(\log{\lambda_{\rm edd}}) {\rm d}\log{\lambda_{\rm edd}},
\end{align}
where $C$ is $4 \pi G m_{\rm p}c / \sigma_{\rm T}$ (as in Eq.  (\ref{eq:edd})) and we have assumed that SMBHs with $\log{\lambda_{\rm edd}}>-4$ are active (which also defines the duty-cycle). The duty-cycle can be determined by making the normalization of the predicted luminosity function consistent with observations at the bright end. We note that the parameterization of the ERDF and the simple constant duty-cycle assumed here are purely for ``a proof of concept'', with the intention to check whether predictions from the seeding model can be reconciled with observations with some level of tuning of the model for SMBH growth. We do not try to argue for a specific  model of SMBH growth through the study here.

The bolometric luminosity of quasars is the integrated luminosity over the entire spectrum, representing the total energy output. However, in observations, the luminosity function measurements are performed in certain photometric bands (commonly far-UV and X-ray for quasars at high redshift) covering restricted parts of the quasar spectral energy distribution and are subject to corrections for dust and neutral hydrogen extinction, survey completeness, and selection biases. Ref. \cite{Shen:2020obl} has updated the constraints on the bolometric quasar luminosity function at high redshift based on the latest compilation of observations in far-UV, X-ray, and infrared. The observational binned estimations from compiled observations are converted onto the bolometric plane, taking account of the extinction and bolometric corrections. 

\begin{figure}[h!]
\centering
\includegraphics[width=0.49\textwidth]{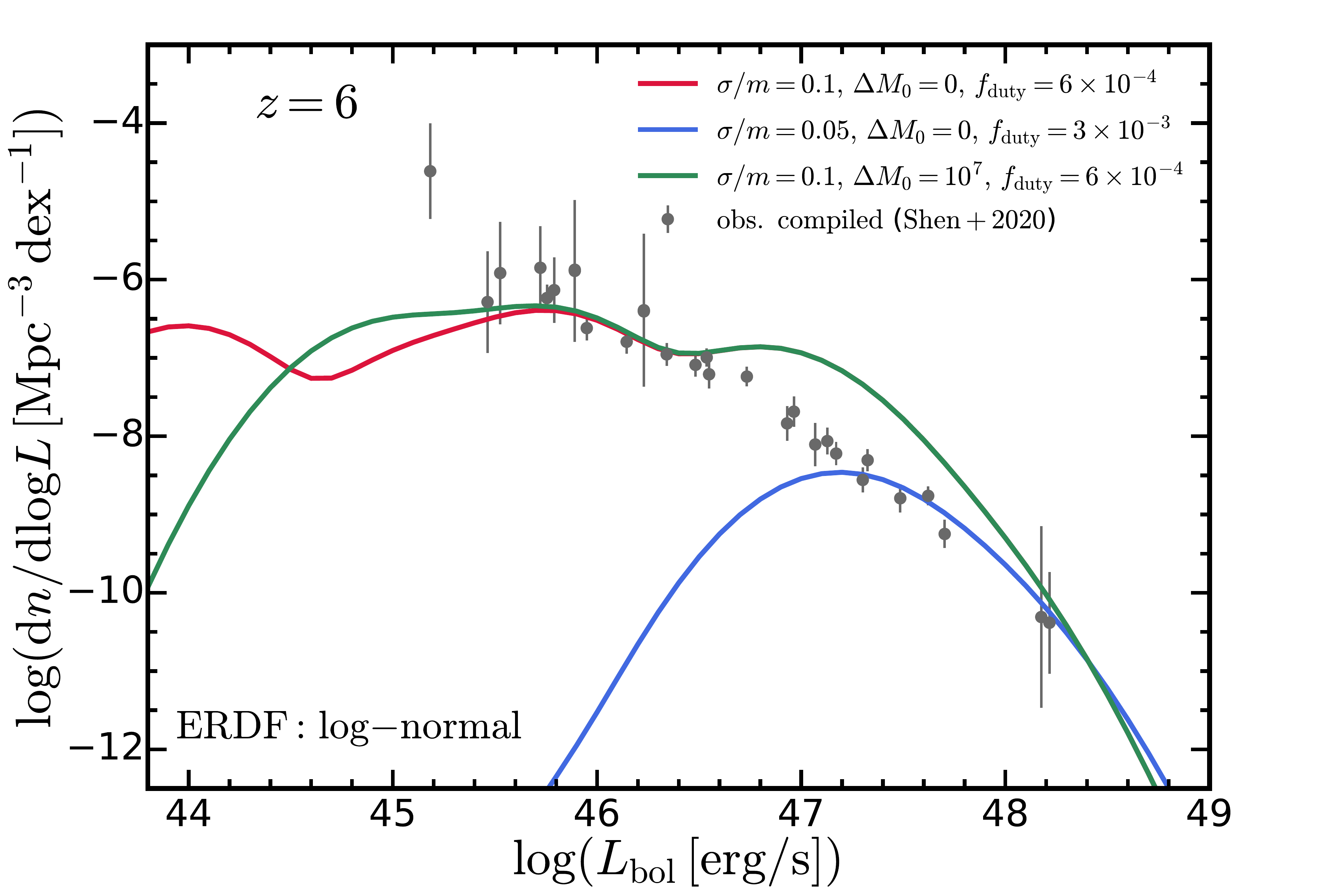}
\includegraphics[width=0.49\textwidth]{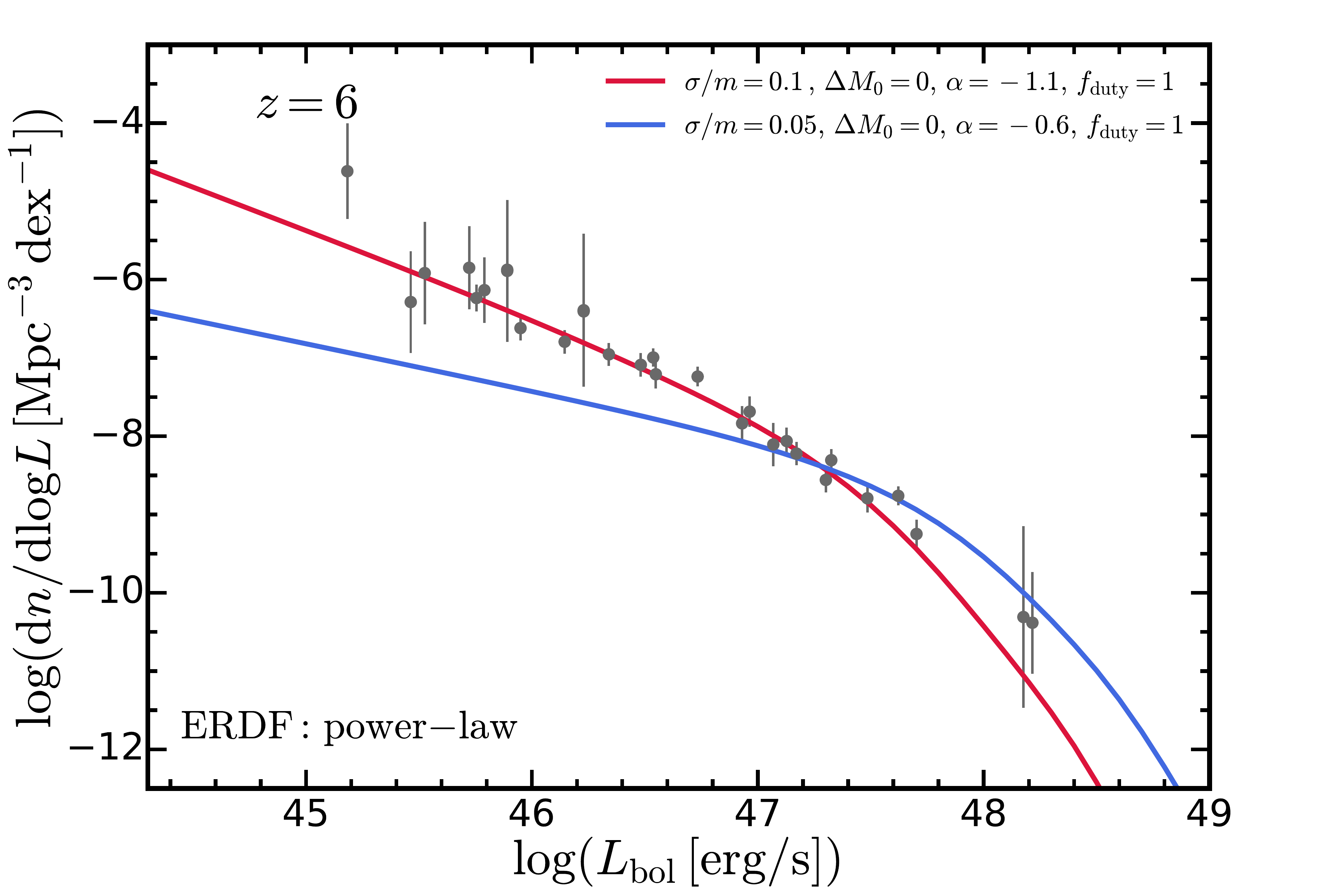}
\caption{\textbf{Bolometric quasar luminosity function at $z=6$.} {\it Top:} Model predictions, varying $\sigma/m$ and $\Delta M_{\rm 0}$. The predictions are derived by convolving the SMBH mass function with a log-normal ERDF, tuning the duty-cycle to match the abundance of luminous quasars. The solid circles represent observational constraints compiled in \cite{Shen:2020obl}. The prediction assuming $\sigma/m = 0.05\cpm$ is compatible with the observations and produces the observed abundance of luminous quasars at $z=6$, assuming a relatively low duty-cycle. On the other hand, the model with $\sigma/m = 0.05\cpm$ will overproduce quasars of $L_{\rm bol} \sim 10^{47}\erg/{\rm s}$. {\it Bottom:} We show the predictions with a cut-off power-law as the ERDF. The duty-cycle is assumed to be unity. The faint-end slope of the ERDF ($\alpha$) is tuned to make the predicted quasar luminosity function close to observations. Both models can agree well with the luminous quasar abundances in observations. But the model $\sigma/m = 0.05\cpm$ does not fit perfectly with the faint end luminosity function regardless of the $\alpha$ adopted.}
\label{fig:lmf}
\end{figure}

In the top panel of Fig.~\ref{fig:lmf}, we show the predicted bolometric quasar luminosity function at $z=6$ from the merger trees, assuming a log-normal ERDF. The results are compared to the observational constraints compiled in \cite{Shen:2020obl}. With a duty-cycle of $3\times 10^{-3}$ ($6\times 10^{-4}$), the predicted abundance of the most luminous quasars in the model with $\sigma/m = 0.05$ ($0.1$) $\cpm$ can match the observed abundances. The prediction assuming $\sigma/m = 0.1\cpm$ gives better agreement at faint luminosities ($L_{\rm bol} \lesssim 10^{46.5} \erg/{\rm s}$) but over-predicts the quasars at intermediate luminosities ($L_{\rm bol} \sim 10^{47} \erg/{\rm s}$). The prediction with $\sigma/m = 0.05\cpm$ agrees with observations at the luminous end ($L_{\rm bol} \gtrsim 10^{47.5} \erg/{\rm s}$) and is not in tension with observations at intermediate and faint luminosities. Acknowledging that other seeding mechanisms could still be responsible for the formation of low-mass and faint quasars, the prediction with $\sigma/m = 0.05\cpm$ is compatible with observations. In terms of the duty-cycle, some observational studies of quasar clustering \cite{Shen2007,White2008,Shankar2010} have suggested that the duty-cycle of high redshift AGNs in the most massive halos may approach unity at $z\simeq 6$. That duty-cycle is much larger than the median value required for our models, especially for the $\sigma/m = 0.1\cpm$ case, to not overproduce the abundance of luminous quasars. However, if the Eddington ratio of SMBH has a strong positive dependence on the host halo mass or environment, the averaged duty-cycle of all SMBHs could be much smaller than inferred from the clustering of currently observed luminous quasars. It is still debated observationally whether high redshift quasars have order unity duty-cycles, or we are observing the tip-of-the-iceberg of the SMBH population. Some studies \cite{Chen2018,Tucci:2016tyc} have instead argued for a low duty-cycle of the quasar population at $z\gtrsim 6$. We examine this possibility by using the cut-off power-law function defined in Eq. (\ref{eq:erdf-power-law}) as the ERDF, which essentially includes a power-law tail of SMBHs with low Eddington ratios. The quasar luminosity functions predicted from this ERDF are shown in the lower panel of Fig.~\ref{fig:lmf}. We have set $D=1$ (since we have already considered quasars with low activity with this ERDF) and tuned $\alpha$ in order to best match the observational constraints. The model with $\sigma/m = 0.1\cpm$ and $\alpha=-1.1$ is in perfect agreement with observations at all luminosities. The model with $\sigma/m = 0.05 \cpm$ with $\alpha=-0.6$ can produce the correct abundance of bright quasars but predicts a shallower faint-end slope. We note that this discrepancy cannot be alleviated by tuning $\alpha$ and $D$, since further decreasing $\alpha$ will decrease the normalization at the bright end and require an unphysical value $D>1$ to match observations. The comparisons here demonstrate that with a little tuning of parameters of the ERDF, the model can reproduce the observed quasar luminosity function. Meanwhile, despite the detailed functional form we use for the ERDF, our results suggest that if the collapse of dissipative dark matter halo is the dominant seeding mechanism for SMBHs at high redshift, a significant fraction of non-active SMBHs or SMBHs with low Eddington ratios would be expected. Such a feature can be tested with future surveys of high redshift quasars with improved completeness.

\begin{figure}[h!]
\centering
\includegraphics[width=0.8\textwidth]{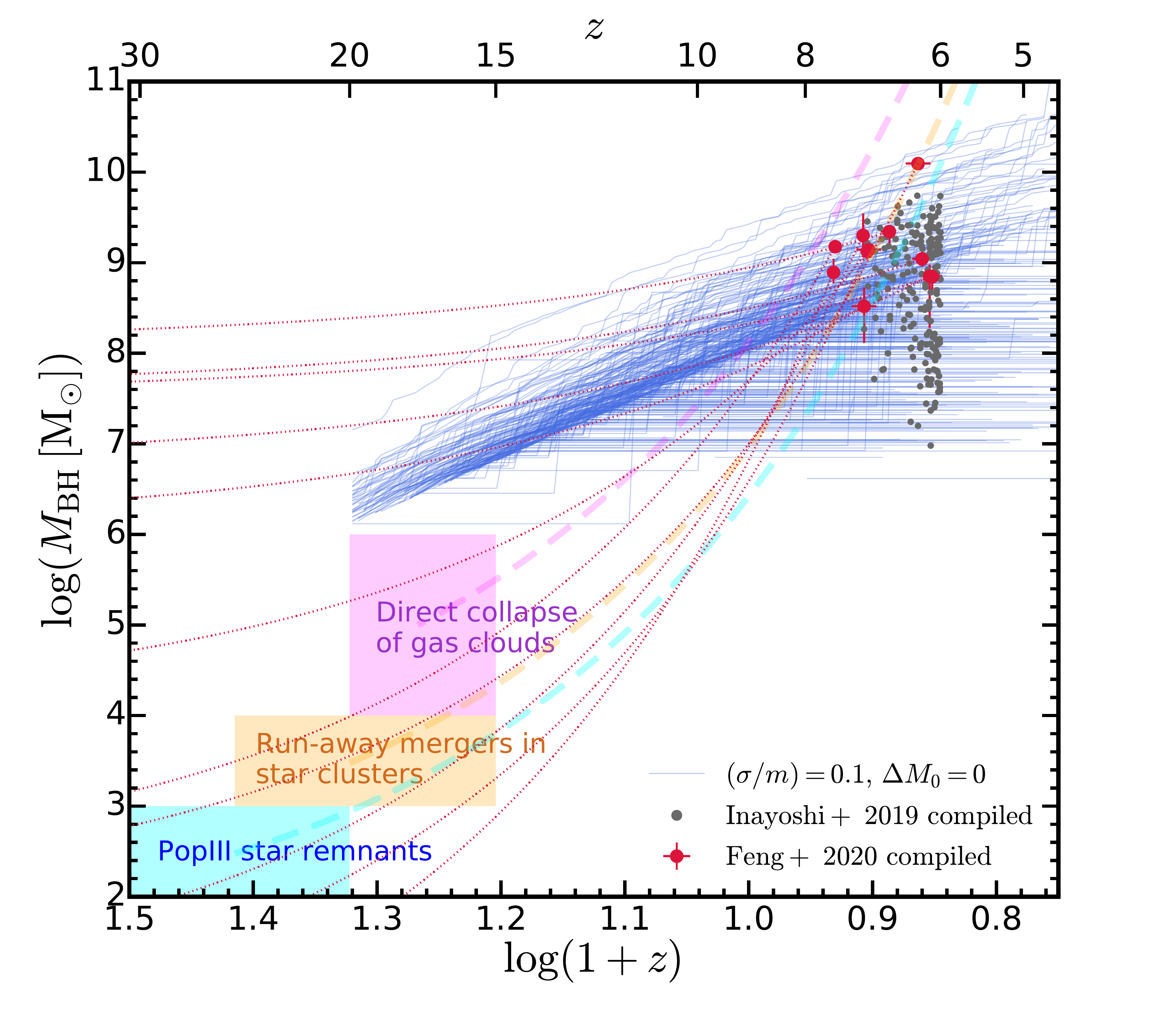}
\caption{\textbf{Mass growth history of SMBHs.} The blue solid lines show the mass of SMBHs as a function of redshift in our model assuming $\sigma/m=0.1\cpm$. These SMBHs are selected from merger trees with $M_{\rm BH} \leq 10^{10} \msun$. The red points are the observed massive quasars at $z\gtrsim 6$ compiled in \cite{feng2020seeding,Wang2019} with the mass estimated using the virial method. The gray points are a more complete set of $196$ quasars at $z\gtrsim 6$ compiled in \cite{Inayoshi_2020}, with the mass estimated indirectly from UV luminosity. The red dotted lines indicate the growth history of the observed quasars assuming they exhibit the same Eddington ratio as the measured value at the redshift of discovery. The typical mass and formation redshift of SMBH seeds from classical seeding mechanisms are shown in shaded regions, with the Eddington-limit growth tracks of these seeds in dashed lines for reference. Seeds formed in canonical mechanisms need to accrete at rates near the Eddington limit in order to produce billion solar mass SMBHs at $z\simeq 6-8$. This is in tension with the low Eddington ratios of some observed quasars, which require seed masses of $\sim 10^{8}M_{\odot}$ implied by their observed Eddington ratio. However, such quasars can be accommodated in our seeding model.}
\label{fig:BHgrowh}
\end{figure}

\begin{figure}[h!]
\centering
\includegraphics[width=0.7\textwidth]{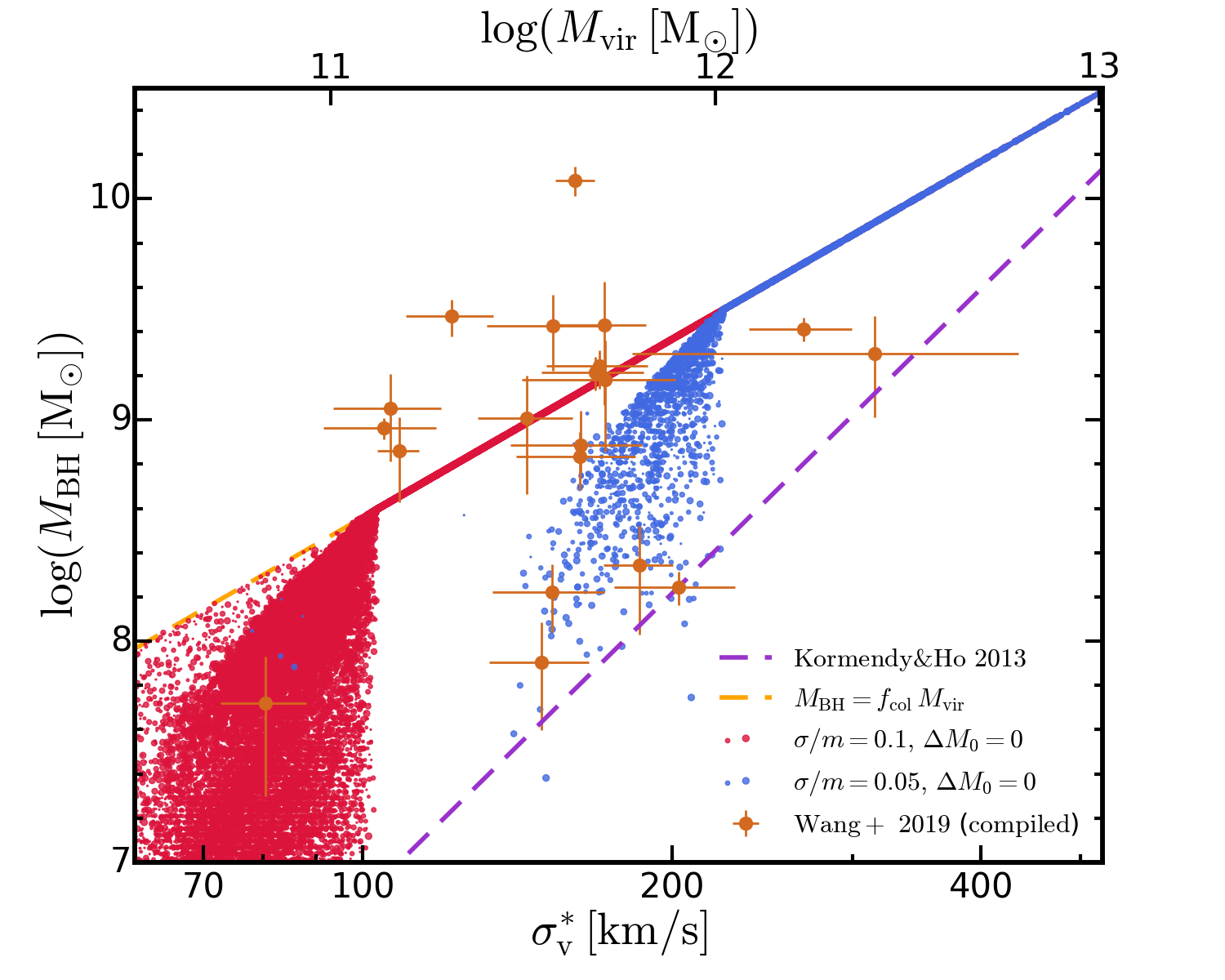}
\caption{\textbf{The $M_{\rm BH}-\sigma^{\ast}_{\rm v}$ relation of high redshift SMBHs.} We show SMBHs in the merger trees selected at $z=7$ in solid circles, with the marker size scaling with the statistical weight. Red and blue circles correspond to the model with $\sigma/m=0.1$ and $0.05\cpm$, respectively. The local $M_{\rm BH}-\sigma^{\ast}_{\rm v}$ relation \cite{Kormendy2013} is shown with the purple dashed line. The orange dashed line shows the relation $M_{\rm BH} \sim f_{\rm col}M_{\rm vir}$, assuming the relation between $M_{\rm vir}$ and $\sigma^{\ast}_{\rm v}$ (Eq. (\ref{eq:Vc-sigma}) and Eq. (\ref{eq:Vc-Mhalo})) holds. Observational samples based on the [C \rom{2}] line observations of the quasar host galaxies compiled in \cite{Wang2019} (originally from \cite{Decarli2018}) are shown in orange circles.}
\label{fig:Msigma}
\end{figure}

Since the most important implication of the model is the existence of extremely massive SMBHs, we explicitly track the mass growth history of $\sim 300$ randomly selected massive SMBHs with $\log{M_{\rm BH}}\leq 10$ at $z=7$ in the merger trees. The results of the model with $\sigma/m=0.1\,\cpm$ are shown in Fig.~\ref{fig:BHgrowh} and compared to the mass measurements of high redshift quasars in the Ref. \cite{feng2020seeding} compilation, including observations from Refs. \cite{Mortlock2011,Wu2015,Banados2018,Wang2018,Matsuoka2019,Onoue2019,Yang2020}. The masses were measured using the virial method based on the broad line emission from quasars. The recent measurement of a $z\sim 7$ quasar \cite{Wang2020} is added to this compilation. For these quasars, we show their mass growth history assuming they have the same Eddington ratio as the measured value at the redshift of discovery. In addition, we show a more complete set of $196$ quasars at $z\gtrsim 6$ compiled by \cite{Inayoshi_2020}, where the SMBH mass was inferred from the UV luminosity with bolometric corrections and assuming $\lambda_{\rm edd}=1$. The massive quasars observed at $z\simeq 6\operatorname{-}8$ with relatively low measured Eddington ratios are hard to reconcile with the canonical seeding mechanisms since the seeds need to continuously accrete at the Eddington limit to reach more than a billion solar mass at the redshift of discovery. On the other hand, in our model, the masses of selected SMBH seeds are in agreement with the massive quasars revealed by observations at $z \simeq 6-8$. Among these seeds, the relatively massive ones are still coupled to the seeding mechanism down to $z\simeq 6$ and have their mass growth following the growth of host halo mass. These seeds are already very massive ($M_{\rm BH} \gtrsim 10^{6}\msun$) when initially seeded at $z\gtrsim 15$ and the mass growth is dominated by the accretion of dissipative dark matter, so the observed low Eddington ratios can be tolerated. Such a picture is consistent with the large fraction of in-active quasars constrained above in the discussion of quasar luminosity functions. In addition, recent observational studies found that a few objects have extremely small proximity zone sizes that imply UV-luminous quasar lifetimes of $\lesssim$ 100,000 yr \cite{Eilers:2020htq}. The short lifetimes of these quasars also pose challenges to canonical black hole formation models which require a much longer period of seed accretion to reach the SMBH mass at the redshift of discovery. However, these young quasars can be accommodated in our seeding model, where the mass growth of SMBHs is dominated by dissipative dark matter accretion with no impact on the ambient intergalactic medium.

In Fig.~\ref{fig:Msigma}, we show the $M_{\rm BH}-\sigma^{\ast}_{\rm v}$ relation of $z\gtrsim 6$ quasars. The SMBHs in the merger trees at $z=7$ are shown in this plane for comparison to observational results. We convert the host halo mass to the bulge velocity dispersion using Eq.~(\ref{eq:Vc-sigma}) and Eq.~(\ref{eq:Vc-Mhalo}), assuming that the locally observed scaling relations can be applied to high redshift galaxies. The SMBHs in the merger trees tightly follow the $M_{\rm BH} \sim f_{\rm col}M_{\rm vir}$ relation in massive host galaxies, and start to scatter toward lower $M_{\rm BH}$ at the mass when the halo decouples from the seeding mechanism. We compare our results with the observational constraints compiled in Ref.~\cite{Wang2019}, based on the [C \rom{2}] line observations of the quasar host galaxies compiled in Ref.~\cite{Decarli2018}. Observations \cite{Decarli2018,Wang2019,Neeleman2021} indicate that the host galaxies of massive, luminous quasars at $z\gtrsim 6$ have halo dynamical masses and velocity dispersions at least an order of magnitude lower than expected from the local $M_{\rm BH}-M_{\rm bulge}$ and $M_{\rm BH}-\sigma^{\ast}_{\rm v}$ relations. However, as shown in Fig.~\ref{fig:Msigma}, SMBHs seeded by tdSIDM, which exhibit a much larger SMBH-to-halo mass ratio than local constraints, are in better agreement with these measurements. At the massive end, SMBHs in this model cluster around a straight line fixed by the $V_{\rm c}-\sigma^{\ast}_{\rm v}$ relation (Eq.~(\ref{eq:Vc-sigma})) we assumed. The statistical scatter of the relation is not reflected here. The typical uncertainty of the normalization of the relation measured at low redshift is $\sim 0.1-0.2\,{\rm dex}$ in $V_{\rm c}(\sigma^{\ast}_{\rm v})$ \cite{Ferrarese:2002ct}, which roughly corresponds to $\sim 0.2-0.6\,{\rm dex}$ in $M_{\rm vir}$. At the low-mass end, SMBHs decouple from the linear relation and the scatter at the tail is due to variations in the merger histories of host halos. Such an $M_{\rm BH}-\sigma^{\ast}_{\rm v}$ relation predicted at high redshift will still be consistent with the relation measured at low redshift, since the SMBHs below $10^{10}\msun$ will have already decoupled from the seeding mechanism and have their mass growth dominated by baryonic accretion.

\begin{figure}[h!]
\centering
\includegraphics[width=0.7\textwidth]{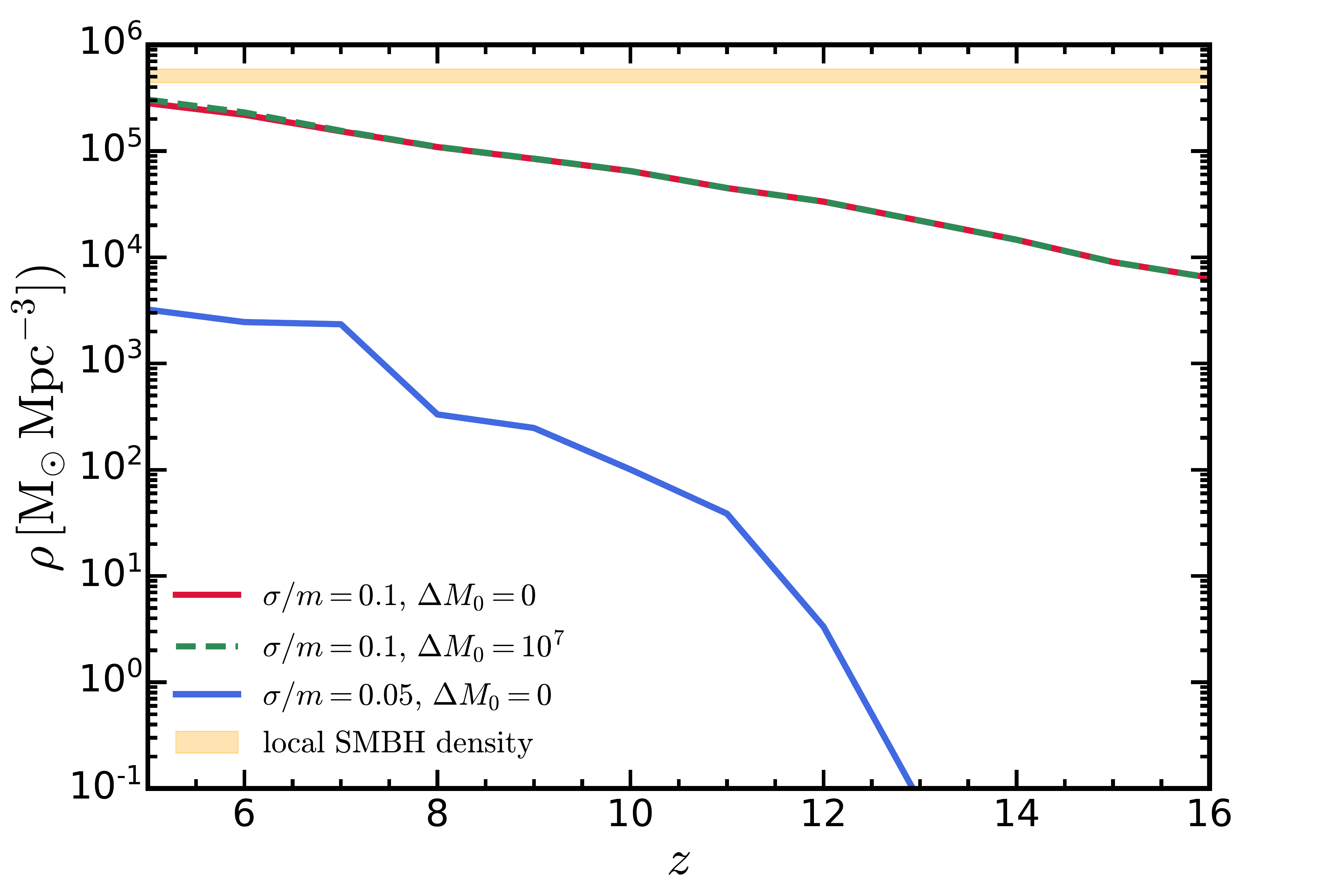}
\caption{\textbf{The comoving SMBH mass density in the Universe versus redshift.} The cumulative mass density of SMBHs integrated over the mass function. The results with different model parameters are shown as labelled and compared to the local SMBH mass density, $4.4-5.9\times 10^{5}\msun \Mpc^{-3}$~\cite{Graham2007}, as indicated by the horizontal line. The mass density from the model with $\sigma/m = 0.1\cpm$ approaches the local mass density already at $z\simeq 6$, which is potentially problematic since the integrated quasar luminosity density matches the local SMBH mass density \cite{Shen:2020obl} (at $0.5\,\mathrm{dex}$ level, assuming $\epsilon_{\rm r} = 0.1$). Therefore, the mass density at high redshift needs to be significantly lower than the local value in order to be consistent with the observation of quasar luminosity functions, unless $\epsilon_{\rm r}$ is larger ({\em i.e.} SMBHs are rapidly rotating).}
\label{fig:bhmass_density}
\end{figure}

In Figure~\ref{fig:bhmass_density}, we show the cosmic SMBH mass density as a function of redshift predicted by our seeding mechanism and compare it to the local SMBH mass density \cite{Graham2007}, which poses an upper limit. The mass density from the model with $\sigma/m = 0.1\cpm$ is close to the local value already at $z\simeq 6$. The mass density is quite sensitive to the self-interaction cross-section and the model with $\sigma/m = 0.05\cpm$ predicts about two orders of magnitude lower mass density at $z\simeq 6$. On the other hand, baryonic accretion has little impact on the SMBH mass density in our model. Quasar surveys indicate that the integrated quasar luminosity density matches the local SMBH mass density \cite{Shen:2020obl} (at $0.5\,\mathrm{dex}$ level, assuming $\epsilon_{\rm r} = 0.1$). Therefore, the SMBH mass density at high redshift has to be significantly lower than the local value in order to be consistent with the observation of quasar luminosity functions, unless $\epsilon_{\rm r}$ is larger (i.e. SMBHs are rapidly rotating). The model with $\sigma/m = 0.1\cpm$ is thus potentially in tension with the observations while the model with  $\sigma/m = 0.05\cpm$ is still consistent with observations. Meanwhile, since the mass growth of the seeds is dominated by accretion of dissipative dark matter rather than baryonic matter, our model predicts that the integrated luminosity density of quasars (which reflects baryonic accretion) at high redshift will be significantly smaller than the change in SMBH mass density at high redshift. Future surveys of high redshift quasars with the next-generation instruments, such as the Nancy Grace Roman Space Telescope, the Rubin Observatory Legacy Survey of Space and Time (LSST), and the James Webb Space Telescope (JWST), may be able to further test our seeding mechanism. 

As mentioned above, the high redshift predictions and the comparisons with observations presented in this section are affected by many astrophysical uncertainties. These come from both modelling the seeding/growth of SMBHs and connecting them to the observed quasars. We have compared the predictions with the observationally inferred bolometric quasar luminosity functions at $z=6$, where bolometric luminosities are affected by uncertainties in bolometric and extinction corrections (see discussions in \cite{Shen:2020obl}). Towards higher redshift, the measurements of quasar luminosity functions have been limited by the survey volume with respect to the vastly decreasing quasar number density (e.g. \cite{Fiore2012, Jiang2016, Matsuoka2018, Wang2019b, Shen:2020obl}). Meanwhile, in modelling the seeding, we have ignored the baryon content of early galaxies. If the baryons have a non-neglibible contribution to the central gravitational potential, the collapse of halo into compact objects was shown to be accelerated \cite{feng2020seeding}. However, the bursty star formation and feedback from the condensed baryon matter could compete with dissipative collapse of dark matter (e.g. \cite{Governato2010,Pontzen2012,Madau2014}). Moreover, in modelling SMBH growth, we have adopted scaling relations in connecting SMBH growth rates to host galaxy bulge properties and host halo properties, while these relations are largely based on low redshift observations. Finally, the largest uncertainty comes from the fuelling model to connect SMBHs to observed quasars, for which there is limited observational constraints even at moderate redshift. We essentially allow the ERDF and the quasar duty-cycle as free parameterized inputs. We expect that none of these uncertainties will likely overturn the general viability of the tdSIDM model, but improved constraints on the astrophysical inputs will certainly help pin point the working tdSIDM parameters more precisely. This will be explored in follow-up studies.

\section{Consistency with low redshift SMBHs}\label{sec:conssistency_check}

There are two branches of halos that are most likely to host SMBH seeds, as illustrated in Fig. \ref{fig:massz}. The first branch consists of rare, massive halos at high redshift that can seed SMBHs shortly after they formed. These rare halos typically have low concentrations (usually below $4$ as shown by most halo mass-concentration relations {\em e.g.} \cite{Dutton_2014,Diemer_2019}). However, the central dark matter density in these halos is still very high since they form at unusually high redshift, as indicated by Eq.~(\ref{eq:collapse_time}), leading to efficient SMBH formation. For this branch, according to Eq.~(\ref{eq:collapse_time}), $t_{\rm col}$ depends on redshift as $\rho^{-7/6}_{\rm crit}(z) \sim (1+z)^{-7/2}$, when $M_{\rm vir}$ and $c_{\rm vir}$ are fixed. At high redshift, when the dark energy is subdominant to matter, $t_{\rm H}$ depends on redshift as $(1+z)^{-3/2}$. Therefore, the ratio $t_{\rm diss}/t_{\rm H}$ of this branch has a simple redshift dependence as $(1+z)^{-2}$, indicating that the seeding is more likely to happen at earlier times assuming a fixed halo mass and concentration. The second branch consists of normal mass halos at low redshift with early assembly times, in which SMBH seeds do not form immediately but when they evolve to low redshift. These halos inherit high central dark matter densities at formation, which manifests as high halo concentrations after they accrete matter at late times. By first order approximation, the central densities of such halos are roughly constant after the majority of their mass is assembled, and the redshift evolution of the ratio $t_{\rm diss}/t_{\rm H}$ is dominated by the evolution of the $t_{\rm H}$ term assuming a fixed halo mass, which approximately gives a redshift dependence $(1+z)^{3/2}$. This indicates that this branch of halos will more likely seed SMBHs at low redshift.

\begin{figure*}
    \centering
    \includegraphics[width = 0.49\textwidth]{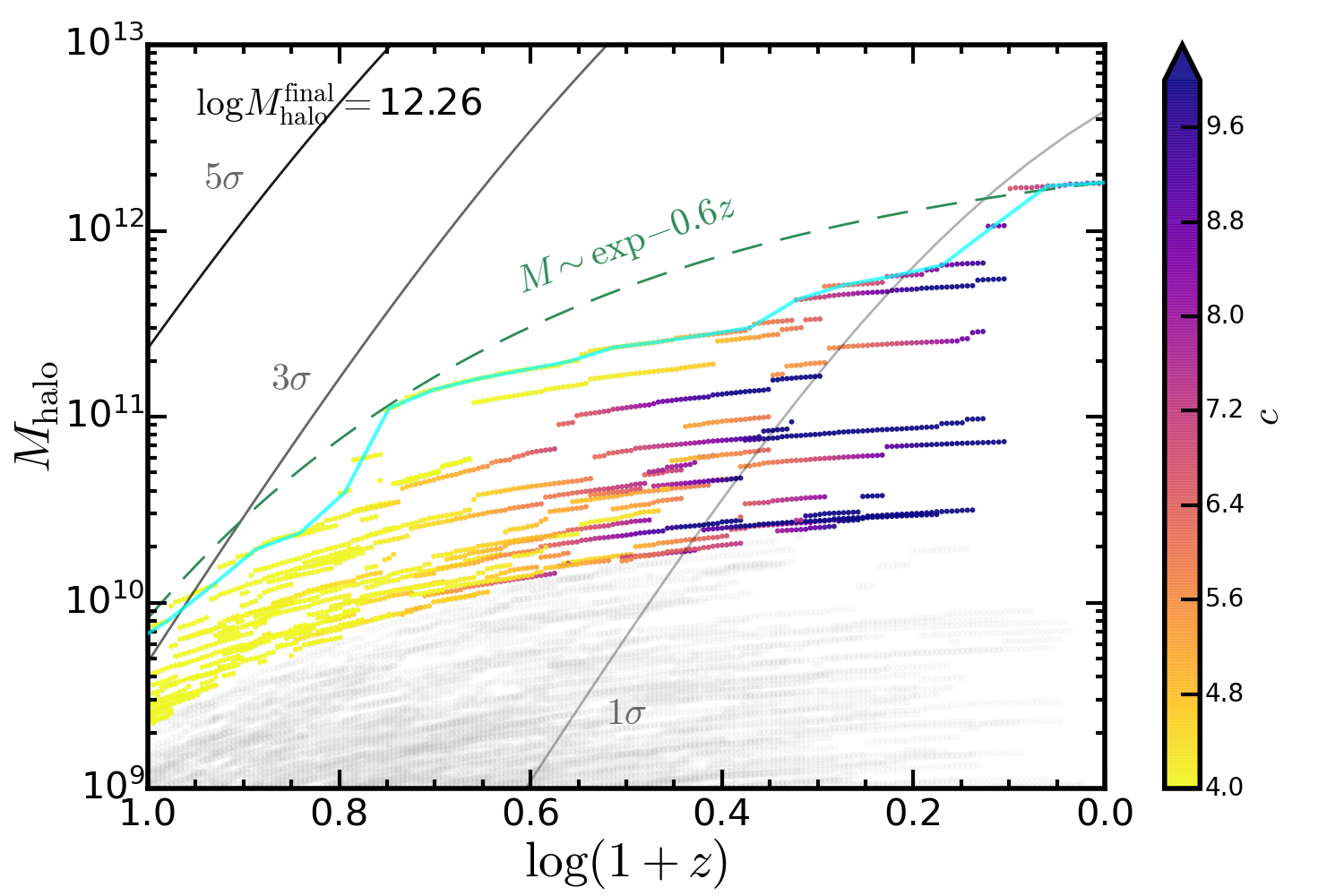}
    \includegraphics[width = 0.49\textwidth]{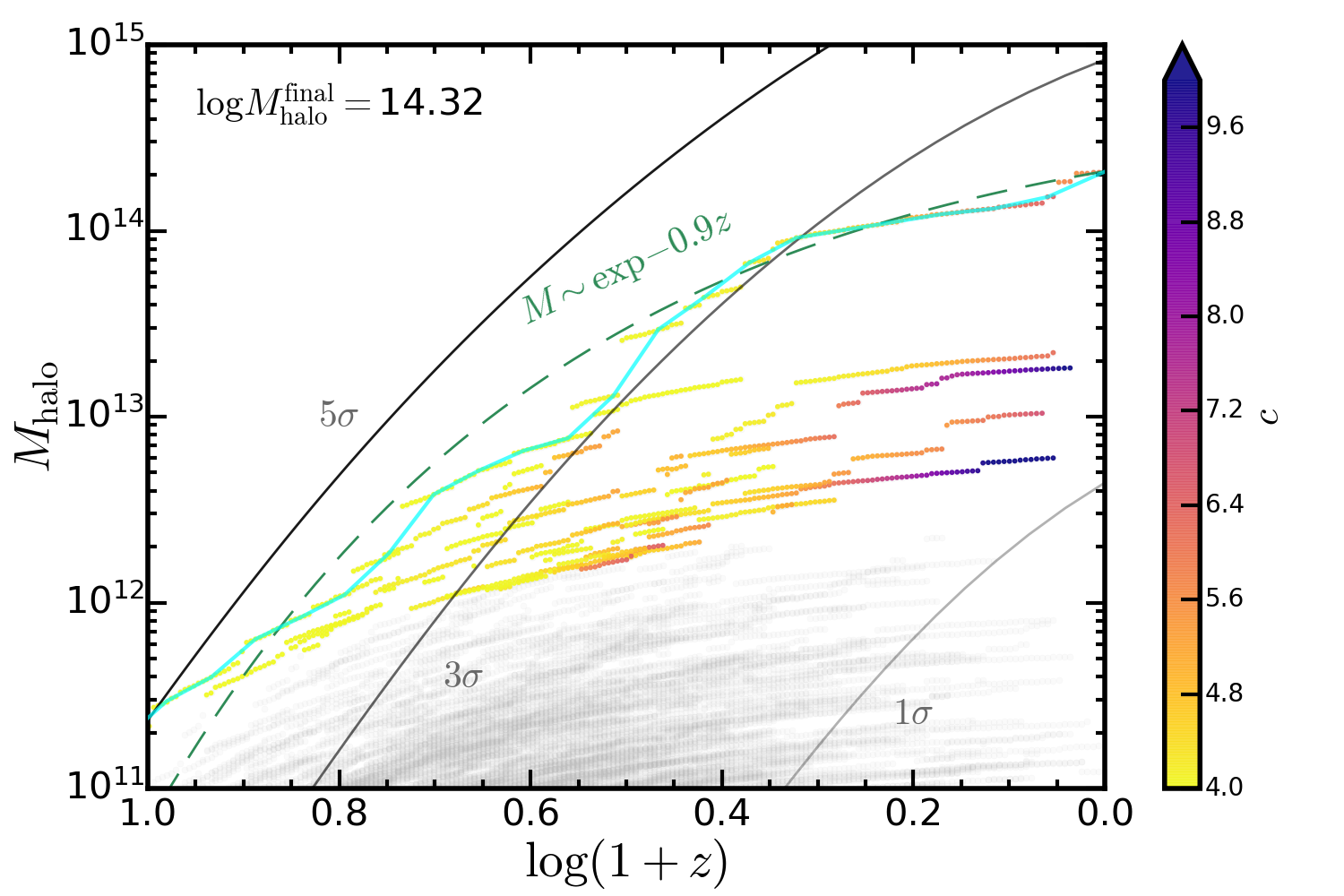}
    \includegraphics[width = 0.49\textwidth]{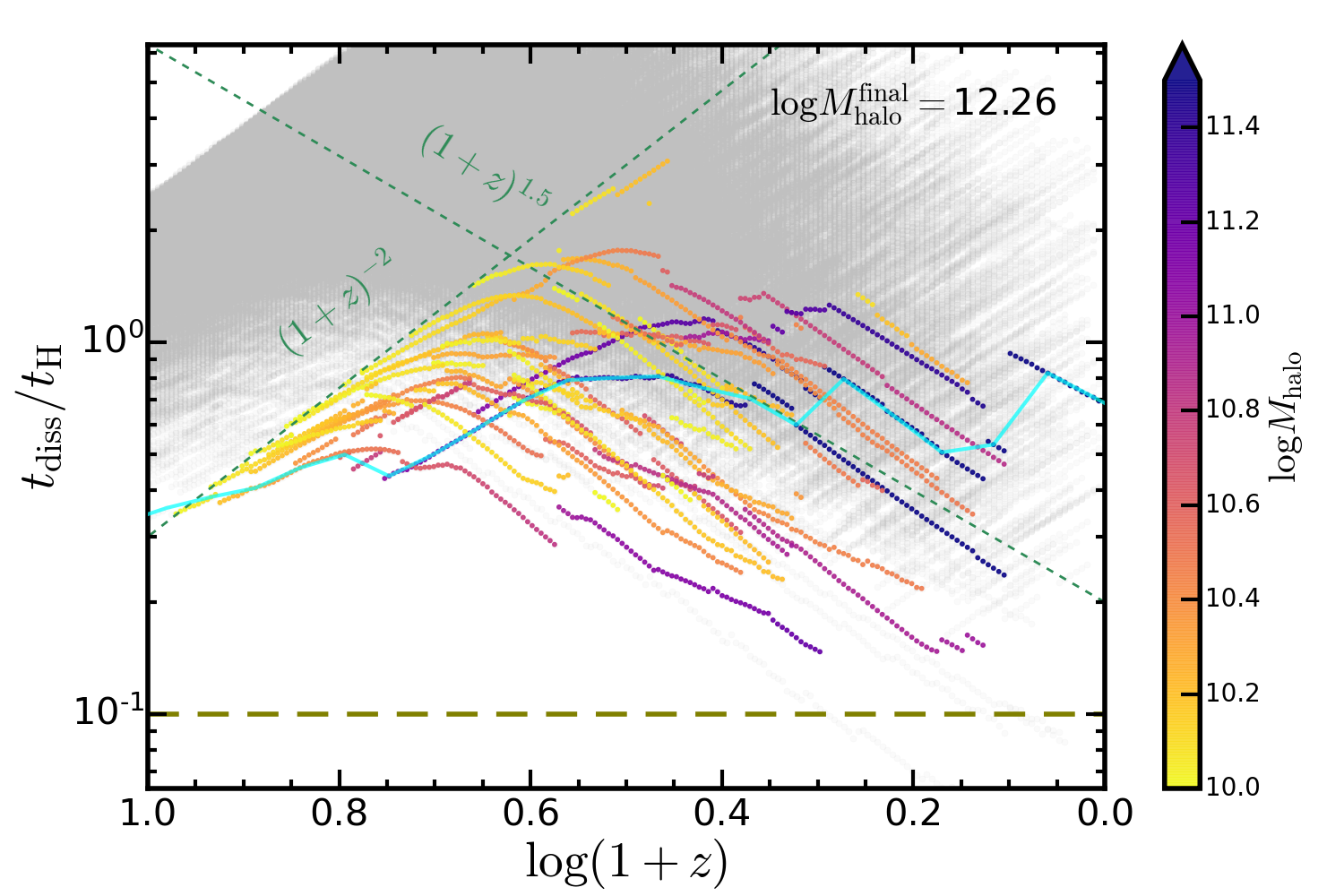}
    \includegraphics[width = 0.49\textwidth]{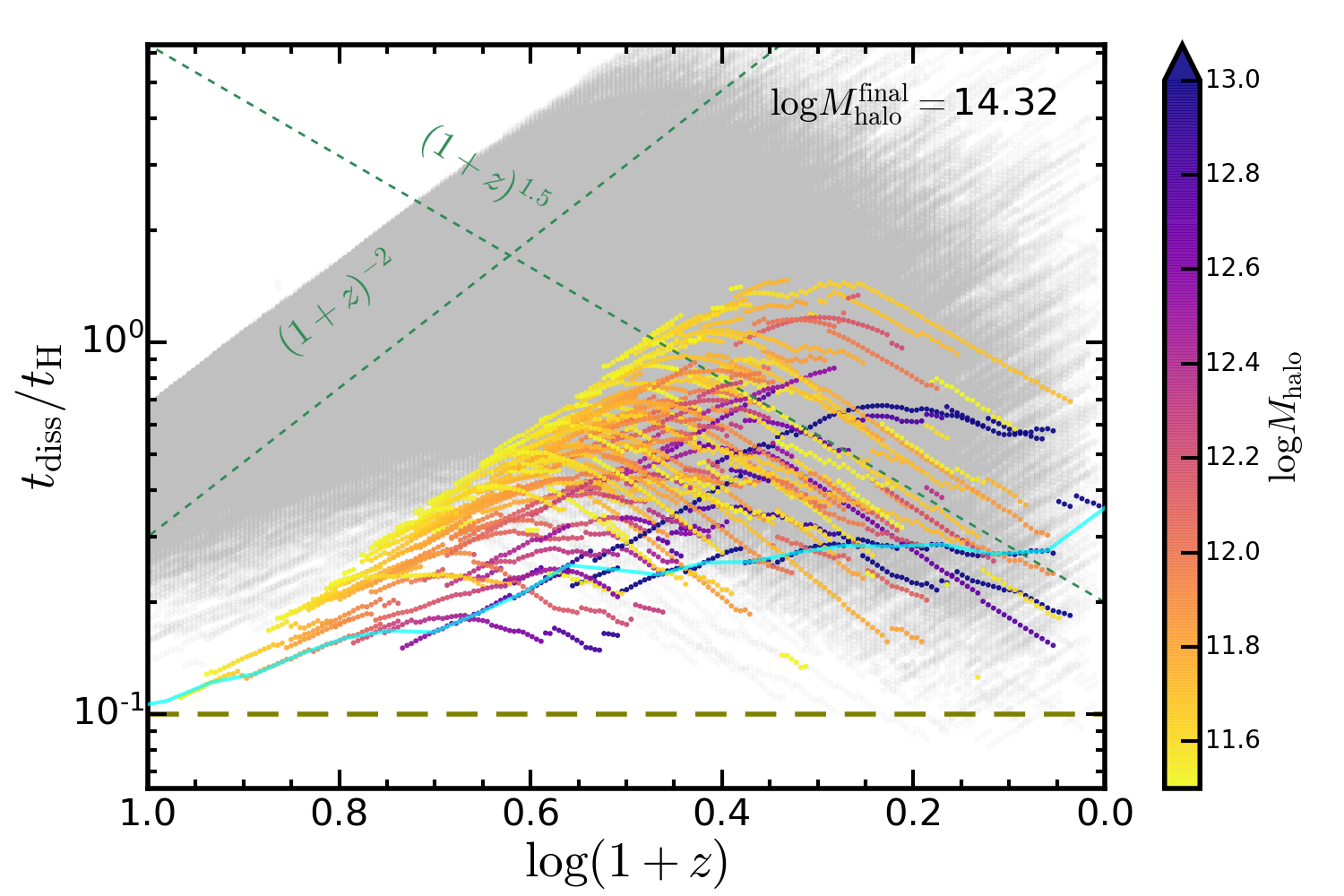}
    \caption{ {\it Top}: \textbf{Mass growth history of halo progenitors.} The growth tracks of relatively massive progenitors are color coded by their concentrations. Low mass progenitors are shown by the gray cloud. The main progenitor is indicated by the cyan solid line. The green dashed line shows an analytic model for the main progenitor mass growth history \cite{Wechsler_2002}. The gray solid lines show the mass of the halo corresponding to a certain rareness of fluctuations. {\it Bottom}: \textbf{Ratio $t_{\rm diss}/t_{\rm H}$ of halo progenitors versus redshift.} The cross-section $(\sigma/m)=0.05\cpm$ is assumed here. Progenitors that are more massive than $10^{10}\msun$ are color coded by their halo masses. The labelling is the same as the top row. The green dashed lines show analytic expectations for the timescales of the low and high redshift branches (as discussed in the main text). The horizontal dashed line indicates the threshold where SMBH seeding will occur assuming $(\sigma/m)=0.05\cpm$ and $\epsilon=0.1$. }
    \label{fig:tree-lowz}
\end{figure*}

The model predicts the SMBH-to-halo mass ratio to be $3\times 10^{-3}$ at seeding, which is apparently much larger than that of local SMBHs in observations. Therefore, the second branch must be checked for the formation of overly massive SMBHs at low redshift. The distinct halos in the Local Universe (with halo masses as large as $10^{14-15}\msun$) with median concentration could evade the seeding criterion at low redshift and avoid hosting an overly massive SMBH. However, given the strong dependence of the collapse timescale on halo concentration, a highly concentrated progenitor (assembled early in cosmic time) could still seed an SMBH, which is later merged into the main progenitor. To investigate the SMBH seeds formed in this scenario and check the consistency of the model with local SMBHs, we generate a second set of merger trees, sampling $5$ halos of mass $\sim 10^{12}\msun$ and $5$ halos of mass $\sim 10^{14}\msun$ at $z=0$, which correspond to the Milky Way mass and cluster mass galaxies in the Local Universe. The mass resolution and highest redshift they trace are the same as the first set. We explicitly track the mass growth history of all the progenitors in these trees to $z=0$ and check if they are able to host SMBH seeds. We will show that the model with small cross-sections ($\sigma\lesssim 0.1\rm cm^2/g$) can stay consistent with low redshift observations while explaining the massive high redshift SMBHs. 

In the top row of Fig.~\ref{fig:tree-lowz}, we show the mass growth history of the progenitors of a Milky Way mass halo (left) and a cluster mass halo (right). The evolution tracks of progenitors are color coded by their halo concentration and end when the progenitors merge. The mass growth history of the main progenitor is well described by the analytical model $M\propto e^{-\alpha z}$ \cite{Wechsler_2002}, with the $\alpha$ values consistent with the ones found therein for both halos. Apparently, for both halos, there exists a population of halos with early assembly times and with limited mass accretion at late times. These halos become much more concentrated than expected from a median mass-concentration relation. Such halos are more abundant in the Milky Way mass halo than in the cluster mass halo, due to the later assembly time of the cluster mass halo and its progenitor (i.e. larger $\alpha$ values). In the bottom row of Fig.~\ref{fig:tree-lowz}, we show the ratio $t_{\rm diss}/t_{\rm H}$ of halo progenitors as a function of redshift and compare it to the seeding threshold (assuming the fiducial choice of cross-section $\sigma/m = 0.05\cpm$) indicated by the dashed line. It is obvious that there are two branches of halos that are close to the seeding threshold, with the redshift dependence of the timescale as expected from the analytic estimations above. For a Milky Way mass halo, the low redshift branch is closer to the seeding threshold. These highly concentrated, massive progenitors have their central mass densities almost preserved towards low redshift before they merge into the main progenitor and the dissipation timescale is almost a constant in these halos. We note that in most of the Milky Way mass halo merger trees, under the choice of cross-section here, no progenitor can cross the seeding threshold. Occasionally, as indicated by the example in Fig.~\ref{fig:tree-lowz}), a low-mass progenitor could cross the seeding threshold, but the mass of the SMBH seed formed and its statistics are still compatible with the observed local SMBHs. Such a low-mass seed (compared to the main progenitor) may take too long time to sink to the halo center under dynamical friction to cause any real issues ({\em e.g.} \cite{Ma2021}). Whether this branch of SMBHs can fully explain the local SMBH populations requires more careful modelling of the late time evolution of SMBHs and galaxies, which is beyond the scope of this paper. For the cluster mass halo, the high redshift branch is closer to the seeding threshold. Although the entire population of progenitors is closer to the seeding threshold, the low redshift branch stays at roughly the same position as those in the Milky Way mass halo, primarily due to the late formation times and low halo concentrations. Again, the low redshift branch can hardly cross the seeding threshold. The high redshift branch in the cluster mass halo is at the edge of the seeding threshold such that seeding is most likely to happen in the main progenitor. It is expected that, for more massive halos, the seeding will continue favoring the high redshift branch and eventually SMBH seeds may form in the main progenitor at high redshift. This is exactly the SMBH population discussed in previous sections.

The discussion here demonstrates that there exists a parameter space of dissipative dark matter where the predictions are consistent with observations at both high and low redshift. The model with $\sigma/m = 0.05\cpm$ (or $\epsilon \sigma/m = 0.005\cpm$ if we make $\epsilon$ free) can give rise to the correct abundance of luminous quasars at high redshift while not producing overly massive SMBH in the low redshift Universe. Note again the seeding criterion depends on the product of $\sigma/m$ and $\epsilon$, so uncertainty in $\epsilon$ is degenerate with uncertainty in $\sigma/m$. If a generic dSIDM model with a constant dissipation fraction $f$ is considered, $f$ is degenerate with the cross-section in the seeding criterion and the relevant parameter is $f \epsilon \sigma/m$. 

\section{Observational constraints of tdSIDM}

Most of the observational constraints for SIDM come from studies of the elastic case, with the stringent ones $(\sigma /m) \lesssim 0.3-1 \cpm$ from merging galaxy clusters (e.g. \cite{Randall_2008,Kahlhoefer2015,Harvey2015,Wittman2018}). The tdSIDM models with $(\sigma /m) \sim 0.05-0.1 \cpm$ considered in this work are consistent with these constraints, although it is not clear whether dissipation will create any distinct signatures at cluster scale compared to the elastic case. 

Specifically, at dwarf scale, dSIDM has been considered in some recent studies. For instance, dSIDM with a constant energy dissipation per collision, $E_{\rm loss}\equiv m \nu^{2}_{\rm loss}$, has been studied in \cite{Essig2019} through semi-analytic modelling of dwarf galaxies. The central densities of dwarf galaxies was found to be significantly enhanced by dissipation--accelerated gravothermal collapse, which confronted with the observed local dwarfs led to constraints on dSIDM. The constraints they derived can be roughly translated to our model when the constant energy loss is comparable to the kinetic energy of dark matter particles. It is roughly equivalent to the $f\sim 0.5$ case, if $E_{\rm loss} \sim E^{\rm rel}_{\rm k} \sim m \langle v^{2}_{\rm rel} \rangle / 4$, which is equivalent to $\nu_{\rm loss} \sim 2/\sqrt{\pi}\sigma_{\rm v}$ assuming the Maxwell-Boltzmann velocity distribution. Considering the typical one-dimensional velocity dispersion of the dwarfs they used, we get $f (\sigma/m) \lesssim 0.15\cpm$ approximately from their constraints. On the other hand, the dSIDM model with fractional energy dissipation, which is of the same family as the tdSIDM model, has been studied in \cite{Shen2021} via hydrodynamical simulations of galaxies. Assuming a lower disspation fraction of $f=0.5$, they found that dSIDM with $(\sigma/m) \gtrsim 0.1\cpm$ could lead to cuspy and power-law like central density profiles of dwarf galaxies at sub-kpc scale. The cuspy profiles are potentially in tension with the kinematic and rotation curve measurements of Local dwarf galaxies (this aspect is expected be analyzed in more detail in the follow-up work Shen {\it et al.} [in prep., 2021]). Further investigations are required to consolidate these constraints. Nevertheless, the favored tdSIDM models in this work are still consistent with these low-redshift studies. 

In addition, dissipative dark matter has potential impacts on halo substructures and corresponding strong lensing signals, which remains an appealing aspect to explore. The condensation of dSIDM has implications in explaining the excess of small-scale gravitational lenses recently found in galaxy clusters \cite{Meneghetti2020} as well as the unexpected concentration of some substructures \cite{Minor2020}. 

\section{Conclusions}

In this paper, we have studied a mechanism to seed high redshift SMBHs via the collapse of totally dissipative self-interacting dark matter (tdSIDM) halos, where the dark matter particle loses nearly all its kinetic energy during a single collision. The study is motivated by the existence of billion solar mass SMBHs observed in the early Universe ($z\gtrsim 6$), which are in tension with canonical seeding mechanisms. We develop an analytical model for the collapse criteria and timescale of tdSIDM halos, calibrated based on numerical N-body simulations of isolated halos, and then apply this model to Monte-Carlo halo merger trees to make predictions of SMBHs and observed quasars in the cosmological context. Our findings can be summarized as:

\begin{itemize}
    \item  We have performed N-body simulations of isolated, rare halos at high redshift initialized with the Navarro–Frenk–White (NFW) profile, with the inclusion of dissipative dark matter self-interactions. We find that a constant fraction, $f_{\rm col}\simeq 3\times 10^{-3}$, of the halo mass will eventually collapse to the scale below the spatial resolution of the simulations. Surprisingly, the collapsed fraction is insensitive to the mass, size, spin and redshift of the sampled halo. An analytic description of the collapse criteria and timescale is developed and calibrated based on these simulations. This analytic prescription can be applied to halos with various masses, concentrations, formation redshifts and in different cosmological models. 
    
    \item The unique feature of our seeding mechanism is the rapid formation of SMBHs seeds with an SMBH-to-halo mass ratio of $\sim 3\times 10^{-3}$. The SMBHs directly seeded from the catastrophic collapse of tdSIDM halos are massive enough to explain the high mass end of SMBHs at $z\gtrsim6$. The rapid formation of SMBHs in our model implies the existence of very young quasars at high redshift, which is consistent with recent studies that attempt to measure the lifetimes of quasars \cite{Eilers:2020htq}. Such a young population of quasars is difficult to explain in standard scenarios where SMBHs have to live long enough to grow at some modest multiple of the Eddington limit from much smaller masses.
    
    \item  We trace the seeding and growth of SMBHs via halo merger trees and derive predictions for the cosmological abundance of SMBHs. With little tuning of the fueling model of SMBHs (the ERDF and the quasar duty-cycle), our model with $\sigma/m=0.05/0.1\cpm$ (or $\epsilon \sigma/m = 0.005/0.01\cpm$ if we make $\epsilon$ free) successfully reproduces the observed quasar luminosity functions at high redshift, particularly at the bright end. The tuned ERDF and duty-cycle imply that a significant fraction of SMBHs seeded in this way must have low quasar activity, which will hopefully be tested by future quasar surveys.
    
    \item SMBHs seeded directly from tdSIDM halos exhibit much larger SMBH-to-halo mass ratios than local SMBHs and lie systematically above the local $M_{\rm BH}-\sigma_{\rm v}^{\ast}$ relation. This feature is in better agreement with [C \rom{2}] gas velocity dispersion and host galaxy dynamical mass measured for high redshift massive quasars. 
    
    \item We compare the cosmic SMBH mass density predicted in our model to the observed SMBH mass density in the Local Universe. We find that the model with $\sigma/m=0.1\cpm$ (or $\epsilon \sigma/m = 0.01\cpm$) is potentially in tension with observations, since the mass density in this model approaches the local value already at $z\sim 6$, requiring large radiative efficiency to remain consistent with low redshift data. The model with $(\sigma/m)=0.05\cpm$ (or $\epsilon \sigma/m = 0.005\cpm$) is still compatible with observations. In addition, we find that the growth of SMBHs at high redshift is dominated by dissipative dark matter rather than baryonic matter, predicting that the integrated luminosity density of quasars (which reflects baryonic accretion) will be significantly smaller than the change in SMBH mass density at high redshift, which is a testable feature of our seeding mechanism. 
    
    \item While the large SMBH-to-halo mass ratio ($3\times 10^{-3}$) found in our N-body simulations can easily explain the most massive SMBHs at $z\gtrsim 6$, which are the most difficult to understand in the standard scenario, one must check with consistency at low redshift, particularly if halos with mass $M\gtrsim 10^{15}M_{\odot}$ also collapse to form overly massive SMBHs. We show this does not occur because the dissipation timescale sensitively depends on the halo central density, which is relatively low for those massive halos at $z\sim 0$. Therefore our seeding model based on dissipative self-interacting dark matter is capable of producing SMBHs that are challenging to explain in standard scenarios while remaining consistent with low redshift observations. Though this work focused on explaining the population of high redshift SMBHs, tdSIDM may also explain the origin of SMBHs in Milky Way mass halos. As shown in Fig.~\ref{fig:massz} and Fig.~\ref{fig:tree-lowz}, Milky Way mass halos may contain progenitors that are formed from rare fluctuations at high redshift. Such rare progenitors have a large central density and are more likely to collapse compared to other progenitors. 
    
\end{itemize}

Our model prefers a cross-section of $\sigma/m\sim0.05\rm cm^2/g$ (or $\epsilon \sigma/m \sim 0.005\cpm$) to explain the quasar luminosity function at high redshift while remaining consistent with low redshift observations. Such a model is testable in the future once the quasar luminosity function is measured at more redshifts. In the future, quasar surveys conducted with the Nancy Grace Roman Space Telescope, the Rubin Observatory Legacy Survey of Space and Time (LSST) and the James Webb Space Telescope (JWST) can further test our predictions of the quasar luminosity function and the density change of SMBHs at high redshift.

\acknowledgments
We thank Fangzhou Jiang for useful discussions regarding halo merger trees with the SatGen code. We thank Moira Gresham for useful discussions.
Numerical calculations were run on the Caltech compute cluster ``Wheeler,'' allocations FTA-Hopkins/AST20016 supported by the NSF and TACC, and NASA HEC SMD-16-7592. Support for PFH and XS was provided by NSF Research Grants 1911233 \&\ 20009234, NSF CAREER grant 1455342, NASA grants 80NSSC18K0562, HST-AR-15800.001-A.
HX is supported in part by the U.S. Department of Energy under grant number DE-SC0011637 and the Kenneth K. Young Chair in Physics. KZ is supported by the U.S. Department of Energy, Office of Science, Office of High Energy Physics, under Award Number DE-SC0021431 and a Simons Investigator award.

\appendix
\section{Convergence Testing}\label{append:convergence_test}
This appendix investigates whether our primary results for isolated NFW halos are sensitive to our choice of gravitational softening
length. The worry is that the physics of SMBH formation is significantly different from structure formation, and simulations with different gravitational softening lengths may lead to very different results. We compare our fiducial run to a simulation with different particle number and gravitational softening length and show that the central regions of dark matter halos still collapse at the same timescale.

\begin{figure}[h!]
\centering
\includegraphics[width=0.7\textwidth]{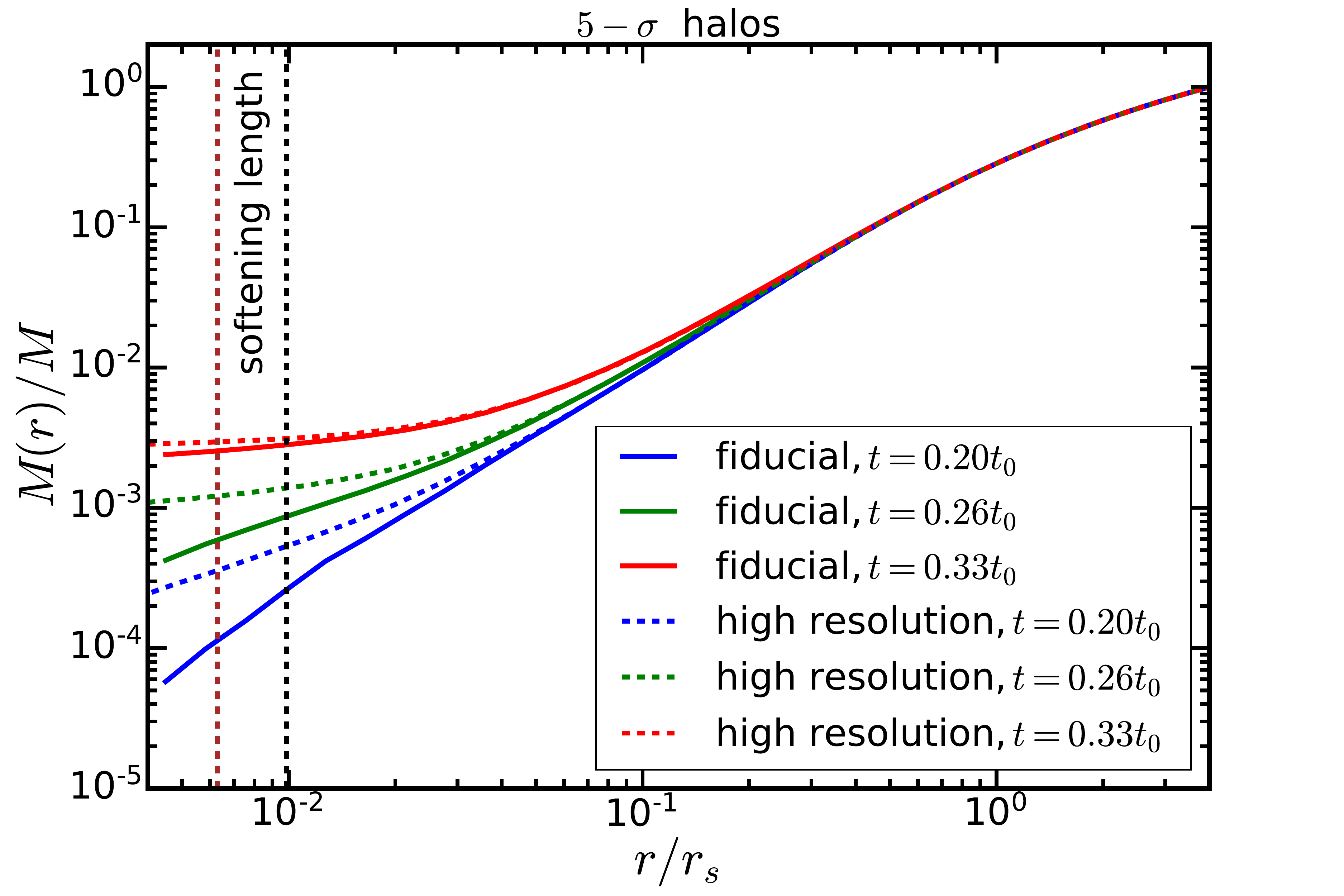}
\caption{Enclosed mass profile of two $5-\sigma$ halos at $z=10$ with the same NFW parameters but different simulation parameters. Solid curves represent the mass profile at different times for the fiducial run, while the dashed curves are for a run with a factor of 4 improved resolution (improving both mass and force resolution accordingly). The vertical dashed lines indicate the gravitational force softening length for both the fiducial run and the high resolution run. }
\label{fig:convergence_test}
\end{figure}
In a cosmological N-body simulation, the gravitational softening length is often taken to be the $d/30$, where $d$ is the particle mean separation in the simulation box. However, our simulations with isolated NFW halos are different from a cosmological simulation. Our focus in this work is the SMBH formation process at the halo center. Therefore, we are more interested in the particle separation length in the region where catastrophic collapse happens. We take our gravitational softening length to be $2d_0$ where $d_0$ is the particle mean separation within radius $0.07 r_{\rm s}$ at the beginning of the simulation. 

In our fiducial run for various halo masses, the particle number is chosen to be $6\times 10^6$ and the simulation box size is fixed to be 1000~pc. The fiducial run simulated the collapse of $5-\sigma$ halos from $z=4-10$. To test for convergence, we select a $5-\sigma$ halo at $z=10$ with NFW parameter $r_{\rm s}=3.3$ kpc and $\rho_0=0.316\,M_{\odot}\rm pc^{-3}$ with gravitational softening length 0.033 kpc. We then run another simulation with improved mass resolution and correspondingly improved force resolution. The particle number in the new run is taken to be $2.4\times 10^7$ and the gravitational softening length is still $2d_0$, corresponding to 0.021~kpc.

As shown in Fig.~\ref{fig:convergence_test}, the enclosed mass profiles $M(r)/M$ converge very well when the time approaches the collapse time $0.35 t_0$. Even though the simulation with an improved resolution has a larger $M(r)/M$ for small $r$ before the catastrophic collapse, their final predictions for the collapse timescale and the SMBH-to-halo mass ratio do converge. Therefore, we conclude that our fiducial simulations reliably predict the collapse timescale to form an SMBH seed and the SMBH-to-halo mass ratio. 

\section{Considerations in Centrifugal Barrier and Fragmentation}\label{append:barrier_fragmentation}
The goal of this Appendix is to demonstrate that the halo angular momentum is not an important consideration for SMBH seeding with tdSIDM, justifying the neglect of angular momentum in the bulk of the analysis.

\subsection{Centrifugal Barrier}

The collapse of a realistic halo with non-zero spin may be halted by the centrifugal barrier~\cite[e.g.,][]{Mo:1998}. The scale of the centrifugal barrier ($\sim \lambda R_{\rm vir}$) is much larger than the physical scale of SMBH seed formation. 

Similar to the seeding mechanism in pristine gas disks~\cite{Bromm2003,Koushiappas2004,Begelman2006,Lodato:2006,Ferrara2014,Pacucci2015,Valiante2016}, we first note that the non-axial-symmetric structures originating from global gravitational instability transfer angular momentum outward and enable further collapse of the halo. As the halo center becomes denser, instability builds, triggering a further collapse of the halo. Run-away collapse to compact objects is realized in this way, even when there is no microscopic physical mechanism to transfer angular momentum outward. Following \cite{Koushiappas2004,Begelman2006,Lodato:2006}, we consider the configuration of the system as a spherical isothermal dark matter halo of virial mass $M_{\rm vir}$, with a constant circular velocity $V_{\rm c}$ and some of the dark matter condensed to a thick dark disk having mass $m_{\rm d}M_{\rm vir}$. The surface density of the dark disk is assumed to be
\begin{equation}
    \Sigma(r) = \Sigma_{\rm 0} e^{-r/R_{\rm d}},
\end{equation}
where $\Sigma_{\rm 0}$ is the normalization of the surface density, $R_{\rm d}$ is the scale length of the disk. Note that the qualitative conclusion is not sensitive to the density profile assumed here. The instability of the dark disk is evaluated by the ``Toomre Q'' parameter~\cite{Toomre:1964}, defined as
\begin{equation}
    Q = \dfrac{c_{\rm s} \kappa }{\pi G \Sigma} = \sqrt{2}\dfrac{\sigma_{\rm v} V_{\rm c}}{\pi G \Sigma_{\rm 0} R_{\rm d}},
    \label{eq:ToomreQ}
\end{equation}
where we have replaced the sound speed $c_{\rm s}$ with the one-dimensional velocity dispersion of dark matter in the dark disk $\sigma_{\rm v}$, $\kappa = \sqrt{2}V_{\rm c}/R_{\rm d}$ is the epicyclic frequency and we use $\Sigma_{\rm 0}$ and $R_{\rm d}$ as a representative surface density and disk scale. The disk is considered unstable when $Q$ drops below a critical value $Q_{\rm c}$ of order unity. Since the spherical halo plus dark disk we consider here is only a crude approximation of the dissipative dark matter configuration, the detailed value of $Q_{\rm c}$ is uncertain and is left as a free parameter.

If we assume that some mass, $m_{\rm a}M$, is accreted at the center of the halo and the remaining mass in the disk is $(m_{\rm d}-m_{\rm a})M$, $\Sigma_{\rm 0}$ and $R_{\rm d}$ are related with
\begin{equation}
    (m_{\rm d}-m_{\rm a})M = 2\pi \Sigma_{\rm 0} R_{\rm d}^2.
    \label{eq:Sigma-Rd}
\end{equation}
We assume that the dark disk has angular momentum $J_{\rm d}=j_{\rm d} J$, where $J$ is the total angular momentum of the halo. $J$ is related to the spin parameter $\lambda$ of the halo \cite{Mo:1998}
\begin{equation}
    J = \dfrac{\lambda G M^{5/2}}{|E|^{1/2}} = \sqrt{2} \dfrac{\lambda G M^{2}}{V_{\rm c}},
    \label{eq:Jd1}
\end{equation}
where $E$ is the total energy of the halo, and we have assumed that the halo takes an isothermal distribution of matter (circular velocity is a constant). Taking the condensed dark disk to have the same circular velocity as the halo, we obtain
\begin{align}
    J_{\rm d} &= \int V_{\rm c} \Sigma_{\rm 0} e^{-r/R_{\rm d}} (2\pi r) r {\rm d}r \nonumber \\
    & = 4\pi V_{\rm c} \Sigma_{\rm 0} R_{\rm d}^3 \nonumber \\
    & = 2 (m_{\rm d}-m_{\rm a}) M R_{\rm d} V_{\rm c}.
    \label{eq:Jd2}
\end{align}
Combining Eqs.~(\ref{eq:Jd1}) and (\ref{eq:Jd2}), we obtain the disk scale length as
\begin{equation}
    R_{\rm d} = \dfrac{1}{\sqrt{2}} \lambda \Big( \dfrac{j_{\rm d}}{m_{\rm d}} \Big) \Big( \dfrac{1}{1 - m_{\rm a}/m_{\rm d}} \Big) \dfrac{GM}{V_{\rm c}^2}. 
\end{equation}
Inserting this into Eq. (\ref{eq:Sigma-Rd}), we obtain $\Sigma_{\rm 0}$, and further substituting into Eq.~(\ref{eq:ToomreQ}) gives
\begin{equation}
    Q = \dfrac{2\lambda}{m_{\rm d}} \Big( \dfrac{j_{\rm d}}{m_{\rm d}} \Big) \dfrac{1}{(1 - m_{\rm a}/m_{\rm d})^2} \dfrac{\sigma_{\rm v}}{V_{\rm c}}.
\end{equation}
At the end of accretion and collapse, the configuration of the system is marginally stable, so that the accreted/collapsed mass $m_{\rm a}$ can be derived replacing $Q$ with $Q_{\rm c}$:
\begin{equation}
    \dfrac{m_{\rm a}}{m_{\rm d}} = 1 - \sqrt{ \dfrac{2\lambda}{m_{\rm d} Q_{\rm c}} \Big( \dfrac{j_{\rm d}}{m_{\rm d}} \Big) \Big( \dfrac{\sigma_{\rm v}}{V_{\rm c}} \Big) }.
\end{equation}
If we neglect angular momentum transfer and the dark disk is formed adiabatically, $j_{\rm d}/m_{\rm d}$ should be $1$. In the absence of halo spin, the final SMBH seed mass is $m_{\rm d}M$, so we replace $m_{\rm d}$ with the collapse fraction $f_{\rm col}$ of a zero-spin halo. Finally, since $m_{\rm a}/m_{\rm d}$ cannot exceed unity, we obtain the instability criterion that collapse only occurs when
\begin{equation}
    \lambda < \lambda_{\rm max} = \dfrac{Q_{\rm c} f_{\rm col}}{2} \dfrac{V_{\rm c}}{\sigma_{\rm v}}.
\end{equation}
The corresponding SMBH seed mass fraction is therefore
\begin{equation}
    f = f_{\rm col} \Bigg(   1 - \sqrt{ \dfrac{2\lambda}{f_{\rm col} Q_{\rm c}} \Big( \dfrac{\sigma_{\rm v}}{V_{\rm c}} \Big) }  \, \Bigg).
\end{equation}
If we approximate $\sigma_{\rm v}$ as $\sigma_{\rm v}(0.07 r_{\rm s})$ of an NFW halo given by Eq.~(\ref{eq:velocity-dispersion}), and calculate $V_{\rm c}$ with Eq.~(\ref{eq:Vc-Mhalo}), the ratio $V_{\rm c}/\sigma_{\rm v}$ will be a constant $\sqrt{f(2.15)/2.15/F(0.07)} \simeq 1.9$, and the angular momentum barrier for seed formation will thus be independent of halo mass. Under these assumptions, we obtain $\lambda_{\rm max}\simeq 0.003$ when $Q_{\rm c}=1$. However, in simulations, we have found that halos with much larger spin parameters still collapse under dissipative dark matter self-interactions at a similar collapse timescale.

An alternative to the picture discussed above is angular momentum transfer through microscopic physical processes. In our case, the viscosity from dark matter self-interactions  transports angular momentum through the dark disk. The viscosity of SIDM in the long-mean-free-path regime~\footnote{The mean free path of dark matter particles is much longer than the gravitational scale height of the system. For the model studied in this paper with $\sigma/m \lesssim 0.1 \cpm$, the requirement is satisfied.} can be written as
\begin{align}
    \eta & = C \rho \dfrac{H^2}{t_{\rm r}} \nonumber \\
    & \simeq \dfrac{\rho (\sigma/m) \sigma_{\rm v}^{3}}{4\pi G},
    \label{eq:viscosity}
\end{align}
where $C$ is a numerical constant of order unity, and $H=\sqrt{\sigma^2_{\rm v}/4\pi G \rho}$ is the gravitational scale height. Similar to the theory of accretion disks, the typical timescale for angular momentum to be transported over a length scale $L$ is
\begin{equation}
    t_{\rm v} = \dfrac{\rho L^2}{\eta} = \dfrac{4\pi G L^{2}}{(\sigma/m) \sigma_{\rm v}^3},
\end{equation}
where we have used Eq. (\ref{eq:viscosity}) in the second line. If we assume that the typical length scale $L$ for angular momentum transport is the collapse radius $\sim 0.07 r_{\rm s}$ found in our simulations, and approximate $\sigma_{\rm v}$ with $\sigma_{\rm v}(0.07 r_{\rm s})$ of a NFW halo given by Eq. (\ref{eq:velocity-dispersion}), we obtain
\begin{equation}
    t_{\rm v} = \dfrac{0.07^{2}}{(\sigma/m)F^{3/2}(0.07)} \dfrac{1}{\sqrt{4\pi G \rho_{0}^{3} r^{2}_{\rm s}}}.
\end{equation}
The viscous timescale has exactly the same scaling behavior as the dissipation timescale in Eq.~(\ref{eq:dissipation_timescale}). The ratio between them can be estimated as
\begin{align}
    \dfrac{t_{\rm v}}{t_{\rm diss}} & \simeq \dfrac{t_{\rm v}}{1/\rho(0.07 r_{\rm s}) (\sigma/m) \sigma_{\rm v}(0.07 r_{\rm s})} \nonumber \\
    & = \dfrac{0.07}{(1+0.07) F(0.07)} \sim O(1).
    \label{eq:tv_to_tdiss}
\end{align}
This suggests that the viscous timescale is comparable to the dissipation timescale. In this case, angular momentum is transported efficiently, and the central collapse mimics the zero spin case. This is the reason why we do not observe the effect of the centrifugal barrier in the simulations. 

\subsection{Fragmentation limit}
Another criterion is that the dissipation timescale remains larger than the dynamical timescale at the center of the halos, such that local fragmentation does not occur, preventing the formation of a single SMBH seed \cite{2006MNRAS.371.1813L}. If fragmentation does occur, the concentration of the largest amount of mass in the center will be suppressed and small clumps will form instead.
The dynamical time within a collapse radius is defined as $t_{\rm dyn}=1/\sqrt{4\pi G \rho_{\rm col}}$, where $\rho_{\rm col}$ is the average density of dark matter halo within collapse radius $0.07r_{\rm s}$. We find the ratio of dissipation time to dynamical time is
\begin{align}
\frac{t_{\rm diss}}{t_{\rm dyn}} & \approx 1.8 \left(\frac{4}{c}\right)^{2}\left(\frac{10}{1+z}\right)^{2}\left[{\rm ln}(1+c) - \frac{c}{1+c}\right] \nonumber \\
& \times\left(\frac{\rm 1 cm^2/g}{\sigma/m}\right)
\left(\frac{10^{12}M_{\odot}}{M}\right)^{1/3}.
\end{align}
For rare halos that can seed SMBHs, the dissipation timescale is always larger than the dynamical time when $\sigma/m\lesssim{\rm 1 cm^2/g}$. As we show in Sec.~\ref{sec:cosmological_evolution} and Sec.~\ref{sec:conssistency_check}, the preferred cross-section for seeding SMBHs at high redshift, while maintaining consistency with low redshift observations, is $\sigma/m=0.05\rm cm^2/g$. In such cases, the dissipation time scale is always an order of magnitude larger than the dynamical timescale, preventing the fragmentation of the dark matter halo.

\begin{figure}[h!]
\centering
\includegraphics[width=0.7\textwidth]{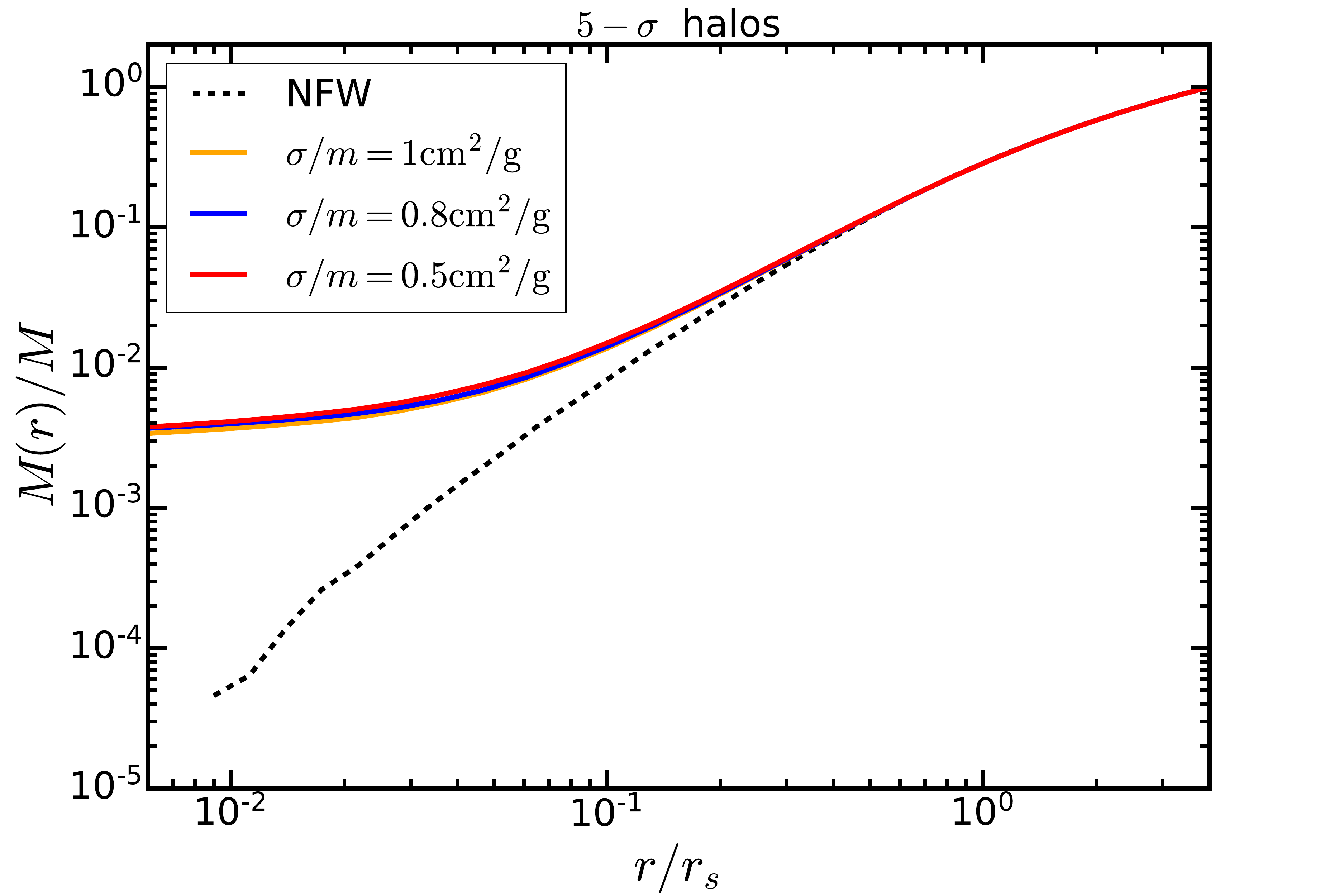}
\caption{Enclosed mass profile of three collapsed $5-\sigma$ halos at $z=10$ with the same NFW profile parameters. The collapse of those halos is simulated with different cross-sections but the same mass and force resolution. The enclosed profiles are flat at small radii, suggesting the formation of SMBHs. Therefore, the universality of the collapse fraction is not violated by changing the cross-section of tdSIDM.}
\label{fig:universality_test}
\end{figure}

\section{Consideration of the universal collapse fraction}\label{append:col_frac}
In the paper, we find a universal collapse fraction, $f_{\rm col}\simeq 3\times 10^{-3}$, of tdSIDM halos that is independent of halo mass, size, spin, and formation redshift. This universal collapse fraction corresponds to a collapse radius of $r_0 \simeq 0.07r_{\rm s}$. To further confirm the universality of the collapse fraction numerically, we run N-body simulations with different dark matter self-interaction cross-sections for the same $5-\sigma$ halo formed at $z=10$. The enclosed profiles of the collapsed dark matter halos, as shown in Fig.~\ref{fig:universality_test}, suggest that the collapse fraction is also universal for different cross-sections. The goal of the following is to explain the universal collapse fraction from the theoretical perspective.

The characteristic length scale of the gravitational collapse of gas clouds against thermal pressure support is the Jeans length
\begin{equation}
    \lambda_{\rm J} = c_{\rm s} \sqrt{\dfrac{\pi}{G \rho}},
\end{equation}
where $c_{\rm s}$ is the sound speed and $\rho$ is the mass density. Applying the concept to weakly collisional dSIDM fluid, we replace $c_{\rm s}$ with the one-dimensional velocity dispersion $\sigma_{\rm v}$ and calculate $\rho$ as the averaged mass density within a radius $r$. Thus, we obtain
\begin{equation}
    \frac{r}{\lambda_{\rm J}(r)}\propto\sqrt{\frac{GM(r)/r^{2}}{\sigma_{\rm v}^{2}(r)/r}},
\end{equation}
where $\sigma_{\rm v}(r)$ is the velocity dispersion given by Eq.~(\ref{eq:velocity-dispersion}), $M(r)$ is the enclosed halo mass within radius $r$. When the ratio $r/\lambda_{\rm J}$ is at its maximum, it reaches the point with the maximum gravitational instability. If the enclosed halo mass is given by the NFW profile, one can obtain that $r/\lambda_{\rm J}$ reaches maximum at $r\simeq 0.06r_{\rm s}$, which is close to the collapse radius $0.07r_{\rm s}$ we found in our simulations. The surprising coincidence suggests that the size of the initially collapsed region is likely related to the gravitational instability.

Furthermore, after the mass within a radius $r$ collapse to a point mass, the boundary of spherical accretion of the surrounding medium is given by the Bondi-Hoyle-Lyttleton (BHL) radius \cite{Hoyle1939,Bondi1944,Bondi1952,Shima1985}
\begin{equation}
    R_{\rm B}(r) = \dfrac{2 G M(r)}{c_{\rm s}^2} = \dfrac{2GM(r)}{\sigma_{\rm v}^{2}(r)},
\end{equation}
where the point mass has been assumed to be stationary with respect to the surrounding medium and we have substituted $c_{\rm s}$ with $\sigma_{\rm v}(r)$ again. It is worth noting that the ratio of the BHL radius to $r$ is proportional to $(r/{\lambda_{\rm J}})^2$. Therefore, the ratio $R_{\rm B}/r$ also reaches its maximum at $r\simeq 0.06r_{\rm s}$ and the numeric value of the maximum is actually close to unity. This indicates that the accretion of surrounding dark matter is strongest at this universal radius and will be less effective when collapse extends to larger radii since $R_{\rm B}(r)$ will quickly drop below $r$.

Although accretion will be prohibited when $R_{\rm B}(r)$ drops below $r$, dissipative self-interactions can continually lower the kinetic energy of dark matter, decrease the velocity dispersion and enlarge the BHL radius, which will restore accretion again. But the accretion also relies on mechanisms to transfer angular momentum outward. As discussed in Section~\ref{append:barrier_fragmentation}, in the system considered here, two important mechanisms would be torques of non-axial-symmetric structures originating from gravitational instability and viscous angular momentum transfer. As found earlier this section, the gravitational instability becomes weaker at larger radius beyond $r\simeq 0.06r_{\rm s}$. For the viscosity, we can compute the ratio between $t_{\rm v}$ and $t_{\rm diss}$ as in Eq.(\ref{eq:tv_to_tdiss})
\begin{align}
    t_{\rm v}/t_{\rm diss} & \simeq  \dfrac{4\pi G r^{2}}{(\sigma/m) \sigma_{\rm v}^3(r)} \Big/  \dfrac{1}{ \overline{\rho}(r) (\sigma/m) \sigma_{\rm v}(r)} \nonumber \\
    & = \dfrac{3 G M(r)}{\sigma_{\rm v}^2(r) r}
\end{align}
where we have used $r$ as the characteristic length scale for angular momentum transfer and used the averaged mass density within radius $r$ for the calculation of $t_{\rm diss}$. It is surprising that the ratio $t_{\rm v}/t_{\rm diss}$ is proportional to $R_{\rm B}/r$ as well as $(r/{\lambda_{\rm J}})^2$. The ratio also reaches its maximum at $r\simeq 0.06r_{\rm s}$ and takes an order unity value at its maximum. Beyond the radius $0.06r_{\rm s}$, viscous angular momentum transfer will also quickly become ineffective. Therefore, the collapse of the central halo eventually stagnates at the universal radius $r\simeq 0.06r_{\rm s}$. {These arguments we discussed above should work for generic dissipative dark matter models. For instance, the fractional kinetic energy loss $f$ does not change the form of the collapse timescale and the Jeans length. If the dissipation is velocity dependent, the arguments of the Jeans length and Bondi radius still applies. 
Therefore, even if the fractional kinetic energy loss $f$ is significantly different from the current value or varies with velocities, one would still expect the collapse fraction to be universal. Baryonic physics can potentially affect the collapse fraction, which we leave for future work. }

\bibliographystyle{apsrev4-2}
\bibliography{BHseed}

\begin{thebibliography}{156}%
\makeatletter
\providecommand \@ifxundefined [1]{%
 \@ifx{#1\undefined}
}%
\providecommand \@ifnum [1]{%
 \ifnum #1\expandafter \@firstoftwo
 \else \expandafter \@secondoftwo
 \fi
}%
\providecommand \@ifx [1]{%
 \ifx #1\expandafter \@firstoftwo
 \else \expandafter \@secondoftwo
 \fi
}%
\providecommand \natexlab [1]{#1}%
\providecommand \enquote  [1]{``#1''}%
\providecommand \bibnamefont  [1]{#1}%
\providecommand \bibfnamefont [1]{#1}%
\providecommand \citenamefont [1]{#1}%
\providecommand \href@noop [0]{\@secondoftwo}%
\providecommand \href [0]{\begingroup \@sanitize@url \@href}%
\providecommand \@href[1]{\@@startlink{#1}\@@href}%
\providecommand \@@href[1]{\endgroup#1\@@endlink}%
\providecommand \@sanitize@url [0]{\catcode `\\12\catcode `\$12\catcode
  `\&12\catcode `\#12\catcode `\^12\catcode `\_12\catcode `\%12\relax}%
\providecommand \@@startlink[1]{}%
\providecommand \@@endlink[0]{}%
\providecommand \url  [0]{\begingroup\@sanitize@url \@url }%
\providecommand \@url [1]{\endgroup\@href {#1}{\urlprefix }}%
\providecommand \urlprefix  [0]{URL }%
\providecommand \Eprint [0]{\href }%
\providecommand \doibase [0]{https://doi.org/}%
\providecommand \selectlanguage [0]{\@gobble}%
\providecommand \bibinfo  [0]{\@secondoftwo}%
\providecommand \bibfield  [0]{\@secondoftwo}%
\providecommand \translation [1]{[#1]}%
\providecommand \BibitemOpen [0]{}%
\providecommand \bibitemStop [0]{}%
\providecommand \bibitemNoStop [0]{.\EOS\space}%
\providecommand \EOS [0]{\spacefactor3000\relax}%
\providecommand \BibitemShut  [1]{\csname bibitem#1\endcsname}%
\let\auto@bib@innerbib\@empty
\bibitem [{\citenamefont {Matsuoka}\ \emph {et~al.}(2016)\citenamefont
  {Matsuoka}, \citenamefont {Onoue}, \citenamefont {Kashikawa}, \citenamefont
  {Iwasawa}, \citenamefont {Strauss}, \citenamefont {Nagao}, \citenamefont
  {Imanishi}, \citenamefont {Niida}, \citenamefont {Toba}, \citenamefont
  {Akiyama},\ and\ \citenamefont {et~al.}}]{Matsuoka_2016}%
  \BibitemOpen
  \bibfield  {author} {\bibinfo {author} {\bibfnamefont {Y.}~\bibnamefont
  {Matsuoka}}, \bibinfo {author} {\bibfnamefont {M.}~\bibnamefont {Onoue}},
  \bibinfo {author} {\bibfnamefont {N.}~\bibnamefont {Kashikawa}}, \bibinfo
  {author} {\bibfnamefont {K.}~\bibnamefont {Iwasawa}}, \bibinfo {author}
  {\bibfnamefont {M.~A.}\ \bibnamefont {Strauss}}, \bibinfo {author}
  {\bibfnamefont {T.}~\bibnamefont {Nagao}}, \bibinfo {author} {\bibfnamefont
  {M.}~\bibnamefont {Imanishi}}, \bibinfo {author} {\bibfnamefont
  {M.}~\bibnamefont {Niida}}, \bibinfo {author} {\bibfnamefont
  {Y.}~\bibnamefont {Toba}}, \bibinfo {author} {\bibfnamefont {M.}~\bibnamefont
  {Akiyama}},\ and\ \bibinfo {author} {\bibnamefont {et~al.}},\ }\href
  {https://doi.org/10.3847/0004-637x/828/1/26} {\bibfield  {journal} {\bibinfo
  {journal} {The Astrophysical Journal}\ }\textbf {\bibinfo {volume} {828}},\
  \bibinfo {pages} {26} (\bibinfo {year} {2016})}\BibitemShut {NoStop}%
\bibitem [{\citenamefont {Bañados}\ \emph {et~al.}(2017)\citenamefont
  {Bañados}, \citenamefont {Venemans}, \citenamefont {Mazzucchelli},
  \citenamefont {Farina}, \citenamefont {Walter}, \citenamefont {Wang},
  \citenamefont {Decarli}, \citenamefont {Stern}, \citenamefont {Fan},
  \citenamefont {Davies},\ and\ \citenamefont {et~al.}}]{Ba_ados_2017}%
  \BibitemOpen
  \bibfield  {author} {\bibinfo {author} {\bibfnamefont {E.}~\bibnamefont
  {Bañados}}, \bibinfo {author} {\bibfnamefont {B.~P.}\ \bibnamefont
  {Venemans}}, \bibinfo {author} {\bibfnamefont {C.}~\bibnamefont
  {Mazzucchelli}}, \bibinfo {author} {\bibfnamefont {E.~P.}\ \bibnamefont
  {Farina}}, \bibinfo {author} {\bibfnamefont {F.}~\bibnamefont {Walter}},
  \bibinfo {author} {\bibfnamefont {F.}~\bibnamefont {Wang}}, \bibinfo {author}
  {\bibfnamefont {R.}~\bibnamefont {Decarli}}, \bibinfo {author} {\bibfnamefont
  {D.}~\bibnamefont {Stern}}, \bibinfo {author} {\bibfnamefont
  {X.}~\bibnamefont {Fan}}, \bibinfo {author} {\bibfnamefont {F.~B.}\
  \bibnamefont {Davies}},\ and\ \bibinfo {author} {\bibnamefont {et~al.}},\
  }\href {https://doi.org/10.1038/nature25180} {\bibfield  {journal} {\bibinfo
  {journal} {Nature}\ }\textbf {\bibinfo {volume} {553}},\ \bibinfo {pages}
  {473–476} (\bibinfo {year} {2017})}\BibitemShut {NoStop}%
\bibitem [{\citenamefont {Wang}\ \emph {et~al.}(2018)\citenamefont {Wang},
  \citenamefont {Yang}, \citenamefont {Fan}, \citenamefont {Yue}, \citenamefont
  {Wu}, \citenamefont {Schindler}, \citenamefont {Bian}, \citenamefont {Li},
  \citenamefont {Farina}, \citenamefont {Bañados},\ and\ \citenamefont
  {et~al.}}]{Wang_2018}%
  \BibitemOpen
  \bibfield  {author} {\bibinfo {author} {\bibfnamefont {F.}~\bibnamefont
  {Wang}}, \bibinfo {author} {\bibfnamefont {J.}~\bibnamefont {Yang}}, \bibinfo
  {author} {\bibfnamefont {X.}~\bibnamefont {Fan}}, \bibinfo {author}
  {\bibfnamefont {M.}~\bibnamefont {Yue}}, \bibinfo {author} {\bibfnamefont
  {X.-B.}\ \bibnamefont {Wu}}, \bibinfo {author} {\bibfnamefont {J.-T.}\
  \bibnamefont {Schindler}}, \bibinfo {author} {\bibfnamefont {F.}~\bibnamefont
  {Bian}}, \bibinfo {author} {\bibfnamefont {J.-T.}\ \bibnamefont {Li}},
  \bibinfo {author} {\bibfnamefont {E.~P.}\ \bibnamefont {Farina}}, \bibinfo
  {author} {\bibfnamefont {E.}~\bibnamefont {Bañados}},\ and\ \bibinfo
  {author} {\bibnamefont {et~al.}},\ }\href
  {https://doi.org/10.3847/2041-8213/aaf1d2} {\bibfield  {journal} {\bibinfo
  {journal} {The Astrophysical Journal}\ }\textbf {\bibinfo {volume} {869}},\
  \bibinfo {pages} {L9} (\bibinfo {year} {2018})}\BibitemShut {NoStop}%
\bibitem [{\citenamefont {Matsuoka}\ \emph {et~al.}(2019)\citenamefont
  {Matsuoka}, \citenamefont {Onoue}, \citenamefont {Kashikawa}, \citenamefont
  {Strauss}, \citenamefont {Iwasawa}, \citenamefont {Lee}, \citenamefont
  {Imanishi}, \citenamefont {Nagao}, \citenamefont {Akiyama}, \citenamefont
  {Asami}, \citenamefont {Bosch}, \citenamefont {Furusawa}, \citenamefont
  {Goto}, \citenamefont {Gunn}, \citenamefont {Harikane}, \citenamefont
  {Ikeda}, \citenamefont {Izumi}, \citenamefont {Kawaguchi}, \citenamefont
  {Kato}, \citenamefont {Kikuta}, \citenamefont {Kohno}, \citenamefont
  {Komiyama}, \citenamefont {Koyama}, \citenamefont {Lupton}, \citenamefont
  {Minezaki}, \citenamefont {Miyazaki}, \citenamefont {Murayama}, \citenamefont
  {Niida}, \citenamefont {Nishizawa}, \citenamefont {Noboriguchi},
  \citenamefont {Oguri}, \citenamefont {Ono}, \citenamefont {Ouchi},
  \citenamefont {Price}, \citenamefont {Sameshima}, \citenamefont {Schulze},
  \citenamefont {Shirakata}, \citenamefont {Silverman}, \citenamefont
  {Sugiyama}, \citenamefont {Tait}, \citenamefont {Takada}, \citenamefont
  {Takata}, \citenamefont {Tanaka}, \citenamefont {Tang}, \citenamefont {Toba},
  \citenamefont {Utsumi}, \citenamefont {Wang},\ and\ \citenamefont
  {Yamashita}}]{Matsuoka_2019}%
  \BibitemOpen
  \bibfield  {author} {\bibinfo {author} {\bibfnamefont {Y.}~\bibnamefont
  {Matsuoka}}, \bibinfo {author} {\bibfnamefont {M.}~\bibnamefont {Onoue}},
  \bibinfo {author} {\bibfnamefont {N.}~\bibnamefont {Kashikawa}}, \bibinfo
  {author} {\bibfnamefont {M.~A.}\ \bibnamefont {Strauss}}, \bibinfo {author}
  {\bibfnamefont {K.}~\bibnamefont {Iwasawa}}, \bibinfo {author} {\bibfnamefont
  {C.-H.}\ \bibnamefont {Lee}}, \bibinfo {author} {\bibfnamefont
  {M.}~\bibnamefont {Imanishi}}, \bibinfo {author} {\bibfnamefont
  {T.}~\bibnamefont {Nagao}}, \bibinfo {author} {\bibfnamefont
  {M.}~\bibnamefont {Akiyama}}, \bibinfo {author} {\bibfnamefont
  {N.}~\bibnamefont {Asami}}, \bibinfo {author} {\bibfnamefont
  {J.}~\bibnamefont {Bosch}}, \bibinfo {author} {\bibfnamefont
  {H.}~\bibnamefont {Furusawa}}, \bibinfo {author} {\bibfnamefont
  {T.}~\bibnamefont {Goto}}, \bibinfo {author} {\bibfnamefont {J.~E.}\
  \bibnamefont {Gunn}}, \bibinfo {author} {\bibfnamefont {Y.}~\bibnamefont
  {Harikane}}, \bibinfo {author} {\bibfnamefont {H.}~\bibnamefont {Ikeda}},
  \bibinfo {author} {\bibfnamefont {T.}~\bibnamefont {Izumi}}, \bibinfo
  {author} {\bibfnamefont {T.}~\bibnamefont {Kawaguchi}}, \bibinfo {author}
  {\bibfnamefont {N.}~\bibnamefont {Kato}}, \bibinfo {author} {\bibfnamefont
  {S.}~\bibnamefont {Kikuta}}, \bibinfo {author} {\bibfnamefont
  {K.}~\bibnamefont {Kohno}}, \bibinfo {author} {\bibfnamefont
  {Y.}~\bibnamefont {Komiyama}}, \bibinfo {author} {\bibfnamefont
  {S.}~\bibnamefont {Koyama}}, \bibinfo {author} {\bibfnamefont {R.~H.}\
  \bibnamefont {Lupton}}, \bibinfo {author} {\bibfnamefont {T.}~\bibnamefont
  {Minezaki}}, \bibinfo {author} {\bibfnamefont {S.}~\bibnamefont {Miyazaki}},
  \bibinfo {author} {\bibfnamefont {H.}~\bibnamefont {Murayama}}, \bibinfo
  {author} {\bibfnamefont {M.}~\bibnamefont {Niida}}, \bibinfo {author}
  {\bibfnamefont {A.~J.}\ \bibnamefont {Nishizawa}}, \bibinfo {author}
  {\bibfnamefont {A.}~\bibnamefont {Noboriguchi}}, \bibinfo {author}
  {\bibfnamefont {M.}~\bibnamefont {Oguri}}, \bibinfo {author} {\bibfnamefont
  {Y.}~\bibnamefont {Ono}}, \bibinfo {author} {\bibfnamefont {M.}~\bibnamefont
  {Ouchi}}, \bibinfo {author} {\bibfnamefont {P.~A.}\ \bibnamefont {Price}},
  \bibinfo {author} {\bibfnamefont {H.}~\bibnamefont {Sameshima}}, \bibinfo
  {author} {\bibfnamefont {A.}~\bibnamefont {Schulze}}, \bibinfo {author}
  {\bibfnamefont {H.}~\bibnamefont {Shirakata}}, \bibinfo {author}
  {\bibfnamefont {J.~D.}\ \bibnamefont {Silverman}}, \bibinfo {author}
  {\bibfnamefont {N.}~\bibnamefont {Sugiyama}}, \bibinfo {author}
  {\bibfnamefont {P.~J.}\ \bibnamefont {Tait}}, \bibinfo {author}
  {\bibfnamefont {M.}~\bibnamefont {Takada}}, \bibinfo {author} {\bibfnamefont
  {T.}~\bibnamefont {Takata}}, \bibinfo {author} {\bibfnamefont
  {M.}~\bibnamefont {Tanaka}}, \bibinfo {author} {\bibfnamefont {J.-J.}\
  \bibnamefont {Tang}}, \bibinfo {author} {\bibfnamefont {Y.}~\bibnamefont
  {Toba}}, \bibinfo {author} {\bibfnamefont {Y.}~\bibnamefont {Utsumi}},
  \bibinfo {author} {\bibfnamefont {S.-Y.}\ \bibnamefont {Wang}},\ and\
  \bibinfo {author} {\bibfnamefont {T.}~\bibnamefont {Yamashita}},\ }\href
  {https://doi.org/10.3847/2041-8213/ab0216} {\bibfield  {journal} {\bibinfo
  {journal} {The Astrophysical Journal}\ }\textbf {\bibinfo {volume} {872}},\
  \bibinfo {pages} {L2} (\bibinfo {year} {2019})}\BibitemShut {NoStop}%
\bibitem [{\citenamefont {Yang}\ \emph {et~al.}(2020)\citenamefont {Yang},
  \citenamefont {Wang}, \citenamefont {Fan}, \citenamefont {Hennawi},
  \citenamefont {Davies}, \citenamefont {Yue}, \citenamefont {Banados},
  \citenamefont {Wu}, \citenamefont {Venemans}, \citenamefont {Barth},
  \citenamefont {Bian}, \citenamefont {Boutsia}, \citenamefont {Decarli},
  \citenamefont {Farina}, \citenamefont {Green}, \citenamefont {Jiang},
  \citenamefont {Li}, \citenamefont {Mazzucchelli},\ and\ \citenamefont
  {Walter}}]{Yang_2020}%
  \BibitemOpen
  \bibfield  {author} {\bibinfo {author} {\bibfnamefont {J.}~\bibnamefont
  {Yang}}, \bibinfo {author} {\bibfnamefont {F.}~\bibnamefont {Wang}}, \bibinfo
  {author} {\bibfnamefont {X.}~\bibnamefont {Fan}}, \bibinfo {author}
  {\bibfnamefont {J.~F.}\ \bibnamefont {Hennawi}}, \bibinfo {author}
  {\bibfnamefont {F.~B.}\ \bibnamefont {Davies}}, \bibinfo {author}
  {\bibfnamefont {M.}~\bibnamefont {Yue}}, \bibinfo {author} {\bibfnamefont
  {E.}~\bibnamefont {Banados}}, \bibinfo {author} {\bibfnamefont {X.-B.}\
  \bibnamefont {Wu}}, \bibinfo {author} {\bibfnamefont {B.}~\bibnamefont
  {Venemans}}, \bibinfo {author} {\bibfnamefont {A.~J.}\ \bibnamefont {Barth}},
  \bibinfo {author} {\bibfnamefont {F.}~\bibnamefont {Bian}}, \bibinfo {author}
  {\bibfnamefont {K.}~\bibnamefont {Boutsia}}, \bibinfo {author} {\bibfnamefont
  {R.}~\bibnamefont {Decarli}}, \bibinfo {author} {\bibfnamefont {E.~P.}\
  \bibnamefont {Farina}}, \bibinfo {author} {\bibfnamefont {R.}~\bibnamefont
  {Green}}, \bibinfo {author} {\bibfnamefont {L.}~\bibnamefont {Jiang}},
  \bibinfo {author} {\bibfnamefont {J.-T.}\ \bibnamefont {Li}}, \bibinfo
  {author} {\bibfnamefont {C.}~\bibnamefont {Mazzucchelli}},\ and\ \bibinfo
  {author} {\bibfnamefont {F.}~\bibnamefont {Walter}},\ }\href
  {https://doi.org/10.3847/2041-8213/ab9c26} {\bibfield  {journal} {\bibinfo
  {journal} {The Astrophysical Journal}\ }\textbf {\bibinfo {volume} {897}},\
  \bibinfo {pages} {L14} (\bibinfo {year} {2020})}\BibitemShut {NoStop}%
\bibitem [{\citenamefont {{Madau}}\ and\ \citenamefont
  {{Rees}}(2001)}]{Madau2001}%
  \BibitemOpen
  \bibfield  {author} {\bibinfo {author} {\bibfnamefont {P.}~\bibnamefont
  {{Madau}}}\ and\ \bibinfo {author} {\bibfnamefont {M.~J.}\ \bibnamefont
  {{Rees}}},\ }\href {https://doi.org/10.1086/319848} {\bibfield  {journal}
  {\bibinfo  {journal} {\apjl}\ }\textbf {\bibinfo {volume} {551}},\ \bibinfo
  {pages} {L27} (\bibinfo {year} {2001})},\ \Eprint
  {https://arxiv.org/abs/astro-ph/0101223} {arXiv:astro-ph/0101223 [astro-ph]}
  \BibitemShut {NoStop}%
\bibitem [{\citenamefont {{Abel}}\ \emph {et~al.}(2002)\citenamefont {{Abel}},
  \citenamefont {{Bryan}},\ and\ \citenamefont {{Norman}}}]{Abel2002}%
  \BibitemOpen
  \bibfield  {author} {\bibinfo {author} {\bibfnamefont {T.}~\bibnamefont
  {{Abel}}}, \bibinfo {author} {\bibfnamefont {G.~L.}\ \bibnamefont
  {{Bryan}}},\ and\ \bibinfo {author} {\bibfnamefont {M.~L.}\ \bibnamefont
  {{Norman}}},\ }\href {https://doi.org/10.1126/science.295.5552.93} {\bibfield
   {journal} {\bibinfo  {journal} {Science}\ }\textbf {\bibinfo {volume}
  {295}},\ \bibinfo {pages} {93} (\bibinfo {year} {2002})},\ \Eprint
  {https://arxiv.org/abs/astro-ph/0112088} {arXiv:astro-ph/0112088 [astro-ph]}
  \BibitemShut {NoStop}%
\bibitem [{\citenamefont {{Bromm}}\ \emph {et~al.}(2002)\citenamefont
  {{Bromm}}, \citenamefont {{Coppi}},\ and\ \citenamefont
  {{Larson}}}]{Bromm2002}%
  \BibitemOpen
  \bibfield  {author} {\bibinfo {author} {\bibfnamefont {V.}~\bibnamefont
  {{Bromm}}}, \bibinfo {author} {\bibfnamefont {P.~S.}\ \bibnamefont
  {{Coppi}}},\ and\ \bibinfo {author} {\bibfnamefont {R.~B.}\ \bibnamefont
  {{Larson}}},\ }\href {https://doi.org/10.1086/323947} {\bibfield  {journal}
  {\bibinfo  {journal} {\apj}\ }\textbf {\bibinfo {volume} {564}},\ \bibinfo
  {pages} {23} (\bibinfo {year} {2002})},\ \Eprint
  {https://arxiv.org/abs/astro-ph/0102503} {arXiv:astro-ph/0102503 [astro-ph]}
  \BibitemShut {NoStop}%
\bibitem [{\citenamefont {{O'Shea}}\ and\ \citenamefont
  {{Norman}}(2007)}]{OShea2007}%
  \BibitemOpen
  \bibfield  {author} {\bibinfo {author} {\bibfnamefont {B.~W.}\ \bibnamefont
  {{O'Shea}}}\ and\ \bibinfo {author} {\bibfnamefont {M.~L.}\ \bibnamefont
  {{Norman}}},\ }\href {https://doi.org/10.1086/509250} {\bibfield  {journal}
  {\bibinfo  {journal} {\apj}\ }\textbf {\bibinfo {volume} {654}},\ \bibinfo
  {pages} {66} (\bibinfo {year} {2007})},\ \Eprint
  {https://arxiv.org/abs/astro-ph/0607013} {arXiv:astro-ph/0607013 [astro-ph]}
  \BibitemShut {NoStop}%
\bibitem [{\citenamefont {{Turk}}\ \emph {et~al.}(2009)\citenamefont {{Turk}},
  \citenamefont {{Abel}},\ and\ \citenamefont {{O'Shea}}}]{Turk2009}%
  \BibitemOpen
  \bibfield  {author} {\bibinfo {author} {\bibfnamefont {M.~J.}\ \bibnamefont
  {{Turk}}}, \bibinfo {author} {\bibfnamefont {T.}~\bibnamefont {{Abel}}},\
  and\ \bibinfo {author} {\bibfnamefont {B.}~\bibnamefont {{O'Shea}}},\ }\href
  {https://doi.org/10.1126/science.1173540} {\bibfield  {journal} {\bibinfo
  {journal} {Science}\ }\textbf {\bibinfo {volume} {325}},\ \bibinfo {pages}
  {601} (\bibinfo {year} {2009})},\ \Eprint {https://arxiv.org/abs/0907.2919}
  {arXiv:0907.2919 [astro-ph.CO]} \BibitemShut {NoStop}%
\bibitem [{\citenamefont {{Tanaka}}\ and\ \citenamefont
  {{Haiman}}(2009)}]{Tanaka2009}%
  \BibitemOpen
  \bibfield  {author} {\bibinfo {author} {\bibfnamefont {T.}~\bibnamefont
  {{Tanaka}}}\ and\ \bibinfo {author} {\bibfnamefont {Z.}~\bibnamefont
  {{Haiman}}},\ }\href {https://doi.org/10.1088/0004-637X/696/2/1798}
  {\bibfield  {journal} {\bibinfo  {journal} {\apj}\ }\textbf {\bibinfo
  {volume} {696}},\ \bibinfo {pages} {1798} (\bibinfo {year} {2009})},\ \Eprint
  {https://arxiv.org/abs/0807.4702} {arXiv:0807.4702 [astro-ph]} \BibitemShut
  {NoStop}%
\bibitem [{\citenamefont {{Greif}}\ \emph {et~al.}(2012)\citenamefont
  {{Greif}}, \citenamefont {{Bromm}}, \citenamefont {{Clark}}, \citenamefont
  {{Glover}}, \citenamefont {{Smith}}, \citenamefont {{Klessen}}, \citenamefont
  {{Yoshida}},\ and\ \citenamefont {{Springel}}}]{Greif2012}%
  \BibitemOpen
  \bibfield  {author} {\bibinfo {author} {\bibfnamefont {T.~H.}\ \bibnamefont
  {{Greif}}}, \bibinfo {author} {\bibfnamefont {V.}~\bibnamefont {{Bromm}}},
  \bibinfo {author} {\bibfnamefont {P.~C.}\ \bibnamefont {{Clark}}}, \bibinfo
  {author} {\bibfnamefont {S.~C.~O.}\ \bibnamefont {{Glover}}}, \bibinfo
  {author} {\bibfnamefont {R.~J.}\ \bibnamefont {{Smith}}}, \bibinfo {author}
  {\bibfnamefont {R.~S.}\ \bibnamefont {{Klessen}}}, \bibinfo {author}
  {\bibfnamefont {N.}~\bibnamefont {{Yoshida}}},\ and\ \bibinfo {author}
  {\bibfnamefont {V.}~\bibnamefont {{Springel}}},\ }\href
  {https://doi.org/10.1111/j.1365-2966.2012.21212.x} {\bibfield  {journal}
  {\bibinfo  {journal} {\mnras}\ }\textbf {\bibinfo {volume} {424}},\ \bibinfo
  {pages} {399} (\bibinfo {year} {2012})}\BibitemShut {NoStop}%
\bibitem [{\citenamefont {{Valiante}}\ \emph {et~al.}(2016)\citenamefont
  {{Valiante}}, \citenamefont {{Schneider}}, \citenamefont {{Volonteri}},\ and\
  \citenamefont {{Omukai}}}]{Valiante2016}%
  \BibitemOpen
  \bibfield  {author} {\bibinfo {author} {\bibfnamefont {R.}~\bibnamefont
  {{Valiante}}}, \bibinfo {author} {\bibfnamefont {R.}~\bibnamefont
  {{Schneider}}}, \bibinfo {author} {\bibfnamefont {M.}~\bibnamefont
  {{Volonteri}}},\ and\ \bibinfo {author} {\bibfnamefont {K.}~\bibnamefont
  {{Omukai}}},\ }\href {https://doi.org/10.1093/mnras/stw225} {\bibfield
  {journal} {\bibinfo  {journal} {\mnras}\ }\textbf {\bibinfo {volume} {457}},\
  \bibinfo {pages} {3356} (\bibinfo {year} {2016})},\ \Eprint
  {https://arxiv.org/abs/1601.07915} {arXiv:1601.07915 [astro-ph.GA]}
  \BibitemShut {NoStop}%
\bibitem [{\citenamefont {Hirano}\ \emph {et~al.}(2014)\citenamefont {Hirano},
  \citenamefont {Hosokawa}, \citenamefont {Yoshida}, \citenamefont {Umeda},
  \citenamefont {Omukai}, \citenamefont {Chiaki},\ and\ \citenamefont
  {Yorke}}]{Hirano_2014}%
  \BibitemOpen
  \bibfield  {author} {\bibinfo {author} {\bibfnamefont {S.}~\bibnamefont
  {Hirano}}, \bibinfo {author} {\bibfnamefont {T.}~\bibnamefont {Hosokawa}},
  \bibinfo {author} {\bibfnamefont {N.}~\bibnamefont {Yoshida}}, \bibinfo
  {author} {\bibfnamefont {H.}~\bibnamefont {Umeda}}, \bibinfo {author}
  {\bibfnamefont {K.}~\bibnamefont {Omukai}}, \bibinfo {author} {\bibfnamefont
  {G.}~\bibnamefont {Chiaki}},\ and\ \bibinfo {author} {\bibfnamefont {H.~W.}\
  \bibnamefont {Yorke}},\ }\href {https://doi.org/10.1088/0004-637x/781/2/60}
  {\bibfield  {journal} {\bibinfo  {journal} {The Astrophysical Journal}\
  }\textbf {\bibinfo {volume} {781}},\ \bibinfo {pages} {60} (\bibinfo {year}
  {2014})}\BibitemShut {NoStop}%
\bibitem [{\citenamefont {Johnson}\ and\ \citenamefont
  {Bromm}(2007)}]{10.1111/j.1365-2966.2006.11275.x}%
  \BibitemOpen
  \bibfield  {author} {\bibinfo {author} {\bibfnamefont {J.~L.}\ \bibnamefont
  {Johnson}}\ and\ \bibinfo {author} {\bibfnamefont {V.}~\bibnamefont
  {Bromm}},\ }\href {https://doi.org/10.1111/j.1365-2966.2006.11275.x}
  {\bibfield  {journal} {\bibinfo  {journal} {Monthly Notices of the Royal
  Astronomical Society}\ }\textbf {\bibinfo {volume} {374}},\ \bibinfo {pages}
  {1557} (\bibinfo {year} {2007})},\ \Eprint
  {https://arxiv.org/abs/https://academic.oup.com/mnras/article-pdf/374/4/1557/2877186/mnras0374-1557.pdf}
  {https://academic.oup.com/mnras/article-pdf/374/4/1557/2877186/mnras0374-1557.pdf}
  \BibitemShut {NoStop}%
\bibitem [{\citenamefont {Whalen}\ \emph {et~al.}(2008)\citenamefont {Whalen},
  \citenamefont {van Veelen}, \citenamefont {O'Shea},\ and\ \citenamefont
  {Norman}}]{Whalen_2008}%
  \BibitemOpen
  \bibfield  {author} {\bibinfo {author} {\bibfnamefont {D.}~\bibnamefont
  {Whalen}}, \bibinfo {author} {\bibfnamefont {B.}~\bibnamefont {van Veelen}},
  \bibinfo {author} {\bibfnamefont {B.~W.}\ \bibnamefont {O'Shea}},\ and\
  \bibinfo {author} {\bibfnamefont {M.~L.}\ \bibnamefont {Norman}},\ }\href
  {https://doi.org/10.1086/589643} {\bibfield  {journal} {\bibinfo  {journal}
  {The Astrophysical Journal}\ }\textbf {\bibinfo {volume} {682}},\ \bibinfo
  {pages} {49} (\bibinfo {year} {2008})}\BibitemShut {NoStop}%
\bibitem [{\citenamefont {Milosavljevi{\'{c}}}\ \emph
  {et~al.}(2009)\citenamefont {Milosavljevi{\'{c}}}, \citenamefont {Couch},\
  and\ \citenamefont {Bromm}}]{Milosavljevi__2009}%
  \BibitemOpen
  \bibfield  {author} {\bibinfo {author} {\bibfnamefont {M.}~\bibnamefont
  {Milosavljevi{\'{c}}}}, \bibinfo {author} {\bibfnamefont {S.~M.}\
  \bibnamefont {Couch}},\ and\ \bibinfo {author} {\bibfnamefont
  {V.}~\bibnamefont {Bromm}},\ }\href
  {https://doi.org/10.1088/0004-637x/696/2/l146} {\bibfield  {journal}
  {\bibinfo  {journal} {The Astrophysical Journal}\ }\textbf {\bibinfo {volume}
  {696}},\ \bibinfo {pages} {L146} (\bibinfo {year} {2009})}\BibitemShut
  {NoStop}%
\bibitem [{\citenamefont {Alvarez}\ \emph {et~al.}(2009)\citenamefont
  {Alvarez}, \citenamefont {Wise},\ and\ \citenamefont {Abel}}]{Alvarez_2009}%
  \BibitemOpen
  \bibfield  {author} {\bibinfo {author} {\bibfnamefont {M.~A.}\ \bibnamefont
  {Alvarez}}, \bibinfo {author} {\bibfnamefont {J.~H.}\ \bibnamefont {Wise}},\
  and\ \bibinfo {author} {\bibfnamefont {T.}~\bibnamefont {Abel}},\ }\href
  {https://doi.org/10.1088/0004-637x/701/2/l133} {\bibfield  {journal}
  {\bibinfo  {journal} {The Astrophysical Journal}\ }\textbf {\bibinfo {volume}
  {701}},\ \bibinfo {pages} {L133} (\bibinfo {year} {2009})}\BibitemShut
  {NoStop}%
\bibitem [{\citenamefont {Inayoshi}\ \emph {et~al.}(2020)\citenamefont
  {Inayoshi}, \citenamefont {Visbal},\ and\ \citenamefont
  {Haiman}}]{Inayoshi_2020}%
  \BibitemOpen
  \bibfield  {author} {\bibinfo {author} {\bibfnamefont {K.}~\bibnamefont
  {Inayoshi}}, \bibinfo {author} {\bibfnamefont {E.}~\bibnamefont {Visbal}},\
  and\ \bibinfo {author} {\bibfnamefont {Z.}~\bibnamefont {Haiman}},\ }\href
  {https://doi.org/10.1146/annurev-astro-120419-014455} {\bibfield  {journal}
  {\bibinfo  {journal} {Annual Review of Astronomy and Astrophysics}\ }\textbf
  {\bibinfo {volume} {58}},\ \bibinfo {pages} {27–97} (\bibinfo {year}
  {2020})}\BibitemShut {NoStop}%
\bibitem [{\citenamefont {Brightman}\ \emph {et~al.}(2019)\citenamefont
  {Brightman}, \citenamefont {Bachetti}, \citenamefont {Earnshaw},
  \citenamefont {Fürst}, \citenamefont {García}, \citenamefont
  {Grefenstette}, \citenamefont {Heida}, \citenamefont {Kara}, \citenamefont
  {Madsen}, \citenamefont {Middleton}, \citenamefont {Stern}, \citenamefont
  {Tombesi},\ and\ \citenamefont {Walton}}]{brightman2019breaking}%
  \BibitemOpen
  \bibfield  {author} {\bibinfo {author} {\bibfnamefont {M.}~\bibnamefont
  {Brightman}}, \bibinfo {author} {\bibfnamefont {M.}~\bibnamefont {Bachetti}},
  \bibinfo {author} {\bibfnamefont {H.~P.}\ \bibnamefont {Earnshaw}}, \bibinfo
  {author} {\bibfnamefont {F.}~\bibnamefont {Fürst}}, \bibinfo {author}
  {\bibfnamefont {J.}~\bibnamefont {García}}, \bibinfo {author} {\bibfnamefont
  {B.}~\bibnamefont {Grefenstette}}, \bibinfo {author} {\bibfnamefont
  {M.}~\bibnamefont {Heida}}, \bibinfo {author} {\bibfnamefont
  {E.}~\bibnamefont {Kara}}, \bibinfo {author} {\bibfnamefont {K.~K.}\
  \bibnamefont {Madsen}}, \bibinfo {author} {\bibfnamefont {M.~J.}\
  \bibnamefont {Middleton}}, \bibinfo {author} {\bibfnamefont {D.}~\bibnamefont
  {Stern}}, \bibinfo {author} {\bibfnamefont {F.}~\bibnamefont {Tombesi}},\
  and\ \bibinfo {author} {\bibfnamefont {D.~J.}\ \bibnamefont {Walton}},\
  }\href@noop {} {\bibinfo {title} {Breaking the limit: Super-eddington
  accretion onto black holes and neutron stars}} (\bibinfo {year} {2019}),\
  \Eprint {https://arxiv.org/abs/1903.06844} {arXiv:1903.06844 [astro-ph.HE]}
  \BibitemShut {NoStop}%
\bibitem [{\citenamefont {Madau}\ \emph {et~al.}(2014)\citenamefont {Madau},
  \citenamefont {Haardt},\ and\ \citenamefont {Dotti}}]{Madau_2014}%
  \BibitemOpen
  \bibfield  {author} {\bibinfo {author} {\bibfnamefont {P.}~\bibnamefont
  {Madau}}, \bibinfo {author} {\bibfnamefont {F.}~\bibnamefont {Haardt}},\ and\
  \bibinfo {author} {\bibfnamefont {M.}~\bibnamefont {Dotti}},\ }\href
  {https://doi.org/10.1088/2041-8205/784/2/l38} {\bibfield  {journal} {\bibinfo
   {journal} {The Astrophysical Journal}\ }\textbf {\bibinfo {volume} {784}},\
  \bibinfo {pages} {L38} (\bibinfo {year} {2014})}\BibitemShut {NoStop}%
\bibitem [{\citenamefont {{Bromm}}\ and\ \citenamefont
  {{Loeb}}(2003)}]{Bromm2003}%
  \BibitemOpen
  \bibfield  {author} {\bibinfo {author} {\bibfnamefont {V.}~\bibnamefont
  {{Bromm}}}\ and\ \bibinfo {author} {\bibfnamefont {A.}~\bibnamefont
  {{Loeb}}},\ }\href {https://doi.org/10.1086/377529} {\bibfield  {journal}
  {\bibinfo  {journal} {\apj}\ }\textbf {\bibinfo {volume} {596}},\ \bibinfo
  {pages} {34} (\bibinfo {year} {2003})},\ \Eprint
  {https://arxiv.org/abs/astro-ph/0212400} {arXiv:astro-ph/0212400 [astro-ph]}
  \BibitemShut {NoStop}%
\bibitem [{\citenamefont {{Koushiappas}}\ \emph {et~al.}(2004)\citenamefont
  {{Koushiappas}}, \citenamefont {{Bullock}},\ and\ \citenamefont
  {{Dekel}}}]{Koushiappas2004}%
  \BibitemOpen
  \bibfield  {author} {\bibinfo {author} {\bibfnamefont {S.~M.}\ \bibnamefont
  {{Koushiappas}}}, \bibinfo {author} {\bibfnamefont {J.~S.}\ \bibnamefont
  {{Bullock}}},\ and\ \bibinfo {author} {\bibfnamefont {A.}~\bibnamefont
  {{Dekel}}},\ }\href {https://doi.org/10.1111/j.1365-2966.2004.08190.x}
  {\bibfield  {journal} {\bibinfo  {journal} {\mnras}\ }\textbf {\bibinfo
  {volume} {354}},\ \bibinfo {pages} {292} (\bibinfo {year} {2004})},\ \Eprint
  {https://arxiv.org/abs/astro-ph/0311487} {arXiv:astro-ph/0311487 [astro-ph]}
  \BibitemShut {NoStop}%
\bibitem [{\citenamefont {{Begelman}}\ \emph {et~al.}(2006)\citenamefont
  {{Begelman}}, \citenamefont {{Volonteri}},\ and\ \citenamefont
  {{Rees}}}]{Begelman2006}%
  \BibitemOpen
  \bibfield  {author} {\bibinfo {author} {\bibfnamefont {M.~C.}\ \bibnamefont
  {{Begelman}}}, \bibinfo {author} {\bibfnamefont {M.}~\bibnamefont
  {{Volonteri}}},\ and\ \bibinfo {author} {\bibfnamefont {M.~J.}\ \bibnamefont
  {{Rees}}},\ }\href {https://doi.org/10.1111/j.1365-2966.2006.10467.x}
  {\bibfield  {journal} {\bibinfo  {journal} {\mnras}\ }\textbf {\bibinfo
  {volume} {370}},\ \bibinfo {pages} {289} (\bibinfo {year} {2006})},\ \Eprint
  {https://arxiv.org/abs/astro-ph/0602363} {arXiv:astro-ph/0602363 [astro-ph]}
  \BibitemShut {NoStop}%
\bibitem [{\citenamefont {{Lodato}}\ and\ \citenamefont
  {{Natarajan}}(2006{\natexlab{a}})}]{Lodato:2006}%
  \BibitemOpen
  \bibfield  {author} {\bibinfo {author} {\bibfnamefont {G.}~\bibnamefont
  {{Lodato}}}\ and\ \bibinfo {author} {\bibfnamefont {P.}~\bibnamefont
  {{Natarajan}}},\ }\href {https://doi.org/10.1111/j.1365-2966.2006.10801.x}
  {\bibfield  {journal} {\bibinfo  {journal} {\mnras}\ }\textbf {\bibinfo
  {volume} {371}},\ \bibinfo {pages} {1813} (\bibinfo {year}
  {2006}{\natexlab{a}})},\ \Eprint {https://arxiv.org/abs/astro-ph/0606159}
  {arXiv:astro-ph/0606159 [astro-ph]} \BibitemShut {NoStop}%
\bibitem [{\citenamefont {{Ferrara}}\ \emph {et~al.}(2014)\citenamefont
  {{Ferrara}}, \citenamefont {{Salvadori}}, \citenamefont {{Yue}},\ and\
  \citenamefont {{Schleicher}}}]{Ferrara2014}%
  \BibitemOpen
  \bibfield  {author} {\bibinfo {author} {\bibfnamefont {A.}~\bibnamefont
  {{Ferrara}}}, \bibinfo {author} {\bibfnamefont {S.}~\bibnamefont
  {{Salvadori}}}, \bibinfo {author} {\bibfnamefont {B.}~\bibnamefont {{Yue}}},\
  and\ \bibinfo {author} {\bibfnamefont {D.}~\bibnamefont {{Schleicher}}},\
  }\href {https://doi.org/10.1093/mnras/stu1280} {\bibfield  {journal}
  {\bibinfo  {journal} {\mnras}\ }\textbf {\bibinfo {volume} {443}},\ \bibinfo
  {pages} {2410} (\bibinfo {year} {2014})},\ \Eprint
  {https://arxiv.org/abs/1406.6685} {arXiv:1406.6685 [astro-ph.GA]}
  \BibitemShut {NoStop}%
\bibitem [{\citenamefont {{Pacucci}}\ \emph {et~al.}(2015)\citenamefont
  {{Pacucci}}, \citenamefont {{Volonteri}},\ and\ \citenamefont
  {{Ferrara}}}]{Pacucci2015}%
  \BibitemOpen
  \bibfield  {author} {\bibinfo {author} {\bibfnamefont {F.}~\bibnamefont
  {{Pacucci}}}, \bibinfo {author} {\bibfnamefont {M.}~\bibnamefont
  {{Volonteri}}},\ and\ \bibinfo {author} {\bibfnamefont {A.}~\bibnamefont
  {{Ferrara}}},\ }\href {https://doi.org/10.1093/mnras/stv1465} {\bibfield
  {journal} {\bibinfo  {journal} {\mnras}\ }\textbf {\bibinfo {volume} {452}},\
  \bibinfo {pages} {1922} (\bibinfo {year} {2015})},\ \Eprint
  {https://arxiv.org/abs/1506.04750} {arXiv:1506.04750 [astro-ph.GA]}
  \BibitemShut {NoStop}%
\bibitem [{\citenamefont {{Matsuoka}}\ \emph {et~al.}(2019)\citenamefont
  {{Matsuoka}}, \citenamefont {{Onoue}}, \citenamefont {{Kashikawa}},
  \citenamefont {{Strauss}}, \citenamefont {{Iwasawa}}, \citenamefont {{Lee}},
  \citenamefont {{Imanishi}}, \citenamefont {{Nagao}}, \citenamefont
  {{Akiyama}}, \citenamefont {{Asami}}, \citenamefont {{Bosch}}, \citenamefont
  {{Furusawa}}, \citenamefont {{Goto}}, \citenamefont {{Gunn}}, \citenamefont
  {{Harikane}}, \citenamefont {{Ikeda}}, \citenamefont {{Izumi}}, \citenamefont
  {{Kawaguchi}}, \citenamefont {{Kato}}, \citenamefont {{Kikuta}},
  \citenamefont {{Kohno}}, \citenamefont {{Komiyama}}, \citenamefont
  {{Koyama}}, \citenamefont {{Lupton}}, \citenamefont {{Minezaki}},
  \citenamefont {{Miyazaki}}, \citenamefont {{Murayama}}, \citenamefont
  {{Niida}}, \citenamefont {{Nishizawa}}, \citenamefont {{Noboriguchi}},
  \citenamefont {{Oguri}}, \citenamefont {{Ono}}, \citenamefont {{Ouchi}},
  \citenamefont {{Price}}, \citenamefont {{Sameshima}}, \citenamefont
  {{Schulze}}, \citenamefont {{Shirakata}}, \citenamefont {{Silverman}},
  \citenamefont {{Sugiyama}}, \citenamefont {{Tait}}, \citenamefont {{Takada}},
  \citenamefont {{Takata}}, \citenamefont {{Tanaka}}, \citenamefont {{Tang}},
  \citenamefont {{Toba}}, \citenamefont {{Utsumi}}, \citenamefont {{Wang}},\
  and\ \citenamefont {{Yamashita}}}]{Matsuoka2019}%
  \BibitemOpen
  \bibfield  {author} {\bibinfo {author} {\bibfnamefont {Y.}~\bibnamefont
  {{Matsuoka}}}, \bibinfo {author} {\bibfnamefont {M.}~\bibnamefont {{Onoue}}},
  \bibinfo {author} {\bibfnamefont {N.}~\bibnamefont {{Kashikawa}}}, \bibinfo
  {author} {\bibfnamefont {M.~A.}\ \bibnamefont {{Strauss}}}, \bibinfo {author}
  {\bibfnamefont {K.}~\bibnamefont {{Iwasawa}}}, \bibinfo {author}
  {\bibfnamefont {C.-H.}\ \bibnamefont {{Lee}}}, \bibinfo {author}
  {\bibfnamefont {M.}~\bibnamefont {{Imanishi}}}, \bibinfo {author}
  {\bibfnamefont {T.}~\bibnamefont {{Nagao}}}, \bibinfo {author} {\bibfnamefont
  {M.}~\bibnamefont {{Akiyama}}}, \bibinfo {author} {\bibfnamefont
  {N.}~\bibnamefont {{Asami}}}, \bibinfo {author} {\bibfnamefont
  {J.}~\bibnamefont {{Bosch}}}, \bibinfo {author} {\bibfnamefont
  {H.}~\bibnamefont {{Furusawa}}}, \bibinfo {author} {\bibfnamefont
  {T.}~\bibnamefont {{Goto}}}, \bibinfo {author} {\bibfnamefont {J.~E.}\
  \bibnamefont {{Gunn}}}, \bibinfo {author} {\bibfnamefont {Y.}~\bibnamefont
  {{Harikane}}}, \bibinfo {author} {\bibfnamefont {H.}~\bibnamefont {{Ikeda}}},
  \bibinfo {author} {\bibfnamefont {T.}~\bibnamefont {{Izumi}}}, \bibinfo
  {author} {\bibfnamefont {T.}~\bibnamefont {{Kawaguchi}}}, \bibinfo {author}
  {\bibfnamefont {N.}~\bibnamefont {{Kato}}}, \bibinfo {author} {\bibfnamefont
  {S.}~\bibnamefont {{Kikuta}}}, \bibinfo {author} {\bibfnamefont
  {K.}~\bibnamefont {{Kohno}}}, \bibinfo {author} {\bibfnamefont
  {Y.}~\bibnamefont {{Komiyama}}}, \bibinfo {author} {\bibfnamefont
  {S.}~\bibnamefont {{Koyama}}}, \bibinfo {author} {\bibfnamefont {R.~H.}\
  \bibnamefont {{Lupton}}}, \bibinfo {author} {\bibfnamefont {T.}~\bibnamefont
  {{Minezaki}}}, \bibinfo {author} {\bibfnamefont {S.}~\bibnamefont
  {{Miyazaki}}}, \bibinfo {author} {\bibfnamefont {H.}~\bibnamefont
  {{Murayama}}}, \bibinfo {author} {\bibfnamefont {M.}~\bibnamefont {{Niida}}},
  \bibinfo {author} {\bibfnamefont {A.~J.}\ \bibnamefont {{Nishizawa}}},
  \bibinfo {author} {\bibfnamefont {A.}~\bibnamefont {{Noboriguchi}}}, \bibinfo
  {author} {\bibfnamefont {M.}~\bibnamefont {{Oguri}}}, \bibinfo {author}
  {\bibfnamefont {Y.}~\bibnamefont {{Ono}}}, \bibinfo {author} {\bibfnamefont
  {M.}~\bibnamefont {{Ouchi}}}, \bibinfo {author} {\bibfnamefont {P.~A.}\
  \bibnamefont {{Price}}}, \bibinfo {author} {\bibfnamefont {H.}~\bibnamefont
  {{Sameshima}}}, \bibinfo {author} {\bibfnamefont {A.}~\bibnamefont
  {{Schulze}}}, \bibinfo {author} {\bibfnamefont {H.}~\bibnamefont
  {{Shirakata}}}, \bibinfo {author} {\bibfnamefont {J.~D.}\ \bibnamefont
  {{Silverman}}}, \bibinfo {author} {\bibfnamefont {N.}~\bibnamefont
  {{Sugiyama}}}, \bibinfo {author} {\bibfnamefont {P.~J.}\ \bibnamefont
  {{Tait}}}, \bibinfo {author} {\bibfnamefont {M.}~\bibnamefont {{Takada}}},
  \bibinfo {author} {\bibfnamefont {T.}~\bibnamefont {{Takata}}}, \bibinfo
  {author} {\bibfnamefont {M.}~\bibnamefont {{Tanaka}}}, \bibinfo {author}
  {\bibfnamefont {J.-J.}\ \bibnamefont {{Tang}}}, \bibinfo {author}
  {\bibfnamefont {Y.}~\bibnamefont {{Toba}}}, \bibinfo {author} {\bibfnamefont
  {Y.}~\bibnamefont {{Utsumi}}}, \bibinfo {author} {\bibfnamefont {S.-Y.}\
  \bibnamefont {{Wang}}},\ and\ \bibinfo {author} {\bibfnamefont
  {T.}~\bibnamefont {{Yamashita}}},\ }\href
  {https://doi.org/10.3847/2041-8213/ab0216} {\bibfield  {journal} {\bibinfo
  {journal} {\apjl}\ }\textbf {\bibinfo {volume} {872}},\ \bibinfo {eid} {L2}
  (\bibinfo {year} {2019})},\ \Eprint {https://arxiv.org/abs/1901.10487}
  {arXiv:1901.10487 [astro-ph.GA]} \BibitemShut {NoStop}%
\bibitem [{\citenamefont {{Onoue}}\ \emph {et~al.}(2019)\citenamefont
  {{Onoue}}, \citenamefont {{Kashikawa}}, \citenamefont {{Matsuoka}},
  \citenamefont {{Kato}}, \citenamefont {{Izumi}}, \citenamefont {{Nagao}},
  \citenamefont {{Strauss}}, \citenamefont {{Harikane}}, \citenamefont
  {{Imanishi}}, \citenamefont {{Ito}}, \citenamefont {{Iwasawa}}, \citenamefont
  {{Kawaguchi}}, \citenamefont {{Lee}}, \citenamefont {{Noboriguchi}},
  \citenamefont {{Suh}}, \citenamefont {{Tanaka}},\ and\ \citenamefont
  {{Toba}}}]{Onoue2019}%
  \BibitemOpen
  \bibfield  {author} {\bibinfo {author} {\bibfnamefont {M.}~\bibnamefont
  {{Onoue}}}, \bibinfo {author} {\bibfnamefont {N.}~\bibnamefont
  {{Kashikawa}}}, \bibinfo {author} {\bibfnamefont {Y.}~\bibnamefont
  {{Matsuoka}}}, \bibinfo {author} {\bibfnamefont {N.}~\bibnamefont {{Kato}}},
  \bibinfo {author} {\bibfnamefont {T.}~\bibnamefont {{Izumi}}}, \bibinfo
  {author} {\bibfnamefont {T.}~\bibnamefont {{Nagao}}}, \bibinfo {author}
  {\bibfnamefont {M.~A.}\ \bibnamefont {{Strauss}}}, \bibinfo {author}
  {\bibfnamefont {Y.}~\bibnamefont {{Harikane}}}, \bibinfo {author}
  {\bibfnamefont {M.}~\bibnamefont {{Imanishi}}}, \bibinfo {author}
  {\bibfnamefont {K.}~\bibnamefont {{Ito}}}, \bibinfo {author} {\bibfnamefont
  {K.}~\bibnamefont {{Iwasawa}}}, \bibinfo {author} {\bibfnamefont
  {T.}~\bibnamefont {{Kawaguchi}}}, \bibinfo {author} {\bibfnamefont {C.-H.}\
  \bibnamefont {{Lee}}}, \bibinfo {author} {\bibfnamefont {A.}~\bibnamefont
  {{Noboriguchi}}}, \bibinfo {author} {\bibfnamefont {H.}~\bibnamefont
  {{Suh}}}, \bibinfo {author} {\bibfnamefont {M.}~\bibnamefont {{Tanaka}}},\
  and\ \bibinfo {author} {\bibfnamefont {Y.}~\bibnamefont {{Toba}}},\ }\href
  {https://doi.org/10.3847/1538-4357/ab29e9} {\bibfield  {journal} {\bibinfo
  {journal} {\apj}\ }\textbf {\bibinfo {volume} {880}},\ \bibinfo {eid} {77}
  (\bibinfo {year} {2019})},\ \Eprint {https://arxiv.org/abs/1904.07278}
  {arXiv:1904.07278 [astro-ph.GA]} \BibitemShut {NoStop}%
\bibitem [{\citenamefont {Eilers}\ \emph {et~al.}(2020)\citenamefont {Eilers}
  \emph {et~al.}}]{Eilers:2020htq}%
  \BibitemOpen
  \bibfield  {author} {\bibinfo {author} {\bibfnamefont {A.-C.}\ \bibnamefont
  {Eilers}} \emph {et~al.},\ }\href {https://doi.org/10.3847/1538-4357/aba52e}
  {\bibfield  {journal} {\bibinfo  {journal} {Astrophys. J.}\ }\textbf
  {\bibinfo {volume} {900}},\ \bibinfo {pages} {37} (\bibinfo {year} {2020})},\
  \Eprint {https://arxiv.org/abs/2002.01811} {arXiv:2002.01811 [astro-ph.GA]}
  \BibitemShut {NoStop}%
\bibitem [{\citenamefont {{Eilers}}\ \emph {et~al.}(2017)\citenamefont
  {{Eilers}}, \citenamefont {{Davies}}, \citenamefont {{Hennawi}},
  \citenamefont {{Prochaska}}, \citenamefont {{Luki{\'c}}},\ and\ \citenamefont
  {{Mazzucchelli}}}]{2017ApJ...840...24E}%
  \BibitemOpen
  \bibfield  {author} {\bibinfo {author} {\bibfnamefont {A.-C.}\ \bibnamefont
  {{Eilers}}}, \bibinfo {author} {\bibfnamefont {F.~B.}\ \bibnamefont
  {{Davies}}}, \bibinfo {author} {\bibfnamefont {J.~F.}\ \bibnamefont
  {{Hennawi}}}, \bibinfo {author} {\bibfnamefont {J.~X.}\ \bibnamefont
  {{Prochaska}}}, \bibinfo {author} {\bibfnamefont {Z.}~\bibnamefont
  {{Luki{\'c}}}},\ and\ \bibinfo {author} {\bibfnamefont {C.}~\bibnamefont
  {{Mazzucchelli}}},\ }\href {https://doi.org/10.3847/1538-4357/aa6c60}
  {\bibfield  {journal} {\bibinfo  {journal} {\apj}\ }\textbf {\bibinfo
  {volume} {840}},\ \bibinfo {eid} {24} (\bibinfo {year} {2017})},\ \Eprint
  {https://arxiv.org/abs/1703.02539} {arXiv:1703.02539 [astro-ph.GA]}
  \BibitemShut {NoStop}%
\bibitem [{\citenamefont {Eilers}\ \emph {et~al.}(2018)\citenamefont {Eilers},
  \citenamefont {Hennawi},\ and\ \citenamefont {Davies}}]{Eilers_2018}%
  \BibitemOpen
  \bibfield  {author} {\bibinfo {author} {\bibfnamefont {A.-C.}\ \bibnamefont
  {Eilers}}, \bibinfo {author} {\bibfnamefont {J.~F.}\ \bibnamefont
  {Hennawi}},\ and\ \bibinfo {author} {\bibfnamefont {F.~B.}\ \bibnamefont
  {Davies}},\ }\href {https://doi.org/10.3847/1538-4357/aae081} {\bibfield
  {journal} {\bibinfo  {journal} {The Astrophysical Journal}\ }\textbf
  {\bibinfo {volume} {867}},\ \bibinfo {pages} {30} (\bibinfo {year}
  {2018})}\BibitemShut {NoStop}%
\bibitem [{\citenamefont {Davies}\ \emph {et~al.}(2019)\citenamefont {Davies},
  \citenamefont {Hennawi},\ and\ \citenamefont {Eilers}}]{Davies_2019}%
  \BibitemOpen
  \bibfield  {author} {\bibinfo {author} {\bibfnamefont {F.~B.}\ \bibnamefont
  {Davies}}, \bibinfo {author} {\bibfnamefont {J.~F.}\ \bibnamefont
  {Hennawi}},\ and\ \bibinfo {author} {\bibfnamefont {A.-C.}\ \bibnamefont
  {Eilers}},\ }\href {https://doi.org/10.1093/mnras/stz3303} {\bibfield
  {journal} {\bibinfo  {journal} {Monthly Notices of the Royal Astronomical
  Society}\ }\textbf {\bibinfo {volume} {493}},\ \bibinfo {pages} {1330–1343}
  (\bibinfo {year} {2019})}\BibitemShut {NoStop}%
\bibitem [{\citenamefont {Andika}\ \emph {et~al.}(2020)\citenamefont {Andika},
  \citenamefont {Jahnke}, \citenamefont {Onoue}, \citenamefont {Bañados},
  \citenamefont {Mazzucchelli}, \citenamefont {Novak}, \citenamefont {Eilers},
  \citenamefont {Venemans}, \citenamefont {Schindler}, \citenamefont {Walter},\
  and\ \citenamefont {et~al.}}]{Andika_2020}%
  \BibitemOpen
  \bibfield  {author} {\bibinfo {author} {\bibfnamefont {I.~T.}\ \bibnamefont
  {Andika}}, \bibinfo {author} {\bibfnamefont {K.}~\bibnamefont {Jahnke}},
  \bibinfo {author} {\bibfnamefont {M.}~\bibnamefont {Onoue}}, \bibinfo
  {author} {\bibfnamefont {E.}~\bibnamefont {Bañados}}, \bibinfo {author}
  {\bibfnamefont {C.}~\bibnamefont {Mazzucchelli}}, \bibinfo {author}
  {\bibfnamefont {M.}~\bibnamefont {Novak}}, \bibinfo {author} {\bibfnamefont
  {A.-C.}\ \bibnamefont {Eilers}}, \bibinfo {author} {\bibfnamefont {B.~P.}\
  \bibnamefont {Venemans}}, \bibinfo {author} {\bibfnamefont {J.-T.}\
  \bibnamefont {Schindler}}, \bibinfo {author} {\bibfnamefont {F.}~\bibnamefont
  {Walter}},\ and\ \bibinfo {author} {\bibnamefont {et~al.}},\ }\href
  {https://doi.org/10.3847/1538-4357/abb9a6} {\bibfield  {journal} {\bibinfo
  {journal} {The Astrophysical Journal}\ }\textbf {\bibinfo {volume} {903}},\
  \bibinfo {pages} {34} (\bibinfo {year} {2020})}\BibitemShut {NoStop}%
\bibitem [{\citenamefont {Balberg}\ \emph {et~al.}(2002)\citenamefont
  {Balberg}, \citenamefont {Shapiro},\ and\ \citenamefont
  {Inagaki}}]{Balberg:2002ue}%
  \BibitemOpen
  \bibfield  {author} {\bibinfo {author} {\bibfnamefont {S.}~\bibnamefont
  {Balberg}}, \bibinfo {author} {\bibfnamefont {S.~L.}\ \bibnamefont
  {Shapiro}},\ and\ \bibinfo {author} {\bibfnamefont {S.}~\bibnamefont
  {Inagaki}},\ }\href {https://doi.org/10.1086/339038} {\bibfield  {journal}
  {\bibinfo  {journal} {Astrophys. J.}\ }\textbf {\bibinfo {volume} {568}},\
  \bibinfo {pages} {475} (\bibinfo {year} {2002})},\ \Eprint
  {https://arxiv.org/abs/astro-ph/0110561} {arXiv:astro-ph/0110561}
  \BibitemShut {NoStop}%
\bibitem [{\citenamefont {Pollack}\ \emph {et~al.}(2015)\citenamefont
  {Pollack}, \citenamefont {Spergel},\ and\ \citenamefont
  {Steinhardt}}]{Pollack:2014rja}%
  \BibitemOpen
  \bibfield  {author} {\bibinfo {author} {\bibfnamefont {J.}~\bibnamefont
  {Pollack}}, \bibinfo {author} {\bibfnamefont {D.~N.}\ \bibnamefont
  {Spergel}},\ and\ \bibinfo {author} {\bibfnamefont {P.~J.}\ \bibnamefont
  {Steinhardt}},\ }\href {https://doi.org/10.1088/0004-637X/804/2/131}
  {\bibfield  {journal} {\bibinfo  {journal} {Astrophys. J.}\ }\textbf
  {\bibinfo {volume} {804}},\ \bibinfo {pages} {131} (\bibinfo {year}
  {2015})},\ \Eprint {https://arxiv.org/abs/1501.00017} {arXiv:1501.00017
  [astro-ph.CO]} \BibitemShut {NoStop}%
\bibitem [{\citenamefont {Hu}\ \emph {et~al.}(2006)\citenamefont {Hu},
  \citenamefont {Shen}, \citenamefont {Lou},\ and\ \citenamefont
  {Zhang}}]{Hu_2006}%
  \BibitemOpen
  \bibfield  {author} {\bibinfo {author} {\bibfnamefont {J.}~\bibnamefont
  {Hu}}, \bibinfo {author} {\bibfnamefont {Y.}~\bibnamefont {Shen}}, \bibinfo
  {author} {\bibfnamefont {Y.-Q.}\ \bibnamefont {Lou}},\ and\ \bibinfo {author}
  {\bibfnamefont {S.}~\bibnamefont {Zhang}},\ }\href
  {https://doi.org/10.1111/j.1365-2966.2005.09712.x} {\bibfield  {journal}
  {\bibinfo  {journal} {Monthly Notices of the Royal Astronomical Society}\
  }\textbf {\bibinfo {volume} {365}},\ \bibinfo {pages} {345–351} (\bibinfo
  {year} {2006})}\BibitemShut {NoStop}%
\bibitem [{\citenamefont {Padilla}\ \emph {et~al.}(2020)\citenamefont
  {Padilla}, \citenamefont {Rindler-Daller}, \citenamefont {Shapiro},
  \citenamefont {Matos},\ and\ \citenamefont {Vázquez}}]{padilla2020corehalo}%
  \BibitemOpen
  \bibfield  {author} {\bibinfo {author} {\bibfnamefont {L.~E.}\ \bibnamefont
  {Padilla}}, \bibinfo {author} {\bibfnamefont {T.}~\bibnamefont
  {Rindler-Daller}}, \bibinfo {author} {\bibfnamefont {P.~R.}\ \bibnamefont
  {Shapiro}}, \bibinfo {author} {\bibfnamefont {T.}~\bibnamefont {Matos}},\
  and\ \bibinfo {author} {\bibfnamefont {J.~A.}\ \bibnamefont {Vázquez}},\
  }\href@noop {} {\bibinfo {title} {On the core-halo mass relation in scalar
  field dark matter models and its consequences for the formation of
  supermassive black holes}} (\bibinfo {year} {2020}),\ \Eprint
  {https://arxiv.org/abs/2010.12716} {arXiv:2010.12716 [astro-ph.GA]}
  \BibitemShut {NoStop}%
\bibitem [{\citenamefont {{Koda}}\ and\ \citenamefont
  {{Shapiro}}(2011)}]{Koda2011}%
  \BibitemOpen
  \bibfield  {author} {\bibinfo {author} {\bibfnamefont {J.}~\bibnamefont
  {{Koda}}}\ and\ \bibinfo {author} {\bibfnamefont {P.~R.}\ \bibnamefont
  {{Shapiro}}},\ }\href {https://doi.org/10.1111/j.1365-2966.2011.18684.x}
  {\bibfield  {journal} {\bibinfo  {journal} {\mnras}\ }\textbf {\bibinfo
  {volume} {415}},\ \bibinfo {pages} {1125} (\bibinfo {year} {2011})},\ \Eprint
  {https://arxiv.org/abs/1101.3097} {arXiv:1101.3097 [astro-ph.CO]}
  \BibitemShut {NoStop}%
\bibitem [{\citenamefont {{Lynden-Bell}}\ and\ \citenamefont
  {{Wood}}(1968)}]{Bell1968}%
  \BibitemOpen
  \bibfield  {author} {\bibinfo {author} {\bibfnamefont {D.}~\bibnamefont
  {{Lynden-Bell}}}\ and\ \bibinfo {author} {\bibfnamefont {R.}~\bibnamefont
  {{Wood}}},\ }\href {https://doi.org/10.1093/mnras/138.4.495} {\bibfield
  {journal} {\bibinfo  {journal} {\mnras}\ }\textbf {\bibinfo {volume} {138}},\
  \bibinfo {pages} {495} (\bibinfo {year} {1968})}\BibitemShut {NoStop}%
\bibitem [{\citenamefont {{Lynden-Bell}}\ and\ \citenamefont
  {{Eggleton}}(1980)}]{Bell1980}%
  \BibitemOpen
  \bibfield  {author} {\bibinfo {author} {\bibfnamefont {D.}~\bibnamefont
  {{Lynden-Bell}}}\ and\ \bibinfo {author} {\bibfnamefont {P.~P.}\ \bibnamefont
  {{Eggleton}}},\ }\href {https://doi.org/10.1093/mnras/191.3.483} {\bibfield
  {journal} {\bibinfo  {journal} {\mnras}\ }\textbf {\bibinfo {volume} {191}},\
  \bibinfo {pages} {483} (\bibinfo {year} {1980})}\BibitemShut {NoStop}%
\bibitem [{\citenamefont {{Burkert}}(2000)}]{Burkert2000}%
  \BibitemOpen
  \bibfield  {author} {\bibinfo {author} {\bibfnamefont {A.}~\bibnamefont
  {{Burkert}}},\ }\href {https://doi.org/10.1086/312674} {\bibfield  {journal}
  {\bibinfo  {journal} {\apjl}\ }\textbf {\bibinfo {volume} {534}},\ \bibinfo
  {pages} {L143} (\bibinfo {year} {2000})},\ \Eprint
  {https://arxiv.org/abs/astro-ph/0002409} {arXiv:astro-ph/0002409 [astro-ph]}
  \BibitemShut {NoStop}%
\bibitem [{\citenamefont {{Kochanek}}\ and\ \citenamefont
  {{White}}(2000)}]{Kochanek2000}%
  \BibitemOpen
  \bibfield  {author} {\bibinfo {author} {\bibfnamefont {C.~S.}\ \bibnamefont
  {{Kochanek}}}\ and\ \bibinfo {author} {\bibfnamefont {M.}~\bibnamefont
  {{White}}},\ }\href {https://doi.org/10.1086/317149} {\bibfield  {journal}
  {\bibinfo  {journal} {\apj}\ }\textbf {\bibinfo {volume} {543}},\ \bibinfo
  {pages} {514} (\bibinfo {year} {2000})},\ \Eprint
  {https://arxiv.org/abs/astro-ph/0003483} {arXiv:astro-ph/0003483 [astro-ph]}
  \BibitemShut {NoStop}%
\bibitem [{\citenamefont {{Col{\'\i}n}}\ \emph {et~al.}(2002)\citenamefont
  {{Col{\'\i}n}}, \citenamefont {{Avila-Reese}}, \citenamefont {{Valenzuela}},\
  and\ \citenamefont {{Firmani}}}]{Colin2002}%
  \BibitemOpen
  \bibfield  {author} {\bibinfo {author} {\bibfnamefont {P.}~\bibnamefont
  {{Col{\'\i}n}}}, \bibinfo {author} {\bibfnamefont {V.}~\bibnamefont
  {{Avila-Reese}}}, \bibinfo {author} {\bibfnamefont {O.}~\bibnamefont
  {{Valenzuela}}},\ and\ \bibinfo {author} {\bibfnamefont {C.}~\bibnamefont
  {{Firmani}}},\ }\href {https://doi.org/10.1086/344259} {\bibfield  {journal}
  {\bibinfo  {journal} {\apj}\ }\textbf {\bibinfo {volume} {581}},\ \bibinfo
  {pages} {777} (\bibinfo {year} {2002})},\ \Eprint
  {https://arxiv.org/abs/astro-ph/0205322} {arXiv:astro-ph/0205322 [astro-ph]}
  \BibitemShut {NoStop}%
\bibitem [{\citenamefont {{Vogelsberger}}\ \emph {et~al.}(2012)\citenamefont
  {{Vogelsberger}}, \citenamefont {{Zavala}},\ and\ \citenamefont
  {{Loeb}}}]{Vogelsberger2012}%
  \BibitemOpen
  \bibfield  {author} {\bibinfo {author} {\bibfnamefont {M.}~\bibnamefont
  {{Vogelsberger}}}, \bibinfo {author} {\bibfnamefont {J.}~\bibnamefont
  {{Zavala}}},\ and\ \bibinfo {author} {\bibfnamefont {A.}~\bibnamefont
  {{Loeb}}},\ }\href {https://doi.org/10.1111/j.1365-2966.2012.21182.x}
  {\bibfield  {journal} {\bibinfo  {journal} {\mnras}\ }\textbf {\bibinfo
  {volume} {423}},\ \bibinfo {pages} {3740} (\bibinfo {year} {2012})},\ \Eprint
  {https://arxiv.org/abs/1201.5892} {arXiv:1201.5892 [astro-ph.CO]}
  \BibitemShut {NoStop}%
\bibitem [{\citenamefont {Randall}\ \emph {et~al.}(2008)\citenamefont
  {Randall}, \citenamefont {Markevitch}, \citenamefont {Clowe}, \citenamefont
  {Gonzalez},\ and\ \citenamefont {Bradač}}]{Randall_2008}%
  \BibitemOpen
  \bibfield  {author} {\bibinfo {author} {\bibfnamefont {S.~W.}\ \bibnamefont
  {Randall}}, \bibinfo {author} {\bibfnamefont {M.}~\bibnamefont {Markevitch}},
  \bibinfo {author} {\bibfnamefont {D.}~\bibnamefont {Clowe}}, \bibinfo
  {author} {\bibfnamefont {A.~H.}\ \bibnamefont {Gonzalez}},\ and\ \bibinfo
  {author} {\bibfnamefont {M.}~\bibnamefont {Bradač}},\ }\href
  {https://doi.org/10.1086/587859} {\bibfield  {journal} {\bibinfo  {journal}
  {The Astrophysical Journal}\ }\textbf {\bibinfo {volume} {679}},\ \bibinfo
  {pages} {1173–1180} (\bibinfo {year} {2008})}\BibitemShut {NoStop}%
\bibitem [{\citenamefont {Choquette}\ \emph {et~al.}(2019)\citenamefont
  {Choquette}, \citenamefont {Cline},\ and\ \citenamefont
  {Cornell}}]{Choquette:2018lvq}%
  \BibitemOpen
  \bibfield  {author} {\bibinfo {author} {\bibfnamefont {J.}~\bibnamefont
  {Choquette}}, \bibinfo {author} {\bibfnamefont {J.~M.}\ \bibnamefont
  {Cline}},\ and\ \bibinfo {author} {\bibfnamefont {J.~M.}\ \bibnamefont
  {Cornell}},\ }\href {https://doi.org/10.1088/1475-7516/2019/07/036}
  {\bibfield  {journal} {\bibinfo  {journal} {JCAP}\ }\textbf {\bibinfo
  {volume} {07}},\ \bibinfo {pages} {036}},\ \Eprint
  {https://arxiv.org/abs/1812.05088} {arXiv:1812.05088 [astro-ph.CO]}
  \BibitemShut {NoStop}%
\bibitem [{\citenamefont {Feng}\ \emph {et~al.}(2020)\citenamefont {Feng},
  \citenamefont {Yu},\ and\ \citenamefont {Zhong}}]{feng2020seeding}%
  \BibitemOpen
  \bibfield  {author} {\bibinfo {author} {\bibfnamefont {W.-X.}\ \bibnamefont
  {Feng}}, \bibinfo {author} {\bibfnamefont {H.-B.}\ \bibnamefont {Yu}},\ and\
  \bibinfo {author} {\bibfnamefont {Y.-M.}\ \bibnamefont {Zhong}},\ }\href@noop
  {} {\bibinfo {title} {Seeding supermassive black holes with self-interacting
  dark matter}} (\bibinfo {year} {2020}),\ \Eprint
  {https://arxiv.org/abs/2010.15132} {arXiv:2010.15132 [astro-ph.CO]}
  \BibitemShut {NoStop}%
\bibitem [{\citenamefont {Huo}\ \emph {et~al.}(2020)\citenamefont {Huo},
  \citenamefont {Yu},\ and\ \citenamefont {Zhong}}]{Huo:2019yhk}%
  \BibitemOpen
  \bibfield  {author} {\bibinfo {author} {\bibfnamefont {R.}~\bibnamefont
  {Huo}}, \bibinfo {author} {\bibfnamefont {H.-B.}\ \bibnamefont {Yu}},\ and\
  \bibinfo {author} {\bibfnamefont {Y.-M.}\ \bibnamefont {Zhong}},\ }\href
  {https://doi.org/10.1088/1475-7516/2020/06/051} {\bibfield  {journal}
  {\bibinfo  {journal} {JCAP}\ }\textbf {\bibinfo {volume} {06}},\ \bibinfo
  {pages} {051}},\ \Eprint {https://arxiv.org/abs/1912.06757} {arXiv:1912.06757
  [astro-ph.CO]} \BibitemShut {NoStop}%
\bibitem [{\citenamefont {{Essig}}\ \emph {et~al.}(2019)\citenamefont
  {{Essig}}, \citenamefont {{McDermott}}, \citenamefont {{Yu}},\ and\
  \citenamefont {{Zhong}}}]{Essig2019}%
  \BibitemOpen
  \bibfield  {author} {\bibinfo {author} {\bibfnamefont {R.}~\bibnamefont
  {{Essig}}}, \bibinfo {author} {\bibfnamefont {S.~D.}\ \bibnamefont
  {{McDermott}}}, \bibinfo {author} {\bibfnamefont {H.-B.}\ \bibnamefont
  {{Yu}}},\ and\ \bibinfo {author} {\bibfnamefont {Y.-M.}\ \bibnamefont
  {{Zhong}}},\ }\href {https://doi.org/10.1103/PhysRevLett.123.121102}
  {\bibfield  {journal} {\bibinfo  {journal} {\prl}\ }\textbf {\bibinfo
  {volume} {123}},\ \bibinfo {eid} {121102} (\bibinfo {year} {2019})},\ \Eprint
  {https://arxiv.org/abs/1809.01144} {arXiv:1809.01144 [hep-ph]} \BibitemShut
  {NoStop}%
\bibitem [{\citenamefont {Gresham}\ \emph
  {et~al.}(2018{\natexlab{a}})\citenamefont {Gresham}, \citenamefont {Lou},\
  and\ \citenamefont {Zurek}}]{Gresham:2018anj}%
  \BibitemOpen
  \bibfield  {author} {\bibinfo {author} {\bibfnamefont {M.~I.}\ \bibnamefont
  {Gresham}}, \bibinfo {author} {\bibfnamefont {H.~K.}\ \bibnamefont {Lou}},\
  and\ \bibinfo {author} {\bibfnamefont {K.~M.}\ \bibnamefont {Zurek}},\ }\href
  {https://doi.org/10.1103/PhysRevD.98.096001} {\bibfield  {journal} {\bibinfo
  {journal} {Phys. Rev. D}\ }\textbf {\bibinfo {volume} {98}},\ \bibinfo
  {pages} {096001} (\bibinfo {year} {2018}{\natexlab{a}})},\ \Eprint
  {https://arxiv.org/abs/1805.04512} {arXiv:1805.04512 [hep-ph]} \BibitemShut
  {NoStop}%
\bibitem [{\citenamefont {Gresham}\ \emph
  {et~al.}(2018{\natexlab{b}})\citenamefont {Gresham}, \citenamefont {Lou},\
  and\ \citenamefont {Zurek}}]{Gresham:2017cvl}%
  \BibitemOpen
  \bibfield  {author} {\bibinfo {author} {\bibfnamefont {M.~I.}\ \bibnamefont
  {Gresham}}, \bibinfo {author} {\bibfnamefont {H.~K.}\ \bibnamefont {Lou}},\
  and\ \bibinfo {author} {\bibfnamefont {K.~M.}\ \bibnamefont {Zurek}},\ }\href
  {https://doi.org/10.1103/PhysRevD.97.036003} {\bibfield  {journal} {\bibinfo
  {journal} {Phys. Rev. D}\ }\textbf {\bibinfo {volume} {97}},\ \bibinfo
  {pages} {036003} (\bibinfo {year} {2018}{\natexlab{b}})},\ \Eprint
  {https://arxiv.org/abs/1707.02316} {arXiv:1707.02316 [hep-ph]} \BibitemShut
  {NoStop}%
\bibitem [{\citenamefont {Gresham}\ \emph {et~al.}(2017)\citenamefont
  {Gresham}, \citenamefont {Lou},\ and\ \citenamefont
  {Zurek}}]{Gresham:2017zqi}%
  \BibitemOpen
  \bibfield  {author} {\bibinfo {author} {\bibfnamefont {M.~I.}\ \bibnamefont
  {Gresham}}, \bibinfo {author} {\bibfnamefont {H.~K.}\ \bibnamefont {Lou}},\
  and\ \bibinfo {author} {\bibfnamefont {K.~M.}\ \bibnamefont {Zurek}},\ }\href
  {https://doi.org/10.1103/PhysRevD.96.096012} {\bibfield  {journal} {\bibinfo
  {journal} {Phys. Rev. D}\ }\textbf {\bibinfo {volume} {96}},\ \bibinfo
  {pages} {096012} (\bibinfo {year} {2017})},\ \Eprint
  {https://arxiv.org/abs/1707.02313} {arXiv:1707.02313 [hep-ph]} \BibitemShut
  {NoStop}%
\bibitem [{\citenamefont {Wise}\ and\ \citenamefont
  {Zhang}(2014)}]{Wise:2014jva}%
  \BibitemOpen
  \bibfield  {author} {\bibinfo {author} {\bibfnamefont {M.~B.}\ \bibnamefont
  {Wise}}\ and\ \bibinfo {author} {\bibfnamefont {Y.}~\bibnamefont {Zhang}},\
  }\href {https://doi.org/10.1103/PhysRevD.90.055030} {\bibfield  {journal}
  {\bibinfo  {journal} {Phys. Rev. D}\ }\textbf {\bibinfo {volume} {90}},\
  \bibinfo {pages} {055030} (\bibinfo {year} {2014})},\ \bibinfo {note}
  {[Erratum: Phys.Rev.D 91, 039907 (2015)]},\ \Eprint
  {https://arxiv.org/abs/1407.4121} {arXiv:1407.4121 [hep-ph]} \BibitemShut
  {NoStop}%
\bibitem [{\citenamefont {Wise}\ and\ \citenamefont
  {Zhang}(2015)}]{Wise:2014ola}%
  \BibitemOpen
  \bibfield  {author} {\bibinfo {author} {\bibfnamefont {M.~B.}\ \bibnamefont
  {Wise}}\ and\ \bibinfo {author} {\bibfnamefont {Y.}~\bibnamefont {Zhang}},\
  }\href {https://doi.org/10.1007/JHEP02(2015)023} {\bibfield  {journal}
  {\bibinfo  {journal} {JHEP}\ }\textbf {\bibinfo {volume} {02}},\ \bibinfo
  {pages} {023}},\ \bibinfo {note} {[Erratum: JHEP 10, 165 (2015)]},\ \Eprint
  {https://arxiv.org/abs/1411.1772} {arXiv:1411.1772 [hep-ph]} \BibitemShut
  {NoStop}%
\bibitem [{\citenamefont {Fan}\ \emph {et~al.}(2013{\natexlab{a}})\citenamefont
  {Fan}, \citenamefont {Katz}, \citenamefont {Randall},\ and\ \citenamefont
  {Reece}}]{Fan:2013tia}%
  \BibitemOpen
  \bibfield  {author} {\bibinfo {author} {\bibfnamefont {J.}~\bibnamefont
  {Fan}}, \bibinfo {author} {\bibfnamefont {A.}~\bibnamefont {Katz}}, \bibinfo
  {author} {\bibfnamefont {L.}~\bibnamefont {Randall}},\ and\ \bibinfo {author}
  {\bibfnamefont {M.}~\bibnamefont {Reece}},\ }\href
  {https://doi.org/10.1103/PhysRevLett.110.211302} {\bibfield  {journal}
  {\bibinfo  {journal} {Phys. Rev. Lett.}\ }\textbf {\bibinfo {volume} {110}},\
  \bibinfo {pages} {211302} (\bibinfo {year} {2013}{\natexlab{a}})},\ \Eprint
  {https://arxiv.org/abs/1303.3271} {arXiv:1303.3271 [hep-ph]} \BibitemShut
  {NoStop}%
\bibitem [{\citenamefont {Fan}\ \emph {et~al.}(2013{\natexlab{b}})\citenamefont
  {Fan}, \citenamefont {Katz}, \citenamefont {Randall},\ and\ \citenamefont
  {Reece}}]{Fan:2013yva}%
  \BibitemOpen
  \bibfield  {author} {\bibinfo {author} {\bibfnamefont {J.}~\bibnamefont
  {Fan}}, \bibinfo {author} {\bibfnamefont {A.}~\bibnamefont {Katz}}, \bibinfo
  {author} {\bibfnamefont {L.}~\bibnamefont {Randall}},\ and\ \bibinfo {author}
  {\bibfnamefont {M.}~\bibnamefont {Reece}},\ }\href
  {https://doi.org/10.1016/j.dark.2013.07.001} {\bibfield  {journal} {\bibinfo
  {journal} {Phys. Dark Univ.}\ }\textbf {\bibinfo {volume} {2}},\ \bibinfo
  {pages} {139} (\bibinfo {year} {2013}{\natexlab{b}})},\ \Eprint
  {https://arxiv.org/abs/1303.1521} {arXiv:1303.1521 [astro-ph.CO]}
  \BibitemShut {NoStop}%
\bibitem [{\citenamefont {Foot}\ and\ \citenamefont
  {Vagnozzi}(2015)}]{Foot_2015}%
  \BibitemOpen
  \bibfield  {author} {\bibinfo {author} {\bibfnamefont {R.}~\bibnamefont
  {Foot}}\ and\ \bibinfo {author} {\bibfnamefont {S.}~\bibnamefont
  {Vagnozzi}},\ }\bibfield  {journal} {\bibinfo  {journal} {Physical Review D}\
  }\textbf {\bibinfo {volume} {91}},\ \href
  {https://doi.org/10.1103/physrevd.91.023512} {10.1103/physrevd.91.023512}
  (\bibinfo {year} {2015})\BibitemShut {NoStop}%
\bibitem [{\citenamefont {Boddy}\ \emph {et~al.}(2014)\citenamefont {Boddy},
  \citenamefont {Feng}, \citenamefont {Kaplinghat},\ and\ \citenamefont
  {Tait}}]{Boddy_2014}%
  \BibitemOpen
  \bibfield  {author} {\bibinfo {author} {\bibfnamefont {K.~K.}\ \bibnamefont
  {Boddy}}, \bibinfo {author} {\bibfnamefont {J.~L.}\ \bibnamefont {Feng}},
  \bibinfo {author} {\bibfnamefont {M.}~\bibnamefont {Kaplinghat}},\ and\
  \bibinfo {author} {\bibfnamefont {T.~M.}\ \bibnamefont {Tait}},\ }\bibfield
  {journal} {\bibinfo  {journal} {Physical Review D}\ }\textbf {\bibinfo
  {volume} {89}},\ \href {https://doi.org/10.1103/physrevd.89.115017}
  {10.1103/physrevd.89.115017} (\bibinfo {year} {2014})\BibitemShut {NoStop}%
\bibitem [{\citenamefont {Kaplan}\ \emph {et~al.}(2010)\citenamefont {Kaplan},
  \citenamefont {Krnjaic}, \citenamefont {Rehermann},\ and\ \citenamefont
  {Wells}}]{Kaplan_2010}%
  \BibitemOpen
  \bibfield  {author} {\bibinfo {author} {\bibfnamefont {D.~E.}\ \bibnamefont
  {Kaplan}}, \bibinfo {author} {\bibfnamefont {G.~Z.}\ \bibnamefont {Krnjaic}},
  \bibinfo {author} {\bibfnamefont {K.~R.}\ \bibnamefont {Rehermann}},\ and\
  \bibinfo {author} {\bibfnamefont {C.~M.}\ \bibnamefont {Wells}},\ }\href
  {https://doi.org/10.1088/1475-7516/2010/05/021} {\bibfield  {journal}
  {\bibinfo  {journal} {Journal of Cosmology and Astroparticle Physics}\
  }\textbf {\bibinfo {volume} {2010}}\bibinfo  {number} { (05)},\ \bibinfo
  {pages} {021–021}}\BibitemShut {NoStop}%
\bibitem [{\citenamefont {Cyr-Racine}\ and\ \citenamefont
  {Sigurdson}(2013)}]{CyrRacine:2012fz}%
  \BibitemOpen
\bibfield  {number} {  }\bibfield  {author} {\bibinfo {author} {\bibfnamefont
  {F.-Y.}\ \bibnamefont {Cyr-Racine}}\ and\ \bibinfo {author} {\bibfnamefont
  {K.}~\bibnamefont {Sigurdson}},\ }\href
  {https://doi.org/10.1103/PhysRevD.87.103515} {\bibfield  {journal} {\bibinfo
  {journal} {Phys. Rev. D}\ }\textbf {\bibinfo {volume} {87}},\ \bibinfo
  {pages} {103515} (\bibinfo {year} {2013})},\ \Eprint
  {https://arxiv.org/abs/1209.5752} {arXiv:1209.5752 [astro-ph.CO]}
  \BibitemShut {NoStop}%
\bibitem [{\citenamefont {Cline}\ \emph {et~al.}(2014)\citenamefont {Cline},
  \citenamefont {Liu}, \citenamefont {Moore},\ and\ \citenamefont
  {Xue}}]{Cline:2013pca}%
  \BibitemOpen
  \bibfield  {author} {\bibinfo {author} {\bibfnamefont {J.~M.}\ \bibnamefont
  {Cline}}, \bibinfo {author} {\bibfnamefont {Z.}~\bibnamefont {Liu}}, \bibinfo
  {author} {\bibfnamefont {G.}~\bibnamefont {Moore}},\ and\ \bibinfo {author}
  {\bibfnamefont {W.}~\bibnamefont {Xue}},\ }\href
  {https://doi.org/10.1103/PhysRevD.89.043514} {\bibfield  {journal} {\bibinfo
  {journal} {Phys. Rev. D}\ }\textbf {\bibinfo {volume} {89}},\ \bibinfo
  {pages} {043514} (\bibinfo {year} {2014})},\ \Eprint
  {https://arxiv.org/abs/1311.6468} {arXiv:1311.6468 [hep-ph]} \BibitemShut
  {NoStop}%
\bibitem [{\citenamefont {Finkbeiner}\ and\ \citenamefont
  {Weiner}(2016)}]{Finkbeiner:2014sja}%
  \BibitemOpen
  \bibfield  {author} {\bibinfo {author} {\bibfnamefont {D.~P.}\ \bibnamefont
  {Finkbeiner}}\ and\ \bibinfo {author} {\bibfnamefont {N.}~\bibnamefont
  {Weiner}},\ }\href {https://doi.org/10.1103/PhysRevD.94.083002} {\bibfield
  {journal} {\bibinfo  {journal} {Phys. Rev. D}\ }\textbf {\bibinfo {volume}
  {94}},\ \bibinfo {pages} {083002} (\bibinfo {year} {2016})},\ \Eprint
  {https://arxiv.org/abs/1402.6671} {arXiv:1402.6671 [hep-ph]} \BibitemShut
  {NoStop}%
\bibitem [{\citenamefont {Boddy}\ \emph {et~al.}(2016)\citenamefont {Boddy},
  \citenamefont {Kaplinghat}, \citenamefont {Kwa},\ and\ \citenamefont
  {Peter}}]{Boddy:2016bbu}%
  \BibitemOpen
  \bibfield  {author} {\bibinfo {author} {\bibfnamefont {K.~K.}\ \bibnamefont
  {Boddy}}, \bibinfo {author} {\bibfnamefont {M.}~\bibnamefont {Kaplinghat}},
  \bibinfo {author} {\bibfnamefont {A.}~\bibnamefont {Kwa}},\ and\ \bibinfo
  {author} {\bibfnamefont {A.~H.~G.}\ \bibnamefont {Peter}},\ }\href
  {https://doi.org/10.1103/PhysRevD.94.123017} {\bibfield  {journal} {\bibinfo
  {journal} {Phys. Rev. D}\ }\textbf {\bibinfo {volume} {94}},\ \bibinfo
  {pages} {123017} (\bibinfo {year} {2016})},\ \Eprint
  {https://arxiv.org/abs/1609.03592} {arXiv:1609.03592 [hep-ph]} \BibitemShut
  {NoStop}%
\bibitem [{\citenamefont {Schutz}\ and\ \citenamefont
  {Slatyer}(2015)}]{Schutz:2014nka}%
  \BibitemOpen
  \bibfield  {author} {\bibinfo {author} {\bibfnamefont {K.}~\bibnamefont
  {Schutz}}\ and\ \bibinfo {author} {\bibfnamefont {T.~R.}\ \bibnamefont
  {Slatyer}},\ }\href {https://doi.org/10.1088/1475-7516/2015/01/021}
  {\bibfield  {journal} {\bibinfo  {journal} {JCAP}\ }\textbf {\bibinfo
  {volume} {01}},\ \bibinfo {pages} {021}},\ \Eprint
  {https://arxiv.org/abs/1409.2867} {arXiv:1409.2867 [hep-ph]} \BibitemShut
  {NoStop}%
\bibitem [{\citenamefont {Das}\ and\ \citenamefont
  {Dasgupta}(2018)}]{Das:2017fyl}%
  \BibitemOpen
  \bibfield  {author} {\bibinfo {author} {\bibfnamefont {A.}~\bibnamefont
  {Das}}\ and\ \bibinfo {author} {\bibfnamefont {B.}~\bibnamefont {Dasgupta}},\
  }\href {https://doi.org/10.1103/PhysRevD.97.023002} {\bibfield  {journal}
  {\bibinfo  {journal} {Phys. Rev. D}\ }\textbf {\bibinfo {volume} {97}},\
  \bibinfo {pages} {023002} (\bibinfo {year} {2018})},\ \Eprint
  {https://arxiv.org/abs/1709.06577} {arXiv:1709.06577 [hep-ph]} \BibitemShut
  {NoStop}%
\bibitem [{\citenamefont {Jo}\ \emph {et~al.}(2020)\citenamefont {Jo},
  \citenamefont {Kim}, \citenamefont {Kim},\ and\ \citenamefont
  {Shin}}]{jo2020exploring}%
  \BibitemOpen
  \bibfield  {author} {\bibinfo {author} {\bibfnamefont {B.}~\bibnamefont
  {Jo}}, \bibinfo {author} {\bibfnamefont {H.}~\bibnamefont {Kim}}, \bibinfo
  {author} {\bibfnamefont {H.~D.}\ \bibnamefont {Kim}},\ and\ \bibinfo {author}
  {\bibfnamefont {C.~S.}\ \bibnamefont {Shin}},\ }\href@noop {} {\bibinfo
  {title} {Exploring the universe with dark light scalars}} (\bibinfo {year}
  {2020}),\ \Eprint {https://arxiv.org/abs/2010.10880} {arXiv:2010.10880
  [hep-ph]} \BibitemShut {NoStop}%
\bibitem [{\citenamefont {Hopkins}(2015)}]{Hopkins_2015}%
  \BibitemOpen
  \bibfield  {author} {\bibinfo {author} {\bibfnamefont {P.~F.}\ \bibnamefont
  {Hopkins}},\ }\href {https://doi.org/10.1093/mnras/stv195} {\bibfield
  {journal} {\bibinfo  {journal} {Monthly Notices of the Royal Astronomical
  Society}\ }\textbf {\bibinfo {volume} {450}},\ \bibinfo {pages} {53–110}
  (\bibinfo {year} {2015})}\BibitemShut {NoStop}%
\bibitem [{\citenamefont {{Herpich}}\ \emph {et~al.}(2017)\citenamefont
  {{Herpich}}, \citenamefont {{Stinson}}, \citenamefont {{Rix}}, \citenamefont
  {{Martig}},\ and\ \citenamefont {{Dutton}}}]{2017MNRAS.470.4941H}%
  \BibitemOpen
  \bibfield  {author} {\bibinfo {author} {\bibfnamefont {J.}~\bibnamefont
  {{Herpich}}}, \bibinfo {author} {\bibfnamefont {G.~S.}\ \bibnamefont
  {{Stinson}}}, \bibinfo {author} {\bibfnamefont {H.~W.}\ \bibnamefont
  {{Rix}}}, \bibinfo {author} {\bibfnamefont {M.}~\bibnamefont {{Martig}}},\
  and\ \bibinfo {author} {\bibfnamefont {A.~A.}\ \bibnamefont {{Dutton}}},\
  }\href {https://doi.org/10.1093/mnras/stx1511} {\bibfield  {journal}
  {\bibinfo  {journal} {\mnras}\ }\textbf {\bibinfo {volume} {470}},\ \bibinfo
  {pages} {4941} (\bibinfo {year} {2017})},\ \Eprint
  {https://arxiv.org/abs/1511.04442} {arXiv:1511.04442 [astro-ph.GA]}
  \BibitemShut {NoStop}%
\bibitem [{\citenamefont {{Springel}}(2005)}]{Springel2005}%
  \BibitemOpen
  \bibfield  {author} {\bibinfo {author} {\bibfnamefont {V.}~\bibnamefont
  {{Springel}}},\ }\href {https://doi.org/10.1111/j.1365-2966.2005.09655.x}
  {\bibfield  {journal} {\bibinfo  {journal} {\mnras}\ }\textbf {\bibinfo
  {volume} {364}},\ \bibinfo {pages} {1105} (\bibinfo {year} {2005})},\ \Eprint
  {https://arxiv.org/abs/astro-ph/0505010} {arXiv:astro-ph/0505010 [astro-ph]}
  \BibitemShut {NoStop}%
\bibitem [{\citenamefont {{Rocha}}\ \emph {et~al.}(2013)\citenamefont
  {{Rocha}}, \citenamefont {{Peter}}, \citenamefont {{Bullock}}, \citenamefont
  {{Kaplinghat}}, \citenamefont {{Garrison-Kimmel}}, \citenamefont
  {{O{\~n}orbe}},\ and\ \citenamefont {{Moustakas}}}]{Rocha2013}%
  \BibitemOpen
  \bibfield  {author} {\bibinfo {author} {\bibfnamefont {M.}~\bibnamefont
  {{Rocha}}}, \bibinfo {author} {\bibfnamefont {A.~H.~G.}\ \bibnamefont
  {{Peter}}}, \bibinfo {author} {\bibfnamefont {J.~S.}\ \bibnamefont
  {{Bullock}}}, \bibinfo {author} {\bibfnamefont {M.}~\bibnamefont
  {{Kaplinghat}}}, \bibinfo {author} {\bibfnamefont {S.}~\bibnamefont
  {{Garrison-Kimmel}}}, \bibinfo {author} {\bibfnamefont {J.}~\bibnamefont
  {{O{\~n}orbe}}},\ and\ \bibinfo {author} {\bibfnamefont {L.~A.}\ \bibnamefont
  {{Moustakas}}},\ }\href {https://doi.org/10.1093/mnras/sts514} {\bibfield
  {journal} {\bibinfo  {journal} {\mnras}\ }\textbf {\bibinfo {volume} {430}},\
  \bibinfo {pages} {81} (\bibinfo {year} {2013})},\ \Eprint
  {https://arxiv.org/abs/1208.3025} {arXiv:1208.3025 [astro-ph.CO]}
  \BibitemShut {NoStop}%
\bibitem [{\citenamefont {{Power}}\ \emph {et~al.}(2003)\citenamefont
  {{Power}}, \citenamefont {{Navarro}}, \citenamefont {{Jenkins}},
  \citenamefont {{Frenk}}, \citenamefont {{White}}, \citenamefont {{Springel}},
  \citenamefont {{Stadel}},\ and\ \citenamefont {{Quinn}}}]{Power2003}%
  \BibitemOpen
  \bibfield  {author} {\bibinfo {author} {\bibfnamefont {C.}~\bibnamefont
  {{Power}}}, \bibinfo {author} {\bibfnamefont {J.~F.}\ \bibnamefont
  {{Navarro}}}, \bibinfo {author} {\bibfnamefont {A.}~\bibnamefont
  {{Jenkins}}}, \bibinfo {author} {\bibfnamefont {C.~S.}\ \bibnamefont
  {{Frenk}}}, \bibinfo {author} {\bibfnamefont {S.~D.~M.}\ \bibnamefont
  {{White}}}, \bibinfo {author} {\bibfnamefont {V.}~\bibnamefont {{Springel}}},
  \bibinfo {author} {\bibfnamefont {J.}~\bibnamefont {{Stadel}}},\ and\
  \bibinfo {author} {\bibfnamefont {T.}~\bibnamefont {{Quinn}}},\ }\href
  {https://doi.org/10.1046/j.1365-8711.2003.05925.x} {\bibfield  {journal}
  {\bibinfo  {journal} {\mnras}\ }\textbf {\bibinfo {volume} {338}},\ \bibinfo
  {pages} {14} (\bibinfo {year} {2003})},\ \Eprint
  {https://arxiv.org/abs/astro-ph/0201544} {arXiv:astro-ph/0201544 [astro-ph]}
  \BibitemShut {NoStop}%
\bibitem [{\citenamefont {{Boylan-Kolchin}}\ \emph {et~al.}(2009)\citenamefont
  {{Boylan-Kolchin}}, \citenamefont {{Springel}}, \citenamefont {{White}},
  \citenamefont {{Jenkins}},\ and\ \citenamefont {{Lemson}}}]{MBK2009}%
  \BibitemOpen
  \bibfield  {author} {\bibinfo {author} {\bibfnamefont {M.}~\bibnamefont
  {{Boylan-Kolchin}}}, \bibinfo {author} {\bibfnamefont {V.}~\bibnamefont
  {{Springel}}}, \bibinfo {author} {\bibfnamefont {S.~D.~M.}\ \bibnamefont
  {{White}}}, \bibinfo {author} {\bibfnamefont {A.}~\bibnamefont {{Jenkins}}},\
  and\ \bibinfo {author} {\bibfnamefont {G.}~\bibnamefont {{Lemson}}},\ }\href
  {https://doi.org/10.1111/j.1365-2966.2009.15191.x} {\bibfield  {journal}
  {\bibinfo  {journal} {\mnras}\ }\textbf {\bibinfo {volume} {398}},\ \bibinfo
  {pages} {1150} (\bibinfo {year} {2009})},\ \Eprint
  {https://arxiv.org/abs/0903.3041} {arXiv:0903.3041 [astro-ph.CO]}
  \BibitemShut {NoStop}%
\bibitem [{\citenamefont {{Klypin}}\ \emph {et~al.}(2011)\citenamefont
  {{Klypin}}, \citenamefont {{Trujillo-Gomez}},\ and\ \citenamefont
  {{Primack}}}]{Klypin2011}%
  \BibitemOpen
  \bibfield  {author} {\bibinfo {author} {\bibfnamefont {A.~A.}\ \bibnamefont
  {{Klypin}}}, \bibinfo {author} {\bibfnamefont {S.}~\bibnamefont
  {{Trujillo-Gomez}}},\ and\ \bibinfo {author} {\bibfnamefont {J.}~\bibnamefont
  {{Primack}}},\ }\href {https://doi.org/10.1088/0004-637X/740/2/102}
  {\bibfield  {journal} {\bibinfo  {journal} {\apj}\ }\textbf {\bibinfo
  {volume} {740}},\ \bibinfo {eid} {102} (\bibinfo {year} {2011})},\ \Eprint
  {https://arxiv.org/abs/1002.3660} {arXiv:1002.3660 [astro-ph.CO]}
  \BibitemShut {NoStop}%
\bibitem [{\citenamefont {{Hopkins}}\ \emph {et~al.}(2018)\citenamefont
  {{Hopkins}}, \citenamefont {{Wetzel}}, \citenamefont {{Kere{\v{s}}}},
  \citenamefont {{Faucher-Gigu{\`e}re}}, \citenamefont {{Quataert}},
  \citenamefont {{Boylan-Kolchin}}, \citenamefont {{Murray}}, \citenamefont
  {{Hayward}}, \citenamefont {{Garrison-Kimmel}}, \citenamefont {{Hummels}},
  \citenamefont {{Feldmann}}, \citenamefont {{Torrey}}, \citenamefont {{Ma}},
  \citenamefont {{Angl{\'e}s-Alc{\'a}zar}}, \citenamefont {{Su}}, \citenamefont
  {{Orr}}, \citenamefont {{Schmitz}}, \citenamefont {{Escala}}, \citenamefont
  {{Sanderson}}, \citenamefont {{Grudi{\'c}}}, \citenamefont {{Hafen}},
  \citenamefont {{Kim}}, \citenamefont {{Fitts}}, \citenamefont {{Bullock}},
  \citenamefont {{Wheeler}}, \citenamefont {{Chan}}, \citenamefont {{Elbert}},\
  and\ \citenamefont {{Narayanan}}}]{Hopkins2018}%
  \BibitemOpen
  \bibfield  {author} {\bibinfo {author} {\bibfnamefont {P.~F.}\ \bibnamefont
  {{Hopkins}}}, \bibinfo {author} {\bibfnamefont {A.}~\bibnamefont {{Wetzel}}},
  \bibinfo {author} {\bibfnamefont {D.}~\bibnamefont {{Kere{\v{s}}}}}, \bibinfo
  {author} {\bibfnamefont {C.-A.}\ \bibnamefont {{Faucher-Gigu{\`e}re}}},
  \bibinfo {author} {\bibfnamefont {E.}~\bibnamefont {{Quataert}}}, \bibinfo
  {author} {\bibfnamefont {M.}~\bibnamefont {{Boylan-Kolchin}}}, \bibinfo
  {author} {\bibfnamefont {N.}~\bibnamefont {{Murray}}}, \bibinfo {author}
  {\bibfnamefont {C.~C.}\ \bibnamefont {{Hayward}}}, \bibinfo {author}
  {\bibfnamefont {S.}~\bibnamefont {{Garrison-Kimmel}}}, \bibinfo {author}
  {\bibfnamefont {C.}~\bibnamefont {{Hummels}}}, \bibinfo {author}
  {\bibfnamefont {R.}~\bibnamefont {{Feldmann}}}, \bibinfo {author}
  {\bibfnamefont {P.}~\bibnamefont {{Torrey}}}, \bibinfo {author}
  {\bibfnamefont {X.}~\bibnamefont {{Ma}}}, \bibinfo {author} {\bibfnamefont
  {D.}~\bibnamefont {{Angl{\'e}s-Alc{\'a}zar}}}, \bibinfo {author}
  {\bibfnamefont {K.-Y.}\ \bibnamefont {{Su}}}, \bibinfo {author}
  {\bibfnamefont {M.}~\bibnamefont {{Orr}}}, \bibinfo {author} {\bibfnamefont
  {D.}~\bibnamefont {{Schmitz}}}, \bibinfo {author} {\bibfnamefont
  {I.}~\bibnamefont {{Escala}}}, \bibinfo {author} {\bibfnamefont
  {R.}~\bibnamefont {{Sanderson}}}, \bibinfo {author} {\bibfnamefont {M.~Y.}\
  \bibnamefont {{Grudi{\'c}}}}, \bibinfo {author} {\bibfnamefont
  {Z.}~\bibnamefont {{Hafen}}}, \bibinfo {author} {\bibfnamefont {J.-H.}\
  \bibnamefont {{Kim}}}, \bibinfo {author} {\bibfnamefont {A.}~\bibnamefont
  {{Fitts}}}, \bibinfo {author} {\bibfnamefont {J.~S.}\ \bibnamefont
  {{Bullock}}}, \bibinfo {author} {\bibfnamefont {C.}~\bibnamefont
  {{Wheeler}}}, \bibinfo {author} {\bibfnamefont {T.~K.}\ \bibnamefont
  {{Chan}}}, \bibinfo {author} {\bibfnamefont {O.~D.}\ \bibnamefont
  {{Elbert}}},\ and\ \bibinfo {author} {\bibfnamefont {D.}~\bibnamefont
  {{Narayanan}}},\ }\href {https://doi.org/10.1093/mnras/sty1690} {\bibfield
  {journal} {\bibinfo  {journal} {\mnras}\ }\textbf {\bibinfo {volume} {480}},\
  \bibinfo {pages} {800} (\bibinfo {year} {2018})},\ \Eprint
  {https://arxiv.org/abs/1702.06148} {arXiv:1702.06148 [astro-ph.GA]}
  \BibitemShut {NoStop}%
\bibitem [{\citenamefont {{Pillepich}}\ \emph {et~al.}(2018)\citenamefont
  {{Pillepich}}, \citenamefont {{Springel}}, \citenamefont {{Nelson}},
  \citenamefont {{Genel}}, \citenamefont {{Naiman}}, \citenamefont {{Pakmor}},
  \citenamefont {{Hernquist}}, \citenamefont {{Torrey}}, \citenamefont
  {{Vogelsberger}}, \citenamefont {{Weinberger}},\ and\ \citenamefont
  {{Marinacci}}}]{Pillepich2018}%
  \BibitemOpen
  \bibfield  {author} {\bibinfo {author} {\bibfnamefont {A.}~\bibnamefont
  {{Pillepich}}}, \bibinfo {author} {\bibfnamefont {V.}~\bibnamefont
  {{Springel}}}, \bibinfo {author} {\bibfnamefont {D.}~\bibnamefont
  {{Nelson}}}, \bibinfo {author} {\bibfnamefont {S.}~\bibnamefont {{Genel}}},
  \bibinfo {author} {\bibfnamefont {J.}~\bibnamefont {{Naiman}}}, \bibinfo
  {author} {\bibfnamefont {R.}~\bibnamefont {{Pakmor}}}, \bibinfo {author}
  {\bibfnamefont {L.}~\bibnamefont {{Hernquist}}}, \bibinfo {author}
  {\bibfnamefont {P.}~\bibnamefont {{Torrey}}}, \bibinfo {author}
  {\bibfnamefont {M.}~\bibnamefont {{Vogelsberger}}}, \bibinfo {author}
  {\bibfnamefont {R.}~\bibnamefont {{Weinberger}}},\ and\ \bibinfo {author}
  {\bibfnamefont {F.}~\bibnamefont {{Marinacci}}},\ }\href
  {https://doi.org/10.1093/mnras/stx2656} {\bibfield  {journal} {\bibinfo
  {journal} {\mnras}\ }\textbf {\bibinfo {volume} {473}},\ \bibinfo {pages}
  {4077} (\bibinfo {year} {2018})},\ \Eprint {https://arxiv.org/abs/1703.02970}
  {arXiv:1703.02970 [astro-ph.GA]} \BibitemShut {NoStop}%
\bibitem [{\citenamefont {{Vogelsberger}}\ \emph {et~al.}(2016)\citenamefont
  {{Vogelsberger}}, \citenamefont {{Zavala}}, \citenamefont {{Cyr-Racine}},
  \citenamefont {{Pfrommer}}, \citenamefont {{Bringmann}},\ and\ \citenamefont
  {{Sigurdson}}}]{Vogelsberger2016}%
  \BibitemOpen
  \bibfield  {author} {\bibinfo {author} {\bibfnamefont {M.}~\bibnamefont
  {{Vogelsberger}}}, \bibinfo {author} {\bibfnamefont {J.}~\bibnamefont
  {{Zavala}}}, \bibinfo {author} {\bibfnamefont {F.-Y.}\ \bibnamefont
  {{Cyr-Racine}}}, \bibinfo {author} {\bibfnamefont {C.}~\bibnamefont
  {{Pfrommer}}}, \bibinfo {author} {\bibfnamefont {T.}~\bibnamefont
  {{Bringmann}}},\ and\ \bibinfo {author} {\bibfnamefont {K.}~\bibnamefont
  {{Sigurdson}}},\ }\href {https://doi.org/10.1093/mnras/stw1076} {\bibfield
  {journal} {\bibinfo  {journal} {\mnras}\ }\textbf {\bibinfo {volume} {460}},\
  \bibinfo {pages} {1399} (\bibinfo {year} {2016})},\ \Eprint
  {https://arxiv.org/abs/1512.05349} {arXiv:1512.05349 [astro-ph.CO]}
  \BibitemShut {NoStop}%
\bibitem [{\citenamefont {{Lovell}}\ \emph {et~al.}(2018)\citenamefont
  {{Lovell}}, \citenamefont {{Zavala}}, \citenamefont {{Vogelsberger}},
  \citenamefont {{Shen}}, \citenamefont {{Cyr-Racine}}, \citenamefont
  {{Pfrommer}}, \citenamefont {{Sigurdson}}, \citenamefont {{Boylan-Kolchin}},\
  and\ \citenamefont {{Pillepich}}}]{Lovell2018}%
  \BibitemOpen
  \bibfield  {author} {\bibinfo {author} {\bibfnamefont {M.~R.}\ \bibnamefont
  {{Lovell}}}, \bibinfo {author} {\bibfnamefont {J.}~\bibnamefont {{Zavala}}},
  \bibinfo {author} {\bibfnamefont {M.}~\bibnamefont {{Vogelsberger}}},
  \bibinfo {author} {\bibfnamefont {X.}~\bibnamefont {{Shen}}}, \bibinfo
  {author} {\bibfnamefont {F.-Y.}\ \bibnamefont {{Cyr-Racine}}}, \bibinfo
  {author} {\bibfnamefont {C.}~\bibnamefont {{Pfrommer}}}, \bibinfo {author}
  {\bibfnamefont {K.}~\bibnamefont {{Sigurdson}}}, \bibinfo {author}
  {\bibfnamefont {M.}~\bibnamefont {{Boylan-Kolchin}}},\ and\ \bibinfo {author}
  {\bibfnamefont {A.}~\bibnamefont {{Pillepich}}},\ }\href
  {https://doi.org/10.1093/mnras/sty818} {\bibfield  {journal} {\bibinfo
  {journal} {\mnras}\ }\textbf {\bibinfo {volume} {477}},\ \bibinfo {pages}
  {2886} (\bibinfo {year} {2018})},\ \Eprint {https://arxiv.org/abs/1711.10497}
  {arXiv:1711.10497 [astro-ph.CO]} \BibitemShut {NoStop}%
\bibitem [{\citenamefont {Macciò}\ \emph {et~al.}(2008)\citenamefont
  {Macciò}, \citenamefont {Dutton},\ and\ \citenamefont {van~den
  Bosch}}]{Macci__2008}%
  \BibitemOpen
  \bibfield  {author} {\bibinfo {author} {\bibfnamefont {A.~V.}\ \bibnamefont
  {Macciò}}, \bibinfo {author} {\bibfnamefont {A.~A.}\ \bibnamefont
  {Dutton}},\ and\ \bibinfo {author} {\bibfnamefont {F.~C.}\ \bibnamefont
  {van~den Bosch}},\ }\href {https://doi.org/10.1111/j.1365-2966.2008.14029.x}
  {\bibfield  {journal} {\bibinfo  {journal} {Monthly Notices of the Royal
  Astronomical Society}\ }\textbf {\bibinfo {volume} {391}},\ \bibinfo {pages}
  {1940–1954} (\bibinfo {year} {2008})}\BibitemShut {NoStop}%
\bibitem [{\citenamefont {Bullock}\ \emph {et~al.}(2001)\citenamefont
  {Bullock}, \citenamefont {Kolatt}, \citenamefont {Sigad}, \citenamefont
  {Somerville}, \citenamefont {Kravtsov}, \citenamefont {Klypin}, \citenamefont
  {Primack},\ and\ \citenamefont {Dekel}}]{Bullock_2001}%
  \BibitemOpen
  \bibfield  {author} {\bibinfo {author} {\bibfnamefont {J.~S.}\ \bibnamefont
  {Bullock}}, \bibinfo {author} {\bibfnamefont {T.~S.}\ \bibnamefont {Kolatt}},
  \bibinfo {author} {\bibfnamefont {Y.}~\bibnamefont {Sigad}}, \bibinfo
  {author} {\bibfnamefont {R.~S.}\ \bibnamefont {Somerville}}, \bibinfo
  {author} {\bibfnamefont {A.~V.}\ \bibnamefont {Kravtsov}}, \bibinfo {author}
  {\bibfnamefont {A.~A.}\ \bibnamefont {Klypin}}, \bibinfo {author}
  {\bibfnamefont {J.~R.}\ \bibnamefont {Primack}},\ and\ \bibinfo {author}
  {\bibfnamefont {A.}~\bibnamefont {Dekel}},\ }\href
  {https://doi.org/10.1046/j.1365-8711.2001.04068.x} {\bibfield  {journal}
  {\bibinfo  {journal} {Monthly Notices of the Royal Astronomical Society}\
  }\textbf {\bibinfo {volume} {321}},\ \bibinfo {pages} {559–575} (\bibinfo
  {year} {2001})}\BibitemShut {NoStop}%
\bibitem [{\citenamefont {Navarro}\ \emph {et~al.}(1996)\citenamefont
  {Navarro}, \citenamefont {Frenk},\ and\ \citenamefont
  {White}}]{Navarro_1996}%
  \BibitemOpen
  \bibfield  {author} {\bibinfo {author} {\bibfnamefont {J.~F.}\ \bibnamefont
  {Navarro}}, \bibinfo {author} {\bibfnamefont {C.~S.}\ \bibnamefont {Frenk}},\
  and\ \bibinfo {author} {\bibfnamefont {S.~D.~M.}\ \bibnamefont {White}},\
  }\href {https://doi.org/10.1086/177173} {\bibfield  {journal} {\bibinfo
  {journal} {The Astrophysical Journal}\ }\textbf {\bibinfo {volume} {462}},\
  \bibinfo {pages} {563} (\bibinfo {year} {1996})}\BibitemShut {NoStop}%
\bibitem [{\citenamefont {Zhao}\ \emph {et~al.}(2003)\citenamefont {Zhao},
  \citenamefont {Jing}, \citenamefont {Mo},\ and\ \citenamefont
  {Brner}}]{Zhao_2003}%
  \BibitemOpen
  \bibfield  {author} {\bibinfo {author} {\bibfnamefont {D.~H.}\ \bibnamefont
  {Zhao}}, \bibinfo {author} {\bibfnamefont {Y.~P.}\ \bibnamefont {Jing}},
  \bibinfo {author} {\bibfnamefont {H.~J.}\ \bibnamefont {Mo}},\ and\ \bibinfo
  {author} {\bibfnamefont {G.}~\bibnamefont {Brner}},\ }\href
  {https://doi.org/10.1086/379734} {\bibfield  {journal} {\bibinfo  {journal}
  {The Astrophysical Journal}\ }\textbf {\bibinfo {volume} {597}},\ \bibinfo
  {pages} {L9} (\bibinfo {year} {2003})}\BibitemShut {NoStop}%
\bibitem [{\citenamefont {Diemer}\ \emph {et~al.}(2013)\citenamefont {Diemer},
  \citenamefont {More},\ and\ \citenamefont {Kravtsov}}]{Diemer_2013}%
  \BibitemOpen
  \bibfield  {author} {\bibinfo {author} {\bibfnamefont {B.}~\bibnamefont
  {Diemer}}, \bibinfo {author} {\bibfnamefont {S.}~\bibnamefont {More}},\ and\
  \bibinfo {author} {\bibfnamefont {A.~V.}\ \bibnamefont {Kravtsov}},\ }\href
  {https://doi.org/10.1088/0004-637x/766/1/25} {\bibfield  {journal} {\bibinfo
  {journal} {The Astrophysical Journal}\ }\textbf {\bibinfo {volume} {766}},\
  \bibinfo {pages} {25} (\bibinfo {year} {2013})}\BibitemShut {NoStop}%
\bibitem [{\citenamefont {{Lemze}}\ \emph {et~al.}(2012)\citenamefont
  {{Lemze}}, \citenamefont {{Wagner}}, \citenamefont {{Rephaeli}},
  \citenamefont {{Sadeh}}, \citenamefont {{Norman}}, \citenamefont {{Barkana}},
  \citenamefont {{Broadhurst}}, \citenamefont {{Ford}},\ and\ \citenamefont
  {{Postman}}}]{Lemze2012}%
  \BibitemOpen
  \bibfield  {author} {\bibinfo {author} {\bibfnamefont {D.}~\bibnamefont
  {{Lemze}}}, \bibinfo {author} {\bibfnamefont {R.}~\bibnamefont {{Wagner}}},
  \bibinfo {author} {\bibfnamefont {Y.}~\bibnamefont {{Rephaeli}}}, \bibinfo
  {author} {\bibfnamefont {S.}~\bibnamefont {{Sadeh}}}, \bibinfo {author}
  {\bibfnamefont {M.~L.}\ \bibnamefont {{Norman}}}, \bibinfo {author}
  {\bibfnamefont {R.}~\bibnamefont {{Barkana}}}, \bibinfo {author}
  {\bibfnamefont {T.}~\bibnamefont {{Broadhurst}}}, \bibinfo {author}
  {\bibfnamefont {H.}~\bibnamefont {{Ford}}},\ and\ \bibinfo {author}
  {\bibfnamefont {M.}~\bibnamefont {{Postman}}},\ }\href
  {https://doi.org/10.1088/0004-637X/752/2/141} {\bibfield  {journal} {\bibinfo
   {journal} {\apj}\ }\textbf {\bibinfo {volume} {752}},\ \bibinfo {eid} {141}
  (\bibinfo {year} {2012})},\ \Eprint {https://arxiv.org/abs/1106.6048}
  {arXiv:1106.6048 [astro-ph.CO]} \BibitemShut {NoStop}%
\bibitem [{\citenamefont {{Sparre}}\ and\ \citenamefont
  {{Hansen}}(2012)}]{Sparre2012}%
  \BibitemOpen
  \bibfield  {author} {\bibinfo {author} {\bibfnamefont {M.}~\bibnamefont
  {{Sparre}}}\ and\ \bibinfo {author} {\bibfnamefont {S.~H.}\ \bibnamefont
  {{Hansen}}},\ }\href {https://doi.org/10.1088/1475-7516/2012/10/049}
  {\bibfield  {journal} {\bibinfo  {journal} {\jcap}\ }\textbf {\bibinfo
  {volume} {2012}},\ \bibinfo {eid} {049} (\bibinfo {year} {2012})},\ \Eprint
  {https://arxiv.org/abs/1210.2392} {arXiv:1210.2392 [astro-ph.CO]}
  \BibitemShut {NoStop}%
\bibitem [{\citenamefont {{Wojtak}}\ \emph {et~al.}(2013)\citenamefont
  {{Wojtak}}, \citenamefont {{Gottl{\"o}ber}},\ and\ \citenamefont
  {{Klypin}}}]{Wojtak2013}%
  \BibitemOpen
  \bibfield  {author} {\bibinfo {author} {\bibfnamefont {R.}~\bibnamefont
  {{Wojtak}}}, \bibinfo {author} {\bibfnamefont {S.}~\bibnamefont
  {{Gottl{\"o}ber}}},\ and\ \bibinfo {author} {\bibfnamefont {A.}~\bibnamefont
  {{Klypin}}},\ }\href {https://doi.org/10.1093/mnras/stt1113} {\bibfield
  {journal} {\bibinfo  {journal} {\mnras}\ }\textbf {\bibinfo {volume} {434}},\
  \bibinfo {pages} {1576} (\bibinfo {year} {2013})},\ \Eprint
  {https://arxiv.org/abs/1303.2056} {arXiv:1303.2056 [astro-ph.CO]}
  \BibitemShut {NoStop}%
\bibitem [{\citenamefont {{{\L}okas}}\ and\ \citenamefont
  {{Mamon}}(2001)}]{Lokas2001}%
  \BibitemOpen
  \bibfield  {author} {\bibinfo {author} {\bibfnamefont {E.~L.}\ \bibnamefont
  {{{\L}okas}}}\ and\ \bibinfo {author} {\bibfnamefont {G.~A.}\ \bibnamefont
  {{Mamon}}},\ }\href {https://doi.org/10.1046/j.1365-8711.2001.04007.x}
  {\bibfield  {journal} {\bibinfo  {journal} {\mnras}\ }\textbf {\bibinfo
  {volume} {321}},\ \bibinfo {pages} {155} (\bibinfo {year} {2001})},\ \Eprint
  {https://arxiv.org/abs/astro-ph/0002395} {arXiv:astro-ph/0002395 [astro-ph]}
  \BibitemShut {NoStop}%
\bibitem [{\citenamefont {{Shen}}\ \emph {et~al.}(2021)\citenamefont {{Shen}},
  \citenamefont {{Hopkins}}, \citenamefont {{Necib}}, \citenamefont {{Jiang}},
  \citenamefont {{Boylan-Kolchin}},\ and\ \citenamefont {{Wetzel}}}]{Shen2021}%
  \BibitemOpen
  \bibfield  {author} {\bibinfo {author} {\bibfnamefont {X.}~\bibnamefont
  {{Shen}}}, \bibinfo {author} {\bibfnamefont {P.~F.}\ \bibnamefont
  {{Hopkins}}}, \bibinfo {author} {\bibfnamefont {L.}~\bibnamefont {{Necib}}},
  \bibinfo {author} {\bibfnamefont {F.}~\bibnamefont {{Jiang}}}, \bibinfo
  {author} {\bibfnamefont {M.}~\bibnamefont {{Boylan-Kolchin}}},\ and\ \bibinfo
  {author} {\bibfnamefont {A.}~\bibnamefont {{Wetzel}}},\ }\href@noop {}
  {\bibfield  {journal} {\bibinfo  {journal} {arXiv e-prints}\ ,\ \bibinfo
  {eid} {arXiv:2102.09580}} (\bibinfo {year} {2021})},\ \Eprint
  {https://arxiv.org/abs/2102.09580} {arXiv:2102.09580 [astro-ph.GA]}
  \BibitemShut {NoStop}%
\bibitem [{\citenamefont {Diemer}\ and\ \citenamefont
  {Joyce}(2019)}]{Diemer_2019}%
  \BibitemOpen
  \bibfield  {author} {\bibinfo {author} {\bibfnamefont {B.}~\bibnamefont
  {Diemer}}\ and\ \bibinfo {author} {\bibfnamefont {M.}~\bibnamefont {Joyce}},\
  }\href {https://doi.org/10.3847/1538-4357/aafad6} {\bibfield  {journal}
  {\bibinfo  {journal} {The Astrophysical Journal}\ }\textbf {\bibinfo {volume}
  {871}},\ \bibinfo {pages} {168} (\bibinfo {year} {2019})}\BibitemShut
  {NoStop}%
\bibitem [{\citenamefont {Diemer}\ and\ \citenamefont
  {Kravtsov}(2015)}]{Diemer_2015}%
  \BibitemOpen
  \bibfield  {author} {\bibinfo {author} {\bibfnamefont {B.}~\bibnamefont
  {Diemer}}\ and\ \bibinfo {author} {\bibfnamefont {A.~V.}\ \bibnamefont
  {Kravtsov}},\ }\href {https://doi.org/10.1088/0004-637x/799/1/108} {\bibfield
   {journal} {\bibinfo  {journal} {The Astrophysical Journal}\ }\textbf
  {\bibinfo {volume} {799}},\ \bibinfo {pages} {108} (\bibinfo {year}
  {2015})}\BibitemShut {NoStop}%
\bibitem [{\citenamefont {Ishiyama}\ \emph {et~al.}(2020)\citenamefont
  {Ishiyama}, \citenamefont {Prada}, \citenamefont {Klypin}, \citenamefont
  {Sinha}, \citenamefont {Metcalf}, \citenamefont {Jullo}, \citenamefont
  {Altieri}, \citenamefont {Cora}, \citenamefont {Croton}, \citenamefont {de~la
  Torre}, \citenamefont {Millán-Calero}, \citenamefont {Oogi}, \citenamefont
  {Ruedas},\ and\ \citenamefont {Vega-Martínez}}]{ishiyama2020uchuu}%
  \BibitemOpen
  \bibfield  {author} {\bibinfo {author} {\bibfnamefont {T.}~\bibnamefont
  {Ishiyama}}, \bibinfo {author} {\bibfnamefont {F.}~\bibnamefont {Prada}},
  \bibinfo {author} {\bibfnamefont {A.~A.}\ \bibnamefont {Klypin}}, \bibinfo
  {author} {\bibfnamefont {M.}~\bibnamefont {Sinha}}, \bibinfo {author}
  {\bibfnamefont {R.~B.}\ \bibnamefont {Metcalf}}, \bibinfo {author}
  {\bibfnamefont {E.}~\bibnamefont {Jullo}}, \bibinfo {author} {\bibfnamefont
  {B.}~\bibnamefont {Altieri}}, \bibinfo {author} {\bibfnamefont {S.~A.}\
  \bibnamefont {Cora}}, \bibinfo {author} {\bibfnamefont {D.}~\bibnamefont
  {Croton}}, \bibinfo {author} {\bibfnamefont {S.}~\bibnamefont {de~la Torre}},
  \bibinfo {author} {\bibfnamefont {D.~E.}\ \bibnamefont {Millán-Calero}},
  \bibinfo {author} {\bibfnamefont {T.}~\bibnamefont {Oogi}}, \bibinfo {author}
  {\bibfnamefont {J.}~\bibnamefont {Ruedas}},\ and\ \bibinfo {author}
  {\bibfnamefont {C.~A.}\ \bibnamefont {Vega-Martínez}},\ }\href@noop {}
  {\bibinfo {title} {The uchuu simulations: Data release 1 and dark matter halo
  concentrations}} (\bibinfo {year} {2020}),\ \Eprint
  {https://arxiv.org/abs/2007.14720} {arXiv:2007.14720 [astro-ph.CO]}
  \BibitemShut {NoStop}%
\bibitem [{\citenamefont {{Bullock}}\ \emph {et~al.}(2001)\citenamefont
  {{Bullock}}, \citenamefont {{Kolatt}}, \citenamefont {{Sigad}}, \citenamefont
  {{Somerville}}, \citenamefont {{Kravtsov}}, \citenamefont {{Klypin}},
  \citenamefont {{Primack}},\ and\ \citenamefont
  {{Dekel}}}]{2001MNRAS.321..559B}%
  \BibitemOpen
  \bibfield  {author} {\bibinfo {author} {\bibfnamefont {J.~S.}\ \bibnamefont
  {{Bullock}}}, \bibinfo {author} {\bibfnamefont {T.~S.}\ \bibnamefont
  {{Kolatt}}}, \bibinfo {author} {\bibfnamefont {Y.}~\bibnamefont {{Sigad}}},
  \bibinfo {author} {\bibfnamefont {R.~S.}\ \bibnamefont {{Somerville}}},
  \bibinfo {author} {\bibfnamefont {A.~V.}\ \bibnamefont {{Kravtsov}}},
  \bibinfo {author} {\bibfnamefont {A.~A.}\ \bibnamefont {{Klypin}}}, \bibinfo
  {author} {\bibfnamefont {J.~R.}\ \bibnamefont {{Primack}}},\ and\ \bibinfo
  {author} {\bibfnamefont {A.}~\bibnamefont {{Dekel}}},\ }\href
  {https://doi.org/10.1046/j.1365-8711.2001.04068.x} {\bibfield  {journal}
  {\bibinfo  {journal} {\mnras}\ }\textbf {\bibinfo {volume} {321}},\ \bibinfo
  {pages} {559} (\bibinfo {year} {2001})},\ \Eprint
  {https://arxiv.org/abs/astro-ph/9908159} {arXiv:astro-ph/9908159 [astro-ph]}
  \BibitemShut {NoStop}%
\bibitem [{\citenamefont {Wechsler}\ \emph {et~al.}(2002)\citenamefont
  {Wechsler}, \citenamefont {Bullock}, \citenamefont {Primack}, \citenamefont
  {Kravtsov},\ and\ \citenamefont {Dekel}}]{Wechsler_2002}%
  \BibitemOpen
  \bibfield  {author} {\bibinfo {author} {\bibfnamefont {R.~H.}\ \bibnamefont
  {Wechsler}}, \bibinfo {author} {\bibfnamefont {J.~S.}\ \bibnamefont
  {Bullock}}, \bibinfo {author} {\bibfnamefont {J.~R.}\ \bibnamefont
  {Primack}}, \bibinfo {author} {\bibfnamefont {A.~V.}\ \bibnamefont
  {Kravtsov}},\ and\ \bibinfo {author} {\bibfnamefont {A.}~\bibnamefont
  {Dekel}},\ }\href {https://doi.org/10.1086/338765} {\bibfield  {journal}
  {\bibinfo  {journal} {The Astrophysical Journal}\ }\textbf {\bibinfo {volume}
  {568}},\ \bibinfo {pages} {52–70} (\bibinfo {year} {2002})}\BibitemShut
  {NoStop}%
\bibitem [{\citenamefont {{Mortlock}}\ \emph {et~al.}(2011)\citenamefont
  {{Mortlock}}, \citenamefont {{Warren}}, \citenamefont {{Venemans}},
  \citenamefont {{Patel}}, \citenamefont {{Hewett}}, \citenamefont {{McMahon}},
  \citenamefont {{Simpson}}, \citenamefont {{Theuns}}, \citenamefont
  {{Gonz{\'a}les-Solares}}, \citenamefont {{Adamson}}, \citenamefont {{Dye}},
  \citenamefont {{Hambly}}, \citenamefont {{Hirst}}, \citenamefont {{Irwin}},
  \citenamefont {{Kuiper}}, \citenamefont {{Lawrence}},\ and\ \citenamefont
  {{R{\"o}ttgering}}}]{Mortlock2011}%
  \BibitemOpen
  \bibfield  {author} {\bibinfo {author} {\bibfnamefont {D.~J.}\ \bibnamefont
  {{Mortlock}}}, \bibinfo {author} {\bibfnamefont {S.~J.}\ \bibnamefont
  {{Warren}}}, \bibinfo {author} {\bibfnamefont {B.~P.}\ \bibnamefont
  {{Venemans}}}, \bibinfo {author} {\bibfnamefont {M.}~\bibnamefont {{Patel}}},
  \bibinfo {author} {\bibfnamefont {P.~C.}\ \bibnamefont {{Hewett}}}, \bibinfo
  {author} {\bibfnamefont {R.~G.}\ \bibnamefont {{McMahon}}}, \bibinfo {author}
  {\bibfnamefont {C.}~\bibnamefont {{Simpson}}}, \bibinfo {author}
  {\bibfnamefont {T.}~\bibnamefont {{Theuns}}}, \bibinfo {author}
  {\bibfnamefont {E.~A.}\ \bibnamefont {{Gonz{\'a}les-Solares}}}, \bibinfo
  {author} {\bibfnamefont {A.}~\bibnamefont {{Adamson}}}, \bibinfo {author}
  {\bibfnamefont {S.}~\bibnamefont {{Dye}}}, \bibinfo {author} {\bibfnamefont
  {N.~C.}\ \bibnamefont {{Hambly}}}, \bibinfo {author} {\bibfnamefont
  {P.}~\bibnamefont {{Hirst}}}, \bibinfo {author} {\bibfnamefont {M.~J.}\
  \bibnamefont {{Irwin}}}, \bibinfo {author} {\bibfnamefont {E.}~\bibnamefont
  {{Kuiper}}}, \bibinfo {author} {\bibfnamefont {A.}~\bibnamefont
  {{Lawrence}}},\ and\ \bibinfo {author} {\bibfnamefont {H.~J.~A.}\
  \bibnamefont {{R{\"o}ttgering}}},\ }\href
  {https://doi.org/10.1038/nature10159} {\bibfield  {journal} {\bibinfo
  {journal} {\nat}\ }\textbf {\bibinfo {volume} {474}},\ \bibinfo {pages} {616}
  (\bibinfo {year} {2011})},\ \Eprint {https://arxiv.org/abs/1106.6088}
  {arXiv:1106.6088 [astro-ph.CO]} \BibitemShut {NoStop}%
\bibitem [{\citenamefont {{Venemans}}\ \emph {et~al.}(2013)\citenamefont
  {{Venemans}}, \citenamefont {{Findlay}}, \citenamefont {{Sutherland}},
  \citenamefont {{De Rosa}}, \citenamefont {{McMahon}}, \citenamefont
  {{Simcoe}}, \citenamefont {{Gonz{\'a}lez-Solares}}, \citenamefont
  {{Kuijken}},\ and\ \citenamefont {{Lewis}}}]{Venemans2013}%
  \BibitemOpen
  \bibfield  {author} {\bibinfo {author} {\bibfnamefont {B.~P.}\ \bibnamefont
  {{Venemans}}}, \bibinfo {author} {\bibfnamefont {J.~R.}\ \bibnamefont
  {{Findlay}}}, \bibinfo {author} {\bibfnamefont {W.~J.}\ \bibnamefont
  {{Sutherland}}}, \bibinfo {author} {\bibfnamefont {G.}~\bibnamefont {{De
  Rosa}}}, \bibinfo {author} {\bibfnamefont {R.~G.}\ \bibnamefont {{McMahon}}},
  \bibinfo {author} {\bibfnamefont {R.}~\bibnamefont {{Simcoe}}}, \bibinfo
  {author} {\bibfnamefont {E.~A.}\ \bibnamefont {{Gonz{\'a}lez-Solares}}},
  \bibinfo {author} {\bibfnamefont {K.}~\bibnamefont {{Kuijken}}},\ and\
  \bibinfo {author} {\bibfnamefont {J.~R.}\ \bibnamefont {{Lewis}}},\ }\href
  {https://doi.org/10.1088/0004-637X/779/1/24} {\bibfield  {journal} {\bibinfo
  {journal} {\apj}\ }\textbf {\bibinfo {volume} {779}},\ \bibinfo {eid} {24}
  (\bibinfo {year} {2013})},\ \Eprint {https://arxiv.org/abs/1311.3666}
  {arXiv:1311.3666 [astro-ph.CO]} \BibitemShut {NoStop}%
\bibitem [{\citenamefont {{Wu}}\ \emph {et~al.}(2015)\citenamefont {{Wu}},
  \citenamefont {{Wang}}, \citenamefont {{Fan}}, \citenamefont {{Yi}},
  \citenamefont {{Zuo}}, \citenamefont {{Bian}}, \citenamefont {{Jiang}},
  \citenamefont {{McGreer}}, \citenamefont {{Wang}}, \citenamefont {{Yang}},
  \citenamefont {{Yang}}, \citenamefont {{Thompson}},\ and\ \citenamefont
  {{Beletsky}}}]{Wu2015}%
  \BibitemOpen
  \bibfield  {author} {\bibinfo {author} {\bibfnamefont {X.-B.}\ \bibnamefont
  {{Wu}}}, \bibinfo {author} {\bibfnamefont {F.}~\bibnamefont {{Wang}}},
  \bibinfo {author} {\bibfnamefont {X.}~\bibnamefont {{Fan}}}, \bibinfo
  {author} {\bibfnamefont {W.}~\bibnamefont {{Yi}}}, \bibinfo {author}
  {\bibfnamefont {W.}~\bibnamefont {{Zuo}}}, \bibinfo {author} {\bibfnamefont
  {F.}~\bibnamefont {{Bian}}}, \bibinfo {author} {\bibfnamefont
  {L.}~\bibnamefont {{Jiang}}}, \bibinfo {author} {\bibfnamefont {I.~D.}\
  \bibnamefont {{McGreer}}}, \bibinfo {author} {\bibfnamefont {R.}~\bibnamefont
  {{Wang}}}, \bibinfo {author} {\bibfnamefont {J.}~\bibnamefont {{Yang}}},
  \bibinfo {author} {\bibfnamefont {Q.}~\bibnamefont {{Yang}}}, \bibinfo
  {author} {\bibfnamefont {D.}~\bibnamefont {{Thompson}}},\ and\ \bibinfo
  {author} {\bibfnamefont {Y.}~\bibnamefont {{Beletsky}}},\ }\href
  {https://doi.org/10.1038/nature14241} {\bibfield  {journal} {\bibinfo
  {journal} {\nat}\ }\textbf {\bibinfo {volume} {518}},\ \bibinfo {pages} {512}
  (\bibinfo {year} {2015})},\ \Eprint {https://arxiv.org/abs/1502.07418}
  {arXiv:1502.07418 [astro-ph.GA]} \BibitemShut {NoStop}%
\bibitem [{\citenamefont {{Mazzucchelli}}\ \emph {et~al.}(2017)\citenamefont
  {{Mazzucchelli}}, \citenamefont {{Ba{\~n}ados}}, \citenamefont {{Venemans}},
  \citenamefont {{Decarli}}, \citenamefont {{Farina}}, \citenamefont
  {{Walter}}, \citenamefont {{Eilers}}, \citenamefont {{Rix}}, \citenamefont
  {{Simcoe}}, \citenamefont {{Stern}}, \citenamefont {{Fan}}, \citenamefont
  {{Schlafly}}, \citenamefont {{De Rosa}}, \citenamefont {{Hennawi}},
  \citenamefont {{Chambers}}, \citenamefont {{Greiner}}, \citenamefont
  {{Burgett}}, \citenamefont {{Draper}}, \citenamefont {{Kaiser}},
  \citenamefont {{Kudritzki}}, \citenamefont {{Magnier}}, \citenamefont
  {{Metcalfe}}, \citenamefont {{Waters}},\ and\ \citenamefont
  {{Wainscoat}}}]{Mazzucchelli2017}%
  \BibitemOpen
  \bibfield  {author} {\bibinfo {author} {\bibfnamefont {C.}~\bibnamefont
  {{Mazzucchelli}}}, \bibinfo {author} {\bibfnamefont {E.}~\bibnamefont
  {{Ba{\~n}ados}}}, \bibinfo {author} {\bibfnamefont {B.~P.}\ \bibnamefont
  {{Venemans}}}, \bibinfo {author} {\bibfnamefont {R.}~\bibnamefont
  {{Decarli}}}, \bibinfo {author} {\bibfnamefont {E.~P.}\ \bibnamefont
  {{Farina}}}, \bibinfo {author} {\bibfnamefont {F.}~\bibnamefont {{Walter}}},
  \bibinfo {author} {\bibfnamefont {A.~C.}\ \bibnamefont {{Eilers}}}, \bibinfo
  {author} {\bibfnamefont {H.~W.}\ \bibnamefont {{Rix}}}, \bibinfo {author}
  {\bibfnamefont {R.}~\bibnamefont {{Simcoe}}}, \bibinfo {author}
  {\bibfnamefont {D.}~\bibnamefont {{Stern}}}, \bibinfo {author} {\bibfnamefont
  {X.}~\bibnamefont {{Fan}}}, \bibinfo {author} {\bibfnamefont
  {E.}~\bibnamefont {{Schlafly}}}, \bibinfo {author} {\bibfnamefont
  {G.}~\bibnamefont {{De Rosa}}}, \bibinfo {author} {\bibfnamefont
  {J.}~\bibnamefont {{Hennawi}}}, \bibinfo {author} {\bibfnamefont {K.~C.}\
  \bibnamefont {{Chambers}}}, \bibinfo {author} {\bibfnamefont
  {J.}~\bibnamefont {{Greiner}}}, \bibinfo {author} {\bibfnamefont
  {W.}~\bibnamefont {{Burgett}}}, \bibinfo {author} {\bibfnamefont {P.~W.}\
  \bibnamefont {{Draper}}}, \bibinfo {author} {\bibfnamefont {N.}~\bibnamefont
  {{Kaiser}}}, \bibinfo {author} {\bibfnamefont {R.~P.}\ \bibnamefont
  {{Kudritzki}}}, \bibinfo {author} {\bibfnamefont {E.}~\bibnamefont
  {{Magnier}}}, \bibinfo {author} {\bibfnamefont {N.}~\bibnamefont
  {{Metcalfe}}}, \bibinfo {author} {\bibfnamefont {C.}~\bibnamefont
  {{Waters}}},\ and\ \bibinfo {author} {\bibfnamefont {R.~J.}\ \bibnamefont
  {{Wainscoat}}},\ }\href {https://doi.org/10.3847/1538-4357/aa9185} {\bibfield
   {journal} {\bibinfo  {journal} {\apj}\ }\textbf {\bibinfo {volume} {849}},\
  \bibinfo {eid} {91} (\bibinfo {year} {2017})},\ \Eprint
  {https://arxiv.org/abs/1710.01251} {arXiv:1710.01251 [astro-ph.GA]}
  \BibitemShut {NoStop}%
\bibitem [{\citenamefont {{Ba{\~n}ados}}\ \emph {et~al.}(2018)\citenamefont
  {{Ba{\~n}ados}}, \citenamefont {{Venemans}}, \citenamefont {{Mazzucchelli}},
  \citenamefont {{Farina}}, \citenamefont {{Walter}}, \citenamefont {{Wang}},
  \citenamefont {{Decarli}}, \citenamefont {{Stern}}, \citenamefont {{Fan}},
  \citenamefont {{Davies}}, \citenamefont {{Hennawi}}, \citenamefont
  {{Simcoe}}, \citenamefont {{Turner}}, \citenamefont {{Rix}}, \citenamefont
  {{Yang}}, \citenamefont {{Kelson}}, \citenamefont {{Rudie}},\ and\
  \citenamefont {{Winters}}}]{Banados2018}%
  \BibitemOpen
  \bibfield  {author} {\bibinfo {author} {\bibfnamefont {E.}~\bibnamefont
  {{Ba{\~n}ados}}}, \bibinfo {author} {\bibfnamefont {B.~P.}\ \bibnamefont
  {{Venemans}}}, \bibinfo {author} {\bibfnamefont {C.}~\bibnamefont
  {{Mazzucchelli}}}, \bibinfo {author} {\bibfnamefont {E.~P.}\ \bibnamefont
  {{Farina}}}, \bibinfo {author} {\bibfnamefont {F.}~\bibnamefont {{Walter}}},
  \bibinfo {author} {\bibfnamefont {F.}~\bibnamefont {{Wang}}}, \bibinfo
  {author} {\bibfnamefont {R.}~\bibnamefont {{Decarli}}}, \bibinfo {author}
  {\bibfnamefont {D.}~\bibnamefont {{Stern}}}, \bibinfo {author} {\bibfnamefont
  {X.}~\bibnamefont {{Fan}}}, \bibinfo {author} {\bibfnamefont {F.~B.}\
  \bibnamefont {{Davies}}}, \bibinfo {author} {\bibfnamefont {J.~F.}\
  \bibnamefont {{Hennawi}}}, \bibinfo {author} {\bibfnamefont {R.~A.}\
  \bibnamefont {{Simcoe}}}, \bibinfo {author} {\bibfnamefont {M.~L.}\
  \bibnamefont {{Turner}}}, \bibinfo {author} {\bibfnamefont {H.-W.}\
  \bibnamefont {{Rix}}}, \bibinfo {author} {\bibfnamefont {J.}~\bibnamefont
  {{Yang}}}, \bibinfo {author} {\bibfnamefont {D.~D.}\ \bibnamefont
  {{Kelson}}}, \bibinfo {author} {\bibfnamefont {G.~C.}\ \bibnamefont
  {{Rudie}}},\ and\ \bibinfo {author} {\bibfnamefont {J.~M.}\ \bibnamefont
  {{Winters}}},\ }\href {https://doi.org/10.1038/nature25180} {\bibfield
  {journal} {\bibinfo  {journal} {\nat}\ }\textbf {\bibinfo {volume} {553}},\
  \bibinfo {pages} {473} (\bibinfo {year} {2018})},\ \Eprint
  {https://arxiv.org/abs/1712.01860} {arXiv:1712.01860 [astro-ph.GA]}
  \BibitemShut {NoStop}%
\bibitem [{\citenamefont {{Wang}}\ \emph {et~al.}(2018)\citenamefont {{Wang}},
  \citenamefont {{Yang}}, \citenamefont {{Fan}}, \citenamefont {{Yue}},
  \citenamefont {{Wu}}, \citenamefont {{Schindler}}, \citenamefont {{Bian}},
  \citenamefont {{Li}}, \citenamefont {{Farina}}, \citenamefont
  {{Ba{\~n}ados}}, \citenamefont {{Davies}}, \citenamefont {{Decarli}},
  \citenamefont {{Green}}, \citenamefont {{Jiang}}, \citenamefont {{Hennawi}},
  \citenamefont {{Huang}}, \citenamefont {{Mazzucchelli}}, \citenamefont
  {{McGreer}}, \citenamefont {{Venemans}}, \citenamefont {{Walter}},\ and\
  \citenamefont {{Beletsky}}}]{Wang2018}%
  \BibitemOpen
  \bibfield  {author} {\bibinfo {author} {\bibfnamefont {F.}~\bibnamefont
  {{Wang}}}, \bibinfo {author} {\bibfnamefont {J.}~\bibnamefont {{Yang}}},
  \bibinfo {author} {\bibfnamefont {X.}~\bibnamefont {{Fan}}}, \bibinfo
  {author} {\bibfnamefont {M.}~\bibnamefont {{Yue}}}, \bibinfo {author}
  {\bibfnamefont {X.-B.}\ \bibnamefont {{Wu}}}, \bibinfo {author}
  {\bibfnamefont {J.-T.}\ \bibnamefont {{Schindler}}}, \bibinfo {author}
  {\bibfnamefont {F.}~\bibnamefont {{Bian}}}, \bibinfo {author} {\bibfnamefont
  {J.-T.}\ \bibnamefont {{Li}}}, \bibinfo {author} {\bibfnamefont {E.~P.}\
  \bibnamefont {{Farina}}}, \bibinfo {author} {\bibfnamefont {E.}~\bibnamefont
  {{Ba{\~n}ados}}}, \bibinfo {author} {\bibfnamefont {F.~B.}\ \bibnamefont
  {{Davies}}}, \bibinfo {author} {\bibfnamefont {R.}~\bibnamefont {{Decarli}}},
  \bibinfo {author} {\bibfnamefont {R.}~\bibnamefont {{Green}}}, \bibinfo
  {author} {\bibfnamefont {L.}~\bibnamefont {{Jiang}}}, \bibinfo {author}
  {\bibfnamefont {J.~F.}\ \bibnamefont {{Hennawi}}}, \bibinfo {author}
  {\bibfnamefont {Y.-H.}\ \bibnamefont {{Huang}}}, \bibinfo {author}
  {\bibfnamefont {C.}~\bibnamefont {{Mazzucchelli}}}, \bibinfo {author}
  {\bibfnamefont {I.~D.}\ \bibnamefont {{McGreer}}}, \bibinfo {author}
  {\bibfnamefont {B.}~\bibnamefont {{Venemans}}}, \bibinfo {author}
  {\bibfnamefont {F.}~\bibnamefont {{Walter}}},\ and\ \bibinfo {author}
  {\bibfnamefont {Y.}~\bibnamefont {{Beletsky}}},\ }\href
  {https://doi.org/10.3847/2041-8213/aaf1d2} {\bibfield  {journal} {\bibinfo
  {journal} {\apjl}\ }\textbf {\bibinfo {volume} {869}},\ \bibinfo {eid} {L9}
  (\bibinfo {year} {2018})},\ \Eprint {https://arxiv.org/abs/1810.11925}
  {arXiv:1810.11925 [astro-ph.GA]} \BibitemShut {NoStop}%
\bibitem [{\citenamefont {{Yang}}\ \emph {et~al.}(2020)\citenamefont {{Yang}},
  \citenamefont {{Wang}}, \citenamefont {{Fan}}, \citenamefont {{Hennawi}},
  \citenamefont {{Davies}}, \citenamefont {{Yue}}, \citenamefont {{Banados}},
  \citenamefont {{Wu}}, \citenamefont {{Venemans}}, \citenamefont {{Barth}},
  \citenamefont {{Bian}}, \citenamefont {{Boutsia}}, \citenamefont {{Decarli}},
  \citenamefont {{Farina}}, \citenamefont {{Green}}, \citenamefont {{Jiang}},
  \citenamefont {{Li}}, \citenamefont {{Mazzucchelli}},\ and\ \citenamefont
  {{Walter}}}]{Yang2020}%
  \BibitemOpen
  \bibfield  {author} {\bibinfo {author} {\bibfnamefont {J.}~\bibnamefont
  {{Yang}}}, \bibinfo {author} {\bibfnamefont {F.}~\bibnamefont {{Wang}}},
  \bibinfo {author} {\bibfnamefont {X.}~\bibnamefont {{Fan}}}, \bibinfo
  {author} {\bibfnamefont {J.~F.}\ \bibnamefont {{Hennawi}}}, \bibinfo {author}
  {\bibfnamefont {F.~B.}\ \bibnamefont {{Davies}}}, \bibinfo {author}
  {\bibfnamefont {M.}~\bibnamefont {{Yue}}}, \bibinfo {author} {\bibfnamefont
  {E.}~\bibnamefont {{Banados}}}, \bibinfo {author} {\bibfnamefont {X.-B.}\
  \bibnamefont {{Wu}}}, \bibinfo {author} {\bibfnamefont {B.}~\bibnamefont
  {{Venemans}}}, \bibinfo {author} {\bibfnamefont {A.~J.}\ \bibnamefont
  {{Barth}}}, \bibinfo {author} {\bibfnamefont {F.}~\bibnamefont {{Bian}}},
  \bibinfo {author} {\bibfnamefont {K.}~\bibnamefont {{Boutsia}}}, \bibinfo
  {author} {\bibfnamefont {R.}~\bibnamefont {{Decarli}}}, \bibinfo {author}
  {\bibfnamefont {E.~P.}\ \bibnamefont {{Farina}}}, \bibinfo {author}
  {\bibfnamefont {R.}~\bibnamefont {{Green}}}, \bibinfo {author} {\bibfnamefont
  {L.}~\bibnamefont {{Jiang}}}, \bibinfo {author} {\bibfnamefont {J.-T.}\
  \bibnamefont {{Li}}}, \bibinfo {author} {\bibfnamefont {C.}~\bibnamefont
  {{Mazzucchelli}}},\ and\ \bibinfo {author} {\bibfnamefont {F.}~\bibnamefont
  {{Walter}}},\ }\href {https://doi.org/10.3847/2041-8213/ab9c26} {\bibfield
  {journal} {\bibinfo  {journal} {\apjl}\ }\textbf {\bibinfo {volume} {897}},\
  \bibinfo {eid} {L14} (\bibinfo {year} {2020})},\ \Eprint
  {https://arxiv.org/abs/2006.13452} {arXiv:2006.13452 [astro-ph.GA]}
  \BibitemShut {NoStop}%
\bibitem [{\citenamefont {Zhao}\ \emph {et~al.}(2009)\citenamefont {Zhao},
  \citenamefont {Jing}, \citenamefont {Mo},\ and\ \citenamefont
  {Börner}}]{Zhao_2009}%
  \BibitemOpen
  \bibfield  {author} {\bibinfo {author} {\bibfnamefont {D.~H.}\ \bibnamefont
  {Zhao}}, \bibinfo {author} {\bibfnamefont {Y.~P.}\ \bibnamefont {Jing}},
  \bibinfo {author} {\bibfnamefont {H.~J.}\ \bibnamefont {Mo}},\ and\ \bibinfo
  {author} {\bibfnamefont {G.}~\bibnamefont {Börner}},\ }\href
  {https://doi.org/10.1088/0004-637x/707/1/354} {\bibfield  {journal} {\bibinfo
   {journal} {The Astrophysical Journal}\ }\textbf {\bibinfo {volume} {707}},\
  \bibinfo {pages} {354} (\bibinfo {year} {2009})}\BibitemShut {NoStop}%
\bibitem [{\citenamefont {{Jiang}}\ \emph {et~al.}(2020)\citenamefont
  {{Jiang}}, \citenamefont {{Dekel}}, \citenamefont {{Freundlich}},
  \citenamefont {{van den Bosch}}, \citenamefont {{Green}}, \citenamefont
  {{Hopkins}}, \citenamefont {{Benson}},\ and\ \citenamefont
  {{Du}}}]{Jiang2020}%
  \BibitemOpen
  \bibfield  {author} {\bibinfo {author} {\bibfnamefont {F.}~\bibnamefont
  {{Jiang}}}, \bibinfo {author} {\bibfnamefont {A.}~\bibnamefont {{Dekel}}},
  \bibinfo {author} {\bibfnamefont {J.}~\bibnamefont {{Freundlich}}}, \bibinfo
  {author} {\bibfnamefont {F.~C.}\ \bibnamefont {{van den Bosch}}}, \bibinfo
  {author} {\bibfnamefont {S.~B.}\ \bibnamefont {{Green}}}, \bibinfo {author}
  {\bibfnamefont {P.~F.}\ \bibnamefont {{Hopkins}}}, \bibinfo {author}
  {\bibfnamefont {A.}~\bibnamefont {{Benson}}},\ and\ \bibinfo {author}
  {\bibfnamefont {X.}~\bibnamefont {{Du}}},\ }\href@noop {} {\bibfield
  {journal} {\bibinfo  {journal} {arXiv e-prints}\ ,\ \bibinfo {eid}
  {arXiv:2005.05974}} (\bibinfo {year} {2020})},\ \Eprint
  {https://arxiv.org/abs/2005.05974} {arXiv:2005.05974 [astro-ph.GA]}
  \BibitemShut {NoStop}%
\bibitem [{\citenamefont {{Lacey}}\ and\ \citenamefont
  {{Cole}}(1993)}]{Lacey1993}%
  \BibitemOpen
  \bibfield  {author} {\bibinfo {author} {\bibfnamefont {C.}~\bibnamefont
  {{Lacey}}}\ and\ \bibinfo {author} {\bibfnamefont {S.}~\bibnamefont
  {{Cole}}},\ }\href {https://doi.org/10.1093/mnras/262.3.627} {\bibfield
  {journal} {\bibinfo  {journal} {\mnras}\ }\textbf {\bibinfo {volume} {262}},\
  \bibinfo {pages} {627} (\bibinfo {year} {1993})}\BibitemShut {NoStop}%
\bibitem [{\citenamefont {{Parkinson}}\ \emph {et~al.}(2008)\citenamefont
  {{Parkinson}}, \citenamefont {{Cole}},\ and\ \citenamefont
  {{Helly}}}]{Parkinson2008}%
  \BibitemOpen
  \bibfield  {author} {\bibinfo {author} {\bibfnamefont {H.}~\bibnamefont
  {{Parkinson}}}, \bibinfo {author} {\bibfnamefont {S.}~\bibnamefont
  {{Cole}}},\ and\ \bibinfo {author} {\bibfnamefont {J.}~\bibnamefont
  {{Helly}}},\ }\href {https://doi.org/10.1111/j.1365-2966.2007.12517.x}
  {\bibfield  {journal} {\bibinfo  {journal} {\mnras}\ }\textbf {\bibinfo
  {volume} {383}},\ \bibinfo {pages} {557} (\bibinfo {year} {2008})},\ \Eprint
  {https://arxiv.org/abs/0708.1382} {arXiv:0708.1382 [astro-ph]} \BibitemShut
  {NoStop}%
\bibitem [{\citenamefont {{Benson}}(2017)}]{Benson2017}%
  \BibitemOpen
  \bibfield  {author} {\bibinfo {author} {\bibfnamefont {A.~J.}\ \bibnamefont
  {{Benson}}},\ }\href {https://doi.org/10.1093/mnras/stx343} {\bibfield
  {journal} {\bibinfo  {journal} {\mnras}\ }\textbf {\bibinfo {volume} {467}},\
  \bibinfo {pages} {3454} (\bibinfo {year} {2017})},\ \Eprint
  {https://arxiv.org/abs/1610.01057} {arXiv:1610.01057 [astro-ph.GA]}
  \BibitemShut {NoStop}%
\bibitem [{\citenamefont {{Bryan}}\ and\ \citenamefont
  {{Norman}}(1998)}]{Bryan1998}%
  \BibitemOpen
  \bibfield  {author} {\bibinfo {author} {\bibfnamefont {G.~L.}\ \bibnamefont
  {{Bryan}}}\ and\ \bibinfo {author} {\bibfnamefont {M.~L.}\ \bibnamefont
  {{Norman}}},\ }\href {https://doi.org/10.1086/305262} {\bibfield  {journal}
  {\bibinfo  {journal} {\apj}\ }\textbf {\bibinfo {volume} {495}},\ \bibinfo
  {pages} {80} (\bibinfo {year} {1998})},\ \Eprint
  {https://arxiv.org/abs/astro-ph/9710107} {arXiv:astro-ph/9710107 [astro-ph]}
  \BibitemShut {NoStop}%
\bibitem [{\citenamefont {{Blumenthal}}\ \emph {et~al.}(1986)\citenamefont
  {{Blumenthal}}, \citenamefont {{Faber}}, \citenamefont {{Flores}},\ and\
  \citenamefont {{Primack}}}]{Blumenthal1986}%
  \BibitemOpen
  \bibfield  {author} {\bibinfo {author} {\bibfnamefont {G.~R.}\ \bibnamefont
  {{Blumenthal}}}, \bibinfo {author} {\bibfnamefont {S.~M.}\ \bibnamefont
  {{Faber}}}, \bibinfo {author} {\bibfnamefont {R.}~\bibnamefont {{Flores}}},\
  and\ \bibinfo {author} {\bibfnamefont {J.~R.}\ \bibnamefont {{Primack}}},\
  }\href {https://doi.org/10.1086/163867} {\bibfield  {journal} {\bibinfo
  {journal} {\apj}\ }\textbf {\bibinfo {volume} {301}},\ \bibinfo {pages} {27}
  (\bibinfo {year} {1986})}\BibitemShut {NoStop}%
\bibitem [{\citenamefont {{Ryden}}\ and\ \citenamefont
  {{Gunn}}(1987)}]{Ryden1987}%
  \BibitemOpen
  \bibfield  {author} {\bibinfo {author} {\bibfnamefont {B.~S.}\ \bibnamefont
  {{Ryden}}}\ and\ \bibinfo {author} {\bibfnamefont {J.~E.}\ \bibnamefont
  {{Gunn}}},\ }\href {https://doi.org/10.1086/165349} {\bibfield  {journal}
  {\bibinfo  {journal} {\apj}\ }\textbf {\bibinfo {volume} {318}},\ \bibinfo
  {pages} {15} (\bibinfo {year} {1987})}\BibitemShut {NoStop}%
\bibitem [{\citenamefont {{Murray}}\ \emph {et~al.}(2013)\citenamefont
  {{Murray}}, \citenamefont {{Power}},\ and\ \citenamefont
  {{Robotham}}}]{Murray2013}%
  \BibitemOpen
  \bibfield  {author} {\bibinfo {author} {\bibfnamefont {S.~G.}\ \bibnamefont
  {{Murray}}}, \bibinfo {author} {\bibfnamefont {C.}~\bibnamefont {{Power}}},\
  and\ \bibinfo {author} {\bibfnamefont {A.~S.~G.}\ \bibnamefont
  {{Robotham}}},\ }\href {https://doi.org/10.1016/j.ascom.2013.11.001}
  {\bibfield  {journal} {\bibinfo  {journal} {Astronomy and Computing}\
  }\textbf {\bibinfo {volume} {3}},\ \bibinfo {pages} {23} (\bibinfo {year}
  {2013})},\ \Eprint {https://arxiv.org/abs/1306.6721} {arXiv:1306.6721
  [astro-ph.CO]} \BibitemShut {NoStop}%
\bibitem [{\citenamefont {{Tinker}}\ \emph {et~al.}(2008)\citenamefont
  {{Tinker}}, \citenamefont {{Kravtsov}}, \citenamefont {{Klypin}},
  \citenamefont {{Abazajian}}, \citenamefont {{Warren}}, \citenamefont
  {{Yepes}}, \citenamefont {{Gottl{\"o}ber}},\ and\ \citenamefont
  {{Holz}}}]{Tinker2008}%
  \BibitemOpen
  \bibfield  {author} {\bibinfo {author} {\bibfnamefont {J.}~\bibnamefont
  {{Tinker}}}, \bibinfo {author} {\bibfnamefont {A.~V.}\ \bibnamefont
  {{Kravtsov}}}, \bibinfo {author} {\bibfnamefont {A.}~\bibnamefont
  {{Klypin}}}, \bibinfo {author} {\bibfnamefont {K.}~\bibnamefont
  {{Abazajian}}}, \bibinfo {author} {\bibfnamefont {M.}~\bibnamefont
  {{Warren}}}, \bibinfo {author} {\bibfnamefont {G.}~\bibnamefont {{Yepes}}},
  \bibinfo {author} {\bibfnamefont {S.}~\bibnamefont {{Gottl{\"o}ber}}},\ and\
  \bibinfo {author} {\bibfnamefont {D.~E.}\ \bibnamefont {{Holz}}},\ }\href
  {https://doi.org/10.1086/591439} {\bibfield  {journal} {\bibinfo  {journal}
  {\apj}\ }\textbf {\bibinfo {volume} {688}},\ \bibinfo {pages} {709} (\bibinfo
  {year} {2008})},\ \Eprint {https://arxiv.org/abs/0803.2706} {arXiv:0803.2706
  [astro-ph]} \BibitemShut {NoStop}%
\bibitem [{\citenamefont {Volonteri}\ \emph {et~al.}(2003)\citenamefont
  {Volonteri}, \citenamefont {Haardt},\ and\ \citenamefont
  {Madau}}]{Volonteri:2002vz}%
  \BibitemOpen
  \bibfield  {author} {\bibinfo {author} {\bibfnamefont {M.}~\bibnamefont
  {Volonteri}}, \bibinfo {author} {\bibfnamefont {F.}~\bibnamefont {Haardt}},\
  and\ \bibinfo {author} {\bibfnamefont {P.}~\bibnamefont {Madau}},\ }\href
  {https://doi.org/10.1086/344675} {\bibfield  {journal} {\bibinfo  {journal}
  {Astrophys. J.}\ }\textbf {\bibinfo {volume} {582}},\ \bibinfo {pages} {559}
  (\bibinfo {year} {2003})},\ \Eprint {https://arxiv.org/abs/astro-ph/0207276}
  {arXiv:astro-ph/0207276} \BibitemShut {NoStop}%
\bibitem [{\citenamefont {{Iwasawa}}\ \emph {et~al.}(2006)\citenamefont
  {{Iwasawa}}, \citenamefont {{Funato}},\ and\ \citenamefont
  {{Makino}}}]{Iwasawa2006}%
  \BibitemOpen
  \bibfield  {author} {\bibinfo {author} {\bibfnamefont {M.}~\bibnamefont
  {{Iwasawa}}}, \bibinfo {author} {\bibfnamefont {Y.}~\bibnamefont
  {{Funato}}},\ and\ \bibinfo {author} {\bibfnamefont {J.}~\bibnamefont
  {{Makino}}},\ }\href {https://doi.org/10.1086/507473} {\bibfield  {journal}
  {\bibinfo  {journal} {\apj}\ }\textbf {\bibinfo {volume} {651}},\ \bibinfo
  {pages} {1059} (\bibinfo {year} {2006})},\ \Eprint
  {https://arxiv.org/abs/astro-ph/0511391} {arXiv:astro-ph/0511391 [astro-ph]}
  \BibitemShut {NoStop}%
\bibitem [{\citenamefont {{Hoffman}}\ and\ \citenamefont
  {{Loeb}}(2007)}]{Hoffman2007}%
  \BibitemOpen
  \bibfield  {author} {\bibinfo {author} {\bibfnamefont {L.}~\bibnamefont
  {{Hoffman}}}\ and\ \bibinfo {author} {\bibfnamefont {A.}~\bibnamefont
  {{Loeb}}},\ }\href {https://doi.org/10.1111/j.1365-2966.2007.11694.x}
  {\bibfield  {journal} {\bibinfo  {journal} {\mnras}\ }\textbf {\bibinfo
  {volume} {377}},\ \bibinfo {pages} {957} (\bibinfo {year} {2007})},\ \Eprint
  {https://arxiv.org/abs/astro-ph/0612517} {arXiv:astro-ph/0612517 [astro-ph]}
  \BibitemShut {NoStop}%
\bibitem [{\citenamefont {{Fitchett}}(1983)}]{Fitchett1983}%
  \BibitemOpen
  \bibfield  {author} {\bibinfo {author} {\bibfnamefont {M.~J.}\ \bibnamefont
  {{Fitchett}}},\ }\href {https://doi.org/10.1093/mnras/203.4.1049} {\bibfield
  {journal} {\bibinfo  {journal} {\mnras}\ }\textbf {\bibinfo {volume} {203}},\
  \bibinfo {pages} {1049} (\bibinfo {year} {1983})}\BibitemShut {NoStop}%
\bibitem [{\citenamefont {{Volonteri}}\ \emph {et~al.}(2006)\citenamefont
  {{Volonteri}}, \citenamefont {{Salvaterra}},\ and\ \citenamefont
  {{Haardt}}}]{Volonteri2006}%
  \BibitemOpen
  \bibfield  {author} {\bibinfo {author} {\bibfnamefont {M.}~\bibnamefont
  {{Volonteri}}}, \bibinfo {author} {\bibfnamefont {R.}~\bibnamefont
  {{Salvaterra}}},\ and\ \bibinfo {author} {\bibfnamefont {F.}~\bibnamefont
  {{Haardt}}},\ }\href {https://doi.org/10.1111/j.1365-2966.2006.10976.x}
  {\bibfield  {journal} {\bibinfo  {journal} {\mnras}\ }\textbf {\bibinfo
  {volume} {373}},\ \bibinfo {pages} {121} (\bibinfo {year} {2006})},\ \Eprint
  {https://arxiv.org/abs/astro-ph/0606675} {arXiv:astro-ph/0606675 [astro-ph]}
  \BibitemShut {NoStop}%
\bibitem [{\citenamefont {{Volonteri}}\ \emph {et~al.}(2008)\citenamefont
  {{Volonteri}}, \citenamefont {{Lodato}},\ and\ \citenamefont
  {{Natarajan}}}]{Volonteri2008}%
  \BibitemOpen
  \bibfield  {author} {\bibinfo {author} {\bibfnamefont {M.}~\bibnamefont
  {{Volonteri}}}, \bibinfo {author} {\bibfnamefont {G.}~\bibnamefont
  {{Lodato}}},\ and\ \bibinfo {author} {\bibfnamefont {P.}~\bibnamefont
  {{Natarajan}}},\ }\href {https://doi.org/10.1111/j.1365-2966.2007.12589.x}
  {\bibfield  {journal} {\bibinfo  {journal} {\mnras}\ }\textbf {\bibinfo
  {volume} {383}},\ \bibinfo {pages} {1079} (\bibinfo {year} {2008})},\ \Eprint
  {https://arxiv.org/abs/0709.0529} {arXiv:0709.0529 [astro-ph]} \BibitemShut
  {NoStop}%
\bibitem [{\citenamefont {{Natarajan}}(2014)}]{2014GReGr..46.1702N}%
  \BibitemOpen
  \bibfield  {author} {\bibinfo {author} {\bibfnamefont {P.}~\bibnamefont
  {{Natarajan}}},\ }\href {https://doi.org/10.1007/s10714-014-1702-6}
  {\bibfield  {journal} {\bibinfo  {journal} {General Relativity and
  Gravitation}\ }\textbf {\bibinfo {volume} {46}},\ \bibinfo {eid} {1702}
  (\bibinfo {year} {2014})}\BibitemShut {NoStop}%
\bibitem [{\citenamefont {Ferrarese}(2002)}]{Ferrarese:2002ct}%
  \BibitemOpen
  \bibfield  {author} {\bibinfo {author} {\bibfnamefont {L.}~\bibnamefont
  {Ferrarese}},\ }\href {https://doi.org/10.1086/342308} {\bibfield  {journal}
  {\bibinfo  {journal} {Astrophys. J.}\ }\textbf {\bibinfo {volume} {578}},\
  \bibinfo {pages} {90} (\bibinfo {year} {2002})},\ \Eprint
  {https://arxiv.org/abs/astro-ph/0203469} {arXiv:astro-ph/0203469}
  \BibitemShut {NoStop}%
\bibitem [{\citenamefont {{Kormendy}}\ and\ \citenamefont
  {{Ho}}(2013)}]{Kormendy2013}%
  \BibitemOpen
  \bibfield  {author} {\bibinfo {author} {\bibfnamefont {J.}~\bibnamefont
  {{Kormendy}}}\ and\ \bibinfo {author} {\bibfnamefont {L.~C.}\ \bibnamefont
  {{Ho}}},\ }\href {https://doi.org/10.1146/annurev-astro-082708-101811}
  {\bibfield  {journal} {\bibinfo  {journal} {\araa}\ }\textbf {\bibinfo
  {volume} {51}},\ \bibinfo {pages} {511} (\bibinfo {year} {2013})},\ \Eprint
  {https://arxiv.org/abs/1304.7762} {arXiv:1304.7762 [astro-ph.CO]}
  \BibitemShut {NoStop}%
\bibitem [{\citenamefont {{Klypin}}\ \emph {et~al.}(2001)\citenamefont
  {{Klypin}}, \citenamefont {{Kravtsov}}, \citenamefont {{Bullock}},\ and\
  \citenamefont {{Primack}}}]{Klypin2001}%
  \BibitemOpen
  \bibfield  {author} {\bibinfo {author} {\bibfnamefont {A.}~\bibnamefont
  {{Klypin}}}, \bibinfo {author} {\bibfnamefont {A.~V.}\ \bibnamefont
  {{Kravtsov}}}, \bibinfo {author} {\bibfnamefont {J.~S.}\ \bibnamefont
  {{Bullock}}},\ and\ \bibinfo {author} {\bibfnamefont {J.~R.}\ \bibnamefont
  {{Primack}}},\ }\href {https://doi.org/10.1086/321400} {\bibfield  {journal}
  {\bibinfo  {journal} {\apj}\ }\textbf {\bibinfo {volume} {554}},\ \bibinfo
  {pages} {903} (\bibinfo {year} {2001})},\ \Eprint
  {https://arxiv.org/abs/astro-ph/0006343} {arXiv:astro-ph/0006343 [astro-ph]}
  \BibitemShut {NoStop}%
\bibitem [{\citenamefont {{Trakhtenbrot}}(2020)}]{Trakhtenbrot2020}%
  \BibitemOpen
  \bibfield  {author} {\bibinfo {author} {\bibfnamefont {B.}~\bibnamefont
  {{Trakhtenbrot}}},\ }\href@noop {} {\bibfield  {journal} {\bibinfo  {journal}
  {arXiv e-prints}\ ,\ \bibinfo {eid} {arXiv:2002.00972}} (\bibinfo {year}
  {2020})},\ \Eprint {https://arxiv.org/abs/2002.00972} {arXiv:2002.00972
  [astro-ph.GA]} \BibitemShut {NoStop}%
\bibitem [{\citenamefont {Shen}\ \emph {et~al.}(2020)\citenamefont {Shen},
  \citenamefont {Hopkins}, \citenamefont {Faucher-Giguère}, \citenamefont
  {Alexander}, \citenamefont {Richards}, \citenamefont {Ross},\ and\
  \citenamefont {Hickox}}]{Shen:2020obl}%
  \BibitemOpen
  \bibfield  {author} {\bibinfo {author} {\bibfnamefont {X.}~\bibnamefont
  {Shen}}, \bibinfo {author} {\bibfnamefont {P.~F.}\ \bibnamefont {Hopkins}},
  \bibinfo {author} {\bibfnamefont {C.-A.}\ \bibnamefont {Faucher-Giguère}},
  \bibinfo {author} {\bibfnamefont {D.}~\bibnamefont {Alexander}}, \bibinfo
  {author} {\bibfnamefont {G.~T.}\ \bibnamefont {Richards}}, \bibinfo {author}
  {\bibfnamefont {N.~P.}\ \bibnamefont {Ross}},\ and\ \bibinfo {author}
  {\bibfnamefont {R.}~\bibnamefont {Hickox}},\ }\href
  {https://doi.org/10.1093/mnras/staa1381} {\bibfield  {journal} {\bibinfo
  {journal} {Mon. Not. Roy. Astron. Soc.}\ }\textbf {\bibinfo {volume} {495}},\
  \bibinfo {pages} {3252} (\bibinfo {year} {2020})},\ \Eprint
  {https://arxiv.org/abs/2001.02696} {arXiv:2001.02696 [astro-ph.GA]}
  \BibitemShut {NoStop}%
\bibitem [{\citenamefont {{Kulkarni}}\ \emph {et~al.}(2019)\citenamefont
  {{Kulkarni}}, \citenamefont {{Worseck}},\ and\ \citenamefont
  {{Hennawi}}}]{Kulkarni2019}%
  \BibitemOpen
  \bibfield  {author} {\bibinfo {author} {\bibfnamefont {G.}~\bibnamefont
  {{Kulkarni}}}, \bibinfo {author} {\bibfnamefont {G.}~\bibnamefont
  {{Worseck}}},\ and\ \bibinfo {author} {\bibfnamefont {J.~F.}\ \bibnamefont
  {{Hennawi}}},\ }\href {https://doi.org/10.1093/mnras/stz1493} {\bibfield
  {journal} {\bibinfo  {journal} {\mnras}\ }\textbf {\bibinfo {volume} {488}},\
  \bibinfo {pages} {1035} (\bibinfo {year} {2019})},\ \Eprint
  {https://arxiv.org/abs/1807.09774} {arXiv:1807.09774 [astro-ph.GA]}
  \BibitemShut {NoStop}%
\bibitem [{\citenamefont {{Wang}}\ \emph
  {et~al.}(2019{\natexlab{a}})\citenamefont {{Wang}}, \citenamefont {{Yang}},
  \citenamefont {{Fan}}, \citenamefont {{Wu}}, \citenamefont {{Yue}},
  \citenamefont {{Li}}, \citenamefont {{Bian}}, \citenamefont {{Jiang}},
  \citenamefont {{Ba{\~n}ados}}, \citenamefont {{Schindler}}, \citenamefont
  {{Findlay}}, \citenamefont {{Davies}}, \citenamefont {{Decarli}},
  \citenamefont {{Farina}}, \citenamefont {{Green}}, \citenamefont {{Hennawi}},
  \citenamefont {{Huang}}, \citenamefont {{Mazzuccheli}}, \citenamefont
  {{McGreer}}, \citenamefont {{Venemans}}, \citenamefont {{Walter}},
  \citenamefont {{Dye}}, \citenamefont {{Lyke}}, \citenamefont {{Myers}},\ and\
  \citenamefont {{Haze Nunez}}}]{Wang2019b}%
  \BibitemOpen
  \bibfield  {author} {\bibinfo {author} {\bibfnamefont {F.}~\bibnamefont
  {{Wang}}}, \bibinfo {author} {\bibfnamefont {J.}~\bibnamefont {{Yang}}},
  \bibinfo {author} {\bibfnamefont {X.}~\bibnamefont {{Fan}}}, \bibinfo
  {author} {\bibfnamefont {X.-B.}\ \bibnamefont {{Wu}}}, \bibinfo {author}
  {\bibfnamefont {M.}~\bibnamefont {{Yue}}}, \bibinfo {author} {\bibfnamefont
  {J.-T.}\ \bibnamefont {{Li}}}, \bibinfo {author} {\bibfnamefont
  {F.}~\bibnamefont {{Bian}}}, \bibinfo {author} {\bibfnamefont
  {L.}~\bibnamefont {{Jiang}}}, \bibinfo {author} {\bibfnamefont
  {E.}~\bibnamefont {{Ba{\~n}ados}}}, \bibinfo {author} {\bibfnamefont {J.-T.}\
  \bibnamefont {{Schindler}}}, \bibinfo {author} {\bibfnamefont {J.~R.}\
  \bibnamefont {{Findlay}}}, \bibinfo {author} {\bibfnamefont {F.~B.}\
  \bibnamefont {{Davies}}}, \bibinfo {author} {\bibfnamefont {R.}~\bibnamefont
  {{Decarli}}}, \bibinfo {author} {\bibfnamefont {E.~P.}\ \bibnamefont
  {{Farina}}}, \bibinfo {author} {\bibfnamefont {R.}~\bibnamefont {{Green}}},
  \bibinfo {author} {\bibfnamefont {J.~F.}\ \bibnamefont {{Hennawi}}}, \bibinfo
  {author} {\bibfnamefont {Y.-H.}\ \bibnamefont {{Huang}}}, \bibinfo {author}
  {\bibfnamefont {C.}~\bibnamefont {{Mazzuccheli}}}, \bibinfo {author}
  {\bibfnamefont {I.~D.}\ \bibnamefont {{McGreer}}}, \bibinfo {author}
  {\bibfnamefont {B.}~\bibnamefont {{Venemans}}}, \bibinfo {author}
  {\bibfnamefont {F.}~\bibnamefont {{Walter}}}, \bibinfo {author}
  {\bibfnamefont {S.}~\bibnamefont {{Dye}}}, \bibinfo {author} {\bibfnamefont
  {B.~W.}\ \bibnamefont {{Lyke}}}, \bibinfo {author} {\bibfnamefont {A.~D.}\
  \bibnamefont {{Myers}}},\ and\ \bibinfo {author} {\bibfnamefont
  {E.}~\bibnamefont {{Haze Nunez}}},\ }\href
  {https://doi.org/10.3847/1538-4357/ab2be5} {\bibfield  {journal} {\bibinfo
  {journal} {\apj}\ }\textbf {\bibinfo {volume} {884}},\ \bibinfo {eid} {30}
  (\bibinfo {year} {2019}{\natexlab{a}})},\ \Eprint
  {https://arxiv.org/abs/1810.11926} {arXiv:1810.11926 [astro-ph.GA]}
  \BibitemShut {NoStop}%
\bibitem [{\citenamefont {{Willott}}\ \emph {et~al.}(2010)\citenamefont
  {{Willott}}, \citenamefont {{Albert}}, \citenamefont {{Arzoumanian}},
  \citenamefont {{Bergeron}}, \citenamefont {{Crampton}}, \citenamefont
  {{Delorme}}, \citenamefont {{Hutchings}}, \citenamefont {{Omont}},
  \citenamefont {{Reyl{\'e}}},\ and\ \citenamefont {{Schade}}}]{Willott2010}%
  \BibitemOpen
  \bibfield  {author} {\bibinfo {author} {\bibfnamefont {C.~J.}\ \bibnamefont
  {{Willott}}}, \bibinfo {author} {\bibfnamefont {L.}~\bibnamefont {{Albert}}},
  \bibinfo {author} {\bibfnamefont {D.}~\bibnamefont {{Arzoumanian}}}, \bibinfo
  {author} {\bibfnamefont {J.}~\bibnamefont {{Bergeron}}}, \bibinfo {author}
  {\bibfnamefont {D.}~\bibnamefont {{Crampton}}}, \bibinfo {author}
  {\bibfnamefont {P.}~\bibnamefont {{Delorme}}}, \bibinfo {author}
  {\bibfnamefont {J.~B.}\ \bibnamefont {{Hutchings}}}, \bibinfo {author}
  {\bibfnamefont {A.}~\bibnamefont {{Omont}}}, \bibinfo {author} {\bibfnamefont
  {C.}~\bibnamefont {{Reyl{\'e}}}},\ and\ \bibinfo {author} {\bibfnamefont
  {D.}~\bibnamefont {{Schade}}},\ }\href
  {https://doi.org/10.1088/0004-6256/140/2/546} {\bibfield  {journal} {\bibinfo
   {journal} {\aj}\ }\textbf {\bibinfo {volume} {140}},\ \bibinfo {pages} {546}
  (\bibinfo {year} {2010})},\ \Eprint {https://arxiv.org/abs/1006.1342}
  {arXiv:1006.1342 [astro-ph.CO]} \BibitemShut {NoStop}%
\bibitem [{\citenamefont {{Kelly}}\ and\ \citenamefont
  {{Shen}}(2013)}]{Kelly2013}%
  \BibitemOpen
  \bibfield  {author} {\bibinfo {author} {\bibfnamefont {B.~C.}\ \bibnamefont
  {{Kelly}}}\ and\ \bibinfo {author} {\bibfnamefont {Y.}~\bibnamefont
  {{Shen}}},\ }\href {https://doi.org/10.1088/0004-637X/764/1/45} {\bibfield
  {journal} {\bibinfo  {journal} {\apj}\ }\textbf {\bibinfo {volume} {764}},\
  \bibinfo {eid} {45} (\bibinfo {year} {2013})},\ \Eprint
  {https://arxiv.org/abs/1209.0477} {arXiv:1209.0477 [astro-ph.CO]}
  \BibitemShut {NoStop}%
\bibitem [{\citenamefont {Tucci}\ and\ \citenamefont
  {Volonteri}(2017)}]{Tucci:2016tyc}%
  \BibitemOpen
  \bibfield  {author} {\bibinfo {author} {\bibfnamefont {M.}~\bibnamefont
  {Tucci}}\ and\ \bibinfo {author} {\bibfnamefont {M.}~\bibnamefont
  {Volonteri}},\ }\href {https://doi.org/10.1051/0004-6361/201628419}
  {\bibfield  {journal} {\bibinfo  {journal} {Astron. Astrophys.}\ }\textbf
  {\bibinfo {volume} {600}},\ \bibinfo {pages} {A64} (\bibinfo {year}
  {2017})},\ \Eprint {https://arxiv.org/abs/1603.00823} {arXiv:1603.00823
  [astro-ph.GA]} \BibitemShut {NoStop}%
\bibitem [{\citenamefont {{Shen}}\ \emph {et~al.}(2007)\citenamefont {{Shen}},
  \citenamefont {{Strauss}}, \citenamefont {{Oguri}}, \citenamefont
  {{Hennawi}}, \citenamefont {{Fan}}, \citenamefont {{Richards}}, \citenamefont
  {{Hall}}, \citenamefont {{Gunn}}, \citenamefont {{Schneider}}, \citenamefont
  {{Szalay}}, \citenamefont {{Thakar}}, \citenamefont {{Vanden Berk}},
  \citenamefont {{Anderson}}, \citenamefont {{Bahcall}}, \citenamefont
  {{Connolly}},\ and\ \citenamefont {{Knapp}}}]{Shen2007}%
  \BibitemOpen
  \bibfield  {author} {\bibinfo {author} {\bibfnamefont {Y.}~\bibnamefont
  {{Shen}}}, \bibinfo {author} {\bibfnamefont {M.~A.}\ \bibnamefont
  {{Strauss}}}, \bibinfo {author} {\bibfnamefont {M.}~\bibnamefont {{Oguri}}},
  \bibinfo {author} {\bibfnamefont {J.~F.}\ \bibnamefont {{Hennawi}}}, \bibinfo
  {author} {\bibfnamefont {X.}~\bibnamefont {{Fan}}}, \bibinfo {author}
  {\bibfnamefont {G.~T.}\ \bibnamefont {{Richards}}}, \bibinfo {author}
  {\bibfnamefont {P.~B.}\ \bibnamefont {{Hall}}}, \bibinfo {author}
  {\bibfnamefont {J.~E.}\ \bibnamefont {{Gunn}}}, \bibinfo {author}
  {\bibfnamefont {D.~P.}\ \bibnamefont {{Schneider}}}, \bibinfo {author}
  {\bibfnamefont {A.~S.}\ \bibnamefont {{Szalay}}}, \bibinfo {author}
  {\bibfnamefont {A.~R.}\ \bibnamefont {{Thakar}}}, \bibinfo {author}
  {\bibfnamefont {D.~E.}\ \bibnamefont {{Vanden Berk}}}, \bibinfo {author}
  {\bibfnamefont {S.~F.}\ \bibnamefont {{Anderson}}}, \bibinfo {author}
  {\bibfnamefont {N.~A.}\ \bibnamefont {{Bahcall}}}, \bibinfo {author}
  {\bibfnamefont {A.~J.}\ \bibnamefont {{Connolly}}},\ and\ \bibinfo {author}
  {\bibfnamefont {G.~R.}\ \bibnamefont {{Knapp}}},\ }\href
  {https://doi.org/10.1086/513517} {\bibfield  {journal} {\bibinfo  {journal}
  {\aj}\ }\textbf {\bibinfo {volume} {133}},\ \bibinfo {pages} {2222} (\bibinfo
  {year} {2007})},\ \Eprint {https://arxiv.org/abs/astro-ph/0702214}
  {arXiv:astro-ph/0702214 [astro-ph]} \BibitemShut {NoStop}%
\bibitem [{\citenamefont {{White}}\ \emph {et~al.}(2008)\citenamefont
  {{White}}, \citenamefont {{Martini}},\ and\ \citenamefont
  {{Cohn}}}]{White2008}%
  \BibitemOpen
  \bibfield  {author} {\bibinfo {author} {\bibfnamefont {M.}~\bibnamefont
  {{White}}}, \bibinfo {author} {\bibfnamefont {P.}~\bibnamefont {{Martini}}},\
  and\ \bibinfo {author} {\bibfnamefont {J.~D.}\ \bibnamefont {{Cohn}}},\
  }\href {https://doi.org/10.1111/j.1365-2966.2008.13817.x} {\bibfield
  {journal} {\bibinfo  {journal} {\mnras}\ }\textbf {\bibinfo {volume} {390}},\
  \bibinfo {pages} {1179} (\bibinfo {year} {2008})},\ \Eprint
  {https://arxiv.org/abs/0711.4109} {arXiv:0711.4109 [astro-ph]} \BibitemShut
  {NoStop}%
\bibitem [{\citenamefont {{Shankar}}\ \emph {et~al.}(2010)\citenamefont
  {{Shankar}}, \citenamefont {{Crocce}}, \citenamefont {{Miralda-Escud{\'e}}},
  \citenamefont {{Fosalba}},\ and\ \citenamefont {{Weinberg}}}]{Shankar2010}%
  \BibitemOpen
  \bibfield  {author} {\bibinfo {author} {\bibfnamefont {F.}~\bibnamefont
  {{Shankar}}}, \bibinfo {author} {\bibfnamefont {M.}~\bibnamefont {{Crocce}}},
  \bibinfo {author} {\bibfnamefont {J.}~\bibnamefont {{Miralda-Escud{\'e}}}},
  \bibinfo {author} {\bibfnamefont {P.}~\bibnamefont {{Fosalba}}},\ and\
  \bibinfo {author} {\bibfnamefont {D.~H.}\ \bibnamefont {{Weinberg}}},\ }\href
  {https://doi.org/10.1088/0004-637X/718/1/231} {\bibfield  {journal} {\bibinfo
   {journal} {\apj}\ }\textbf {\bibinfo {volume} {718}},\ \bibinfo {pages}
  {231} (\bibinfo {year} {2010})},\ \Eprint {https://arxiv.org/abs/0810.4919}
  {arXiv:0810.4919 [astro-ph]} \BibitemShut {NoStop}%
\bibitem [{\citenamefont {{Chen}}\ and\ \citenamefont
  {{Gnedin}}(2018)}]{Chen2018}%
  \BibitemOpen
  \bibfield  {author} {\bibinfo {author} {\bibfnamefont {H.}~\bibnamefont
  {{Chen}}}\ and\ \bibinfo {author} {\bibfnamefont {N.~Y.}\ \bibnamefont
  {{Gnedin}}},\ }\href {https://doi.org/10.3847/1538-4357/aae8e8} {\bibfield
  {journal} {\bibinfo  {journal} {\apj}\ }\textbf {\bibinfo {volume} {868}},\
  \bibinfo {eid} {126} (\bibinfo {year} {2018})},\ \Eprint
  {https://arxiv.org/abs/1809.01545} {arXiv:1809.01545 [astro-ph.GA]}
  \BibitemShut {NoStop}%
\bibitem [{\citenamefont {{Wang}}\ \emph
  {et~al.}(2019{\natexlab{b}})\citenamefont {{Wang}}, \citenamefont {{Wang}},
  \citenamefont {{Fan}}, \citenamefont {{Wu}}, \citenamefont {{Yang}},
  \citenamefont {{Neri}},\ and\ \citenamefont {{Yue}}}]{Wang2019}%
  \BibitemOpen
  \bibfield  {author} {\bibinfo {author} {\bibfnamefont {F.}~\bibnamefont
  {{Wang}}}, \bibinfo {author} {\bibfnamefont {R.}~\bibnamefont {{Wang}}},
  \bibinfo {author} {\bibfnamefont {X.}~\bibnamefont {{Fan}}}, \bibinfo
  {author} {\bibfnamefont {X.-B.}\ \bibnamefont {{Wu}}}, \bibinfo {author}
  {\bibfnamefont {J.}~\bibnamefont {{Yang}}}, \bibinfo {author} {\bibfnamefont
  {R.}~\bibnamefont {{Neri}}},\ and\ \bibinfo {author} {\bibfnamefont
  {M.}~\bibnamefont {{Yue}}},\ }\href
  {https://doi.org/10.3847/1538-4357/ab2717} {\bibfield  {journal} {\bibinfo
  {journal} {\apj}\ }\textbf {\bibinfo {volume} {880}},\ \bibinfo {eid} {2}
  (\bibinfo {year} {2019}{\natexlab{b}})},\ \Eprint
  {https://arxiv.org/abs/1906.06801} {arXiv:1906.06801 [astro-ph.GA]}
  \BibitemShut {NoStop}%
\bibitem [{\citenamefont {{Decarli}}\ \emph {et~al.}(2018)\citenamefont
  {{Decarli}}, \citenamefont {{Walter}}, \citenamefont {{Venemans}},
  \citenamefont {{Ba{\~n}ados}}, \citenamefont {{Bertoldi}}, \citenamefont
  {{Carilli}}, \citenamefont {{Fan}}, \citenamefont {{Farina}}, \citenamefont
  {{Mazzucchelli}}, \citenamefont {{Riechers}}, \citenamefont {{Rix}},
  \citenamefont {{Strauss}}, \citenamefont {{Wang}},\ and\ \citenamefont
  {{Yang}}}]{Decarli2018}%
  \BibitemOpen
  \bibfield  {author} {\bibinfo {author} {\bibfnamefont {R.}~\bibnamefont
  {{Decarli}}}, \bibinfo {author} {\bibfnamefont {F.}~\bibnamefont {{Walter}}},
  \bibinfo {author} {\bibfnamefont {B.~P.}\ \bibnamefont {{Venemans}}},
  \bibinfo {author} {\bibfnamefont {E.}~\bibnamefont {{Ba{\~n}ados}}}, \bibinfo
  {author} {\bibfnamefont {F.}~\bibnamefont {{Bertoldi}}}, \bibinfo {author}
  {\bibfnamefont {C.}~\bibnamefont {{Carilli}}}, \bibinfo {author}
  {\bibfnamefont {X.}~\bibnamefont {{Fan}}}, \bibinfo {author} {\bibfnamefont
  {E.~P.}\ \bibnamefont {{Farina}}}, \bibinfo {author} {\bibfnamefont
  {C.}~\bibnamefont {{Mazzucchelli}}}, \bibinfo {author} {\bibfnamefont
  {D.}~\bibnamefont {{Riechers}}}, \bibinfo {author} {\bibfnamefont {H.-W.}\
  \bibnamefont {{Rix}}}, \bibinfo {author} {\bibfnamefont {M.~A.}\ \bibnamefont
  {{Strauss}}}, \bibinfo {author} {\bibfnamefont {R.}~\bibnamefont {{Wang}}},\
  and\ \bibinfo {author} {\bibfnamefont {Y.}~\bibnamefont {{Yang}}},\ }\href
  {https://doi.org/10.3847/1538-4357/aaa5aa} {\bibfield  {journal} {\bibinfo
  {journal} {\apj}\ }\textbf {\bibinfo {volume} {854}},\ \bibinfo {eid} {97}
  (\bibinfo {year} {2018})},\ \Eprint {https://arxiv.org/abs/1801.02641}
  {arXiv:1801.02641 [astro-ph.GA]} \BibitemShut {NoStop}%
\bibitem [{\citenamefont {{Wang}}\ \emph {et~al.}(2020)\citenamefont {{Wang}},
  \citenamefont {{Davies}}, \citenamefont {{Yang}}, \citenamefont {{Hennawi}},
  \citenamefont {{Fan}}, \citenamefont {{Barth}}, \citenamefont {{Jiang}},
  \citenamefont {{Wu}}, \citenamefont {{Mudd}}, \citenamefont {{Ba{\~n}ados}},
  \citenamefont {{Bian}}, \citenamefont {{Decarli}}, \citenamefont {{Eilers}},
  \citenamefont {{Farina}}, \citenamefont {{Venemans}}, \citenamefont
  {{Walter}},\ and\ \citenamefont {{Yue}}}]{Wang2020}%
  \BibitemOpen
  \bibfield  {author} {\bibinfo {author} {\bibfnamefont {F.}~\bibnamefont
  {{Wang}}}, \bibinfo {author} {\bibfnamefont {F.~B.}\ \bibnamefont
  {{Davies}}}, \bibinfo {author} {\bibfnamefont {J.}~\bibnamefont {{Yang}}},
  \bibinfo {author} {\bibfnamefont {J.~F.}\ \bibnamefont {{Hennawi}}}, \bibinfo
  {author} {\bibfnamefont {X.}~\bibnamefont {{Fan}}}, \bibinfo {author}
  {\bibfnamefont {A.~J.}\ \bibnamefont {{Barth}}}, \bibinfo {author}
  {\bibfnamefont {L.}~\bibnamefont {{Jiang}}}, \bibinfo {author} {\bibfnamefont
  {X.-B.}\ \bibnamefont {{Wu}}}, \bibinfo {author} {\bibfnamefont {D.~M.}\
  \bibnamefont {{Mudd}}}, \bibinfo {author} {\bibfnamefont {E.}~\bibnamefont
  {{Ba{\~n}ados}}}, \bibinfo {author} {\bibfnamefont {F.}~\bibnamefont
  {{Bian}}}, \bibinfo {author} {\bibfnamefont {R.}~\bibnamefont {{Decarli}}},
  \bibinfo {author} {\bibfnamefont {A.-C.}\ \bibnamefont {{Eilers}}}, \bibinfo
  {author} {\bibfnamefont {E.~P.}\ \bibnamefont {{Farina}}}, \bibinfo {author}
  {\bibfnamefont {B.}~\bibnamefont {{Venemans}}}, \bibinfo {author}
  {\bibfnamefont {F.}~\bibnamefont {{Walter}}},\ and\ \bibinfo {author}
  {\bibfnamefont {M.}~\bibnamefont {{Yue}}},\ }\href
  {https://doi.org/10.3847/1538-4357/ab8c45} {\bibfield  {journal} {\bibinfo
  {journal} {\apj}\ }\textbf {\bibinfo {volume} {896}},\ \bibinfo {eid} {23}
  (\bibinfo {year} {2020})},\ \Eprint {https://arxiv.org/abs/2004.10877}
  {arXiv:2004.10877 [astro-ph.GA]} \BibitemShut {NoStop}%
\bibitem [{\citenamefont {{Neeleman}}\ \emph {et~al.}(2021)\citenamefont
  {{Neeleman}}, \citenamefont {{Novak}}, \citenamefont {{Venemans}},
  \citenamefont {{Walter}}, \citenamefont {{Decarli}}, \citenamefont
  {{Kaasinen}}, \citenamefont {{Schindler}}, \citenamefont {{Banados}},
  \citenamefont {{Carilli}}, \citenamefont {{Drake}}, \citenamefont {{Fan}},\
  and\ \citenamefont {{Rix}}}]{Neeleman2021}%
  \BibitemOpen
  \bibfield  {author} {\bibinfo {author} {\bibfnamefont {M.}~\bibnamefont
  {{Neeleman}}}, \bibinfo {author} {\bibfnamefont {M.}~\bibnamefont {{Novak}}},
  \bibinfo {author} {\bibfnamefont {B.~P.}\ \bibnamefont {{Venemans}}},
  \bibinfo {author} {\bibfnamefont {F.}~\bibnamefont {{Walter}}}, \bibinfo
  {author} {\bibfnamefont {R.}~\bibnamefont {{Decarli}}}, \bibinfo {author}
  {\bibfnamefont {M.}~\bibnamefont {{Kaasinen}}}, \bibinfo {author}
  {\bibfnamefont {J.-T.}\ \bibnamefont {{Schindler}}}, \bibinfo {author}
  {\bibfnamefont {E.}~\bibnamefont {{Banados}}}, \bibinfo {author}
  {\bibfnamefont {C.~L.}\ \bibnamefont {{Carilli}}}, \bibinfo {author}
  {\bibfnamefont {A.~B.}\ \bibnamefont {{Drake}}}, \bibinfo {author}
  {\bibfnamefont {X.}~\bibnamefont {{Fan}}},\ and\ \bibinfo {author}
  {\bibfnamefont {H.-W.}\ \bibnamefont {{Rix}}},\ }\href@noop {} {\bibfield
  {journal} {\bibinfo  {journal} {arXiv e-prints}\ ,\ \bibinfo {eid}
  {arXiv:2102.05679}} (\bibinfo {year} {2021})},\ \Eprint
  {https://arxiv.org/abs/2102.05679} {arXiv:2102.05679 [astro-ph.GA]}
  \BibitemShut {NoStop}%
\bibitem [{\citenamefont {{Graham}}\ and\ \citenamefont
  {{Driver}}(2007)}]{Graham2007}%
  \BibitemOpen
  \bibfield  {author} {\bibinfo {author} {\bibfnamefont {A.~W.}\ \bibnamefont
  {{Graham}}}\ and\ \bibinfo {author} {\bibfnamefont {S.~P.}\ \bibnamefont
  {{Driver}}},\ }\href {https://doi.org/10.1111/j.1745-3933.2007.00340.x}
  {\bibfield  {journal} {\bibinfo  {journal} {\mnras}\ }\textbf {\bibinfo
  {volume} {380}},\ \bibinfo {pages} {L15} (\bibinfo {year} {2007})},\ \Eprint
  {https://arxiv.org/abs/0705.4505} {arXiv:0705.4505 [astro-ph]} \BibitemShut
  {NoStop}%
\bibitem [{\citenamefont {{Fiore}}\ \emph {et~al.}(2012)\citenamefont
  {{Fiore}}, \citenamefont {{Puccetti}}, \citenamefont {{Grazian}},
  \citenamefont {{Menci}}, \citenamefont {{Shankar}}, \citenamefont
  {{Santini}}, \citenamefont {{Piconcelli}}, \citenamefont {{Koekemoer}},
  \citenamefont {{Fontana}}, \citenamefont {{Boutsia}}, \citenamefont
  {{Castellano}}, \citenamefont {{Lamastra}}, \citenamefont {{Malacaria}},
  \citenamefont {{Feruglio}}, \citenamefont {{Mathur}}, \citenamefont
  {{Miller}},\ and\ \citenamefont {{Pannella}}}]{Fiore2012}%
  \BibitemOpen
  \bibfield  {author} {\bibinfo {author} {\bibfnamefont {F.}~\bibnamefont
  {{Fiore}}}, \bibinfo {author} {\bibfnamefont {S.}~\bibnamefont {{Puccetti}}},
  \bibinfo {author} {\bibfnamefont {A.}~\bibnamefont {{Grazian}}}, \bibinfo
  {author} {\bibfnamefont {N.}~\bibnamefont {{Menci}}}, \bibinfo {author}
  {\bibfnamefont {F.}~\bibnamefont {{Shankar}}}, \bibinfo {author}
  {\bibfnamefont {P.}~\bibnamefont {{Santini}}}, \bibinfo {author}
  {\bibfnamefont {E.}~\bibnamefont {{Piconcelli}}}, \bibinfo {author}
  {\bibfnamefont {A.~M.}\ \bibnamefont {{Koekemoer}}}, \bibinfo {author}
  {\bibfnamefont {A.}~\bibnamefont {{Fontana}}}, \bibinfo {author}
  {\bibfnamefont {K.}~\bibnamefont {{Boutsia}}}, \bibinfo {author}
  {\bibfnamefont {M.}~\bibnamefont {{Castellano}}}, \bibinfo {author}
  {\bibfnamefont {A.}~\bibnamefont {{Lamastra}}}, \bibinfo {author}
  {\bibfnamefont {C.}~\bibnamefont {{Malacaria}}}, \bibinfo {author}
  {\bibfnamefont {C.}~\bibnamefont {{Feruglio}}}, \bibinfo {author}
  {\bibfnamefont {S.}~\bibnamefont {{Mathur}}}, \bibinfo {author}
  {\bibfnamefont {N.}~\bibnamefont {{Miller}}},\ and\ \bibinfo {author}
  {\bibfnamefont {M.}~\bibnamefont {{Pannella}}},\ }\href
  {https://doi.org/10.1051/0004-6361/201117581} {\bibfield  {journal} {\bibinfo
   {journal} {\aap}\ }\textbf {\bibinfo {volume} {537}},\ \bibinfo {eid} {A16}
  (\bibinfo {year} {2012})},\ \Eprint {https://arxiv.org/abs/1109.2888}
  {arXiv:1109.2888 [astro-ph.CO]} \BibitemShut {NoStop}%
\bibitem [{\citenamefont {{Jiang}}\ \emph {et~al.}(2016)\citenamefont
  {{Jiang}}, \citenamefont {{McGreer}}, \citenamefont {{Fan}}, \citenamefont
  {{Strauss}}, \citenamefont {{Ba{\~n}ados}}, \citenamefont {{Becker}},
  \citenamefont {{Bian}}, \citenamefont {{Farnsworth}}, \citenamefont {{Shen}},
  \citenamefont {{Wang}}, \citenamefont {{Wang}}, \citenamefont {{Wang}},
  \citenamefont {{White}}, \citenamefont {{Wu}}, \citenamefont {{Wu}},
  \citenamefont {{Yang}},\ and\ \citenamefont {{Yang}}}]{Jiang2016}%
  \BibitemOpen
  \bibfield  {author} {\bibinfo {author} {\bibfnamefont {L.}~\bibnamefont
  {{Jiang}}}, \bibinfo {author} {\bibfnamefont {I.~D.}\ \bibnamefont
  {{McGreer}}}, \bibinfo {author} {\bibfnamefont {X.}~\bibnamefont {{Fan}}},
  \bibinfo {author} {\bibfnamefont {M.~A.}\ \bibnamefont {{Strauss}}}, \bibinfo
  {author} {\bibfnamefont {E.}~\bibnamefont {{Ba{\~n}ados}}}, \bibinfo {author}
  {\bibfnamefont {R.~H.}\ \bibnamefont {{Becker}}}, \bibinfo {author}
  {\bibfnamefont {F.}~\bibnamefont {{Bian}}}, \bibinfo {author} {\bibfnamefont
  {K.}~\bibnamefont {{Farnsworth}}}, \bibinfo {author} {\bibfnamefont
  {Y.}~\bibnamefont {{Shen}}}, \bibinfo {author} {\bibfnamefont
  {F.}~\bibnamefont {{Wang}}}, \bibinfo {author} {\bibfnamefont
  {R.}~\bibnamefont {{Wang}}}, \bibinfo {author} {\bibfnamefont
  {S.}~\bibnamefont {{Wang}}}, \bibinfo {author} {\bibfnamefont {R.~L.}\
  \bibnamefont {{White}}}, \bibinfo {author} {\bibfnamefont {J.}~\bibnamefont
  {{Wu}}}, \bibinfo {author} {\bibfnamefont {X.-B.}\ \bibnamefont {{Wu}}},
  \bibinfo {author} {\bibfnamefont {J.}~\bibnamefont {{Yang}}},\ and\ \bibinfo
  {author} {\bibfnamefont {Q.}~\bibnamefont {{Yang}}},\ }\href
  {https://doi.org/10.3847/1538-4357/833/2/222} {\bibfield  {journal} {\bibinfo
   {journal} {\apj}\ }\textbf {\bibinfo {volume} {833}},\ \bibinfo {eid} {222}
  (\bibinfo {year} {2016})},\ \Eprint {https://arxiv.org/abs/1610.05369}
  {arXiv:1610.05369 [astro-ph.GA]} \BibitemShut {NoStop}%
\bibitem [{\citenamefont {{Matsuoka}}\ \emph {et~al.}(2018)\citenamefont
  {{Matsuoka}}, \citenamefont {{Strauss}}, \citenamefont {{Kashikawa}},
  \citenamefont {{Onoue}}, \citenamefont {{Iwasawa}}, \citenamefont {{Tang}},
  \citenamefont {{Lee}}, \citenamefont {{Imanishi}}, \citenamefont {{Nagao}},
  \citenamefont {{Akiyama}}, \citenamefont {{Asami}}, \citenamefont {{Bosch}},
  \citenamefont {{Furusawa}}, \citenamefont {{Goto}}, \citenamefont {{Gunn}},
  \citenamefont {{Harikane}}, \citenamefont {{Ikeda}}, \citenamefont {{Izumi}},
  \citenamefont {{Kawaguchi}}, \citenamefont {{Kato}}, \citenamefont
  {{Kikuta}}, \citenamefont {{Kohno}}, \citenamefont {{Komiyama}},
  \citenamefont {{Lupton}}, \citenamefont {{Minezaki}}, \citenamefont
  {{Miyazaki}}, \citenamefont {{Murayama}}, \citenamefont {{Niida}},
  \citenamefont {{Nishizawa}}, \citenamefont {{Noboriguchi}}, \citenamefont
  {{Oguri}}, \citenamefont {{Ono}}, \citenamefont {{Ouchi}}, \citenamefont
  {{Price}}, \citenamefont {{Sameshima}}, \citenamefont {{Schulze}},
  \citenamefont {{Shirakata}}, \citenamefont {{Silverman}}, \citenamefont
  {{Sugiyama}}, \citenamefont {{Tait}}, \citenamefont {{Takada}}, \citenamefont
  {{Takata}}, \citenamefont {{Tanaka}}, \citenamefont {{Toba}}, \citenamefont
  {{Utsumi}}, \citenamefont {{Wang}},\ and\ \citenamefont
  {{Yamashita}}}]{Matsuoka2018}%
  \BibitemOpen
  \bibfield  {author} {\bibinfo {author} {\bibfnamefont {Y.}~\bibnamefont
  {{Matsuoka}}}, \bibinfo {author} {\bibfnamefont {M.~A.}\ \bibnamefont
  {{Strauss}}}, \bibinfo {author} {\bibfnamefont {N.}~\bibnamefont
  {{Kashikawa}}}, \bibinfo {author} {\bibfnamefont {M.}~\bibnamefont
  {{Onoue}}}, \bibinfo {author} {\bibfnamefont {K.}~\bibnamefont {{Iwasawa}}},
  \bibinfo {author} {\bibfnamefont {J.-J.}\ \bibnamefont {{Tang}}}, \bibinfo
  {author} {\bibfnamefont {C.-H.}\ \bibnamefont {{Lee}}}, \bibinfo {author}
  {\bibfnamefont {M.}~\bibnamefont {{Imanishi}}}, \bibinfo {author}
  {\bibfnamefont {T.}~\bibnamefont {{Nagao}}}, \bibinfo {author} {\bibfnamefont
  {M.}~\bibnamefont {{Akiyama}}}, \bibinfo {author} {\bibfnamefont
  {N.}~\bibnamefont {{Asami}}}, \bibinfo {author} {\bibfnamefont
  {J.}~\bibnamefont {{Bosch}}}, \bibinfo {author} {\bibfnamefont
  {H.}~\bibnamefont {{Furusawa}}}, \bibinfo {author} {\bibfnamefont
  {T.}~\bibnamefont {{Goto}}}, \bibinfo {author} {\bibfnamefont {J.~E.}\
  \bibnamefont {{Gunn}}}, \bibinfo {author} {\bibfnamefont {Y.}~\bibnamefont
  {{Harikane}}}, \bibinfo {author} {\bibfnamefont {H.}~\bibnamefont {{Ikeda}}},
  \bibinfo {author} {\bibfnamefont {T.}~\bibnamefont {{Izumi}}}, \bibinfo
  {author} {\bibfnamefont {T.}~\bibnamefont {{Kawaguchi}}}, \bibinfo {author}
  {\bibfnamefont {N.}~\bibnamefont {{Kato}}}, \bibinfo {author} {\bibfnamefont
  {S.}~\bibnamefont {{Kikuta}}}, \bibinfo {author} {\bibfnamefont
  {K.}~\bibnamefont {{Kohno}}}, \bibinfo {author} {\bibfnamefont
  {Y.}~\bibnamefont {{Komiyama}}}, \bibinfo {author} {\bibfnamefont {R.~H.}\
  \bibnamefont {{Lupton}}}, \bibinfo {author} {\bibfnamefont {T.}~\bibnamefont
  {{Minezaki}}}, \bibinfo {author} {\bibfnamefont {S.}~\bibnamefont
  {{Miyazaki}}}, \bibinfo {author} {\bibfnamefont {H.}~\bibnamefont
  {{Murayama}}}, \bibinfo {author} {\bibfnamefont {M.}~\bibnamefont {{Niida}}},
  \bibinfo {author} {\bibfnamefont {A.~J.}\ \bibnamefont {{Nishizawa}}},
  \bibinfo {author} {\bibfnamefont {A.}~\bibnamefont {{Noboriguchi}}}, \bibinfo
  {author} {\bibfnamefont {M.}~\bibnamefont {{Oguri}}}, \bibinfo {author}
  {\bibfnamefont {Y.}~\bibnamefont {{Ono}}}, \bibinfo {author} {\bibfnamefont
  {M.}~\bibnamefont {{Ouchi}}}, \bibinfo {author} {\bibfnamefont {P.~A.}\
  \bibnamefont {{Price}}}, \bibinfo {author} {\bibfnamefont {H.}~\bibnamefont
  {{Sameshima}}}, \bibinfo {author} {\bibfnamefont {A.}~\bibnamefont
  {{Schulze}}}, \bibinfo {author} {\bibfnamefont {H.}~\bibnamefont
  {{Shirakata}}}, \bibinfo {author} {\bibfnamefont {J.~D.}\ \bibnamefont
  {{Silverman}}}, \bibinfo {author} {\bibfnamefont {N.}~\bibnamefont
  {{Sugiyama}}}, \bibinfo {author} {\bibfnamefont {P.~J.}\ \bibnamefont
  {{Tait}}}, \bibinfo {author} {\bibfnamefont {M.}~\bibnamefont {{Takada}}},
  \bibinfo {author} {\bibfnamefont {T.}~\bibnamefont {{Takata}}}, \bibinfo
  {author} {\bibfnamefont {M.}~\bibnamefont {{Tanaka}}}, \bibinfo {author}
  {\bibfnamefont {Y.}~\bibnamefont {{Toba}}}, \bibinfo {author} {\bibfnamefont
  {Y.}~\bibnamefont {{Utsumi}}}, \bibinfo {author} {\bibfnamefont {S.-Y.}\
  \bibnamefont {{Wang}}},\ and\ \bibinfo {author} {\bibfnamefont
  {T.}~\bibnamefont {{Yamashita}}},\ }\href
  {https://doi.org/10.3847/1538-4357/aaee7a} {\bibfield  {journal} {\bibinfo
  {journal} {\apj}\ }\textbf {\bibinfo {volume} {869}},\ \bibinfo {eid} {150}
  (\bibinfo {year} {2018})},\ \Eprint {https://arxiv.org/abs/1811.01963}
  {arXiv:1811.01963 [astro-ph.GA]} \BibitemShut {NoStop}%
\bibitem [{\citenamefont {{Governato}}\ \emph {et~al.}(2010)\citenamefont
  {{Governato}}, \citenamefont {{Brook}}, \citenamefont {{Mayer}},
  \citenamefont {{Brooks}}, \citenamefont {{Rhee}}, \citenamefont {{Wadsley}},
  \citenamefont {{Jonsson}}, \citenamefont {{Willman}}, \citenamefont
  {{Stinson}}, \citenamefont {{Quinn}},\ and\ \citenamefont
  {{Madau}}}]{Governato2010}%
  \BibitemOpen
  \bibfield  {author} {\bibinfo {author} {\bibfnamefont {F.}~\bibnamefont
  {{Governato}}}, \bibinfo {author} {\bibfnamefont {C.}~\bibnamefont
  {{Brook}}}, \bibinfo {author} {\bibfnamefont {L.}~\bibnamefont {{Mayer}}},
  \bibinfo {author} {\bibfnamefont {A.}~\bibnamefont {{Brooks}}}, \bibinfo
  {author} {\bibfnamefont {G.}~\bibnamefont {{Rhee}}}, \bibinfo {author}
  {\bibfnamefont {J.}~\bibnamefont {{Wadsley}}}, \bibinfo {author}
  {\bibfnamefont {P.}~\bibnamefont {{Jonsson}}}, \bibinfo {author}
  {\bibfnamefont {B.}~\bibnamefont {{Willman}}}, \bibinfo {author}
  {\bibfnamefont {G.}~\bibnamefont {{Stinson}}}, \bibinfo {author}
  {\bibfnamefont {T.}~\bibnamefont {{Quinn}}},\ and\ \bibinfo {author}
  {\bibfnamefont {P.}~\bibnamefont {{Madau}}},\ }\href
  {https://doi.org/10.1038/nature08640} {\bibfield  {journal} {\bibinfo
  {journal} {\nat}\ }\textbf {\bibinfo {volume} {463}},\ \bibinfo {pages} {203}
  (\bibinfo {year} {2010})},\ \Eprint {https://arxiv.org/abs/0911.2237}
  {arXiv:0911.2237 [astro-ph.CO]} \BibitemShut {NoStop}%
\bibitem [{\citenamefont {{Pontzen}}\ and\ \citenamefont
  {{Governato}}(2012)}]{Pontzen2012}%
  \BibitemOpen
  \bibfield  {author} {\bibinfo {author} {\bibfnamefont {A.}~\bibnamefont
  {{Pontzen}}}\ and\ \bibinfo {author} {\bibfnamefont {F.}~\bibnamefont
  {{Governato}}},\ }\href {https://doi.org/10.1111/j.1365-2966.2012.20571.x}
  {\bibfield  {journal} {\bibinfo  {journal} {\mnras}\ }\textbf {\bibinfo
  {volume} {421}},\ \bibinfo {pages} {3464} (\bibinfo {year} {2012})},\ \Eprint
  {https://arxiv.org/abs/1106.0499} {arXiv:1106.0499 [astro-ph.CO]}
  \BibitemShut {NoStop}%
\bibitem [{\citenamefont {{Madau}}\ \emph {et~al.}(2014)\citenamefont
  {{Madau}}, \citenamefont {{Shen}},\ and\ \citenamefont
  {{Governato}}}]{Madau2014}%
  \BibitemOpen
  \bibfield  {author} {\bibinfo {author} {\bibfnamefont {P.}~\bibnamefont
  {{Madau}}}, \bibinfo {author} {\bibfnamefont {S.}~\bibnamefont {{Shen}}},\
  and\ \bibinfo {author} {\bibfnamefont {F.}~\bibnamefont {{Governato}}},\
  }\href {https://doi.org/10.1088/2041-8205/789/1/L17} {\bibfield  {journal}
  {\bibinfo  {journal} {\apjl}\ }\textbf {\bibinfo {volume} {789}},\ \bibinfo
  {eid} {L17} (\bibinfo {year} {2014})},\ \Eprint
  {https://arxiv.org/abs/1405.2577} {arXiv:1405.2577 [astro-ph.GA]}
  \BibitemShut {NoStop}%
\bibitem [{\citenamefont {{Dutton}}\ and\ \citenamefont
  {{Macci{\`o}}}(2014)}]{Dutton_2014}%
  \BibitemOpen
  \bibfield  {author} {\bibinfo {author} {\bibfnamefont {A.~A.}\ \bibnamefont
  {{Dutton}}}\ and\ \bibinfo {author} {\bibfnamefont {A.~V.}\ \bibnamefont
  {{Macci{\`o}}}},\ }\href {https://doi.org/10.1093/mnras/stu742} {\bibfield
  {journal} {\bibinfo  {journal} {\mnras}\ }\textbf {\bibinfo {volume} {441}},\
  \bibinfo {pages} {3359} (\bibinfo {year} {2014})},\ \Eprint
  {https://arxiv.org/abs/1402.7073} {arXiv:1402.7073 [astro-ph.CO]}
  \BibitemShut {NoStop}%
\bibitem [{\citenamefont {{Ma}}\ \emph {et~al.}(2021)\citenamefont {{Ma}},
  \citenamefont {{Hopkins}}, \citenamefont {{Ma}}, \citenamefont
  {{Angl{\'e}s-Alc{\'a}zar}}, \citenamefont {{Faucher-Gigu{\`e}re}},\ and\
  \citenamefont {{Kelley}}}]{Ma2021}%
  \BibitemOpen
  \bibfield  {author} {\bibinfo {author} {\bibfnamefont {L.}~\bibnamefont
  {{Ma}}}, \bibinfo {author} {\bibfnamefont {P.~F.}\ \bibnamefont {{Hopkins}}},
  \bibinfo {author} {\bibfnamefont {X.}~\bibnamefont {{Ma}}}, \bibinfo {author}
  {\bibfnamefont {D.}~\bibnamefont {{Angl{\'e}s-Alc{\'a}zar}}}, \bibinfo
  {author} {\bibfnamefont {C.-A.}\ \bibnamefont {{Faucher-Gigu{\`e}re}}},\ and\
  \bibinfo {author} {\bibfnamefont {L.~Z.}\ \bibnamefont {{Kelley}}},\
  }\href@noop {} {\bibfield  {journal} {\bibinfo  {journal} {arXiv e-prints}\
  ,\ \bibinfo {eid} {arXiv:2101.02727}} (\bibinfo {year} {2021})},\ \Eprint
  {https://arxiv.org/abs/2101.02727} {arXiv:2101.02727 [astro-ph.GA]}
  \BibitemShut {NoStop}%
\bibitem [{\citenamefont {{Kahlhoefer}}\ \emph {et~al.}(2015)\citenamefont
  {{Kahlhoefer}}, \citenamefont {{Schmidt-Hoberg}}, \citenamefont {{Kummer}},\
  and\ \citenamefont {{Sarkar}}}]{Kahlhoefer2015}%
  \BibitemOpen
  \bibfield  {author} {\bibinfo {author} {\bibfnamefont {F.}~\bibnamefont
  {{Kahlhoefer}}}, \bibinfo {author} {\bibfnamefont {K.}~\bibnamefont
  {{Schmidt-Hoberg}}}, \bibinfo {author} {\bibfnamefont {J.}~\bibnamefont
  {{Kummer}}},\ and\ \bibinfo {author} {\bibfnamefont {S.}~\bibnamefont
  {{Sarkar}}},\ }\href {https://doi.org/10.1093/mnrasl/slv088} {\bibfield
  {journal} {\bibinfo  {journal} {\mnras}\ }\textbf {\bibinfo {volume} {452}},\
  \bibinfo {pages} {L54} (\bibinfo {year} {2015})},\ \Eprint
  {https://arxiv.org/abs/1504.06576} {arXiv:1504.06576 [astro-ph.CO]}
  \BibitemShut {NoStop}%
\bibitem [{\citenamefont {{Harvey}}\ \emph {et~al.}(2015)\citenamefont
  {{Harvey}}, \citenamefont {{Massey}}, \citenamefont {{Kitching}},
  \citenamefont {{Taylor}},\ and\ \citenamefont {{Tittley}}}]{Harvey2015}%
  \BibitemOpen
  \bibfield  {author} {\bibinfo {author} {\bibfnamefont {D.}~\bibnamefont
  {{Harvey}}}, \bibinfo {author} {\bibfnamefont {R.}~\bibnamefont {{Massey}}},
  \bibinfo {author} {\bibfnamefont {T.}~\bibnamefont {{Kitching}}}, \bibinfo
  {author} {\bibfnamefont {A.}~\bibnamefont {{Taylor}}},\ and\ \bibinfo
  {author} {\bibfnamefont {E.}~\bibnamefont {{Tittley}}},\ }\href
  {https://doi.org/10.1126/science.1261381} {\bibfield  {journal} {\bibinfo
  {journal} {Science}\ }\textbf {\bibinfo {volume} {347}},\ \bibinfo {pages}
  {1462} (\bibinfo {year} {2015})},\ \Eprint {https://arxiv.org/abs/1503.07675}
  {arXiv:1503.07675 [astro-ph.CO]} \BibitemShut {NoStop}%
\bibitem [{\citenamefont {{Wittman}}\ \emph {et~al.}(2018)\citenamefont
  {{Wittman}}, \citenamefont {{Golovich}},\ and\ \citenamefont
  {{Dawson}}}]{Wittman2018}%
  \BibitemOpen
  \bibfield  {author} {\bibinfo {author} {\bibfnamefont {D.}~\bibnamefont
  {{Wittman}}}, \bibinfo {author} {\bibfnamefont {N.}~\bibnamefont
  {{Golovich}}},\ and\ \bibinfo {author} {\bibfnamefont {W.~A.}\ \bibnamefont
  {{Dawson}}},\ }\href {https://doi.org/10.3847/1538-4357/aaee77} {\bibfield
  {journal} {\bibinfo  {journal} {\apj}\ }\textbf {\bibinfo {volume} {869}},\
  \bibinfo {eid} {104} (\bibinfo {year} {2018})},\ \Eprint
  {https://arxiv.org/abs/1701.05877} {arXiv:1701.05877 [astro-ph.CO]}
  \BibitemShut {NoStop}%
\bibitem [{\citenamefont {{Meneghetti}}\ \emph {et~al.}(2020)\citenamefont
  {{Meneghetti}}, \citenamefont {{Davoli}}, \citenamefont {{Bergamini}},
  \citenamefont {{Rosati}}, \citenamefont {{Natarajan}}, \citenamefont
  {{Giocoli}}, \citenamefont {{Caminha}}, \citenamefont {{Metcalf}},
  \citenamefont {{Rasia}}, \citenamefont {{Borgani}}, \citenamefont {{Calura}},
  \citenamefont {{Grillo}}, \citenamefont {{Mercurio}},\ and\ \citenamefont
  {{Vanzella}}}]{Meneghetti2020}%
  \BibitemOpen
  \bibfield  {author} {\bibinfo {author} {\bibfnamefont {M.}~\bibnamefont
  {{Meneghetti}}}, \bibinfo {author} {\bibfnamefont {G.}~\bibnamefont
  {{Davoli}}}, \bibinfo {author} {\bibfnamefont {P.}~\bibnamefont
  {{Bergamini}}}, \bibinfo {author} {\bibfnamefont {P.}~\bibnamefont
  {{Rosati}}}, \bibinfo {author} {\bibfnamefont {P.}~\bibnamefont
  {{Natarajan}}}, \bibinfo {author} {\bibfnamefont {C.}~\bibnamefont
  {{Giocoli}}}, \bibinfo {author} {\bibfnamefont {G.~B.}\ \bibnamefont
  {{Caminha}}}, \bibinfo {author} {\bibfnamefont {R.~B.}\ \bibnamefont
  {{Metcalf}}}, \bibinfo {author} {\bibfnamefont {E.}~\bibnamefont {{Rasia}}},
  \bibinfo {author} {\bibfnamefont {S.}~\bibnamefont {{Borgani}}}, \bibinfo
  {author} {\bibfnamefont {F.}~\bibnamefont {{Calura}}}, \bibinfo {author}
  {\bibfnamefont {C.}~\bibnamefont {{Grillo}}}, \bibinfo {author}
  {\bibfnamefont {A.}~\bibnamefont {{Mercurio}}},\ and\ \bibinfo {author}
  {\bibfnamefont {E.}~\bibnamefont {{Vanzella}}},\ }\href
  {https://doi.org/10.1126/science.aax5164} {\bibfield  {journal} {\bibinfo
  {journal} {Science}\ }\textbf {\bibinfo {volume} {369}},\ \bibinfo {pages}
  {1347} (\bibinfo {year} {2020})},\ \Eprint {https://arxiv.org/abs/2009.04471}
  {arXiv:2009.04471 [astro-ph.GA]} \BibitemShut {NoStop}%
\bibitem [{\citenamefont {{Minor}}\ \emph {et~al.}(2020)\citenamefont
  {{Minor}}, \citenamefont {{Gad-Nasr}}, \citenamefont {{Kaplinghat}},\ and\
  \citenamefont {{Vegetti}}}]{Minor2020}%
  \BibitemOpen
  \bibfield  {author} {\bibinfo {author} {\bibfnamefont {Q.~E.}\ \bibnamefont
  {{Minor}}}, \bibinfo {author} {\bibfnamefont {S.}~\bibnamefont {{Gad-Nasr}}},
  \bibinfo {author} {\bibfnamefont {M.}~\bibnamefont {{Kaplinghat}}},\ and\
  \bibinfo {author} {\bibfnamefont {S.}~\bibnamefont {{Vegetti}}},\ }\href@noop
  {} {\bibfield  {journal} {\bibinfo  {journal} {arXiv e-prints}\ ,\ \bibinfo
  {eid} {arXiv:2011.10627}} (\bibinfo {year} {2020})},\ \Eprint
  {https://arxiv.org/abs/2011.10627} {arXiv:2011.10627 [astro-ph.GA]}
  \BibitemShut {NoStop}%
\bibitem [{\citenamefont {{Mo}}\ \emph {et~al.}(1998)\citenamefont {{Mo}},
  \citenamefont {{Mao}},\ and\ \citenamefont {{White}}}]{Mo:1998}%
  \BibitemOpen
  \bibfield  {author} {\bibinfo {author} {\bibfnamefont {H.~J.}\ \bibnamefont
  {{Mo}}}, \bibinfo {author} {\bibfnamefont {S.}~\bibnamefont {{Mao}}},\ and\
  \bibinfo {author} {\bibfnamefont {S.~D.~M.}\ \bibnamefont {{White}}},\ }\href
  {https://doi.org/10.1046/j.1365-8711.1998.01227.x} {\bibfield  {journal}
  {\bibinfo  {journal} {\mnras}\ }\textbf {\bibinfo {volume} {295}},\ \bibinfo
  {pages} {319} (\bibinfo {year} {1998})},\ \Eprint
  {https://arxiv.org/abs/astro-ph/9707093} {arXiv:astro-ph/9707093 [astro-ph]}
  \BibitemShut {NoStop}%
\bibitem [{\citenamefont {{Toomre}}(1964)}]{Toomre:1964}%
  \BibitemOpen
  \bibfield  {author} {\bibinfo {author} {\bibfnamefont {A.}~\bibnamefont
  {{Toomre}}},\ }\href {https://doi.org/10.1086/147861} {\bibfield  {journal}
  {\bibinfo  {journal} {\apj}\ }\textbf {\bibinfo {volume} {139}},\ \bibinfo
  {pages} {1217} (\bibinfo {year} {1964})}\BibitemShut {NoStop}%
\bibitem [{\citenamefont {{Lodato}}\ and\ \citenamefont
  {{Natarajan}}(2006{\natexlab{b}})}]{2006MNRAS.371.1813L}%
  \BibitemOpen
  \bibfield  {author} {\bibinfo {author} {\bibfnamefont {G.}~\bibnamefont
  {{Lodato}}}\ and\ \bibinfo {author} {\bibfnamefont {P.}~\bibnamefont
  {{Natarajan}}},\ }\href {https://doi.org/10.1111/j.1365-2966.2006.10801.x}
  {\bibfield  {journal} {\bibinfo  {journal} {\mnras}\ }\textbf {\bibinfo
  {volume} {371}},\ \bibinfo {pages} {1813} (\bibinfo {year}
  {2006}{\natexlab{b}})},\ \Eprint {https://arxiv.org/abs/astro-ph/0606159}
  {arXiv:astro-ph/0606159 [astro-ph]} \BibitemShut {NoStop}%
\bibitem [{\citenamefont {{Hoyle}}\ and\ \citenamefont
  {{Lyttleton}}(1939)}]{Hoyle1939}%
  \BibitemOpen
  \bibfield  {author} {\bibinfo {author} {\bibfnamefont {F.}~\bibnamefont
  {{Hoyle}}}\ and\ \bibinfo {author} {\bibfnamefont {R.~A.}\ \bibnamefont
  {{Lyttleton}}},\ }\href {https://doi.org/10.1017/S0305004100021150}
  {\bibfield  {journal} {\bibinfo  {journal} {Proceedings of the Cambridge
  Philosophical Society}\ }\textbf {\bibinfo {volume} {35}},\ \bibinfo {pages}
  {405} (\bibinfo {year} {1939})}\BibitemShut {NoStop}%
\bibitem [{\citenamefont {{Bondi}}\ and\ \citenamefont
  {{Hoyle}}(1944)}]{Bondi1944}%
  \BibitemOpen
  \bibfield  {author} {\bibinfo {author} {\bibfnamefont {H.}~\bibnamefont
  {{Bondi}}}\ and\ \bibinfo {author} {\bibfnamefont {F.}~\bibnamefont
  {{Hoyle}}},\ }\href {https://doi.org/10.1093/mnras/104.5.273} {\bibfield
  {journal} {\bibinfo  {journal} {\mnras}\ }\textbf {\bibinfo {volume} {104}},\
  \bibinfo {pages} {273} (\bibinfo {year} {1944})}\BibitemShut {NoStop}%
\bibitem [{\citenamefont {{Bondi}}(1952)}]{Bondi1952}%
  \BibitemOpen
  \bibfield  {author} {\bibinfo {author} {\bibfnamefont {H.}~\bibnamefont
  {{Bondi}}},\ }\href {https://doi.org/10.1093/mnras/112.2.195} {\bibfield
  {journal} {\bibinfo  {journal} {\mnras}\ }\textbf {\bibinfo {volume} {112}},\
  \bibinfo {pages} {195} (\bibinfo {year} {1952})}\BibitemShut {NoStop}%
\bibitem [{\citenamefont {{Shima}}\ \emph {et~al.}(1985)\citenamefont
  {{Shima}}, \citenamefont {{Matsuda}}, \citenamefont {{Takeda}},\ and\
  \citenamefont {{Sawada}}}]{Shima1985}%
  \BibitemOpen
  \bibfield  {author} {\bibinfo {author} {\bibfnamefont {E.}~\bibnamefont
  {{Shima}}}, \bibinfo {author} {\bibfnamefont {T.}~\bibnamefont {{Matsuda}}},
  \bibinfo {author} {\bibfnamefont {H.}~\bibnamefont {{Takeda}}},\ and\
  \bibinfo {author} {\bibfnamefont {K.}~\bibnamefont {{Sawada}}},\ }\href
  {https://doi.org/10.1093/mnras/217.2.367} {\bibfield  {journal} {\bibinfo
  {journal} {\mnras}\ }\textbf {\bibinfo {volume} {217}},\ \bibinfo {pages}
  {367} (\bibinfo {year} {1985})}\BibitemShut {NoStop}%
\end{thebibliography}%
\end{document}